\def\LL{\mbox{$\:\lambda\lambda $ }}

\def\HA{{H$\alpha$}}

\def\LUM{\:{\rm ergs\:s^{-1}}}
\def\FLUX{\:{\rm ergs\:cm^{-2}\:s^{-1}}}
\def\FLUXARCSEC{\:{\rm ergs\:cm^{-2}\:s^{-1}\:arcsec^{-2}}}
\def\COUNTS{\:{\rm cts\:s^{-1}}}

\def\VEL{\:{\rm km\:s^{-1}}}

\def\sii{[\ion{S}{2}]}
\def\nii{[\ion{N}{2}]}
\def\oiii{[\ion{O}{3}]}
\def\chase{ChASeM33}
\def\chandra{{\it Chandra}}
\def\hii{\ion{H}{2}}

\documentclass[12pt,preprint]{aastex}

\begin{document}

% Additional private definitions that appear to work only inside document

\newcommand{\MSOL}{\mbox{$\:M_{\sun}$}}  

\newcommand{\EXPN}[2]{\mbox{$#1\times 10^{#2}$}}
\newcommand{\EXPU}[3]{\mbox{\rm $#1 \times 10^{#2} \rm\:#3$}}  % exponent with units
\newcommand{\POW}[2]{\mbox{$\rm10^{#1}\rm\:#2$}}
\newcommand{\SING}[2]{#1$\thinspace \lambda $#2}
\newcommand{\MULT}[2]{#1$\thinspace \lambda \lambda $#2}
\newcommand{\CHINU}{\mbox{$\chi_{\nu}^2$}}
\newcommand{\vsini}{\mbox{$v\:\sin{(i)}$}}
\newcommand{\FUSE}{{\it FUSE}}
\newcommand{\HST}{{\it HST}}
\newcommand{\IUE}{{\it IUE}}
\newcommand{\EUVE}{{\it EUVE}}

% End of defining things

\shorttitle{SNRs in M33}
\shortauthors{Long \etal}

%\slugcomment{Could denote version here.}
\title{
% \today\\
The Chandra ACIS Survey of M33: X-ray, Optical and Radio Properties of the Supernova Remnants
}

\author{Knox S. Long\altaffilmark{1},
William P. Blair\altaffilmark{2},
P. Frank Winkler\altaffilmark{3}, \\
Robert H. Becker\altaffilmark{4, 5},
Terrance J. Gaetz\altaffilmark{6},
Parviz Ghavamian\altaffilmark{1},
David J. Helfand\altaffilmark{7},\\
John P. Hughes\altaffilmark{8}, 
Robert P. Kirshner\altaffilmark{6},
Kip D. Kuntz\altaffilmark{2},
Emily K. McNeil\altaffilmark{3, 11},\\
Thomas G. Pannuti\altaffilmark{9},
Paul P. Plucinsky\altaffilmark{6},
Destry Saul \altaffilmark{7}, 
Ralph T\"ullmann \altaffilmark{6},\\ and
Benjamin Williams\altaffilmark{10}
}

\altaffiltext{1}{Space Telescope Science Institute, 3700 San Martin Drive, Baltimore, MD, 21218; long@stsci.edu}
\altaffiltext{2}{The Henry A. Rowland Department of Physics and Astronomy, Johns Hopkins University, 3400 N. Charles
Street, Baltimore, MD, 21218; wpb@pha.jhu.edu}
\altaffiltext{3}{Department of Physics, Middlebury College, Middlebury, VT, 05753; winkler@middlebury.edu}
\altaffiltext{4}{Departmentof Physics, Universityof California, 1 Shields Avenue, Davis, CA 95616}
\altaffiltext{5}{Lawrence Livermore National Laboratory, L-413, Livermore, CA  94550}
\altaffiltext{6}{Harvard-Smithsonian Center for Astrophysics, 60 Garden Street, Cambridge, MA 02138}
\altaffiltext{7}{Columbia Astrophysics Laboratory, 550 W.
$120^{\mathrm{th}}$ St., New York, NY, 10027}
\altaffiltext{8}{Department of Physics and Astronomy, Rutgers University, 136 Frelinghuysen Road, Piscataway, NJ, 08854}
\altaffiltext{9}{Department of Earth and Space Sciences, Space Science Center,235 Martindale Drive, Morehead State
University, Morehead, KY 40351}
\altaffiltext{10}{Astronomy Department, University of Washington, Seattle,
WA 98195}
\altaffiltext{11}{Present address: The Research School of Astronomy and Astrophysics, The Australian National University, Mount Stromlo and Siding Spring Observatories, via Cotter Road, Weston Creek, PO 2611, Australia} 

\begin{abstract}

M33 contains a large number of emission nebulae identified as supernova remnants (SNRs) based on the high \sii:\HA\ ratios characteristic of shocked gas.  Using \chandra\ data from the \chase\ survey with a 0.35-2 keV sensitivity of \EXPU{\sim2}{34}{\LUM}, we have detected 82 of 137 SNR candidates, yielding confirmation of (or at least strongly support for) their SNR identifications. This provides the largest sample of remnants detected at optical and X-ray wavelengths in any galaxy, including the Milky Way.  A spectral analysis of the seven X-ray brightest SNRs reveals that two, G98-31 and G98-35, have spectra that appear to indicate enrichment by ejecta from core-collapse supernova explosions.  In general, the X-ray detected SNRs have soft X-ray spectra compared to the vast majority of sources detected along the line of sight to M33.  It  is unlikely that there are any other undiscovered thermally dominated X-ray SNRs with luminosities in excess of \EXPU{\sim4}{35}{\LUM} in the portions of M33 covered by the \chase\ survey. We have used a combination of new and archival optical and radio observations to attempt to better understand why some objects are detected as X-ray sources and others are not. We have also developed a morphological classification scheme for the optically-identified SNRs, and discuss the efficacy of this scheme as a predictor of X-ray detectability.  Finally, we have compared the SNRs found in M33 to those that have been observed in the Galaxy and the Magellanic Clouds.  There are no close analogs of Cas A,  Kepler's SNR, Tycho's SNR or the Crab Nebula in the regions of M33 surveyed, but we have found an X-ray source with a power law spectrum coincident with a small-diameter radio source that may be the first pulsar-wind nebula recognized in M33.

Subject Headings: galaxies: individual (M33) -- galaxies: ISM -- radio continuum: galaxies --supernova remnants

\end{abstract}
\section{Introduction \label{sec_intro}}

The vast majority of supernova remnants (SNRs) in nearby galaxies have been identified using optical interference-filter imagery where the ratio of \sii\ to \HA\ emission discriminates between SNRs and H II regions.  Pioneered by \cite{mathewson73}, this method is based on the fact that radiative shocks contain cooling zones with temperatures near 10,000K where S$^{+}$ is the dominant ionization stage of S, whereas in (bright) \hii\ regions S$^{++}$ is the dominant ionization state.  As a result, \sii:\HA\ ratios of 0.4 to 1 are common in SNRs whereas in bright H II regions the ratio is typically 0.1 \citep{dodorico76, levenson95}.  

M33, a member of the Local Group, is the nearest late-type spiral galaxy.  At a distance of  817$\pm$58 kpc \citep{freedman01}, a SNR with a diameter of 20 pc would have an angular size of 5.0\arcsec. The galaxy is relatively face-on \cite[$i=56 \pm 1^\circ $,][]{zaritsky89}, lies along a line of sight with low Galactic absorption \cite[$N_H \sim \EXPU{6}{20}{cm^{-2}}$,][]{dickey90, stark92} and, as a result, has been an obvious choice for characterizing the SNR population in nearby galaxies.   \cite{dodorico78} identified the first three SNRs in M33 using interference-filter imagery; other subsequent studies, principally by \cite{dodorico80}, \cite{long90} and \cite{gordon98}, have increased the number of remnant candidates to nearly 100.  

In contrast to M33, the vast majority of SNRs in the Galaxy were first recognized from radio observations, based on their extended non-thermal emission.  Since most SNRs are located near the Galactic plane implying high line-of-sight absorption to most objects, even today a majority of the Galactic SNRs remain undetected in \sii\  and \HA.  Radio searches for SNRs have also been carried out in nearby galaxies; these provide an independent search technique that can detect several classes of SNRs, including Crab-like remnants, very young objects like Cas A, and the so-called Balmer-dominated SNRs, that emit little or no \sii\ and thus are unlikely to be discovered in optical \HA-\sii\ searches.  In the case of M33, \cite{goss80} were the first to detect radio emission from three of the bright optical SNRs, and the number of radio detections has increased with time. \cite{gordon99} used images constructed from data taken with the Very Large Array (VLA) and the Westerbork Synthesis Radio Telescope  (WSRT) to detect 56 of the optical candidates as radio sources. However, no one has  successfully carried out a truly independent survey to find extended, non-thermal radio sources as has been done in the Galaxy; furthermore, no Crab-like SNRs have been found in M33 \citep{reynolds87}.

The $Einstein$ Observatory was the first X-ray experiment with sufficient sensitivity and angular resolution to detect extragalactic SNRs at X-ray wavelengths, although most of these were, not surprisingly, in the Magellanic Clouds \citep{long79, seward81}.   
%Although substantial numbers of SNRs in the Galaxy and the Magellanic Clouds have now been imaged at X-ray wavelengths, few of them have been identified based on such observations,  a notable exceptions being  G347.3-0.5 (RX J1713.7-3946) \citep{pfeffermann96, lazendic04}.  %\cite{markert83}, using the $Einstein$ observatory, appear to have been the first to unambiguously identify a SNR (X-13 = G98-55) in M33  at X-ray wavelengths. 
 \cite{long81}, in the initial $Einstein$ study of M33 suggested that the bright X-ray source M33 X-3, located near NGC~592 was a SNR because of its very soft X-ray spectrum.  \cite{markert83} reported that this source was time-variable, and its nature was, for a time, uncertain. Today we know this object as G98-21, the brightest M33 X-ray SNR \citep{gordon93, gaetz07}.   Later,  using ROSAT, \cite{long96} found X-ray counterparts to 10 optically-identified SNRs. More recently, \cite{misanovic06} reanalyzed the XMM data originally discussed by \cite{pietsch04}, identifying 15 non-variable X-ray sources with optical SNRs and suggesting 11 other X-ray sources might be SNRs based on their X-ray spectra.   \cite{ghavamian05} inspected interference-filter images of the candidates and found that two, XMM068 and XMM270, were located along lines of sight to nebulae with elevated \sii:\HA\  ratios that had been missed by \cite{gordon98} as well as earlier optical observers.\footnote{Here and elsewhere, XMM catalog numbers refer to the source numbers given by \cite{pietsch04}.} 
 
Although XMM has sufficient sensitivity to detect the brighter M33 remnants,  {\it Chandra} is the first X-ray observatory with sufficient sensitivity and spatial resolution that one might expect to discover new SNRs in galaxies as close as M33 using X-ray observations {\it alone}.  With a resolution of 0.5\arcsec\ on-axis, and $<5 \arcsec$ up to 8\arcmin\ off-axis \citep{weisskopf02},  {\it Chandra} should resolve most remnants with diameters $\gtrsim 15$ pc.
%The distance to M33 is 817$\pm$58 kpc \citep{freedman01} and so a SNR with a diameter of 20 pc would have an angular size of 5.0\arcsec, easily resolvable with $Chandra$. 
 \cite{ghavamian05} showed the effectiveness of {\it Chandra} for detecting SNRs, finding a total of 22 X-ray sources that aligned with optically identified SNRs in the data that existed at that time, which had typical exposures of 50 ksec.  For this reason among others, we have undertaken a series of deep observations of M33 with {\it Chandra}, the ChASeM33 survey.  \cite{plucinsky08} provided a ``First Look" at the results from ChASeM33 based on approximately half of the data; there we identified 26 SNRs from the list of \cite{gordon98}.\footnote{Hereafter, we will refer to sources reported by \cite{plucinsky08} as `First Look' sources, and identify sources, e.g., FL175, by their source number in that catalog.}  In addition, \cite{gaetz07} provided a detailed analysis of  G98-21 (M33 X-3), which appears as a 5\arcsec\ shell on the edge of a bright \hii\ region.

The purpose of the present study is to provide a full analysis of the \chase\ survey relevant to SNRs, in combination with supplemental optical and radio data.  A detailed analysis of the survey as a whole, and the point sources that have been detected, will be reported by \cite{tuellmann10}.  The remainder of this paper is organized as follows:   We describe the \chase\ survey and our approach to extracting information about SNRs in \S \ref{sec_obs}.  We then discuss our construction of a set of candidate SNRs (\S \ref{sec_candidates}) and report the detection of more than half this sample as X-ray sources in \S \ref{sec_results}.  We next describe our attempts  to better characterize the sample using a combination of new and archival optical and radio observations of M33 (\S \ref{sec_other_data}) and compare the X-ray, optical and radio properties of the SNRs in \S \ref{sec_compare}.    We explore the question of why we have detected some objects and not others in \S  \ref{sec_real_snrs} and then describe the images and spectra of the brightest  X-ray SNRs in M33 in \S \ref{sec_bright}.  We discuss which Galactic SNRs we would have detected in M33 and compare the samples of SNRs in M33 to those from other galaxies in \S  \ref{sec_discussion}.   We summarize our conclusions in \S \ref{sec_conclusions}.  Finally, in Appendix \ref{sec_notes}, we provide an atlas of X-ray and optical images of M33 SNRs and SNR candidates, including details on why some might have been misclassified.

\section{Observations and Reduction of the \chandra\ Data \label{sec_obs}}

The  strategy for the ChASeM33 survey has been described in detail by \cite{plucinsky08}.  The survey comprises ACIS-I observations of seven overlapping fields which, as shown in Fig.\  \ref{fig_snr_overview}, cover the inner region of M33 to a radius of about 18\arcmin, or 4.3 kpc. Of the 98 SNRs catalogued by \cite{gordon98}, 93 (or about 95\% of the SNRs identified in the last systematic optical-only search  of M33)  lie within the area covered by ChASeM33.  All of the data for the survey have now been obtained.   Each field was observed for a total of $\sim$200 ks, but because of overlaps the total exposure over a substantial portion of the inner galaxy exceeds 400 ks.  As a result, the limiting 0.35-2 keV sensitivity is about \EXPU{2}{34}{\LUM} (2$\sigma$), assuming emission from a thermal plasma with kT of 0.6 keV, a metal abundance of 0.5 times solar, and a line-of-sight absorption of \EXPU{1}{21}{cm^{-2}}.\footnote{The Galactic absorption along the line of sight to M33 is about  \EXPU{6}{20}{cm^{-2}} \citep{dickey90, stark92}, and the average H~I column density through M33 is  \EXPU{1}{21}{cm^{-2}} \citep{newton80}.  Our flux conversion uses a typical value for the total column to the mid-plane of M33. The metallicity of M33 varies with galactocentric radius, but 0.5 is a typical value in the portions of M33 sampled by \chase\ \citep{henry95}.} 
%\footnote{This was calculated with PIMMS  for an Raymond-Smith spectrum and 0.6 solar abundances.  It assumed we needed 16 counts in 400 ks for a detection.  One might do better with XSPEC.} 

The starting point for our analysis is the same screened data that are being used for the creation of the point source catalog; details can be found in \cite{tuellmann10}.  These data include not only those obtained by us as part of ChASeM33 but also ACIS-I data from observations which are available from the $Chandra$ archive (ObsID's 1730 and 2023).  Limited data from chip S3 of the ACIS spectroscopic array exist, but we have not included these in our analysis because this chip has higher background and a very different energy response from that of the I-array CCDs.  The ChASeM33 data were obtained between 2005 September and 2006 November.  Each observation of a field, except that of Field 6, was split into two epochs to enable a more comprehensive search for variable sources. Due to scheduling constraints ObsID 7208, which is part of Field 6 epoch 1, was observed at a different roll angle and has much lower exposure; we analyzed this observation separately. The data used, including the exposure resulting in good data at each position, are summarized in Table \ref{table_obs}.

%\subsection{Procedure}
 
Our goal was to extract information from the X-ray data about all known and candidate SNRs in M33.  For our analysis we used {\tt ACIS Extract} (AE) version 2009-08-12 \citep{broos02},  which is designed specifically to deal with source crowding, to determine source properties like net counts, fluxes, and significances, and to extract spectra simultaneously from multiple observations with overlapping fields of view. The general approach we used to account for the spatial extent of the SNRs, designed for the present case where source extent is of the same order of magnitude as the point spread function  (PSF), is described here.  

We began with separate lists of SNR/SNR candidates and point-source positions. The construction of our SNR/SNR candidate list is discussed in more detail in Section \ref{sec_candidates}, but the list basically consists of all of the objects we and others have suggested are SNRs as well as objects that we felt were plausible candidates based on inspection of the X-ray data and other archival data at our disposal. Although our interest is in the SNRs, the point-source list is also required so that the SNR fluxes can be calculated using background regions that exclude point sources. Since our analysis of the SNRs took place in parallel with the analysis of the entire ChASeM33 survey, our point-source list consisted of a near-final version of the point source catalog, with those sources we have identified with SNR candidates removed.    

We first processed all of the sources together as if they were point sources. The primary purpose of this step was to obtain two sets of region files, one describing the PSF at the position of all of the SNRs in each of the observations and the other describing the areas of each X-ray image affected by point sources.\footnote{For this purpose, in AE parlance, we set {\tt psf\_frac=90\%} for psf generation and {\tt mask\_fraction=0.95 and  mask\_multiplier=1.2} for background mask generation.}  The PSFs and background regions for a given source vary, depending on the roll angle and off-axis angle of each observation.   

Since SNRs in M33 are not point sources but have typical sizes of 3\arcsec\  to $>30$\arcsec, the count rates extracted by AE in the first processing step (which used region files appropriate for point sources) were not expected to yield accurate fluxes (or upper limits) for the SNRs; the region files must be modified to reflect both the physical size of the SNR and the local PSF in each X-ray image. We used the optical interference-filter images discussed in \S \ref{sec_new_opt} as well as (on-axis) X-ray data to estimate the intrinsic size of each SNR or candidate object.  We then used the AE point-source extraction regions at the position of each SNR convolved with these intrinsic sizes to create appropriate region files for each SNR for each of the various X-ray fields in which that SNR was observed.  There were two objects for which a point source was either clearly inside a large-diameter object or was close enough to the  SNR that the expanded region file included the point source in the pointings where the SNR was observed far from the telescope axis (so  the PSF was large).  In these cases we constructed special region files that excluded the point sources from the extraction regions. We then reran AE using these tailored region files to extract count rates for the objects in our SNR list.\footnote{AE determines background count rates by expanding the background region until it contains a certain minimum number of counts.  For our analysis, we set {\tt min\_num\_cts=100}.}

AE is designed to handle the crowding of point sources, and it adjusts extraction regions to reduce the effects of source crowding during the source extraction process.  Inserting our own region files for the extended SNRs interferes with this process.  To assure ourselves that the regions we were using for source extraction and background subtraction were consistent and correct, we created a number of tools to overlay background and region files for the SNRs and nearby point sources on the X-ray images.   Since the M33 field is not exceptionally crowded, there were no problems with AE attempting to resize the extraction regions in most cases.  We did, however, eliminate a few observations from our analysis if there was too much overlap between the SNR and a contaminating point source at large off-axis angles ($>7\arcmin$), or where the SNR was very close to the edge of the field.  As a result of this effort we re-iterated our analysis until we were satisfied that we had obtained accurate count rates for all SNRs and SNR candidates.

\section{A List of SNRs and SNR Candidates for M33 \label{sec_candidates}}

When we began this study there were roughly 100 objects in M33 that had been suggested as SNRs based almost entirely upon optical interference-filter imagery.  Follow-up optical spectroscopy has been carried out for 72 of these objects, and has confirmed that the ratio of \sii:H$\alpha$ is high in essentially all objects observed.  Most (98) of these objects are summarized in the study reported by \cite{gordon98} in which a number of us participated. We used the Gordon list as our initial set of candidates.  Some of these objects are classic spherical shells of high surface brightness, but the majority are not (see \S  \ref{sec_real_snrs}).   

The \cite{gordon98} list is not complete, in part because the images available at the time did not include all of the outer portions of the galaxy, and in part because the identification of a SNR using the \sii:H$\alpha$ ratio at a particular flux or surface brightness limit depends upon the amount of confusion by other sources of emission---\hii\ regions or other diffuse gas in the vicinity.  A few objects, including the two XMM-discovered sources noted earlier, were  simply overlooked.    These facts, together with the existence of better optical and radio data than those used by \cite{gordon98} (some first reported here), have led us to expand the list of candidates based on new interference-filter imagery and on the X-ray data, as we describe in this section.

We were fairly liberal in identifying candidates, since X-ray detection provides strong confirmation that an object is a remnant, and our goal is to construct a complete remnant catalog. This means that some objects in the candidate list may not be SNRs but, rather, related entities such as wind-blown bubbles or superbubbles. The vast majority of objects are,  however {\it bona fide} SNRs. We return to the question of both the completeness of our SNR list and the possibility that some objects are not SNRs in \S \ref{sec_superbubble},  when we have the X-ray count rates and upper limits in hand.

\subsection{New Optically Selected SNR Candidates \label{sec_new_opt}}

We supplemented the \cite{gordon98} list by searching for new candidates using two emission-line surveys of M33:  the Local Group Galaxies Survey (LGGS), carried out by \citet{massey06, massey07} using the 4m Mayall telescope and prime-focus Mosaic detector at KPNO,\footnote{Data from the LGGS are available at http://www.lowell.edu/users/massey/lgsurvey/} and our own survey from 1996 using the 0.6m Burrell Schmidt telescope at KPNO with the S2KA chip at the Newtonian focus  \citep{mcneil06}.   The observational characteristics of the two surveys are summarized in Table \ref{table_optical}.  
Both surveys obtained deep images in \HA, \sii \LL 6716, 6731, and \oiii$\; \lambda 5007$, plus continuum bands that can be used to subtract most of the stellar emission.  For the Schmidt survey, the latter were narrow continuum bands in both red and green; for the LGGS, broad-band R and V images were used.   For the LGGS, five 300 s exposures in each emission line were stacked. 
%For our Schmidt survey the (number) and total exposure  times were: \HA, (6) 5,100 s; \sii, (5) 5,700 s; \oiii, (8) 12,000 s. 
For our Schmidt survey the number and total exposure  times were respectively 6 and 5,100 s for \HA, 5 and 5,700 s for  \sii, and  8 and 12,000 s for \oiii.  
Both surveys covered almost all of M33; the Schmidt + S2KA covered the entire galaxy in a single 61\arcmin\ square field, while the LGGS covered the galaxy in three overlapping 36\arcmin\ square fields.  Both used multiple, dithered frames to paper over chip gaps and defects and to improve signal-to-noise. In both cases, the individual frames were transformed to a standard coordinate system based on USNOB2 stars, and then combined.   To remove, insofar as possible, the stellar continuum from the emission line images, we first used the IRAF task {\it psfmatch} to determine a kernel that we then convolved with the sharper image of a line-continuum pair to produce two matched images.  We then scaled and subtracted the  matched continuum image from each emission line image to
obtain obtain continuum-subtracted emission line images.

To search systematically for SNR candidates, we displayed multiple images in \HA\ and \sii\ (both continuum subtracted), a \sii:\HA\ ratio image, and the B-band image from the LGGS at a large scale (typically a $\sim3\arcmin$\ field), all registered to the same coordinate system.  We searched for shell-like objects and other coherent structures visible on both the \sii\ and \HA\ images with \sii:\HA\ ratio $\gtrsim 0.4$\@.
Objects with a \sii -bright shell surrounding an interior that was \HA-bright and that contained a population of hot stars (as judged from the continuum image) were rejected as primarily photoionized structures.
The two surveys are complementary for SNR searches: the coarser scale ($1.5\arcsec\;{\rm pixel}^{-1}$\ in the combined images) and resolution ($\sim 5\arcsec$) of the Schmidt survey facilitated identification of extended diffuse objects, especially in isolated regions of the galaxy, while the far higher resolution ($\sim 1\arcsec$) of the LGGS was superior for examining the structure and identifying objects in confused regions.  We estimate the sensitivity of the LGGS at 
$\sim 10^{-16}\FLUXARCSEC$, in that coherent structures at this surface brightness could consistently be identified in  \HA\ and/or \sii\  images in all but the most crowded regions of M33.   The Schmidt survey is somewhat deeper; relatively isolated objects with surface brightness  $\sim 4\times 10^{-17}\FLUXARCSEC$\ could be identified.  Regardless of how candidate objects had first been identified, subsequent  flux measurements (\S \ref{sec_opt_flux})  and assessment of their morphology (\S \ref{sec_opt_morph}) were carried out using the higher resolution LGGS images. 

Our search for SNRs revealed 23 new objects that satisfy the usual optical criterion for describing an object as a SNR.  Most, though not all, of the new candidates are in the region covered in the images used by \cite{gordon98}.  The majority of the new candidates are large, low-surface-brightness objects for which the Schmidt survey in particular was more sensitive than earlier surveys. In addition, our inspection of the images suggested to us that three of the \cite{gordon98} objects -- G98-43, G98-57, and G98-97 -- are composite in the sense that the emission is better described as emission from two separate objects in close spatial proximity.\footnote{We have added a designation A and B to the source number in the \cite{gordon98}  list to identify the two separate components we believe are better characterized as separate objects.  See Appendix \ref{sec_notes} for further information about our reasoning. With this accounting, there are 101 separate SNRs in our revised accounting of nebulae identified by \cite{gordon98} as likely SNRs. }

\subsection{X-ray-selected SNR candidates}

At X-ray wavelengths, most SNRs are extended sources, and most have soft spectra dominated by emission from a hot plasma. The bright SNRs described in \S \ref{sec_bright} have both of these characteristics.  Even in the absence of earlier optical or radio observations, these objects would have been recognized as SNRs.  This uniformity suggests the possibility of carrying out a purely X-ray-based search for SNRs in M33. We have attempted such a search while guarding against other types of sources which share some of these characteristics: \hii\ regions and wind-blown bubbles (extended and thermal, but usually centered on one or more young stars), late-type foreground stars (thermal but point-sources), and background clusters of galaxies (extended and thermal but hotter characteristic temperatures than most SNRs).  As outlined below, we have used several criteria to search {\it ab initio} for SNRs in the X-ray data.

\subsubsection{X-ray Spectra}
We examined the X-ray spectra of all of the objects in the ChASeM33 dataset with more than 300 total counts (0.35 - 8 keV) to search for evidence of thermal emission.   Our examination consisted of comparing the fitted parameters for power-law and thermal-plasma model fits to the data and a visual inspection for evidence of line emission.  We chose 300 counts as the minimum based on previous experience, which has shown that a thermal spectrum with this many counts will be fitted with a thermal model of temperature $kT<2.0$~keV, while a power-law model will have a photon index of larger than 3.0. Hence, although the quality of the fit might be similar for the thermal and non-thermal models, we can conclude that the thermal model is more likely to be correct because the power-law index is unrealistically steep. For a thermal spectrum with a characteristic temperature of 0.6 keV and $N_H$ of \EXPU{1}{21} cm$^{-2}$, 300 counts in a typical exposure of 400 ksec corresponds to a 0.35-2 keV luminosity of \EXPU{\sim4}{35}{\LUM}
% Changing from 6e20 to 1e21 was a 4% effect. so did not change the numbers here. .{\bf everywhere else we use either the Galactic value of $6 \times 10^{20}$  or, more appropriate here I think, $10^{21}$}

Of the 95 sources in the full ChASeM33 survey with more than 300 counts, there are 19 sources with spectra that appear to be dominated by emission from a hot plasma, as expected for a SNR.  Of these 19 sources, eleven correspond to well-studied objects in the \cite{gordon98} list, and two are the
objects identified first by \cite{pietsch04} and confirmed by \cite{ghavamian05} as SNRs.   For these 13 SNRs, the best-fit temperatures ranged between 0.30 and 0.76~keV and the best-fit power law indices were all larger than 4.0. 

Of the remaining six sources which were not known to be SNRs, three -- FL175, FL181 and FL362 -- are along lines of sight toward relatively bright foreground stars. None of these three is significantly extended as an X-ray source, and none has associated optical emission with elevated \sii:\HA\  ratios that might indicate an SNR, although FL175 does lie near the edge of a large ring of \HA\  emission. All three likely represent X-ray emission from Galactic stars.

The final three objects --- FL047, FL195, and FL281 --- are more difficult to classify.  Unlike the SNRs discussed above, the spectra of all of these objects are quite hard.  The best-fit temperature for these objects is between 2.6 and 3.8 keV, while the best-fit power law indices are between 2.1 and 2.2, well outside the range of any of the objects known to be SNRs with thermal spectra.  None of these sources are significantly extended in X-rays.  FL047 aligns with a 19.4 magnitude point-like source that could be the counterpart, which suggests the possibility that that this is a neutron-star or black-hole binary in M33.
FL195 is in the outer galaxy, and only a few very faint stellar sources are near the position, with none being an obvious counterpart. FL281 lies within 0.35 arcsec of a slightly extended radio source with a spectral index of $-0.1$ and a flux density of 700 $\mu$Jy at 20 cm.  No optical continuum source is seen, but a very faint, compact emission nebula may be present.  This object is a candidate pulsar wind nebulae (PWN) in M33, and it is discussed further in \S  \ref{sec_radio}.  We did not add any of these objects to our list of SNRs.

\subsubsection{Spatial Extent in X-rays}
We inspected all the sources in the point source catalog to look for evidence of spatial extent. As with the previous test, limitations due to counting statistics mean that we can only measure extent for the brightest sources. Again we conducted both a visual inspection and a quantitative one.   The AE software contains a check for consistency between the observed PSF and that expected for a point source.  A total of 72 sources were flagged as possibly spatially extended based on this comparison.  Of these, 22 were objects that had been previously identified as optical SNRs (20 in the \cite{gordon98} list plus XMM156 and XMM244). With the exception of one or two sources that appeared to be high-surface-brightness peaks in the diffuse emission from an \hii\ region, the remaining objects were at best marginally resolved.  
%{\bf Isn't a high X-ray surface brightness peak in an HII region just what one would expect a young SNR to look like?? --david.  Well yes, but diffuse emission has peaks and in the absence of anything else one could not call it a SNR, and the spatial extent derives from the fact that it is an HII region} 
A significant number of the flagged objects are fairly bright X-ray sources which are identified in the First Look survey with known objects such as foreground stars and the nucleus of M33.  Thirty-nine of the sources were already examined in the high-brightness sample, which as just described were also analyzed for evidence of thermal emission.   We inspected each object in the X-ray image obtained closest to the boresight of the Observatory.  We concluded that it is unlikely that any of these objects is actually extended. None was identified with a radio source.  None was associated with an optical nebula with elevated \sii:\HA\  ratio.  While it is possible that one or more of these sources is a very young SNR in M33 ($D<4$~pc), there was no compelling reason to include any of these objects in a list of SNR candidates. Thus, this test added no new objects to our list of SNRs and SNR candidates.

\subsubsection{Soft X-ray Sources with Optical Counterparts}

We systematically searched for optical counterparts to all soft sources in the point source catalog.  
%Most of the objects observed with Chandra as part of the \chase\ survey have far fewer counts than the brightest six, and as result it is not easy to determine whether the spectra  are thermal or non-thermal, or whether the object is spatially extended.   As a result X-ray searches are bound to reveal fewer SNRs than other techniques, at present.  
%Since, as is discussed in \S \ref{sec_results}, the spectra of most SNRs  are soft compared to the X-ray binaries and background AGN that dominate the point source survey of M33, we inspected optical images of fields of objects with soft X-ray spectra as characterized by the hardness ratios ($<$0; see \S \ref{sec_results} for the exact definition) in our source catalog for M33 for evidence that they might be SNRs.   
As noted earlier, this technique had been used by \citet{ghavamian05} to confirm that two soft X-ray sources suggested as SNRs from XMM surveys \citep{pietsch04, misanovic06} are indeed SNRs based on their association with emission nebulae exhibiting high \sii:\HA\ ratios. 
We used the hardness ratio described in \S \ref{sec_results} and considered all the 160 objects with hardness ratios less than 0 that had not already been identified as SNRs. Each of these sources was displayed in a multi-panel format with aligned coordinates, including LGGS continuum-subtracted \HA, \sii, \oiii,  and a V-band continuum image in addition to the X-ray panel. This search revealed a handful of sources of potential interest for follow-up work, but only three new optical counterparts that passed the \sii:\HA\ criterion:  FL236, FL261, and XMM244.  (The optical counterpart of the source XMM270 was found independently, but this just confirmed the earlier detection by \cite{ghavamian05}.)  The first two of these (plus XMM270) correspond to very faint, small-diameter knots of emission in unconfused regions, while XMM244 is coincident with a more patchy, diffuse nebula in a fairly confused region.  In all four cases, the objects could easily have been missed in previous optical searches.  These have been added to our candidate list.

Finally, several other special cases deserve mention. The soft source FL212 is aligned with a very faint, extended \HA\ region that did not have a counterpart in [S~II], [O~III], or the continuum bands.  Since some young Galactic and Magellanic Cloud SNRs have optical spectra dominated by Balmer line emission, this source was added to the SNR candidate list as a possible remnant of this type.  %FL312 is a soft source that is in a region with no likely stellar counterparts (brighter than about 23rd magnitude) 
Finally, the soft source FL113 (=XMM156) is extended in X-rays, but seems to correspond to a m$_v$=16.7 star.  However, careful inspection of the LGGS [O~III] image indicates an excess of emission that is not consistent with the stellar subtraction residual. A longslit MMT spectrum reported in \S \ref{sec_opt_spec} confirms that this excess emission is real.  This source was also included in our candidate list as a likely SNR.

\subsubsection{Faint, Extended X-ray Emission}

Since our X-ray source detection algorithm was optimized to detect point sources, we conducted a search for faint, extended X-ray sources that would have been missed or rejected in assembling the point source catalog. To do so, we created a mosaic of all the ACIS-I data in the 0.45-1.0 keV band   
and removed all the point sources and all the previously catalogued SNRs. 
We then smoothed the image with a modified square ``Mexican hat" function 
where the positive core had $r=7\farcs87$ and the negative annulus had $15\farcs7<r<23\farcs6$. This convolution favors the detection of sources with sizes comparable to the positive core (e.g., 5\arcsec\ to 15\arcsec) and excludes significantly larger sources such as the \hii\ complexes NGC~604 \citep{tuellmann08} and IC 131 \citep{tuellmann09}. We found 30 significant positive excursions in the diffuse emission map created in this manner. We performed a careful visual inspection of these regions, considering both X-ray and optical emission as in our soft point-source search. We discarded X-ray sources with no extended \HA\ structures as being either background sources or spurious detections.
Six sources had H$\alpha$ emission-line counterparts (typically very faint)
that made them interesting candidates for further study. These are primarily moderate to faint optical shells in the size range 40 - 60 pc, some isolated and some in fairly confused optical regions. Because of their potential association with extended X-ray emission, these objects were added to the SNR candidate list.

One might have expected this search method to produce a significant number of X-ray detections of  \hii\ regions.  Of the giant \hii\ regions, only NGC~604 \citep{tuellmann08}, IC~131 \citep{tuellmann09}, IC~133, IC~142, and NGC~595 appear as distinct objects in X-ray images. Only NGC~604 and IC~131, which have strong cores, appeared in the filtered images described above; the remaining giant \hii\  regions are apparently too large or too faint to pass the filter. Doubling the filter size detected NGC~595, but no additional \hii\ regions or indeed any significant new positive excursions. Of the extended X-ray sources detected through the filtering process and coincident with the myriad of smaller optical emission regions,  most were known SNRs. Only three normal \hii\ regions were seen, indicating that a large majority of the \hii\ regions have X-ray fluxes below the \chase\ detection limit.  

% Calculated from the routine which makes the sample table on 080812
% Ha Luminosity characteristics: 137  9.515880e+34 2.701310e+37  1.859590e+36
% Ha SB characteristics: 137  1.384191e-17 1.505960e-14  2.152531e-16
%Diameters all sources number 137  min 7.920e+00 max 1.790e+02  med 4.429e+01 

\subsection{The Complete SNR List}

Our complete list of 137 SNRs and SNR candidates is presented in Table \ref{tab_sample}.    The objects have diameters ranging from 8 to 179 pc; the median diameter is 44 pc, corresponding to a middle-aged SNR.\footnote{The SNR size was determined primarily from the LGGS images, further informed by the X-ray data.  We defined elliptical regions that best traced the optical shell, expanded slightly in a few cases to embrace X-ray emission that is apparently associated with the SNR (see \S  \ref{sec_opt_flux}).  The ``diameter" as defined and used in this paper is the geometric mean of the major and minor axes of this ellipse.  All of the objects were clearly extended in the LGGS images.  However, the error in measuring the diameter, which for a well-defined object is 0.5-1\arcsec, clearly has a larger fractional effect on the small diameter objects.  For the poorly defined objects in morphology class C (\S \ref{sec_opt_morph}) the size is highly uncertain.} The list includes all 98 of the objects described by \cite{gordon98}, three additional objects based on our assertion that G98-43, G98-57 and G98-97 are multiple SNRs, 23 objects that were selected because of their optical properties, and 13 that were included based on their X-ray emission characteristics.  The distribution of candidate objects in X-ray and optical images of the southern spiral arm of the galaxy is presented in Fig.\  \ref{fig_multiwave}; close inspection shows that many of the candidates are indeed X-ray sources.

\section{X-ray Survey Results \label{sec_results}}

The results of our search for X-ray emitting SNRs in M33 are presented in Table \ref{table_results}. Of the 137 objects in our list of SNR candidates,  131 were in regions observed as part of the ChASeM33 survey of M33, and 82 (58) were detected at significance levels of 2 (3) $\sigma$ or greater.\footnote{The six objects that were not within the region surveyed with \chandra\ were G98-01, G98-03, G98-04, G98-07, G98-43A and G98-43B.  These numbers reflect our opinion that G98-43 comprises two close but distinct SNRs.}
A detailed discussion of the brightest objects appears in \S \ref{sec_bright}. X-ray and optical images and some background information about the remainder of the detected sources are presented in Appendix \ref{sec_notes}. Of the 95 SNRs in our revised accounting of the  \cite{gordon98} list which fall within the region observed by {\it Chandra}, 61 (45) were detected at 2 (3) $\sigma$ or greater.  These numbers include G98-57A, GKL98-57B, GK98-97A and GKL98-97B, all of which were detected at more than 3$\sigma$.\footnote{Had we treated the \cite{gordon98} list as originally described, then the ChASeM33 survey would have covered 93 of the 98 SNRs, and 59 (43) would have been detected at at 2 (3) $\sigma$ or greater.}   Of the 23 new objects in our list of SNR candidates selected from optical interference-filter imagery, only 8 (2) were detected in X-rays. That a smaller fraction of these new objects were detected in X-rays compared to the \cite{gordon98} sample is unsurprising given that they are typically  larger and lower (optical) surface brightness objects; indeed, some may not be SNRs (see Appendix \ref{sec_notes}).  As expected, all of the 13 objects that  initially attracted our attention because of their X-ray properties were detected in the final AE analysis.

Luminosities (and upper limits) for the objects in our list of SNRs and SNR candidates are presented in Fig.\   \ref{fig_lum}  as a function of the diameters determined from our inspection of the optical and X-ray images.  To convert from count rates to luminosities, we assumed a spectrum for a thermal plasma with 0.5 solar abundances, a temperature $kT$ of 0.6 keV and an absorbing column of \EXPU{1}{21}{cm^{-2}}.   The observed luminosities for detected SNRs vary by a factor of 500, from \EXPU{2.4}{34}{\LUM} to \EXPU{1.2}{37}{\LUM} in the 0.35-2.0~keV  band. The brightest SNRs have diameters of 20 to 30 pc.  Smaller-diameter SNRs tend to be fainter, as do larger ones. However, even at relatively small diameters of 20-40 pc,  ten optically-identified SNRs were not detected in X-rays.  Some SNRs were detected in X-rays  throughout the size range from the smallest up to $\sim 100$ pc, though many of the larger ones are quite faint.

% Next paragraph updated to accurate numbers 091228
Four objects with diameters of 80 pc or greater were detected.  Two  of these -- L10-043 (85 pc)  and L10-089 (92 pc) -- were identified initially because of their X-ray properties, but two others -- G98-70 (101 pc),  and L10-122 (111 pc) -- were selected based solely on their optical properties.   Their detection as extended sources supports their identification as SNRs although, if they are isolated SNRs, they would be among the largest such objects known.  An alternative possibility that these are superbubbles in which SNe have exploded is discussed in \S \ref{sec_superbubble}.

We first consider whether any of our X-ray detections represent chance coincidences.  The new point-source catalog will contain about 670 entries, of which 7\% (46) are SNRs.  The \chandra\ survey area covers about 1000 arcmin$^2$.  The summed areas of the region files for SNRs and candidates with diameters less than 60 pc is 2 arcmin$^2$; for candidates with diameters greater than 60 pc the total area is about 5 arcmin$^2$.  Assuming the sources are randomly distributed  on the sky, we thus expect about one chance coincidence for SNRs smaller than 60 pc and about three for larger diameter objects. We do find one point source located within the boundary of one of the SNRs, L10-137; the effect of this point source was removed in calculating the values reported in Table \ref{table_results}.  
%However this simple estimate is overly conservative; as shown in Fig.\ \ref{fig_hardness}, most of the sources in the point source catalog have hard X-ray spectra, whereas the SNRs have soft spectra.  
We conclude that possible point-source contamination to the detections we report is very small.

 Fig.\ \ref{fig_galacto} displays the distribution of objects as a function of diameter and galactocentric radius. The fraction of X-ray detections is roughly constant, apart from the effect that can be attributed to a larger number of large-diameter objects at larger galactocentric radii.

% Calculated from fig_rates.py  080812
% Median diameters - all 44.3 det 37.6 undet 53.9 at 2.000000 sigma
% Median diameters - all 44.3 det 31.7 undet 53.3 at 3.000000 sigma
% Calculaed from fig_lum.  The numbers below are not sensitive to the 2 sigma limit becuause all objects with this L were actually detected at 3sig or greater
% Diameters for sources with L>1e35  number 29  min 1.090e+01 max 5.989e+01  med 3.168e+01 at 2.000000 sig
% Diameters for sources with L>1e36  number  7  min 1.267e+01 max 3.168e+01  med 2.019e+01 at 2.000000 sig

The distribution of SNR diameter is shown in Fig.\ \ref{fig_dia} which compares the distribution for all observed SNR candidates (in black), those detected at  $>2\sigma$ in blue and those detected at $>3\sigma$ in red.  While there are some small ($<$30 pc) diameter objects that are not detected, the fraction of detections is high below 40 pc and gradually falls at larger diameters.   The drop-off in detection frequency is more rapid for the $3\sigma$ limit than at $2\sigma$ as one would expect if large-diameter objects were typically fainter.  A substantial fraction of the objects detected between 2$\sigma$ and 3$\sigma$ have diameters in the 40 to 80 pc range.  The median diameter is 38 (33) pc for objects detected at 2 (3) $\sigma$, compared to the median diameter of 44~pc for the entire sample.  The median diameter for objects detected with $L_{X} >$ \POW{35}{\LUM} is 29 pc; for \POW{36}{\LUM} it is 19 pc.  

Although there are a few SNRs bright enough for detailed spectral analysis (see \S \ref{sec_bright}), most are faint. Nonetheless, it is clear that the spectra of the SNRs are soft compared to other M33 (and background) source populations.  This is illustrated in Fig.\  \ref{fig_hardness}, which shows histograms of the numbers of SNRs and point sources detected in M33 with {\it Chandra} as a function of hardness ratio.  Here, we have defined the hardness ratio, M--S/Total, using the numbers of counts in the 0.35-1.2 keV (S), 1.2-2.6 keV (M), and 0.35-8 keV (Total) bands.  With this choice of bands, the SNRs occupy a region centered around $-0.5$, whereas the point sources are centered near $+0.5$.  The width of the SNR distribution is broad compared to that of the point-source distribution, largely a consequence of the high fraction of the SNRs that are faint, rather than an indication of real spectral variations among the SNRs.  There is a tail of soft sources in the point-source distribution; as discussed in \S  \ref{sec_candidates}, we inspected the positions of all of these sources in optical images and find no evidence that they are SNRs.  A number of them can be identified with foreground stars.  The clear separation between the majority of point sources and the SNRs supports our assertion that we are including few if any chance associations in our list of detected SNRs.

We performed tests to see if there were trends in hardness ratios with the X-ray count rates and/or SNR diameters.  We tested the hardness ratios defined above, as well as hardness ratios based on the bands 0.35-1 keV and 1-2 keV.  One might expect such trends if larger diameter and/or fainter SNRs had lower shock velocities, but none was seen, presumably because of the limited signal-to-noise ratio of the spectra and limited spectral resolution of ACIS I.   There was a general narrowing of the distribution as the number of counts increases, as one would expect from counting statistics.  

% Next paragraph updated to reflect numbers from AE9a
We have detected many more SNRs than the 26 reported by \cite{plucinsky08} in the First Look survey.  There are several reasons for this: (1) we have tailored the extraction regions for the actual sizes of the SNRs; (2) we have selected an energy bandpass of 0.35-2 keV optimized for the relatively soft spectra of SNRs; and (3) we are using the full \chase\ dataset whereas the First Look analysis was carried out only on data obtained during the first year of observations.  Of the 26 sources reported in the First Look survey and associated with SNRs, there are two -- G98-42, and G98-83, identified with FL174, and FL286 -- that do not appear in our list of detections.  In the First Look survey, sources were associated with SNRs by positional coincidence only using a match radius of 10\arcsec.  In these two cases, our refined analysis shows that there were point sources near but outside the region defined by the optical SNR. As noted earlier, \cite{ghavamian05} analyzed a different set of \chandra\ data including some data from the S-array but covering only part of M33; they detected 23 SNRs including 22 from the list of \cite{gordon98}, five of which were not detected in the First Look analysis.  The \cite{ghavamian05} study was specifically focussed on recovering SNRs from the data and used energy bands and extraction regions optimized for SNRs.  Our study confirms that all but one of the SNRs reported by  \cite{ghavamian05} are X-ray emitters.  The one SNR that was not detected by us is G98-83; with 5.1$\pm$5.3 net counts, it falls below our 2$\sigma$ detection threshold.

\section{Optical and Radio Characterization of the SNR Sample \label{sec_other_data}}

In order to better characterize the sample, we have gathered all of the available information on the objects in our list of SNRs and SNR candidates, using data that were already available and, when necessary, acquiring new optical and radio observations.  We describe here the results of these efforts.

\subsection{Optical Flux Measurements \label{sec_opt_flux}}

As discussed in \S \ref{sec_new_opt}, we used interference-filter imagery from Massey et al.'s (2006) LGGS data and our own imagery from the Burrell Schmidt telescope to search for new SNRs in M33.  We used these same data to determine the apparent sizes of SNRs and to measure \HA\  and \sii\ fluxes from the objects in our sample.  We examined each of the SNRs and candidates and created elliptical region files, varying the major and minor axes and the position angle to match the extent of the \HA\ and \sii\ emission.  For full or partial shells where the size could be be estimated with little ambiguity, we chose regions to match.  For the minority of objects with amorphous shapes or where only an individual filament is visible, we simply chose an ellipse that most nearly encloses the visible emission.  For all but a handful of objects, we used identical regions for both optical and X-ray flux extractions, the exceptions being objects that were clearly more extended in one band or the other.   

To measure the optical flux, we used the {\it imcnts} task (part of the IRAF external package {\it xray}) to extract the flux within the elliptical regions defined above from the flux-calibrated\footnote{Emission-line images from both surveys have flux calibrations obtained through measurements of several spectrophotometric standard stars from the list of \citet{massey88}.}, continuum-subtracted images. In a few instances where a nearby bright star had been poorly subtracted, we excluded a small region around the star from the extraction region.  We measured the flux using data from both the LGGS and Schmidt surveys, and in the vast majority of objects obtained results that agreed within $\pm 30\%$\@.  (Instances of disagreement were mostly very small objects where the lower resolution of the Schmidt survey caused much of the flux to spill outside the extraction region.)  

The results are given in Table \ref{tab_sample}, where  the ``diameter" represents the geometric mean between major and minor axes, converted to pc assuming a distance to M33 of 817 kpc.  The surface brightness is the actual mean value within the elliptical region (as reported by {\it imcnts}), and the luminosity $L_{H\alpha}$\ is determined simply by scaling to 817 kpc, with no absorption; both of the latter quantities were measured from the LGGS.  While the actual luminosities will be slightly higher once absorption is taken into account, the absorption is approximately uniform across the face of M33, so the relative values should change little.  Also, it is important to note that the LGGS bandpass of 81 \AA\ passes \nii \LL 6548, 6583, so the ``\HA"  surface brightness and luminosity values include most of the \nii\ emission as well (the relative intensity of which will vary systematically relative to \HA\  with galactocentric radius).

The \HA\  luminosities for the SNRs range from \EXPN{1}{35} to \EXPU{3}{37}{\LUM}, with a median value of \EXPU{1.9}{36}{\LUM}, while the \HA\  surface brightness ranges from \EXPN{1.4}{-17} to \EXPU{1.5}{-14}{ergs~cm^{-2}s^{-1}arcsec^{-2}} with a median value of \EXPU{2.2}{-16}{ergs~cm^{-2}s^{-1}arcsec^{-2}}.  As noted earlier the optical searches for SNRs in M33 are largely surface-brightness limited, as is clear from Fig.\  \ref{fig_ha_lum}, which shows the \HA\  luminosities of objects in our list as a function of diameter.  At smaller diameters, SNRs and candidates have a large range of luminosities, while at large diameters there are no low-luminosity objects since these objects would have such faint average surface brightnesses that they would fall below our detection threshold.

\subsection{Optical Spectroscopy \label{sec_opt_spec}}

Optical spectra are important not only for confirming the SNR identifications through quantitative line ratios, but also for providing physical information such as electron densities in the optically-emitting gas via the \sii\ $\LL 6717,6731$ line ratio.  \cite{gordon98} provide a compendium of SNRs observed spectroscopically up to that time (see their Tables 3 and 9).   In support of the {\it Chandra} SNR analysis, we have embarked on a program to obtain new optical spectroscopy for additional M33 SNR candidates using the MMT. 
%We have observed additional SNR candidates during several observing runs at the MMT observatory in Arizona. 
In 2006 September and 2007 September, we used the longslit CCD spectrograph known as the Blue Channel Spectrograph (BCS) to observe 14 SNR candidates (including some previously observed SNRs for consistency checks). Then, during several runs in autumn 2008, we used the multi-fiber system `Hectospec' \citep{fabricant05} to observe over 130 \chase\  sources, including 12 SNR candidates (again, some observed previously and four new objects).  Both the BCS data and the Hectospec data provided full coverage over the optical window at 3-5 \AA\ resolution, although we only report the red-end data here.  

Standard procedures were used to obtain the data and produce flux-calibrated spectra.  Our principal concern was to obtain accurate line ratios rather than absolute photometry; the optical imagery described in \S \ref{sec_opt_flux}  is better suited for the latter purpose.   In particular, longslit spectra often provide only partial coverage of extended objects, and fiber placement for Hectospec spectra can affect the absolute flux measurements for extended nebulae.

All of the lines were measured using the {\it splot} routine in IRAF. We compared results from the new data sets against previous observations where available, and generally found good agreement within expected errors.  Objects with multiple Hectospec observations sometimes showed  significant differences in the derived H$\alpha$ fluxes, likely resulting from slightly different fiber placements on small-diameter objects, but the derived ratios remained consistent.  This validates the utility of Hectospec observations for making the relevant measurements on these objects.  Indeed, the signal-to-noise  ratio of the Hectospec data were excellent, even on objects we considered to be `faint' in the preliminary analysis. We sorted through all of the available spectroscopic data and selected the best data in each case to derive the flux ratios tabulated in Table \ref{tab_sample}.  The source of the reported spectral information is shown in the `Spec. ref.' column of this table.  Of the 11 newly reported SNR spectra,  10 are confirmed as likely SNRs based on an elevated \sii:\HA\  ratio.  The remaining object, G98-83, had an MMT-BCS spectra exhibiting  a \sii:\HA\  ratio of only 0.28.  This is high relative to most \hii\ regions but below the value of 0.4 usually used to define a likely SNR.  As noted earlier, \cite{ghavamian05} did report the X-ray detection of G98-83, but our survey did not confirm that detection. 

\subsection{High Resolution Radio Imaging \label{sec_radio}}

The most detailed survey of SNRs in M33 at radio wavelengths was carried out by \cite{gordon99}.  Based on a series of 6 and 20 cm observations obtained at the VLA and the WSRT and originally discussed by \cite{duric93}, they constructed a catalog of 186 sources along the line of sight to M33.  They identified 53 of these with objects previously suggested as SNRs based on optical imagery and follow-up spectroscopy.  Their $7\arcsec$ beam was sufficient to resolve the larger remnants ($>30$~pc), but not the smaller ones; in addition, it left their study vulnerable to source confusion, especially in crowded regions near \hii\ complexes.  They did not propose any new SNRs based solely on their radio data.

In order to complement our high-resolution X-ray images, and to search for small-diameter, young remnants and Pulsar Wind Nebulae (PWNe), we undertook new radio observations of M33 using the VLA in its A configuration. Data were taken at both 6~cm and 20~cm in bandwidth-synthesis mode to minimize bandwidth smearing effects at large off-axis angles. At 6~cm we used a mosaic of eight fields to cover 621 arcmin$^2$, roughly co-extensive with the radio coverage of \cite{gordon99} and somewhat less than observed with \chandra; a single pointing at 20~cm covering 721 arcmin$^2$ was also obtained. These images have resolutions of $\sim 0.5^{\prime\prime}$ and $\sim 2^{\prime\prime}$ at 6 and 20~cm, respectively; the typical map noise levels were $15-25 \mu$Jy.  

%Details of the observations and a complete catalog of radio sources can be found in \cite{saul10}.

% We have examined the positions of the 50 reported detections by \cite{gordon99} which lie within the coverage of our new radio images. Twenty-three are detected, 
Of the 53 objects identified as radio-detected SNRs by \cite{gordon99}, 50 lie in the region covered by our new radio images.  Of these, 23 are detected,
with 21 having optical diameters smaller than $10 \arcsec $. This bias toward small-diameter objects results from the fact that our high-resolution data are not sensitive to angular scales that are more than a few times the diameter of the synthesized beam ($\sim 2 \arcsec $ at 20~cm). For another 19 of their detections, some of which were rather weak, we record no emission, although all of these remnants are larger than $8\arcsec $; the one additional small-diameter source from \cite{gordon99} is in a noisy part of our image, and no meaningful limit can be obtained from our data. For the seven remaining sources, we detect radio emission within $<10\arcsec$ of the optical SNR position, but not from the SNR itself; in four cases the emission is coincident with a nearby \hii\ region, and in the other three it is consistent with one or more point sources (most likely background objects). These objects are flagged in Table \ref{tab_sample}.

We also cross-correlated our radio catalog with the complete X-ray source list in order to search for new SNR candidates, particularly small-diameter young PWNe analogous to the Crab Nebula which might have been missed on the grounds of X-ray extent (too small), X-ray spectrum (too hard), or optical data (confused in H~II regions, lacking high [S~II]:H$\alpha$ ratios, etc.). Using a matching radius of $3^{\prime\prime}$ to account for extended sources in both wavelength regimes as well as the larger positional uncertainties of off-axis X-ray detections, we find 34 coincidences. Eleven of these are known SNRs. The majority of the remainder are point-like X-ray sources coincident with point-like radio sources (or, in one case, with a classical double radio source) and are likely to be background AGN. Nine objects, however, show some evidence of radio extent, although most are weak ($\sim 200 \mu$Jy) radio sources and are only marginally larger than the beam. We examined the optical emission at the location of each of these sources and found nothing to suggest any of them are normal shell-type remnants.  Better radio observations are required before any of these can be declared {\it bona fide} radio SNRs.

One source, however, is worthy of further consideration. FL281 
is distinguished by its hard X-ray spectrum approximated by a power law with a photon spectral index of 2.1.\footnote{Following the normal radio and X-ray conventions, we define the radio slope as
an energy index while the X-ray slope is defined in terms of a photon index $\alpha_{x}$. Thus,
$f_\nu\propto \nu^{-\alpha_{radio}}$ while 
$f_\nu\propto \nu^{-\alpha_{x}+1}$.}
It is coincident with a faint H$\alpha$ emission region.  The radio source is just resolved at 6~cm,  with a FWHM of $0.5 \pm 0.1\arcsec$ or $\sim 2$~pc; it also has a flat radio spectrum, with $\alpha_{radio}=0.2$. The X-ray and radio characteristics match those of a young PWN. The field of the candidate object is shown in Fig.\  \ref{fig_pwn}. We discuss this object further in \S 9.

\section{Multiwavelength Properties of the Supernova Remnant Sample \label{sec_compare}}

An obvious question is whether the X-ray properties of SNRs correlate with their observed properties at other wavelengths.  Such comparisons are largely not possible with Galactic samples because of the effects of absorption in the Galactic plane and because there are significant uncertainties in the distances to most Galactic SNRs.  However, all the SNRs in M33 are at essentially the same distance, and the line-of-sight absorption to M33 is quite low.  There are various effects that limit the uniformity and utility of our sample, including spatial differences in X-ray exposure, confusion in crowded regions in the optical, somewhat limited coverage beyond the $D_{25}$ optical elliptical isophot (diameter $\sim\:32\arcmin$), and poor sampling of low spatial frequencies in our radio observations. Nonetheless, the data we have on M33 is likely the most homogenous data set that exists for SNRs today in the sense that effectively the entire galaxy has been covered to uniform depth in each wavelength band and significant numbers of SNRs have been detected in each band. If correlations do exist then one would expect them to appear in the M33 sample.  We have carried out a number of such comparisons and tests to see whether the optical or radio properties of SNRs are good predictors of X-ray emission and vice versa.  In all cases, we have used both the \cite{gordon98} list of SNRs and our expanded list.  The Gordon et al.\ list has the advantage that it was constructed without knowledge of the X-ray properties, while our new list is somewhat larger and presumably more complete, although it contains some objects added as a consequence of their X-ray selection.

% Calculated from the same routine that generates the figure on 090812
% For objects above 1.000000e+35, the median (average ha surb was 2.841337e-16 (1.211364e-15) 
% For objects below 1.000000e+35, the median (average ha surb was 1.635886e-16 (2.762568e-16) 
A comparison of L$_X$ and \HA\  surface brightness is presented in Fig.\ \ref{fig_ha}.  Most of the SNRs in our sample were identified optically, and optical searches in M33 (and other galaxies of comparable distance) are largely surface-brightness-limited, except for smallest diameter objects.  On the other hand, except for the largest ($>$100 pc diameter) objects, X-ray detections of SNRs in the \chase\ survey are primarily limited by flux, because the background count rates are so low.  In some sense then, the comparison of L$_X$ to \HA\  surface brightness is the cleanest from an observational perspective. Fig.\ \ref{fig_ha} shows that the most luminous X-ray objects tend to be the SNRs with the highest \HA\  surface brightness.  These are the objects that are interacting with dense media and hence are the ones evolving on the shortest time scales.  The median (average) \HA\  surface brightness is \EXPU{2.8 \: (15)}{-16}{ergs~cm^{-2}arcsec^{-2}} for SNRs with X-ray luminosities brighter than  \POW{35}{\LUM} and \EXPU{1.6 \: (2.8)}{-16}{ergs~cm^{-2}arcsec^{-2}}  for  SNRs fainter than \POW{35}{\LUM}. Brighter X-ray SNRs tend to have higher \HA\ surface brightness, but the scatter is quite large.

The corresponding comparison of L$_X$ to \HA\  luminosity is presented in Fig. \ref{fig_ha_xray}. Again, the most luminous X-ray SNRs tend to be the SNRs with the highest \HA\   luminosities. The X-ray luminosities of the SNRs in M33 are nearly always less than the optical luminosities as measured by \HA, and in many cases much less.   It is worth remembering in this regard that most of the radiant energy from the SNRs is likely to be in lines in the UV or IR. The large amount of variation in the X-ray-to-optical luminosity ratio is not surprising.  The optical lines originate in relatively dense recombining gas with temperatures of 10,000-20,000 K and short cooling time scales, while the X-ray emission originates in tenuous material with temperatures of a million degrees or more and relatively long cooling time scales.  In Galactic SNRs, where the structure of a SNR is clearly resolvable at both X-ray and optical wavelengths, there is seldom a good detailed correlation between the X-ray and optical emission.

The \sii\ $\lambda\lambda$ 6717,6731 ratio is density-sensitive. Since one anticipates rough pressure equilibrium between the X-ray gas and the gas in the recombination zone where \sii\ is generated, one might anticipate a relatively strong correlation between the \sii\  ratio and X-ray emission, either with $L_x$, or possibly with $\sqrt{L_x/r^3}$.  The available \sii\  line ratios are plotted against remnant diameter in the left panel of Fig. \ref{fig_s2}. Smaller diameter objects are more likely to show departures from the low-density limit than are large-diameter ones.   High density is a good predictor of X-ray detectability;  of the 17 objects with \sii\ line ratios measured to be less than 1.3 (corresponding to a density higher than about 140 cm$^{3}$ \citep{cai93}), 16 were detected in our survey. Furthermore, as is shown in the right panel, in most cases the higher the density in the \sii\ zone the higher the X-ray luminosity. The median (average) line ratio for objects with $L_x > $\POW{35}{\LUM} is 1.31 (1.25) whereas for objects with lower values (including the objects that were not detected), the ratio is 1.42 (1.39).   Once again, however, there are exceptions: G98-21, the highest luminosity object in our sample, has a  \sii\ $\lambda\lambda$ 6717,6731 ratio of 1.39, not far from the low-density limit.

% Updated 100119
While the radio survey of \cite{gordon99} appears to have some deficiencies associated with lack of sufficient spatial resolution as described in \S  \ref{sec_radio}, it remains the most complete survey of SNRs in M33 at radio wavelengths.   The faintest sources they detected have flux densities of 0.2 mJy, low enough to have detected some, but not the majority of, Galactic SNRs at the distance of M33.  For example, RCW86, Puppis A, PKS1209-52 and all of the historical SNe except for SN1006 would have been detected, but the Cygnus Loop would have been missed. Of the 53 objects they identified as radio SNRs based on positional coincidence with an optically-identified SNRs, all but three -- G98-01, G98-03, G98-43 -- are in the regions covered by our X-ray survey; another seven are misidentifications arising from source confusion as noted in \S \ref{sec_radio}, none of which are X-ray detected. Of the (43) real radio-optical matches in our survey region, 29 (26) or 67\% (60\%) were detected at 2 (3) $\sigma$ in our X-ray survey.  This compares to a 64\% (47\%) detection rate of X-ray-counterparts by {\it Chandra} of the sources in the sample of optically-identified SNRs which was available to \cite{gordon99}. One might expect that these radio SNRs would have been preferentially detected in our {\it Chandra} X-ray survey, but the effect is fairly small.  There are  52 objects in our sample derived from the \cite{gordon98} list which are not known (now) to be radio sources\footnote{Our expanded \cite{gordon98} list  consists of 101 objects, 6 of which were not in the \chandra\ fields, and 43 of which have radio emission based on the discussion in \S \ref{sec_radio} leaving 52.}   and consequently whose identification as a SNR excluding the \chase\ observations is based solely on their optical properties.  Of these, 32 (19) or 62\% (37\%) were detected in X-rays. 

Moreover, the radio flux as measured by \cite{gordon99} is not a good predictor of X-ray detectability; the median radio flux for the detected X-ray SNRs (0.6 mJy) is actually lower than the median for the sources that were not detected in X-rays (0.8 mJy). This somewhat surprising finding is due in part to problems associated with identification of the radio sources with SNRs, as was discussed in \S  \ref{sec_radio};  G98-83 for example is bright in \cite{gordon99}, but really is a nearby \hii\ region based on our new radio data.  On the other hand, X-ray brightness is a good predictor of radio detection; the brightest seven X-ray SNRs are all detected by \cite{gordon99}.   A deeper, radio survey with complete $uv$ coverage and scaled-array spectral indices is needed to understand the relationship between the radio and X-ray properties of M33 SNRs.

Our overall conclusion from all of these comparisons of observables in different wavelength bands is that correlations are weak.  It is generally the case that extreme objects are extreme at all wavelengths; if one considers average or median values for the brightest X-ray objects, these tend to have the highest optical surface brightness, the most significant departures from the low-density limit as measured by the \sii\ line ratio, and are the radio brightest.  However, the range of variation is large.  
The inventory of X-ray gas and gas at 10,000 K where the optical emission arises is simply very different in different SNRs.

\section{What Factors Influence X-ray Detectability of SNRs? \label{sec_real_snrs}}

% Update 091230
Of the 131 objects in our sample observed by \chandra, we have detected more than half (62\%).  At least for the 69 X-ray detected objects that were selected initially based on a large \sii:\HA\  ratio and the absence of obvious young star clusters, their detections provide strong support for their identification as SNRs.  
For the 13 X-ray-selected objects,  X-ray detection is not supporting evidence, of course; that comes from the fact that these objects have optical characteristics expected from SNRs.  On the other hand, 49 objects were not detected, including 34 \cite{gordon98} SNR candidates and 15  new candidates we identified from our search of the LGGS and Schmidt plates.  The fraction of undetected objects is higher for objects with larger diameters, but some small-diameter objects were missed as well. What factor(s) determine X-ray detectability?

On an observational level, the answer to this question is simple: many of the objects were detected fairly near our sensitivity limit, and it is clear from Fig. \ref{fig_lum} that SNRs with similar diameters have widely differing luminosities;  had our sensitivity limit been two or three times lower, it is quite likely that more of the objects would have been detected. In addition, while the survey was fairly uniform, objects were observed at varying off-axis angles, as shown in Table \ref{table_results}, and with differing exposure times; a few objects were likely missed because they fell below our sensitivity limit at their positions in the survey fields.  A secondary concern, a possibility we return to in \S \ref{sec_superbubble}, is that a small fraction of the objects may have been misclassified as SNRs.

\subsection{The Importance of Local Gas Density}

On a physical level, the primary answer must surely be that SNe explode in a variety of environments, and that the density of the interstellar medium profoundly affects the appearance of the SNR that results.  Although detailed modeling is required to produce accurate numbers, the basic effects can be seen from the following simple argument:   Consider a ``typical'' SNR in our sample, one with an X-ray luminosity of \POW{35}{\LUM} and a diameter of 30 pc (near the peak of the distribution shown in Fig. \ref{fig_dia}).   The (absorbed) 0.35-2 keV volume emissivity of a plasma in ionization equilibrium with abundances of 0.6 solar, appropriate to M33, peaks at a value of about \EXPU{1.3}{-23}{ergs~cm^{3}s^{-1}} near a temperature $kT\approx 0.7\;$ keV.  The emissivity decreases gradually at higher temperatures, falling by a factor of 2 at about 2 keV.  At lower temperatures it falls by a factor of 2 at 0.2 keV, and plummets rapidly below that. Hence to first order the 0.35-2 keV X-ray luminosity scales as the volume emission measure, until the post-shock temperature drops below 0.2 keV. 

If the typical SNR with $L_x = $\POW{35}{\LUM} is radiating near the peak emissivity, then the volume emission measure for the SNR would be $\int n_e n_H dV \sim$ \EXPU{7.7}{57}{cm^{-3}}.   If we assume a filling factor of 0.25, consistent with compression behind a strong shock, then the emission measure and peak emissivity values above imply a density of 0.27 cm$^{-3}$ for the X-ray emitting plasma, and a rather low value of 0.07 cm$^{-3}$ for the ambient ISM.  The swept-up mass within the 30 pc diameter shell would be $33\,$\MSOL.
In the absence of emission from the ejecta, this SNR would not have have been detected in our survey at a diameter $D \lesssim 15$ pc, since its X-ray luminosity, which scales roughly as $D^3$, would have been \EXPU{\sim1}{34}{\LUM}.   Once it becomes detectable, the X-ray luminosity of our typical remnant would continue to increase with time as the diameter grows, until the shock speed \citep[which for a strong shock is given by $v_s = 900 \sqrt{kT({\rm keV})} \VEL$; see e.g.,][]{hamilton83}  falls below $\sim 400\VEL$.  For a SNR in the Sedov phase, the swept-up mass is $M=83\, E_{51}\,[kT({\rm keV})]^{-1}  \MSOL $, where $E_{51}$ is the SN explosion energy in units of \POW{51}{ergs}.  The luminosity of the SNR would then peak when the swept-up mass was about  400\MSOL, at which point our ``typical' SNR would have a diameter of 70 pc and a luminosity of \EXPU{1.2}{36}{\LUM};  subsequently it would fade from view as the post-shock gas cools below 0.2 keV.    

By this argument, the maximum size at which a SNR would exceed our X-ray detection threshold scales as $n^{-1/3}$.  A SNR, identical in all respects except expanding into a higher density environment of 0.5 cm$^{-3}$ could be observed to a maximum diameter of only $\sim 36\,$pc\@.  Expanding into a denser medium does have an advantage, however:  given the relative constancy of the emissivity,  the luminosity  scales with the emission measure, or  $n^2$, until $v_s$\ falls below 400$\VEL$.  Our SNR expanding into a medium with a density of 0.5 cm$^3$ would have appeared with L$_X=$ \EXPU{2}{34}{\LUM} at a diameter of about 5 pc, would have an L$_X$ of  \EXPU{5}{36} at 30 pc, some 50 times that of our ``typical'' SNR at the same size, and it would reach a peak luminosity of \EXPU{9}{36}{\LUM} before starting to fade.   

Given these arguments, it is quite clear that we should not have expected to detect all the M33 SNRs in our \chandra\ survey.  Objects could fall below the detection threshold because they were either too small to be detected, even if they were expanding into fairly dense media, or because they were too large and could have cooled to a point where the X-ray flux in the 0.35-2 keV band (seen through even just the Galactic $N_H$ of \EXPU{6}{20}{cm^{-2}})  was too low to be detected.  For the large diameter objects, the shock velocity would still be high enough to produce secondary shocks in denser media that would result in optical emission with high \sii:\HA\  ratios.

\subsection{Do Optical Morphology and Environment Influence X-ray Detectability? \label{sec_opt_morph}}

It is clear from an inspection of the optical and X-ray images presented in Appendix \ref{sec_notes} that the objects classified as SNRs on the basis of enhanced \sii:\HA\  are a very heterogeneous set of objects, expanding into a variety of environments.  Some are isolated objects; others lie on or within \hii\ regions.  Some show up at optical and X-ray wavelengths as classical shells, while others have poorly defined morphologies.   We  have therefore attempted to see whether any trends might become apparent if we group SNRs by their optical morphology and/or their general environment.
 
As discussed earlier, emission nebulae in nearby galaxies have historically been characterized as SNRs if the ratio of \sii:\HA\ emission exceeds 0.4.  In more distant galaxies where SNRs are typically unresolved, the morphologies of SNR candidates are largely unknown.  But in M33 the SNRs are resolved, so it is possible, in principle, to define additional criteria to help determine whether an object is really a SNR or not.  In the absence of some independent way of confirming an emission nebula as a SNR, this exercise would have been moot.   But since we now have X-ray detections of about half the SNRs in the optical sample, we have attempted to characterize the objects according to their optical morphology and the environment in which they are found, to see whether objects in some categories are preferentially detected in X-rays.

We classified objects morphologically as follows: $A$, well-defined, nearly complete shells;  $B$, partial shells, whose size can be estimated from the curvature of the visible portion of the shell; and $C$, amorphous or poorly defined objects, sometimes a single filament. As a pragmatic distinction, we defined a small $A^{\prime}$ class, as small (and typically bright) objects that may not show a shell structure but that are nevertheless clearly defined.  In terms of environment, we classified objects as (1) isolated from other nebulosity, (2) within nebulosity, but easily separable from the potential contamination, and (3) confused or difficult to isolate from the surrounding nebulosity.  Our motivation was to explore whether a combination of physical effects and observational biases might lead to differences in the X-ray detection probability for the objects in various morphological and environmental groups.  

Two of our team (WPB and PFW) independently and systematically examined all of the SNR candidates in the sample using the continuum-subtracted versions of the LGGS images \citep{massey06} and classified them according to the two criteria described above.  Following their independent efforts, they compared results.  In most cases they were in agreement, and in the cases where they differed ($\sim 20\%$), the difference was by no more than one sub-grouping in either criterion.  These differences were resolved into a single designation for each object, and these consensus classifications are given in Table \ref{tab_sample}. Comparing the classifications in this Table with  the Figures in Appendix A should provide the reader with a sense of the classification scheme.  

% Updated the numbers 090812 ksl
A summary of the numbers of SNRs that fall into each of the categories is shown in Table \ref{tab_morph}.  Since the detection probabilities for large-diameter objects is lower than that for small-diameter ones, the table presents separately the statistics for objects with diameters greater than and less than 50~pc, as well as the totals.  As is apparent from the table, objects that appear optically as well-formed shells away from confusing nebulosity (class $A1$) constitute 43\% of the entire sample; they are more prominent in the large-diameter subsample (49\%) than in the small-diameter one (38\%).  By contrast, objects of class $C3$, amorphous shapes in confused regions, represent a small fraction (5\%) of the total, and contribute similarly to both the large and small diameter subsamples. 

% paragraph break added by WPB 

It is quite likely that both observational selection effects and physical effects contribute to the distribution of morphologies that are seen.  For example, when one inspects images to look for SNRs, it is easier to focus on isolated shell-like objects that are not confused; furthermore, these can be observed to fainter surface brightness levels.   One likely physical effect is that objects that explode in regions with dense material (and hence appear more confused optically) evolve rapidly compared to objects expanding into the more rarefied interstellar medium; this would reduce the number that would be detected in confused regions if all SNRs could be observed to a fixed age.

Table 5 also summarizes the absolute numbers and the fractional detection rate for objects in our list of SNRs and SNR candidates.  All of the four morphology class $A^{\prime}$\ objects were detected; these have been grouped with class $A$.  As was noted, a higher fraction (74\%) of the objects with diameters less than 50 pc were detected in X-rays compared to the objects with diameters greater than 50 pc (45\%).  This is consistent with the argument made above that the shock speed is more likely to be below that needed to produce X-rays in the large diameter objects, as well the concern that larger diameter SNRs are hard to distinguish from superbubbles or diffuse ionized gas.   Fractionally fewer (40\%) of the amorphous (class $C$)  objects are detected than the better defined full (class $A$) or partial (class $B$) shells (62\% and 84\%. respectively).  This may indicate that some of the amorphous objects identified as candidate SNRs based primarily on their \sii:\HA\ ratio are really something else.  On the other hand, environment does not seem to be a strong discriminator of X-ray detectability.  The percentage of X-ray detections is almost identical for each of the three environment types. One might have guessed that the possibility of mis-identifying a SNR in a confused region was greater than in regions where there was no other confusing emission.  Instead, based on the detections from \chase\ it seems more likely that while it may be harder to pick out SNRs in confused regions, the ones that are identified are just as likely to have sufficient X-ray luminosity to be detected.

\subsection{Supernova Remnants to Superbubbles and Diffuse Ionized Gas \label{sec_superbubble}}

The prototypical SNR is understood to be a nebula produced by a single SN expanding into a medium that is reasonably uniform over the region being encountered by the SN shock expanding with velocities of 100 $\VEL$ or more.  If the medium is not uniform, given that the wind from the progenitor will in many cases have evacuated a region around the progenitor, then at least  the effects of the SN on the medium should dominate everything else.  Most core-collapse SNe occur in regions of active star formation where the environment has been altered by stellar winds of other stars and/or earlier SNe \citep{mccray79}.  If there are enough stars and long enough time, our prototype changes to that of a superbubble in which the nebula expanding at speeds of 20-30 $\VEL$ is best-described in terms of the collective effects from large numbers of stars \citep{maclow88}.  

Observationally, the distinction between a SNR and a superbubble is not so simple, and the boundary between them is fuzzy. The standard optical criterion, the one we used in establishing our list of candidate SNRs, is a \sii:\HA\  line ratio greater than 0.4.  Although there are counterexamples \citep[see, e.g.,][]{oey02}, most superbubbles  do not have high enough line ratios to qualify as SNRs by this criterion.  The ionization balance in most superbubbles is controlled by photoionization from the stars that populate the region.  The expansion velocities are too low to create the radiative shocks that produce large \sii:\HA\ ratios.  This is true despite the fact that superbubbles are known to exist with luminosities well in excess of our luminosity limit of \EXPU{2}{34}{\LUM}.  However, the \sii:\HA\  ratio really discriminates between shocks and photoionized gas.  The most likely source of shocks along the edge of a superbubble is a SN, especially if the SN occurs toward one side of the bubble \citep{chu90}.   Such a SN would create a ``SNR'' (by our optical definition) on one side of the superbubble.  

As the discussion in the previous section has indicated,  SNRs and the types of objects classified as SNRs are an observationally heterogenous set and the amount of supporting information to classify objects is limited.   The morphology class A and environment class 1 objects are very close to the prototypical SNR described above, but the departures become more and more dramatic as one progresses towards morphology class C and environment class 3.  

Given these trends, one has to ask whether there is a substantial number of objects in our sample that should be reclassified as superbubbles rather than as SNRs.   On the whole, the answer to this question is clearly no: high \sii:\HA\  ratios are not common in superbubbles.  Furthermore, the strong concentration of O-stars in a small superbubble should have been apparent in the inspection of the images.  On the other hand, there are some objects for which a strong suspicion is raised. The main candidates for reclassification are large-diameter ($>$ 80 pc) objects, whether isolated or seen as large loops exhibiting elevated \sii:\HA, and large linear structures on the edges of \hii\ regions, where it is difficult to determine the object's size.  It is harder to identify the young stars powering the superbubble in these larger objects, and in any event for older superbubbles the brightest stars may have already destroyed themselves as SNe.  Furthermore, despite the fact that there is no reason an isolated SNR should stop expanding, we do not empricially know of any such objects in the Galaxy or in the Magellanic Clouds that are larger than about 80 - 90 pc. 

There is also a second possibility that could apply to large low-surface-brightness emission regions in M33.  For higher surface brightnesses, there is a fairly well-defined gap between the \sii:\HA\ ratios of objects known to be SNRs and objects known to be \hii\ regions \citep{levenson95}.  In bright \hii\ regions, an abundance of ionizing photons is interacting with the surrounding interstellar medium, and nearly all of the S is ionized to S$^{++}$.  However,  as the density and surface brightness of photoionized regions declines, the ratio of \sii:\HA\ rises.  In NGC~7793, for example, \cite{blair97}  found the \sii:\HA\ ratio regularly exceeds 0.5 for \hii\ regions\footnote{Several such objects were observed serendipitously in longslit observations of SNRs .} with surface brightnesses of less than \POW{-15}{ergs~cm^{-2} s^{-1} arcsec^{-2}}.   Our sample in M33 extends to a surface brightness limit of  $\sim 4\times 10^{-17}\FLUXARCSEC$\ in uncrowded areas, and  to $\sim 10^{-16}~\FLUXARCSEC$\ more generally --- well below the limit observed to cause confusion in NGC~7793.   It is also well-known that for photoionized regions, ionization  tends to drop with increasing distance from the ionizing stars, and the [S~II] lines increase in relative strength with decreasing ionization.  Carrying this to the extreme, the diffuse ionized gas (DIG) in nearby galaxies is known to be enhanced in \sii:\HA, sometimes to the point that it even exceeds the 0.4 dividing line in this ratio \cite[e.g.,][]{blair97, tuellmann00}.  Thus, a relatively bright filament (for the DIG)  with an enhanced ratio could be picked up as a SNR candidate.  Such regions ionized by the general UV background could in principle have been selected as SNR candidates in our sample, although it would be difficult to understand how any could have associated X-ray emission.

%There is also a possibility that some of the objects in our sample are not SNRs but related objects, since we have accepted essentially all objects, including some objects of quite large diameter, into the list of SNR candidates as long as they had evidence for elevated \sii:\HA\  ratios.  
%The primary consideration for this assessment is that the \sii:\HA\ criterion 
%is not an infallible indication of SNR activity, for two reasons. 1) SNe are not the only phenomena 
%that can produce shocks.  Stellar wind shocks, either from an individual massive star or from young 
%associations can in some instances raise the observed \sii:\HA\  even in the presence of 
%(predominantly) photoionized gas.  (A classic example is a superbubble such as N70 in the Large 
%Magellanic Cloud, see REF.)  
%%2) It is also well known that for photoionized regions, ionization  tends to drop with increasing distance from the ionizing stars, and the [S~II] lines increase in strength with decreasing ionization.  Carrying this to the extreme, the diffuse ionized gas (DIG) in nearby galaxies is known to be enhanced in \sii:\HA, sometimes to the point that it even exceeds the 0.4 dividing line in the ratio (e.g. Blair \& Long 1997; other REFs).  Thus, a relatively bright (for the DIG) filament with enhanced ratio could be picked up as a SNR candidate.

\subsection{A Final Evaluation of our Sample \label{sec_final}}

Based on the considerations discussed above, we have systematically reviewed the SNR candidate list in Table \ref{tab_sample}\label{candidates}, using all available data to identify the subset of objects that had been admitted as candidate SNRs based on our fairly inclusive criteria, but that seem in some ways peculiar or unlikely to represent the remnants of single SN explosions.  Our review has identified 26 of these objects in both categories outlined above. 
%{\bf Good statement, but Table 6 does not tell us which we think is which or why -- Also suggest Table 6 have sizes ksl}  
A moderately large shell or diffuse emission region with one or more stars located centrally within the shell at the very least raises suspicion that stellar winds (possibly in conjunction with one or more SNe) may be partially responsible for creating the observed nebula.  A variation on this theme are the large arcs of enhanced \sii:\HA\ emission protruding from bright \hii\ regions which could be `blow-outs' involving stellar winds and/or potentially SNR shocks. Likewise, isolated filaments of enhanced \sii:\HA\  in diffuse regions of emission (without clearly associated shell-like morphology) could be examples of the ``bright DIG'' phenomenon.  Table \ref{table_nonsnr}  provides a summary of this assessment along with an indication of the reason for questioning the veracity of the SNR identification.  These 26 objects constitute 18\% of the sample.  Further details about all the objects, including those in Table \ref{table_nonsnr}, are presented in Appendix \ref{sec_notes}.     

We emphasize that we found no reason to question the identification of the majority of objects in our SNR list.  The bulk of the 137 objects in Table \ref{tab_sample} are certainly SNRs; the fact that 82 of them were detected in X-rays strongly supports the SNR identification for these objects.  
It is also quite possible that some of the objects we have listed as questionable will ultimately turn out to be SNRs; several were detected in X-rays.  Given current capabilities across the electromagnetic spectrum there simply is no way observationally to separate all of the borderline cases cleanly.  

A Venn diagram providing a snapshot of the current status of detections of the M33 SNRs sample is shown in Fig.\ \ref{fig_venn}.  There are 29 objects that have been detected as optical emission nebulae with elevated \sii\ compared to \HA, as X-ray sources, and also as radio sources. The 8 X-ray only objects are associated with some optical emission, either \HA\ or \oiii, but do not appear to show elevated \sii. Similar diagrams are provided by \cite{pannuti07}  for the galaxies discussed briefly in \S \ref{local_group}.  All of these diagrams are strongly affected by the sensitivities of searches at various wavelengths.   In particular, for M33, while \cite{gordon99} did detect  a substantial number of the SNRs that had been identified optically at radio wavelengths, their survey sensitivity and angular resolution were not sufficient for them to identify a radio-only sample of SNRs.  Indeed our higher angular resolution radio observations of M33 showed that seven of the 53 objects that they thought were coincident with optically-identified SNRs were nearby H II regions or background objects. Except for the Galaxy and the Magellanic Clouds, optical searches remain the most effective way to identify SNRs.  The \sii:\HA\ ratio provides a fairly clean diagnostic and photons are relatively plentiful.  

The \chandra\ observations reported here represent  about the best one can hope to do with the current generation of X-ray instrumentation.   Of our total sample of 137 objects, 131 were within our survey region and 82 were detected at X-ray wavelengths. For eight of these, the primary reason for believing they are SNRs is that they are (a) extended and (b) soft X-ray sources.  

The next major advance in the M33 SNR sample is likely to occur with the advent of the EVLA.  This will be important not only for finding more PWNe, but also for finding and detecting additional normal shock-dominated remnants.  It should provide both the angular resolution to construct radio images for all of the known SNRs and the sensitivity and polarimetric fidelity to find the remnants that have been missed in optical surveys; i.e., the PWNe and young remnants without the radiative shocks needed to produce \sii.  At optical wavelengths, the concern as one goes fainter is confusion with diffuse ionized gas masquerading as a shock-heated material; at radio wavelengths it is background galaxies that happen to lie along the line of sight toward \HA\ emission.  Higher quality (or for about one-third of the objects, initial) 
optical spectroscopy would also be very helpful in clarifying which of the objects are bona fide SNRs and what their physical properties are.  Finally,  a new interference-filter optical imaging study of M33 would also be desirable, especially if the observations could be obtained with sub-arcsec seeing conditions; this would allow a more sensitive search for small diameter objects and more accurate measurements of the optical sizes.  The currently available data set at X-ray wavelengths represents close to the limit of what can be expected with the current generation of X-ray observatories, and it is thus unlikely that the X-ray sample in M33 will be increased significantly for some time to come.

\section{The Brightest SNRs \label{sec_bright}}

There are seven SNRs in M33 that stand out in terms of their X-ray brightness -- G98-21 (29 pc diameter) , G98-28 (22 pc), G98-29 (40 pc), G98-31 (17 pc), G98-35 (34 pc), G98-55 (24 pc) , and G98-73 (22 pc).  These remnants are sufficiently bright to allow extraction and fitting of their spectra, and also comparison of their X-ray, optical, and radio morphologies. 
Interestingly, all of these are intermediate-sized objects ($\sim$15-40 pc across).  In addition, most show evidence of significant interaction with surrounding material (e.g., strongly enhanced X-ray emission at localized spots on their boundaries).  There are sufficient counts in the X-ray spectra of these seven SNRs to allow a more detailed characterization of each remnant with more complex spectral models.  In the following, we first discuss the comparison of the X-ray and optical images of each remnant and then discuss results of
our X-ray spectral model fits.

\subsection{Multiwavelength Spatial Analyses}

In order to take advantage of \chandra's imaging quality, we performed 10 iterations of a Lucy-Richardson deconvolution on these objects using the IDL {\tt astron} package {\tt max\_likelihood}.  For each object, we selected a subset of pointings which were closest to on-axis, avoiding cases where the source fell on a detector chip gap.  The data for each object were merged and filtered to include only the 0.35--4 keV band.  For the deconvolution kernel, we constructed a merged PSF by merging ray traces performed at the appropriate off-axis angles, as determined using the CIAO tool {\tt dmcoords}.  The ray traces used {\tt SAOtrace}\footnote{http://jeeves/cxcoptics/Public/SAOTrace/Index} to trace the optics and {\tt MARX}\footnote{http://space.mit.edu/CXC/MARX} to project onto the detector plane.  

Figures \ref{fig_color_bright1} and \ref{fig_color_bright2} show four-panel X-ray/optical/radio comparisons for each of these SNRs.  The leftmost panel in each row is a color composite made from continuum-subtracted \HA\  (red),  \sii\ ( green) image, and  \oiii\ (blue) LGGS images, with superposed radio contours from our VLA observations of M33.  The remaining panels in each row are the deconvolved X-ray image, the raw X-ray image and the PSF for the observations of the fields used.  In addition, the corresponding atlas images of these objects in Appendix A show grayscale renderings (left to right) of the X-ray data, H$\alpha$, \sii, and V-band images.

G98-21 (or L10-025) is the  brightest X-ray SNR in M33, and a thorough study of this object has been published by \cite{gaetz07}.  This remnant is located on the outskirts of the giant \hii\ region NGC~592. As noted earlier, this object was suggested as a SNR by \cite{long81} using {\it Einstein}, and its identification as a SNR was firmly established when \cite{gordon93} showed it to be a  fairly bright, non-thermal radio source. 
 The {\it Chandra}  images, analyzed in detail by \cite{gaetz07}, showed the X-rays to be a nearly circular shell with a diameter of 20 pc (5\arcsec) that is brighter on  the eastern side (i.e., toward the center of NGC~592). 

The optical/X-ray comparison is shown in the top row of Figure \ref{fig_color_bright1} (see also Fig.~\ref{fig_atlas06}, third row).      While a bright star or tight cluster is seen in projection on the eastern limb of the optical shell, the \sii\ image clearly shows the portion of the object dominated by radiative shocks.   In this object, the morphology of the optical shell differs from that of the X-ray SNR, with an arc of optical emission on the south side extending beyond the western edge of the X-ray shell.  The brightest portion of the X-ray ellipse actually lies somewhat east of the most conspicuous optical emission.   This morphology of G98-21 is unusual; while it is common to see differences in the relative surface brightness of features at optical and X-ray wavelengths,  rarely is the overall morphology  so qualitatively different.   If the bright eastern X-ray emission is due to enhanced density in the direction of NGC~592, then we would expect enhanced optical emission there as well, contrary to what Figure \ref{fig_color_bright1} shows.  The \oiii\ emission of the SNR (image not shown) follows a similar pattern to that seen in \sii.  \cite{gaetz07} suggest that photoionization by NGC~592 prevents recombination in the eastern limb of the SNR.  

G98-28 (or L10-036) is shown in the second row of Figure \ref{fig_color_bright1} (see also Fig.~\ref{fig_atlas09}, second row).  This SNR is located away from any obvious local optical nebulosity, although it is about 1\arcmin\ away from the northwestern edge of the giant H~II region NGC~595.  The optical object appears to be a bright partial shell, open to the east-southeast with an enhancement at its southern end.  The X-ray image shows little evidence of a shell, but rather is centered on the very high-surface-brightness optical shell.  The morphology of the object is nearly identical in \HA, \sii, and \oiii.   We inspected images from the $Spitzer$ archive of this region and these show that the SNR is also detected at 24 $\mu$m, indicating the likely presence of warm dust in the radiative shocks.   
Optical echelle spectra  \citep{blair88} show bulk motions of 345 $\VEL$, confirming high velocities.  The appearance of G98-28 in the V-band continuum image is likely due to the transmission of bright [O~III] emission lines through the broadband filter.  The reason for the dominance of emission from the western side of the shell is not clear from these data.

G98-29 (or L10-037), shown in the third row of  Fig.\ \ref{fig_color_bright1} (see also Fig.~\ref{fig_atlas09}, third row) is a large limb-brightened X-ray SNR about 32 pc across. The original and deconvolved images are very similar in this case.  The SNR was detected by \cite{gordon99} at radio wavelengths, but is not contained in our higher resolution radio maps, which is most likely due to the fact that our observations were less sensitive to objects of this size.  Optically, the SNR is relatively isolated from other nebulosity.  The optical emission has a knotty appearance, with only a hint of emission in the NW.  

G98-31 (or L10-039), shown in the bottom row of Fig.\ \ref{fig_color_bright1} (see also Fig.~\ref{fig_atlas10}, top row), is quite small at X-ray wavelengths and the emission appears centrally peaked, with only a hint of shell-like structure extending to the north.  The X-ray emission corresponds to a bright core of optical emission seen in the figure and the high-surface-brightness component of this core is very similar.  The \sii\ line ratio is 0.8 for this object, indicating high density.  Optical echelle spectra \citep{blair88} show velocities of 330 $\VEL$ for this core.  The core also appears strong in [O~III], and it may be this emission being passed by the V-band filter that causes the extended emission in this image.  In a wider view, one can see that the object sits centrally in a larger \HA\  shell with  brighter knots on the east and north sides.

At a stretch that reveals faint structures, the optical emission from G98-31 displays an unusual morphology, with what appear to be two thin arcs or shells extending from the core to the north and east. While somewhat reminiscent of the rings around SN1987A, the physical scale of the thin arcs in G98-31 is nearly 60 times larger, extending some 30~pc.  Longslit spectra confirm that these arcs show elevated \sii:\HA\  ratios, but they do not show the high velocities seen in the core.  The arcs are present but less well-defined in the [O~III] image (not shown).  It is conceivable that the arcs may represent an older SNR located along the same line of sight.

G98-35 (or L10-045), shown in Fig. \ref{fig_color_bright2} top (see also Fig.~\ref{fig_atlas11}, third row), is another complicated object with a bizarre morphology.  A very bright central core of optical emission is surrounded by asymmetrical loops or arcs, one to the east-southeast and one to the northwest.  The eastern lobe is high in \sii\ while the northwestern lobe has lower, but still elevated \sii\ emission.  The [O~III] emission looks quite similar.  
{\it HST} observations with the original WF/PC camera by \cite{blair93} showed the bright core to be a broken shell of very high surface-brightness emission.  This can be seen in the \sii\ panel of the figure, which is scaled to show the highest surface-brightness regions.  Echelle data by \cite{blair88} show bulk motions of 313 km $\rm s^{-1}$ for the core, but the velocities are significantly lower in the portions of the loops measured by their east-west oriented slit.  

Most of the X-ray emission stems from the very bright optical core.  However, there is clearly emission extending beyond the core---especially toward the eastern lobe where faint X-ray emission peaks just inside the optical rim.  Without more information, we define the SNR to include the core and both lobes, which appear to form a coherent structure as seen in the \HA\ panel of the figure, but clearly the core itself is kinematically distinct.

G98-55 (or L10-071), shown in the middle panel of Fig.\ \ref{fig_color_bright2} (see also Fig.~\ref{fig_atlas17}, bottom row),  is another very high-surface-brightness optical SNR on the eastern edge of an \hii\ complex with bright condensations and extensive diffuse emission.  In \sii\ and \oiii, the SNR is distinct from the \hii\  region, and is dramatically brighter on its western side, toward the bulk of the \hii\ region.   The southern edge of a shell is also visible, and the complete shell can be seen with altered contrast settings. The X-ray emission shows a well-developed shell,  brightest in the southwest, where the optical and radio emission are also brightest.  Echelle spectra \citep{blair88} show bulk motions of 272 $\VEL$.  The smudge in the V-band continuum frame is likely due to \oiii\ emission transmitted through the bandpass.

G98-73 (or L10-096), shown in the bottom panel of Fig.\ \ref{fig_color_bright2} (see also Fig.~\ref{fig_atlas23}, bottom row),  is a small, diffuse shell-like SNR showing bright knots on the southeast and northwest limbs.  
%is a small, diffuse shell-like SNR showing bright knots on the northeast and southwest edges.  
A very bright, compact \hii\ region  $\sim6\arcsec$ to the southeast is apparently not related.  The \oiii\ emission (not shown here) reveals a complete shell, with the brighter knots less evident than in the other emission-line images.  Echelle spectra \citep{blair88} show bulk  motions of 185 $\VEL$.  The X-ray emission is barely resolved and appears to fill the shell.  
The compact \hii\ region is a 24 $\mu$m source; the SNR, however, is not.

\subsection{X-ray Spectral Analyses}

Abundance analysis can in principle yield clues to the SN type, while the derived temperature and ionization time scale can yield information on the evolutionary state of the remnant.  The extracted spectrum for each remnant is shown in Figure \ref{fig_pshock}.  Using {\tt XSPEC}, we first fit all of the spectra to simple plane-parallel shock models ({\tt pshock} --  \citep{borkowski2001} in {\tt XSPEC} with {\tt NEIVERS} set to the default of 1.1).  We fixed the Galactic absorption along the line of sight to be \EXPU{6}{20}{cm^{-2}}, using the {\tt tbabs} model \citep{wilms2000} in {\tt XSPEC} with the {\tt abund} parameter set to {\tt wilm} and the cross sections to {\tt vern} \citep{verner1996}.  We fixed the metallicity for absorption in M33 to be 0.5 times solar, but allowed the column in M33 to vary using the {\tt tbvarabs} model in {\tt XSPEC}.  We fit the shock temperatures, the overall metallicities, and the ionization ages for the shocks.  The results are given in Table  \ref{table_pshock}, and the model fits are compared to the data (with residuals plotted below) in Figure \ref{fig_pshock}. The simple plane-parallel shock models reproduce the spectra of G98-21, G98-28, G98-29, G98-55 and G98-73 fairly well.  The metal abundances are sub-solar as expected for M33, varying from 0.18 to 0.66.  The best-fit shock temperatures are of order 0.6 keV (except for G98-29 and G98-73 which have best-fit temperatures of 0.84~keV and 1.05~keV respectively), and \CHINU\ values, while exceeding unity per degree of freedom, are reasonable.  Thus, the fits for these objects are consistent with shocks propagating into M33 ISM material.

The {\tt pshock} fits to the spectra of G98-31, and to a lesser degree G98-35, are not as good.  It is evident from Figure \ref{fig_pshock} that the region around 1.4 keV, which contains the Mg~XI triplet, is fitted particularly poorly.  To improve the fit we considered the {\tt vpshock} model, a model similar to {\tt pshock} but in which individual abundances can vary. In our case, we allowed the abundances of the O, Ne, and Mg to be free, since these elements are expected to emerge from core-collapse SNe and may have a significant signal in our bandpass. We also allowed the abundance of Fe to vary, so that the abundances of O, Ne, and Mg could be compared to Fe. The results are given in Table  \ref{table_vpshock}, and the model fits are compared to the data (again with residuals plotted below) in Figure \ref{fig_vpshock}.  Clearly, the fit around the Mg~XI triplet region is significantly improved in these models, as evidenced by the improvement in the reduced $\chi^2$ from 3.13 to 1.57 for G98-31 and from 1.39 to 0.64 for G98-35.  

For G98-31, the variable abundance fit resulted in [O], [Ne],  [Mg] abundances of 0.21 (0.15-0.31 at 90\% confidence), 0.23 (0.17-0.34), and 0.61 (0.48-0.85), respectively, and an [Fe] abundance of  0.10 (0.09-0.19), where [Z] indicates the abundance of element Z relative to the solar abundance. The other metals were held to the value of  0.19 solar based on the ({\tt pshock}) model fits.  While the {\it absolute} abundances of O, Ne, and Mg are not especially high, the [O]/[Fe] and [Ne]/[Fe] ratios are $\sim2$ and the [Mg]/[Fe] ratio is $\sim6$; we can say with some confidence that O, Ne, and Mg are all enhanced {\it relative to} Fe.   Optically, G98-31 is dominated by swept-up material, so the enhanced abundances in the X-ray analysis indicates enrichment by a significant mass of O, Ne, and Mg from SN ejecta. This argues for a core-collapse explosion of a high-mass progenitor (see \cite{rauscher2002} and references therein). 

For G98-35, the variable abundance fit resulted in  [O], [Ne] and [Mg] abundances of 1.06 (0.32-4.00), 0.99 (0.51-2.64), and 0.93 (0.37-2.26), and an [Fe] abundance of  0.15 (0.07-0.41), while the other metals were held to the value of  0.18. The O, Ne, and Mg appear to be enhanced with respect to Fe ([O]/[Fe], [Ne]/[Fe], and [Mg]/[Fe] are all $\sim6$), and enhanced compared to the mean metallicity derived from the simpler {\tt pshock} model fits of 0.18, but the uncertainties are large.    We consider it likely that G98-35 also resulted from  the core collapse of a high-mass star, and that there is some evidence for this in the abundance pattern derived from the fits, but the case is not as strong as for G98-31.  Interestingly, as with G98-31, there is no clear evidence for peculiar abundances in the optical data for this object.

For both G98-31 and G98-35, we assert that the unusual spectra of these objects are significantly contaminated by ejecta and core-collapse SNe are required since the lower Z elements appear to be overabundant relative to Fe. Beyond that, it is very difficult to be more precise about the nature of the progenitors.  While the spectra for G98-31 and G98-35 are clearly different from those of the remaining objects in our bright sample, the signal-to-noise ratio in our spectra are fairly low, and it is likely that there is a substantial forward-shock contribution to the emission we observe.  Neither of these objects shows evidence of anomalous abundances in optical spectra, although the velocity widths of emission lines are significant (indicating relative youth). Furthermore, in \chandra-based studies of Galactic and Magellanic Cloud SNRs, individual ejecta features are typically isolated and analyzed  whereas our spectra of  G98-31 and G98-35 represent global averages with ejecta contaminated by overlying blast wave emission in each object.   This explains why the abundance ratios in G98-31 and G98-35 are not as extreme as expected from
pure-ejecta material  \citep[see, e.g.,][]{nomoto97}, but it does not help in trying to determine the mass of the progenitor.  The fact that we do not detect the K-shell emission from the higher Z elements, especially Fe, also produces considerable uncertainty.  It is possible that detailed model fits to specific SNR models could improve this picture but that is beyond the scope of this paper.

We can estimate the age, the initial ambient density, the explosion energy and the swept-up mass of these seven brightest M33 remnants if we make the somewhat over-simplified assumption that they are in the Sedov phase of evolution.  Although there is evidence from the X-ray and optical morphology (and in the case of G98-31 and G98-35, X-ray ejecta emission) that these remnants are not yet fully evolved to the Sedov phase, this  is still a useful exercise for a qualitative look at the properties of these seven objects.  For this analysis, we adopted the sizes of the remnants as derived from the X-ray images, the unabsorbed flux in the 0.35--5.0~keV band, and the temperature as obtained from the X-ray spectral fits. For an assumed distance of
817~kpc to M33 and an assumed emissivity of the thermal plasma models {\tt pshock} and {\tt vpshock} appropriate for the best fitted temperature, we can follow the formulation in \cite{cox1972} to derive the parameters of a remnant in the Sedov phase.  For G98-21, G98-28, G98-29, G98-55, and G98-73, we adopted the fitted parameters from the {\tt pshock} models  listed in Table~\ref{table_pshock};  and for G98-31 and G98-35, the fitted parameters from the  {\tt vpshock} models listed in Table~\ref{table_vpshock}. The inferred shock velocity is directly proportional to the square root of the fitted X-ray temperatures,\footnote{The $v_s \propto \sqrt{T}$ dependence assumes electron-ion equilibration, which may not have been fully achieved for these relatively young SNRs.}  and since all of the remnants except G98-29 and G98-73 have an inferred temperature of around 0.6~keV, the inferred shock velocities are all in a narrow range from 600 to 700 $\VEL$.  For G98-29 and G98-73, the X-ray temperatures of 0.84~keV and 1.05 keV give shock velocities of 800$\VEL$ and 900$\VEL$ respectively. 

The inferred ages are about 5,000~yr, with G98-31 the youngest at 3,300~yr and G98-29 the oldest at 9,700~yr. These are most likely overestimates of the true age, since assuming a Sedov model for a remnant which has not yet reached the Sedov phase leads to an overestimate in most cases.  The inferred initial ambient densities  are around a few per ${\rm cm^{-3}}$, ranging from $n_0 \sim 0.20\;{\rm cm^{-3}}$ for G98-29 to $n_0 \sim5\;{\rm cm^{-3}}$ for G98-31. The inferred explosion energies are all in the range 0.3 to  $0.55\times10^{51}\;{\rm ergs}$,  except for G98-21 and G98-29 with energies of $1.5\times10^{51}\;{\rm ergs}$ and $1.4\times10^{51}\;{\rm ergs}$, respectively.  Finally, only two of the remnants appear to have swept-up a mass $\gtrsim 100 M_\sun$: G98-21 and G98-29 have swept up an estimated $250 M_\sun$ and $162 M_\sun$.  As a group, these seven brightest remnants are rather homogeneous in terms of their inferred properties, with G98-21 and G98-29 exhibiting the greatest difference from the others. The picture that emerges is one of fairly young SNRs (ages around 5,000~yr) which went off in fairly dense regions of the ISM  ($n_0 \sim 0.2\; {\rm to}\; 5\: {\rm cm}^{-3}$).  These characteristics are consistent with the fact that these remnants are the most X-ray luminous in our sample. G98-21 and G98-29 may have been overenergetic explosions in a dense medium which has resulted in their high X-ray luminosities.

\section{Discussion \label{sec_discussion}}

\subsection{Comparison with Remnants of Historical Galactic Supernovae}

In our Galaxy, there have been at least five SNe in the last $\sim 1000$ years: SN~1006, SN~1054 (the Crab Nebula), SN~1572 (Tycho), SN~1604 (Kepler) and Cas A.\footnote{A sixth event, SN~1181, is often added to this list, and 3C58 has often been identified as its remnant.  However, recent kinematic data \citep{fesen08}, radio results \citep{bietenholz06}, and X-ray analyses (Gotthelf et al. 2007) indicate that 3C58 is much older.}  All are observed as luminous X-ray sources today.  Of these, the X-ray emission from the Crab Nebula is powered by a pulsar and shows a hard, power-law spectrum, while the remainder have soft X-ray spectra with emission components from a combination of thermal and synchrotron emission behind the forward and reverse shocks.  Would we have detected similar objects in M33?

The Crab Nebula is among the very brightest X-ray sources in the Galaxy in both X-ray and radio emission.  
The Crab's optical filaments  show high \sii:\HA\  ratios \citep{fesen82}; even though the Crab's emitting filaments are photoionized, the character of the photoionizing spectrum from the synchrotron nebula is such that an elevated \sii:\HA\  ratio results.   The Crab itself is an extreme object, and any similar object in M33 would certainly have attracted early notice and extensive investigation at all wavelengths.   We will discuss the possibility of less extreme pulsar wind nebulae in \S 9.3.

Three of the others---Cas A, Tycho, and Kepler---are relatively bright plasma-dominated X-ray sources, all with X-ray luminosities $>10^{36}{\LUM}$---bright enough to place them among the top ten X-ray-detected SNRs in M33.  However, all three of these are rather heavily absorbed,  
%($N_H >$\EXPU{0.5}{22}{cm^{-2}}), 
and all have X-ray spectra that require non-equilibrium ionization modeling of a plasma highly enriched by SN ejecta---considerations which require caution in extrapolating their spectra to lower energies and lower absorption appropriate to M33.  Furthermore, all three are small, with diameters $\lesssim 5\;{\rm pc}\approx 1.3\arcsec$, so their analogs in M33 would be at best marginally resolved with {\it Chandra} or in ground-based imaging.   
%We will consider each in more detail.
In order to assess the detectablity of Tycho, Kepler, and Cas A analogs by the ChASeM33 survey of M33, we fit existing  {\it Chandra}  spectra of these SNRs to nonequilibrium ionization thermal models.  Adjusting their ACIS-I count rates from their actual distances  (2.3 kpc, 5.0 kpc, and 3.4 kpc, respectively) to 817 kpc  resulted in rates (0.5--2 keV band) of \EXPU{4.2}{-4} \COUNTS, \EXPU{8.7}{-4} \COUNTS, and \EXPU{5.5}{-3} \COUNTS, respectively.  For the typical exposure of 400 ks across the area of M33 surveyed, we see that Cas A would produce over 2000 {\it Chandra} counts; Kepler 340, and Tycho 160.  
However, we should additionally correct for the significant column of ISM absorption through the Galaxy for these SNRs:  $N_{\rm H} \approx 1.4 \times 10^{22}$ cm$^{-2}$ to Cas A, and $N_{\rm H} \approx 6 \times 10^{21}$ cm$^{-2}$ to Kepler and Tycho, leading to  much higher predicted numbers of counts.  We now consider each of these in greater detail.   

After correcting for absorption, a Cas A analog would have yielded several thousand counts, even in a minimal 200 ks exposure.  With a spectrum dominated by emission lines at low energy, it would certainly have attracted attention based on its X-ray properties alone; indeed it might well have been identified prior to the launch of {\it Chandra}.
Cas A's optical spectrum is dominated by knots of ejecta \citep[see e.g.,][and references therein]{fesen01}.  In a survey based {\it purely} on optical imaging in \HA\ and \sii, it is not obvious whether or not a Cas A analog in M33 would have been discovered.  Many of the ejecta-dominated knots in Cas A have strong \sii\ emission, while \HA\ is concentrated in the spatially distinct and much fainter quasi-stationary flocculi, as well as  very faint extended \HA\ emission throughout the region surrounding Cas A\@.  If \sii\ was bright enough to attract notice, the \sii:\HA\ ratio would be quite high.  However, the strongest optical emission line from Cas A is \oiii, where we have measured $F_{5007} \approx 5 \times 10^{-12}\FLUX$  through a filter with FWHM = 41 \AA.  (This was measured from the Burrell Schmidt in the same October 1996 run during which we observed M33.)   For a distance of 3.4 kpc \citep{reed95} and foreground absorption $A_V = 6.2$ mag \citep{eriksen09}, this gives a luminosity of $\EXPU{2}{36}{\LUM}$.\footnote{Cas A has very broad lines, so the total \oiii\ flux and luminosity will be higher than these values.  Furthermore, the flux passed by the 55 \AA\ FWHM \oiii\ filter used for the LGGS would certainly have passed more flux than the 41 \AA\ one used in our Schmidt survey, so the values here are quite conservative.}  As noted in \S  \ref{sec_candidates}, objects that  attracted our attention through their soft X-ray and/or extended radio emission, were carefully examined  in {\it all} the LGGS emission-line bands, so a Cas A in M33 would almost certainly have been detected optically as well.
%Furthermore, the older ($\sim$ 3000 yr) galactic O-rich SNR G292.0+1.8 has an \oiii\ luminosity of $1.6 \times 10^{35}\LUM$ and would also probably have been detected in M33 \citep{winkler06}.
%it is somewhat unclear how it would be characterized in an imaging survey based on \HA\  and \sii.  Cas A would show up in \sii\ but not in \HA\ per se, except that Cas A is associated with some more generally extended \HA\ emission, so if the \sii was noticed the [S~II]:\HA\ ratio would be quite high.

Kepler, corrected to the absorption toward M33, would yield $\sim1400$ counts in a 400 ks exposure, with a  0.35--2.0 keV X-ray luminosity of $\EXPU{2.8}{36}\LUM$.    It too would have attracted our attention as a line-dominated soft X-ray source.  Optically,  
Kepler shows significantly elevated \sii:\HA\  emission, marking it as the only one of the four historical shell-type remnants that clearly satisfies the normal criteria for an optical SNR \citep{leibowitz83, blair91}.  It too would easily have been detected in our M33 survey.

Corrected for absorption, Tycho ($L_X\approx \EXPU{1.3}{36}\LUM$) would yield $\sim 500$ counts in 400 ks, certainly enough to attract our attention, and with its soft, line-dominated spectrum would have been flagged as a potential SNR\@.  Optically, 
Tycho (and SN~1006 as well) have spectra that are atypical for SNRs \citep{kirshner87}, showing faint Balmer lines instead of the bright recombination spectra seen in most other SNRs.  The emission arises not from a dense plasma cooling and recombining behind a relatively slow shock, but from a plasma that is being rapidly ionized behind a fast shock.  As a result, such objects would not have been identified in an optical search based on elevated \sii:\HA\  ratios.   But Tycho, like all the other historical remnants, would have been resolved in our 6 cm radio observations, and hence vetted according to the procedures outlined in \S 5.3.  Therefore, we conclude that Tycho, along with Cas A and Kepler, would likely have been detected from our X-ray observations alone, and almost certainly have been recognized from the combination of X-ray, optical and radio data.  We cannot, of course, rule out the possibility that a direct analog of Tycho's SNR could be hidden outside the region we surveyed well with \chandra.

The one remaining remnant among the historical Galactic SNRs, SN~1006, is much fainter, $L_X \approx$  \EXPU{2}{34}{\LUM}---just at our flux threshold, and it is dominated by emission from its bright synchrotron limbs to give a hard power-law integrated X-ray spectrum.  As noted above, its optical emission consists exclusively of Balmer filaments, and even those are faint, so it would not have been picked out in our optical search.   And finally, its radio flux would have fallen a factor of 4 below the sensitivity limit of our 6 cm radio survey.   A SN~1006 analog in M33 would probably have been missing from our remnant catalog, but ones like the other four remnants of Galactic SNe from the past millennium should  have been seen and included.

\subsection{Comparison with the Local Group Remnant Population \label{local_group}}

The samples of SNRs most directly comparable to those in M33 are those in other Local Group galaxies, namely the Large and Small Magellanic Clouds and M31.  In their {\it ROSAT}-based survey of SNRs in the LMC, \cite{williams99} reported detections of 31 SNRs in the LMC using the PSPC and the HRI: this sample includes the extremely young object SN 1987A. \cite{williams99} also identified six other known X-ray-detected SNRs that were not detected or observed by $ROSAT$:  0450-70.9, 0500-702 (N186D), 0513-692, 0524-664 (DEM L175a), 0527-658 (DEM L204) and 0527-658 (DEM L241).  Of these, 0450-70.9 and 0527-658 have been the subjects of recent pointed observations with {\it XMM-Newton} (see \cite{williams04} and \cite{bamba06}, respectively). Therefore, excluding
SN 1987A, a total of 36 LMC SNRs have been detected in X-rays.    The diameter
distribution of the X-ray-detected SNRs peaks near 40 pc, but also includes 0450-70.9 which, with a size of 98 $\times$ 70 pc, is the largest single X-ray emitting SNR known \citep{williams04}. These X-ray-detected SNRs constitute nearly all of the well-studied SNRs in the LMC; \cite{payne08} recently asserted that there are 56 known SNRs and another 20 candidate SNRs in the LMC based on recent radio studies, but very few of the objects not yet detected in X-rays have been confirmed through follow-up optical or X-ray investigation.  In the SMC, 18 SNRs are known to exist \citep{filipovic08}. All have been detected in X-rays; in addition, those
authors identify a new candidate X-ray SNR which was also detected in the radio
but not in the optical, along with a candidate radio SNR detected (weakly) in the optical but not in the X-ray. The diameter distribution of the SMC SNRs peaks near 45 pc. 
% Finally, in the case of M31, while the number of optically-identified SNRs and candidates in the galaxy is large \cite[178;][]{magnier95}, fewer than 30 have been clearly detected in X-rays \citep{pietsch05}. It is possible that because of the high inclination angle at which M31 is observed, significant internal absorption within the galaxy's disk prevents many SNRs from being detected at X-ray energies.
Finally,  M31 has a significant number of optically-identified SNRs and candidates, at least 221 as summarized and discussed by \cite{matonick97}. Most of the sample  is due to  \cite{magnier95}, who reported 178 objects, of which  less than half of these would have passed the [S~II]:H$\alpha$ criterion we have used here for optically-selected SNRs.  Of the M31 SNRs and candidates, fewer than 30 were detected in X-rays by \cite{pietsch05} using XMM.  It is possible that because of the high inclination angle at which M31 is observed, significant internal absorption within the galaxy's disk prevents many SNRs from being detected at X-ray energies. Individual SNRs in M31 have been detected with \chandra\ and discussed \cite[see, e. g.][]{williamsb05}, but a comprehensive study of  SNRs in M31 using all of the available ACIS data has not yet been published.

%\subsection{The X-ray Luminosity Function for Supernova Remnants}

The luminosity function for SNRs in M33 is shown in Fig.\ \ref{fig_lumfunc}.  There are 29 SNRs in M33 with 0.35-2 keV luminosities in excess of \POW{35}{\LUM}, and seven in excess of \POW{36}{\LUM}.  The figure also shows the 0.35-2 keV luminosity function for SNRs in the LMC and SMC.  Despite their much smaller distances, neither the Large nor the Small Magellanic Cloud has been searched to the same limit over the entire face of the galaxy that we have achieved for M33. However, it is at the bright end of the luminosity functions that discrepancies arise. As was pointed out initially by \cite{ghavamian05}, the number of high-luminosity SNRs in the LMC exceeds that for M33. We find only two SNRs with $L_{x} >10^{36.5}{\LUM}$ in M33, whereas the LMC has eight such objects. It is not clear what  accounts for this difference, but it may simply arise from small-number statistics. The bright end of the LMC SNR luminosity function comprises a very heterogeneous collection of objects. The brightest of these, N132D, still shows evidence of SN ejecta from a core-collapse explosion. Some of the other bright remnants have pure Balmer-dominated spectra and are thought on a variety of grounds to arise from Type Ia SNe.  Others like N49 and N63A are interacting with dense material. The brightest SNR in M33, G98-21, is also bright because it is expanding into a dense medium.  G98-31, the second brightest X-ray SNR, is nearly point-like in X-rays and appears to be emitting primarily from shocks in the ejecta.

The star-formation rate (SFR) in M33 is about 0.2 $ \MSOL \: {\rm yr}^{-1}$ \cite[see, e.g.][]{verley07}, similar to that in the LMC  0.19--0.26 $ \MSOL\: {\rm yr}^{-1} $\citep{kennicutt95, whitney08}, and so the rate at which core-collapse SNe occur should be similar. The agreement of the curves at \POW{35}{\LUM}  is comforting in this regard.  By contrast the SFR for the SMC is about 0.05 $\MSOL \:{\rm yr}^{-1}$ \citep{kennicutt95, wilke04}, and, as expected, the number of remnants is about one-third that of the other two galaxies.  At \EXPU{2}{34}{\LUM}, the number of SNRs in M33 exceeds that in the LMC.  Assuming this does not represent incompleteness in the LMC X-ray sample at these luminosities, the easiest way to explain this is to argue that the SNRs in the LMC are evolving faster than they are in M33, and hence fade earlier.  This would seem to require the typical density of the ISM in the LMC to be higher than in M33, but it is not obvious why that would be.   %The idea would be that about the same number are created, but the ones in the LMC are may be expanding  into a somewhat lower density medium.  
Considerable data have been obtained for the LMC in recent years at X-ray, radio, and optical wavelengths, much of which is not yet published.  Given the fact that the Magellanic Clouds are much closer than M33, it should be possible ultimately to create a more complete, better characterized sample of SNRs there than in M33.  It will be interesting to compare the luminosity functions once this is done.

Recently, \cite{pannuti07} examined X-ray SNR populations in five spiral galaxies -- M81, M101, NGC~2403, NGC~4736 (M94), and NGC~6946 -- that are outside the Local Group but still relatively nearby (within 7 Mpc). These five galaxies host a total of 138 objects identified as SNRs based on \sii:\HA\  ratios and 50 objects suggested to be SNRs based on their radio properties (namely, non-thermal spectral indices with \HA\  counterparts). Nine 
of the optically-identified objects and 12 of the radio sources have X-ray counterparts within 2 to 2.5\arcsec, and so most of the associations are likely real. The limiting luminosities for the X-ray observations of the galaxies examined by \cite{pannuti07} was higher (\EXPU{\sim2}{36}{\LUM}) than for M33, but many of the trends were similar. As is the case with M33, the mean diameter of the objects with X-ray counterparts is smaller than that of SNRs without counterparts (24$\pm$4 pc and 62$\pm$6 pc, respectively). 
%{\bf I would question whether or not the following sentences are relevant to this paper -- I would delete DJH.  I agree. --- PFW} They also explore correlations between L$_{H\alpha}  $, \sii:\HA, [N II]:\HA, [O I]:\HA, [O III]:\HB and the \sii\ line ratio, but none of these were significant.  Those authors did find a difference in the radio spectral index, 0.9$\pm$0.1 for detected objects vs 0.6$\pm$0.1 for undetected objects, but no physical explanation was offered for this difference.

\subsection{PWNe in M33}

The Crab Nebula, one of the brightest X-ray and radio sources in the Galaxy, represents the remnant of SN~1054. Its emission is powered by relativistic electrons generated by a young pulsar producing synchrotron radiation from radio to gamma ray frequencies. Over the past three decades, over two dozen other PWNe have been discovered in the Galaxy, some isolated like the Crab, and some inside normal shell-type remnants. Two such objects are known in the LMC, although none are known in any other galaxy. The spread of X-ray and radio luminosities -- and the ratio of the luminosities in these two bands -- span several orders of magnitude, even when the parameters of the pulsars powering these nebulae are similar.

The possible PWN we identified above, FL281, falls well within the ranges of parameters characteristic of Galactic PWNe. Its X-ray and radio spectral indices and its radio diameter are all typical for such objects. Its X-ray luminosity of $\sim 5 \times 10^{35}$ erg s$^{-1}$ is about 2\% that of the Crab and 7\% that of 0540-693, the brightest LMC PWN, but is in the upper quartile of the known Galactic objects, the majority of which fall below our detection threshold in the X-ray, the radio, or both.  The X-ray-to-radio luminosity ratio is about five times lower than that of the Crab, but 40 times greater than this ratio for 3C58, another young Galactic PWN. It is plausible, given the roughly equal star formation rates in the LMC and M33, that we should find $\sim 1$ bright PWN in our survey. While additional PWNe may lurk among our fainter radio sources, the lack of X-ray counterparts and the uncertain optical signatures will make them very difficult to identify. A deep radio image of FL281 with full polarimetric mapping, easily achievable with the EVLA, is required to confirm its identification as the first PWN beyond the Magellanic Clouds, and to reveal additional PWNe candidates.

\section{Conclusions \label{sec_conclusions}}

The \chase\ survey and our supporting multiwavelength analysis has enabled us to identify the largest well-characterized sample of SNRs that exists for any galaxy, except perhaps our own. Of the 137 SNR candidates in this sample, we observed 131 with \chandra\ and  determined the X-ray luminosities for 82, with upper limits on the remaining 49.  The six other objects were outside or at the very edges of the \chase\  survey fields. The detected SNRs have $L_{X}$ ranging from a minimum of \EXPU{2}{34}{\LUM} to a maximum of \EXPU{9}{36}{\LUM}.  From the {\it Chandra} data and our improved measurements of the optical and radio properties of the SNRs --- including a determination of the \HA\ luminosities of the sample, a compilation of the spectroscopic data, and new VLA observations --- we find the following:

\begin{itemize}
\item
There are no missing SNRs dominated by thermal emission brighter than $L_x$\EXPU{\simeq 4}{35}{\LUM} within our survey area.  Our spectroscopic search of the brighter \chandra\ sources in \chase\ should have revealed them all, regardless of whether they had been identified previously.  All of them should have been recognized by their soft, plasma-dominated spectra or, if they were larger than about 10 pc, extended X-ray emission, even if they had been missed in the optical and radio surveys.
\item
There appear to be no direct analogs of Cas A, Kepler, or Tycho in M33, or more precisely in the regions  of M33 that were well-observed with \chandra.  Tycho's SNR, the faintest of these, should have produced $\sim 500$ counts in the energy range 0.35-2 kev in 400 ksec, unless the absorption along the line of sight in M33 to such a SNR is higher than expected.  The only well-known historical SNR that could have been missed is SN~1006, which would be at or below our detection limit.   In FL281, however, we may have discovered the first PWN in M33.

\item
Two of the brightest SNRs in M33 -- G98-31 and G98-35 -- show evidence of abundance anomalies expected if their X-ray emission is contaminated by material produced in the SN explosion. Their spectra are suggestive of a core-collapse SN, which is unsurprising since most of the SNe in M33 should have arisen from core-collapse SNe. However, the abundances are not as extreme as the youngest core-collapse SNRs observed in the LMC and SMC; the resulting weaker lines and  limited signal-to-noise of the spectra do not enable us to place meaningful constraints on progenitor masses.  
% insufficient to determine the mass of the progenitor.  
The optical spectra of these SNRs do not reveal  fast ejecta-dominated filaments, and as a result it is unlikely these SNRs are extremely young objects.

\item
%The results show how heterogenous SNRs are, a heterogeneity that undoubtedly reflects the local conditions.  
There is no characteristic luminosity for a SNR of a certain size, because a SNR expanding into an ISM with a density of 0.1 $\rm cm^{-3}$ (for instance) evolves more slowly and to lower luminosity than one expanding into an ISM with a density of 0.5 $\rm cm^{-3}$.   Unfortunately, without estimates of current expansion velocities, there is no easy way with the existing data to remove or calibrate the effects of local environment on the other properties of SNRs one might want to measure.
  
\item
There are no strong correlations between the X-ray luminosities of M33 SNRs and their properties at other wavelengths, although extreme objects at one wavelength tend to be extreme at other wavelengths.  X-ray emission is not the dominant source of radiative energy loss for most SNRs, at least as measured in the 0.35-2 keV band.  For most SNRs, the \HA\ luminosity exceeds the 0.35-2 keV X-ray luminosity.
\end{itemize}

\acknowledgments{Support for this work was provided by the National Aeronautics and Space Administration through \chandra\ Award Number G06-7073A issued by the Chandra X-ray Observatory 
Center, which is operated by the Smithsonian Astrophysical Observatory for and on behalf of
the National Aeronautics Space Administration under contract NAS8-03060. PPP and TJG acknowledge support under NASA contract NAS8-03060. PFW and EKM acknowledge additional support from the National Science Foundation through grants AST-0307613 and AST-0908566. The work by RHB was partly performed under the auspices of the US Department of Energy by Lawrence Livermore National Laboratoryunder contract DE-AC52-07NA27344. We acknowledge our extensive use of the optical data on M33 from the Local Group Galaxies Survey, and are grateful to P. Massey and his colleagues for obtaining these data and for making them freely available.  This work has made extensive use of SAOImage DS9 \citep{joye03}, developed by the Smithsonian Astrophysical Observatory.  We also acknowledge the heroic efforts of the referee, M. Filipovi{\'c}, for an extremely careful reading of  the original manuscript and numerous comments that have improved this paper.

%  8/7/09: This Appendix is ordered based on increasing RA and so should ease the burden 
%  of adding the "L10-xxx" numbers when these are finally assigned.  I show the (expected)
%  file names for each object, assuming previous conventions for file naming are used.
%  I few of the objects could be "doubled up" in a single figure, but with the exception of
%  GKL57AB I have not assumed that here.  Best candidates for this are: GKL05/06, GKL10/12/13.
%  GKL48/WPB1, GKL51/52,  GKL86/87, and GKL97A/B.
%
%  A few of the objects still need figure work, either because they are outside the LGGS 
%  center field or in one case (GKL93) for some unknown technical glitch.

%  A few editorial comments remain. some pertaining to ill-defined regions.  I am not advocating
%  for changing these, but we might want to track them for future implementation.
%
%   -Bill
%

\appendix

%\begin{appendix}

\section{Notes and Images of SNRs and SNR Candidates \label{sec_notes}}

X-ray and optical images of the SNRs detected in M33 are provided in Figures \ref{fig_atlas01}-\ref{fig_atlas33}, with four SNRs shown in each Figure and four panels per Figure.  The four panels include (from left to right) \chase\ data, continuum-subtracted \HA\ and [S~II] images, and the V-band continuum image of the region, all from the LGGS data.  Where possible, the X-ray images were constructed from the subset of the \chase\ observations where the SNR was within 5\arcmin\ of the center of the \chandra\ field.  A scale bar is shown in the rightmost panel of each four-panel figure.  When appropriate, comments about the appearance in [O~III] $\lambda$5007 are also included, although these data are not shown.  A white contour on the X-ray and V-band panels shows the adopted extraction region we have set to define each object's extent.  The following notes summarize what is known about each object.

\begin{itemize}

\item
L10-001=G98-01  (Fig.\  \ref{fig_atlas01})   %     (fig\_Atlas\_GKL01.eps)

This is a large, irregular nebula somewhat resembling the letter `S' that stands out from the background confusion in the \sii\ image.  No \oiii\ emission is indicated and only a couple of stars are seen in projection; they do not appear to be associated directly with the nebula.  This object was outside the region of the X-ray survey, being on the far western side of the survey region.  Existing optical spectra confirm the elevated \sii:\HA\ ratio and, surprisingly, indicate a density significantly above the low density limit.  The object is either a very old, large SNR or a fossil superbubble whose exciting stars have largely evolved and disappeared.

\item
L10-002=G98-02    (Fig.\  \ref{fig_atlas01})   %    (fig\_Atlas\_GKL02.eps)

This is a faint region of elevated \sii:\HA\ ratio that overlaps a bright H~II region. We can only verify the faint northern portion due to confusion and set the extraction region based on this.  However, the southern extension seen in the [S~II] image is suggestive of a larger shell.  Only an upper limit of \EXPU{7}{34}{\LUM} on the X-ray emission is available.  Optical spectra imply a density somewhat elevated above the low-density limit.

\item
L10-003=G98-03)    (Fig.\  \ref{fig_atlas01})    %   (fig\_Atlas\_GKL03.eps)

G98-03 is located to the south of the region surveyed for ChASeM33 and thus no X-ray data are available.  The object is a large, faint shell in \sii:\HA\ with several bright stars interior to and adjacent  at least in projection.  One compact \hii\ region adjacent to a star is seen centrally located in the shell, something that is sometimes seen in superbubbles such as DEM301/N70 in the LMC.  This combination of features makes a superbubble interpretation most likely for this object.

\item
L10-004=G98-04)   (Fig.\  \ref{fig_atlas01})   %     (fig\_Atlas\_GKL04.eps)

This is a faint, limb-brightened thick shell filled with diffuse emission in \HA\ and [S~II]. No \oiii\ is visible.  No optical spectra are available, and it is outside the X-ray region surveyed by ChASeM33.

\item
L10-005=XMM068  (Fig.\  \ref{fig_atlas02})  %   (fig\_Atlas\_FL016.eps)

This is the object also known as XMM068, reported previously by Ghavamian et al. (2005).  A nearly complete but somewhat irregular broken shell of emission is seen in H$\alpha$ and [S~II], and is also present (but somewhat fainter) in [O~III].  Although the identification is solid from the imagery available, an MMT-BCS longslit spectrum reported in this paper provides a solid confirmation as an SNR, and shows that densities are in the low density limit.  The modest X-ray detection is consistent with a shell with some interior emission as well.

\item
L10-006=G98-05    (Fig.\  \ref{fig_atlas02})  %     (fig\_Atlas\_GKL05.eps)

This is a faint, almost complete and circular limb-brightened shell in \HA\  and [S~II].  [O~III] emission looks similar but is even fainter.  The X-ray remnant appears to have been detected separate from a bright X-ray point source present just east of the SNR.  Densities are in the low density limit of the [S~II] ratio.  Another SNR, G98-06, is centered 20\arcsec\ to the southeast.

\item
L10-007)  (Fig.\  \ref{fig_atlas02})    %  (fig\_Atlas\_Kip-S.eps)

This faint X-ray source was identified as part of the `extended X-ray source' analysis.  Comparing the position against the optical data revealed a large, very faint, patchy circle of emission that was present in H$\alpha$ and [S~II]. The exceedingly faint optical nebulosity has a somewhat enhanced \sii:\HA\ ratio of 0.32-0.36, but given the difficulties of measuring this ratio for such a faint object and the likely association of X-ray emission, it was included in our candidate list.  A few stars are seen in projection against the nebula, but their association with it is not clear. No obvious [O~III] emission is present.   No spectral data are yet available.  The X-ray morphology appears to be center-filled, although the detection significance is low (2.7$\sigma$). 

\item
L10-008=G98-06  (Fig.\  \ref{fig_atlas02})     %    (fig\_Atlas\_GKL06.eps)

This SNR is a nearly circular limb-brightened shell very similar to and only slightly smaller than G98-05 (20\arcsec\ to the northwest).  The SNR is seen most clearly in [S~II] since it overlaps with more extended, structured \HA\  emission on the south and east sides.  Only the brightest part of the \HA\  shell is present in [O~III].  Optical spectra show low densities, and only an upper limit is available in X-rays.

\item
L10-009=G98-07   (Fig.\  \ref{fig_atlas03})    %    (fig\_Atlas\_GKL07.eps)

G98-07 corresponds to an ill-defined patch of emission in [S~II] and \HA, with diffuse, faint emission toward the north and a brighter patch in the south.  A region is simply defined that encompasses the observed emission.  No optical spectrum is yet available. The source was outside the region explored by \chase. % {\bf Note: object is not shown in Table 4, but a `blank" X-ray fits file was available.  Is it on the edge of a FOV? -Bill}

\item
L10-010=G98-08   (Fig.\  \ref{fig_atlas03})   %     (fig\_Atlas\_GKL08.eps)

This is a large, loopy, faint emission structure with elevated \sii:\HA\ ratio.  [O~III] is faintly present from the brightest regions in \HA.  Optical spectra show low densities and only an upper limit is available in X-rays.  Although a couple of stars are seen in projection against this structure, no fossil association is obviously present. This is either an exceeding large SNR or a fossil superbubble whose primary stars have all evolved.

\item
L10-011=G98-09  (Fig.\  \ref{fig_atlas04})    %   7.5     24.3  (fig\_Atlas\_GKL09.eps)

This very high surface brightness optical SNR is present in \HA, [S~II], and [O~III], and has been confirmed spectroscopically to be shock-heated by a number of investigators over the years, including new MMT hectospec data reported here. HST images of the object with the original WF/PC instrument published by \cite{blair93} resolve the bright patch into a partial shell open to the northwest. The X-ray emission arises primarily at the location of the bright optical emission.  Presumably the shock wave from a SN explosion centered northwest of the bright patch must be encountering a significant density enhancement at this location on the shell. The somewhat over-sized extraction region is an attempt to account for the object being larger than the obvious bright emission regions.  Echelle spectra by \cite{blair88} show bulk velocities of order 260 $\VEL$. The [S~II] ratio confirms an elevated electron density for this object.

\item
L10-012=G98-10  (Fig.\  \ref{fig_atlas03})      %   (fig\_Atlas\_GKL10\_12\_13.eps)

This is a faint, broken ring of optical emission with enhanced \sii:\HA\ ratio extending from the western side of a complex H~II region.  [O~III] emission is seen primarily from the western limb of the shell. The X-ray detection implies an X-ray luminosity of about \EXPU{1.2}{35}{\LUM}. Stars present within the H~II region, and even within the nominal shell of the SNR, make it likely that one of the massive stars responsible for the H~II emission exploded as a core-collapse SN. No spectral data exist for this candidate.  Two other \cite{gordon98} SNRs, G98-12 and G98-13, are also associated with this H~II complex, but neither are detected in X-rays.

\item
L10-013=G98-11   (Fig.\  \ref{fig_atlas04})  %    (fig\_Atlas\_GKL11.eps)

This moderately strong X-ray source corresponds with an unremarkable optical SNR candidate consisting of two brighter knots of optical emission surrounded by more diffuse emission that may or may not be related to the SNR.  The [O~III] emission looks similar to the other optical images.  The X-rays give more of a filled shell appearance although some indication of a shell may be present.  In this case, the X-ray emission shows the overall extent of the object better than the optical. The relatively low \sii:\HA\ ratio of 0.47 for this object (compared to many other objects on our list) is still above the nominal threshold for shock heating, and may be artificially low due to potential background subtraction complications.  Interestingly, a moderately high density for the bright knots is indicated by the [S~II] ratio despite the overall low optical surface brightness.

\item
L10-014=G98-12   (Fig.\  \ref{fig_atlas03})    %    (fig\_Atlas\_GKL10\_12\_13.eps)

G98-12 is a half-shell of bright \HA-[S~II] emission on the north side of the H~II region that has also spawned G98-10 and 13.  [O~III] emission is present but very faint from the \HA\  partial shell.  
No optical spectrum is yet available, but X-ray emission appears to have been detected.

\item
L10-015=G98-13 (Fig.\  \ref{fig_atlas03})   %    (fig\_Atlas\_GKL10\_12\_13.eps)

G98-13 is a faint SNR buried in a complex region.  Careful inspection of the \HA\  and [S~II] images shows a faint oval shell with a brighter spot on the southern edge and a fainter enhancement on the northern side.  [O~III] is faint but present and brightest on the southeastern limb of the shell. No optical spectrum is yet available, but the source appears to have been detected at X-ray wavelengths.
\item
L10-016  (Fig.\  \ref{fig_atlas04})   %   (fig\_Atlas\_EM20.eps)

This is a large, oval, diffuse, limb-brightened shell of low surface brightness identified in our Schmidt survey.  In H$\alpha$, the object is surrounded  by faint, patchy emission in the wider field, but the SNR stands out in the [S~II] image.  The shell is barely discernible in [O~III] and brighter on the north and east sides, while H$\alpha$ is brighter on the west.  X-ray emission is confidently detected and seems to fill the shell.  Our new MMT-BCS data on this object confirm elevated [S~II] emission and show densities at the low density limit. 

\item
L10-017=G98-14  (Fig.\  \ref{fig_atlas04})   %   (fig\_Atlas\_GKL14.eps)

G98-14 is a well-formed, limb-brightened shell about 1\arcmin\ northeast of the H~II region that hosts G98-10, G98-12, and G98-13.  [O~III] is present but very faint. No optical spectrum is yet available, and only an upper limit on X-ray emission is available in our data.

\item
L10-018=G98-15  (Fig.\  \ref{fig_atlas04})   %  (fig\_Atlas\_GKL15.eps)

This X-ray source corresponds with a moderately bright, partially-filled shell of emission in H$\alpha$ and [S~II].  Extensions to the north and west of the main shell in \HA\ appear to be associated but do not have elevated [S~II]. Much fainter, diffuse [O~III] emission with a brighter knot on the west side are coincident with the shell.  Oddly, the low-density limit is indicated by the observed [S~II] ratio for this fairly bright optical SNR.  The X-ray emission seems to fill the shell rather than show the outer limb. \cite{blair88} found velocities in excess of 180 km $\rm s^{-1}$ for this object. Based on our inspection of images of M33 in the $Spitzer$ archive, a 24 $\mu$m source appears to coincide with this SNR.

\item
L10-019=G98-16 (Fig.\  \ref{fig_atlas05})  %   (fig\_Atlas\_GKL16.eps)

G98-16 is a large, somewhat oddly shaped limb-brightened shell most clearly delineated in the [S~II] image.  A bright H~II region is visible just to the south, with a stellar association clearly embedded. A faint ridge of emission is present 30\arcsec\ to the east.  The shell is present in [O~III] but more confused with surrounding emission.  No optical spectrum is yet available, and only an upper limit on X-ray emission is available in our data.

\item
L10-020 (Fig.\  \ref{fig_atlas05})  %   (fig\_Atlas\_EM80.eps)

This object was found in our Schmidt survey.  It is on the western edge of a large `ring' H~II region, and its full morphology is somewhat masked in the H$\alpha$ image, but it appears as a limb-brightened shell filled with faint diffuse emission.  [S~II] reveals the SNR somewhat better.  A very faint but complete ring is seen in the [O~III] image. A couple of stellar sources are seen in projection, but are not centrally located in the shell.  The X-ray detection is below 3$\sigma$, but the X-rays appear to be interior to the shell.

\item
L10-021=G98-17   (Fig.\  \ref{fig_atlas05})     %   (fig\_Atlas\_GKL17.eps)

G98-17 is a faint, very oblong ring of [S~II] emission with a brighter spot on its western side.  A corresponding structure is seen in \HA, but is confused by other overlying emission.  An optical spectrum confirms enhanced \sii:\HA\ ratio, and low densities.  Some of this extended structure is visible at very low surface brightness in [O~III].  Our X-ray data only provide an upper limit.   A brighter, more compact SNR (G98-20) is present about 25\arcsec\ to the east, and both objects are in the northern outskirts of the H~II region that hosts the brightest M33 SNR, G98-21.

\item
L10-022=G98-18  (Fig.\  \ref{fig_atlas05})  %   (fig\_Atlas\_GKL18.eps)

This is another example where the X-ray emission better defines the extent of the object than the optical emission does.  This moderately bright X-ray source coincides with a fairly faint, partial shell of optical emission brightest on the northeast side.  X-ray emission seems to fill the shell rather than follow the optical shell.  An [O~III] patch corresponds with the brightest spot of H$\alpha$ and [S~II] emission. Our MMT hectospec data confirm a very high \sii:\HA\ ratio and a low density for the optical nebula.

\item
L10-023=G98-20  (Fig.\  \ref{fig_atlas06})  %  (fig\_Atlas\_GKL20.eps)

This is another example where the X-ray emission better defines the extent of the object than the optical emission does.  This fairly bright X-ray source appears as a limb-brightened shell brightest on the east side.  A moderately bright optical knot of H$\alpha$, [S~II], and [O~III] emission corresponds with the southeastern limb of the SNR, which is located in the northern outskirts if the H~II region that also hosts G98-21 (Gaetz et al. 2007).  Under the best contrast setting, the knot almost resolves into an elongated, partial shell open to the west-northwest. Another larger diameter optical SNR, G98-17, lies $\sim$25\arcsec\ to the southwest but is not apparently detected in X-rays.  A fairly high density is indicated for G98-20 in our MMT hectospec data, consistent with the high optical surface brightness.

\item
L10-024=G98-19 (Fig.\  \ref{fig_atlas06})    %   (fig\_Atlas\_GKL19.eps)

This is a large, loopy shell with a centrally-located bright star or tight association and marginally-enhanced \sii:\HA\ = 0.41 - 0.48 and no evidence for X-ray emission.  Both stellar winds and/or possible SNR shocks may be involved in raising the observed [S~II]:H$\alpha$ ratio, and this appears more consistent with the object being primarily classified as a superbubble.

\item
L10-025=G98-21  (Fig.\  \ref{fig_atlas06}, see also Fig.\ \ref{fig_color_bright1} top row)
This is the brightest M33 X-ray SNR, and has been discussed at length by Gaetz et al. (2007) and above in \S \ref{sec_bright}.  Because of the discrepancy between the physical extent of X-ray and optical regions and the complexity of the surrounding optical emission, the optical extraction region was adjusted from the X-ray region (shown here).

\item
L10-026    (Fig.\  \ref{fig_atlas06})   %    (fig\_Atlas\_Kip-I.eps)

This object was first noticed from the diffuse, extended X-ray search.  A faint, well-formed shell in \HA\  and [S~II] is seen on the east side, and a more complicated structure extends toward the west.  A very bright, compact H~II region appears just to the south.  Stars visible within the shell make a combination of stellar winds and SNRs possible as the source of the shocks, and a superbubble identification likely.  No optical spectrum is available for this newly identified object.

\item
L10-027=G98-22  (Fig.\  \ref{fig_atlas07})      %   (fig\_Atlas\_GKL22.eps)

This is a faint, oddly shaped limb-brightened nebula that stands out best in [S~II].  A star or stars are seen projected within the shell, but a direct association is unclear.  Previous optical spectra confirm elevated \sii:\HA\ ratio and densities in the low density limit. Only an upper limit is available for the X-ray emission.  Another SNR, G98-23, lies $\sim$1\arcmin\ due east.

\item
L10-028  (Fig.\  \ref{fig_atlas07})  %  (fig\_Atlas\_EM22.eps)

This is a very large, faint emission ring brightest is on the southern side and has an image-derived \sii:\HA\  = 0.5 - 0.55. However, a cluster of stars is seen in projection along the major axis of the ellipse, including apparently one stellar object with excess H$\alpha$ emission directly associated.  The physical size of this nebula, $\sim$200 pc, would also seem to make a  superbubble origin more likely for this object.  Stellar winds and/or multiple previous SNe may contribute shocks that have raised the observed \sii:\HA\ ratio.  X-rays were not detected from the object.
\item
L10-029    (Fig.\  \ref{fig_atlas07})    %   (fig\_Atlas\_EM04.eps)

This is a faint, limb-brightened shell, very regular on the south side but somewhat irregular in the north, found as part of the Schmidt survey.  A brighter spot appears on the southwest side of the shell and close to two stars, arguing for a possible association.  An adjacent nebula with very similar characteristics lies just to the west and south.  No optical spectrum is yet available, although the object appears to have been detected at X-ray wavelengths.
\item
L10-030=G98-23    (Fig.\  \ref{fig_atlas07})    %   (fig\_Atlas\_GKL23.eps)

This is a large, faint, limb-brightened shell broken in the east and with a bright spot in the southwest.  The nebula looks identical but fainter in [S~II] compared with \HA, and [O~III] is only marginally detected from the brightest region of \HA.  Several bright stars, including some that form a horseshoe pattern, inhabit the interior, raising the possibility of a superbubble interpretation for this object.  However, no optical spectrum is yet available. The object was not detected in X-rays.

\item
L10-031=G98-24   (Fig.\  \ref{fig_atlas08})     %   (fig\_Atlas\_GKL24.eps)

This is a classic optical `filled smoke ring' SNR in \HA\ and \sii, with comparable \oiii\ emission as well. It is about 24 pc in diameter, making it similar to the Cygnus Loop in our Galaxy.  Our hectospec data confirm the elevated \sii:\HA\ ratio and densities near the low density limit.  Only an upper limit on X-ray emission is available in our data.

\item
L10-032=G98-25 (Fig.\  \ref{fig_atlas08})    %  (fig\_Atlas\_GKL25.eps)

This is an elongated optical SNR with a bright partial shell on the northeast side and fainter, more diffuse emission trailing off to the southwest.  Only faint [O~III] emission is present and it is more uniform.  Optical spectra show a very high \sii:\HA\ ratio and a density somewhat above the low density limit.  The X-ray emission appears to fill the optical shell. The southwestern X-ray source is harder and is apparently a separate source from the SNR.  
%{\bf Give source ID for SW source? -WPB}

\item
L10-033=G98-26 (Fig.\  \ref{fig_atlas08})  %   (fig\_Atlas\_GKL26.eps)

The X-ray luminosity is less than \EXPU{7}{34}{\LUM}.  This optical remnant is unremarkable, primarily standing out as a diffuse, filled circle on the [S~II] image.  Spectra (Table 3) confirm elevated \sii:\HA\ but indicate very low densities.

\item
L10-034=G98-27   (Fig.\  \ref{fig_atlas08})  %  (fig\_Atlas\_GKL27.eps)

This SNR lies on the southern side of a somewhat larger `ring' H~II region, with other emission surrounding to the south.  The SNR shows a fairly diffuse morphology, with a hint of shell-like structure in H$\alpha$ on its northern and southern sides.  Very faint and fairly diffuse emission is also present in [O~III], but is complicated by other photoionized emission in the region.  X-ray emission is well detected and seems to fill the shell rather than showing the limb.  The [S~II] ratio is very nearly at the low density limit in existing spectra.  A number of stars are seen in projection in and near the SNR, but their direct association is not assured. 

%\item
%L10-xxx  (EM81)    (fig\_Atlas\_EM81.eps)
%
% This object is a faint, irregular shell within a fairly complex overall region.  The \sii:\HA\  ratio from imaging analysis is only %marginally enhanced at 0.32, which does not make the nominal cut to be a SNR.   
%The stars seen in projection are not centrally located within the shell and do not appear to be associated.  Since accurate 
%background subtraction in the images is difficult in confused regions such as this, the imaging measurement of the ratio could be questionable.  
%Only an upper limit on X-ray emission is available in our data.  The nature of this object remains uncertain. {\bf This one looks like a dog to me.  WPB}
% Removed due to low ratio as judged by PFW image analysis.  8/11/09 WPB

\item
L10-035=XMM156 (Fig.\  \ref{fig_atlas09})   %    (fig\_Atlas\_XMM156.eps)

This object is a special case, and is potentially an example of a new oxygen-dominated optical SNR coincident with a fairly bright soft extended X-ray source. The X-ray source is clearly extended into an oblong shell brightest in the northwest.  The surrounding region is complex at optical wavelengths, including two adjacent \cite{gordon98} SNRs (G98-26 to the west and G98-30 to the east), knotty and diffuse emission regions, and several bright stellar sources, including one that is essentially coincident with the SNR position.  Careful inspection of the continuum-subtracted emission-line images, and in particular the [O~III] image (not shown here), implied the presence of emission at a level above that expected from an incomplete stellar subtraction.  An MMT-BCS longslit spectrum confirms the presence of strong [O~III] emission with little or no directly corresponding emission in H$\alpha$ and [S~II], from a region directly adjacent to the star, but from a region that is much smaller than the X-ray SNR.  This object could in principle be similar to the object N132D in the LMC, where the optical O-rich knots are concentrated in an interior region much smaller than the extended X-ray shell.  No high velocity emission was detected in our optical spectrum, however.  This object is deserving of further study.

\item
L10-036=G98-28  (Fig.\  \ref{fig_atlas09})  %    (Figure \ref{fig_bright1} right)

This bright, compact optical SNR is discussed in sec. \ref{sec_bright}. Under appropriate contrast settings, this oblong optical source shows indications of a partial shell open to the east (similar to G98-09). \cite{blair88} found velocities in excess of 340 km $\rm s^{-1}$ for this object.  Moderately high densities are indicated by the [S~II] ratio, and a faint 24 $\mu$m source aligns with the optical SNR.

\item
L10-037=G98-29   (Fig.\  \ref{fig_atlas09})  %   (fig\_Atlas\_GKL29.eps)

This is an example where the X-ray emission better defines the extent of the object than the optical emission does, showing a filled circular patch of emission that is brightest on the southeastern side.  The optical emission shows bright knots in the southeast that appear to be the brightest portions of a much fainter shell, also visible in high contrast and on the [O~III] image.  Densities moderately above the low density limit are indicated by existing spectra. A small faint \HA\ patch just west of the shell is not obviously related to the SNR.

\item
L10-038=G98-30    (Fig.\  \ref{fig_atlas09})   %    (fig\_Atlas\_GKL30.eps)

G98-30 is an exceedingly ill-defined patch of enhanced \sii\ emission within extended \HA\ emission.  It is associated with the same regions that include G98-26 and XMM156.  Our new MMT-BCS longslit data show only a marginally enhanced \sii:\HA\ ratio of 0.31, although severe background subtraction problems make this uncertain.  No X-ray detection is claimed, so only an upper limit is reported.  The identification of this source as a SNR is somewhat questionable.

\item
L10-039=G98-31  (Fig.\  \ref{fig_atlas10})   %   (Figure \ref{fig_bright2} left)

This is a very bright X-ray source that aligns with a very high surface brightness optical SNR with a bizarre morphology.  It was discussed in sec. \ref{sec_bright}.  The appearance of apparent diffuse continuum emission at the SNR position could be an artifact of the high surface brightness [O~III] emission lines being passed by the broadband filter image. \cite{blair88} found velocities in excess of 330 km $\rm s^{-1}$ for this object. A strong 24 $\mu$m source is coincident.

\item
L10-040=G98-32  (Fig.\  \ref{fig_atlas10})    %  (fig\_Atlas\_GKL32.eps)

This is a moderately large, nearly complete shell SNR located about 20\arcsec\ east of G98-28, in the outskirts of the H~II region NGC~595.  The X-ray detection is below 3 $\sigma$, but suggests a luminosity of \EXPU{3.7}{34}{\LUM}. Some emission interior to the shell may be present, and densities at the low density limit are indicated by previous spectra.  Any [O~III] emission is very faint and confused in the outer emission from NGC~595.

\item
L10-041   (Fig.\  \ref{fig_atlas10})   %   (fig\_Atlas\_EM18.eps)

This is a large diameter, very low surface brightness diffuse optical nebula found in our Schmidt survey with comparable H$\alpha$ and [S~II].  No [O~III] emission is seen.  While several stellar objects are seen in projection against this object, they are not an organized association and their connection to the nebula is uncertain.  Such an object in a more complicated region of the ISM would almost surely have gone unnoticed. The X-ray image appears to show emission and appears to fill the large shell, but excess is less than 1$\sigma$ and so the object is reported as an upper limit in Table \ref{table_results}.  The association of even faint X-ray emission with such a large, faint nebula is somewhat unusual.

\item
L10-042=G98-33   (Fig.\  \ref{fig_atlas10})    %    (fig\_Atlas\_GKL33.eps)

This is a faint but classic smoke ring SNR about 40 pc in diameter and slightly brighter on its southwest limb.  It is visible just outside the glow from a very bright star  some 40\arcsec\ to the northeast.  No [O~III] emission is seen.  No optical spectrum is yet available, and only an upper limit on X-ray emission is available in our data.

\item
L10-043  (Fig.\  \ref{fig_atlas11})  %   (fig\_Atlas\_Kip-H.eps)

A faint $\sim$90 pc shell with somewhat enhanced \sii:\HA\ = 0.44 - 0.47 is buried in a complex region.  A large, bright H~II region is visible to the southwest and a bright, more compact H~II region is projected onto or within the eastern limb of the shell.  The projection of stars within the shell favor a superbubble interpretation. No optical spectrum is available. However, X-rays are detected with a luminosity of \EXPU{1.2}{35}{\LUM}, which is somewhat unusual for a superbubble.

\item
L10-044=G98-34   (Fig.\  \ref{fig_atlas11})  %  (fig\_Atlas\_GKL34.eps)

This oblong SNR is shell-like, but appears flattened on the northern side.  Very weak [O~III] emission is indicated, and the X-ray emission is weak but seems to be primarily interior to the shell.  Our new hectospec data for this object confirms the elevated \sii:\HA\ ratio for this object and shows that densities are in the low density limit.

\item
L10-045=G98-35  (Fig.\  \ref{fig_atlas11})   %   (Figure \ref{fig_bright2} right)

This is another bright X-ray source coincident with a very high surface brightness optical SNR with bizarre morphology.  It is discussed in sec. \ref{sec_bright}.  Very high electron densities are indicated by optical spectroscopy, including the new MMT spectra presented in this paper. \cite{blair88} found velocities in excess of 310 km $\rm s^{-1}$ for this object.

\item
L10-046=G98-36  (Fig.\  \ref{fig_atlas11})  %    (fig\_Atlas\_GKL36.eps)

This is a fairly diffuse, filled SNR with a shell visible, especially in the north.  Faint [O~III] emission especially along the eastern side is indicated.  The extended X-ray source seems to fill the shell rather than align with the limb.  A high \sii:\HA\ ratio and a density near the low density limit is indicated by our new MMT-BCS optical spectra. 

\item
L10-047=G98-37  (Fig.\  \ref{fig_atlas12})   %    (fig\_Atlas\_GKL37.eps)

This is a faint optical SNR, diffuse and filled but with indications of a partial shell on the west side in \HA\ and [S~II].  Interestingly, it is the east side of the shell that is primarily visible in [O~III] although it is still quite faint.  As with the previous source, the X-ray emission seems to fill the shell rather than showing the outer rim. A high \sii:\HA\ ratio and a density at the low density limit is indicated by previous spectra.

\item
L10-048=G98-38   (Fig.\  \ref{fig_atlas12})   %   (fig\_Atlas\_GKL38.eps)

This is a small enhanced patch of emission at the southern end of a long, faint finger of emission.  Its appearance is complicated by the need to subtract a star that is centrally located (at least in projection).  No appreciable [O~III] emission is seen.  Previously published spectra apparently confirm the \sii:\HA\ ratio, and a density near the low density limit is indicated.  Only an upper limit is available on the X-ray emission from this object.

\item
L10-049=G98-39 (Fig.\  \ref{fig_atlas12})     %   (fig\_Atlas\_GKL39.eps)

This is a faint, clumpy, irregular optical SNR most visible from the surrounding faint contaminating emission in the [S~II] image.  Very faint [O~III] emission is present as well.  Again, the faint X-ray emission seems to fill the region rather than showing a rim.  No optical spectrum is available for this object.

\item
L10-050=G98-41  (Fig.\  \ref{fig_atlas12})   %  (fig\_Atlas\_GKL41.eps)

The long, broken filament identified by \cite{gordon98} with enhanced [S~II] may be the brightest southeastern portion of a fainter, much more extensive emission shell barely visible in the Figure.  A bright H~II region appears coincident with the northern limb of this shell. Some moderately bright stars are seen in projection within the shell, but their association with the shell is unclear.  While stellar winds or even possible SNR shocks may be involved in raising the observed \sii:\HA\ ratio, this appears more consistent with being primarily classified as a superbubble or even a budding supershell.  Only an upper limit is available on X-ray emission.

\item
L10-051=G98-40  (Fig.\  \ref{fig_atlas13})    %   (fig\_Atlas\_GKL40.eps)

This is a faint, moderately large limb-brightened shell SNR.  It is unremarkable in \HA\ and appears as part of a line of faint emission stretching from northeast to southwest, but the shell stands out more conspicuously in \sii. Previous spectra confirm the enhanced \sii:\HA\ and show a density very near the low density limit.  The X-ray detection within the defined contour of the shell is officially an upper limit, but there is a curious clustering of X-rays near the center of the shell that may be significant. 

\item
L10-052=G98-42  (Fig.\  \ref{fig_atlas13})    %   (fig\_Atlas\_GKL42.eps)

G98-42 is a marginally-enhanced smudge of emission extended in a roughly north-south direction within a complex region in the southern spiral arm.   It is not detected in X-rays, but two X-ray sources corresponding with nearby stellar objects are present just to the south of the candidate. No optical spectrum is yet available.  

\item
L10-053=G98-43A   (Fig.\  \ref{fig_atlas13})    %  (fig\_Atlas\_GKL43.eps)

This object is located in the LGGS south field and is outside the region surveyed for ChASeM33.  The shell identified by Gordon et al. (1998) is the brighter shell in the figure, with a fainter, diffuse shell to the southeast that also appears to have enhanced \sii. The nebula Gordon et al. (1998) identified as G98-43 is what we designate as G98-43A.   (The fainter oval region we designate as G98-43B and is discussed next.)  This is either a single SNR with bizarre morphology or two adjacent SNRs that are either physically associated or projected along the line of sight. A spectrum of G98-43A confirms elevated \sii:\HA\ ratio and a \sii\ density in the low density limit. 

\item
L10-054=G98-43B   (Fig.\  \ref{fig_atlas13})    %  (fig\_Atlas\_GKL43.eps)

G98-43B is a faint emission region visible in the LGGS data due to it's greater sensitivity compared with previous surveys.  Interestingly, it is visible (faintly) in Gordon et al.'s figure for this object but was not called out as a separate object or part of the G98-43 SNR candidate.
This object is located in the LGGS south field and is outside the region surveyed for ChASeM33.  This is either a single SNR with bizarre morphology  or two adjacent SNRs that are either physically associated or projected along the line of sight.  The brighter shell to the northeast we designate as G98-43A (see above).  No spectrum is yet available for G98-43B.

\item
L10-055=G98-44  (Fig.\  \ref{fig_atlas13})  %   (fig\_Atlas\_GKL44.eps)

This object may be a more extreme version of the phenomenon seen in G98-41 (L10-050).  The long,  filament with enhanced [S~II] is clearly part of a much larger structure, but unlike G98-41, no larger potential shell-like structure is evident.    No [O~III] is seen from the filament, and it was not detected as an X-ray source. Bright H~II regions are seen to the west.   Stellar winds and/or ancient SNe shocks cannot be eliminated as contributors to the enhanced [S~II], but a clear definition as a superbubble is also not obvious.   Without further information, the origin of this filament cannot be stated.  No spectrum yet exists for G98-44 to confirm the enhanced \sii:\HA\ indicated by imagery. 

\item
L10-056=G98-45   (Fig.\  \ref{fig_atlas14})  %    (fig\_Atlas\_GKL45.eps)

This is a faint partial shell optical SNR just NW of a bright, compact H~II region.  The shell is brightest on the east side, and the X-ray emission seems to be centrally located in the shell.  No obvious [O~III] emission is present for this SNR.  Low densities are indicated by the [S~II] ratio in existing spectra.

\item
L10-057=G98-46 (Fig.\  \ref{fig_atlas14})   %  (fig\_Atlas\_GKL46.eps)

This is a faint, diffuse, oblong filament on the outskirts of a much brighter `ring' H~II region.  No [O~III] emission is indicated, and the X-ray detection is just above 3$\sigma$, with no particular structure seen to the X-ray emission. Existing optical spectra show a \sii:\HA\ ratio of 0.54 and low densities.

\item
L10-058   (Fig.\  \ref{fig_atlas14})    %  (fig\_Atlas\_PFW1.eps)

This is one of the newly identified optical SNR candidates.
The object appears as a moderately large kidney-bean shaped limb-brightened shell.  An indentation with brightening on the east side gives the impression that the shock is encountering denser material there.  However, the object is not detected in X-rays and no optical spectrum is yet available.

\item
L10-059   (Fig.\  \ref{fig_atlas14})   %   (fig\_Atlas\_WPB2.eps)

This is one of the newly identified optical SNR candidates, located in a busy field just west of the galaxy nucleus.  Careful subtraction of the background revealed a faint, perfectly oval diffuse emission region with just a hint of limb-brightening.  Several bright stars are seen projected within the shell, but their association with the nebula is uncertain.  The X-ray image and the excess in counts measured for the object give the impression that the source is detected, but the location of the source within the scattering region from the bright nuclear source makes the measurement somewhat uncertain.  No optical spectrum is available.

\item
L10-060=G98-48  (Fig.\  \ref{fig_atlas15})    %   (fig\_Atlas\_GKL48.eps)

G98-48 is a rather ill-defined patch of enhanced \sii:\HA\ about 20\arcsec\ southeast of L10-059 and hence even closer to the nuclear source.  Nonetheless, the X-ray detection of this source appears to be convincing despite the official statistic showing it below 3$\sigma$. A bright, compact H~II region is directly adjacent to the north.  The object is barely visible in \oiii.   Existing optical spectra confirm the enhanced \sii:\HA\ and show the densities to be in the low density limit.

\item
L10-061=G98-47 (Fig.\  \ref{fig_atlas15})   %  (fig\_Atlas\_GKL47.eps)

This is a fairly large, diffuse emission nebula directly adjacent to a bright, compact H~II region, with significant other surrounding emission in the southern spiral arm.  The region of interest is most visible in the [S~II] image where most of the photoionized gas is fainter.  There are two diffuse but somewhat patchy lobes, one protruding to the east and the other to the northeast from the compact H~II region.  Both lobes show significantly enhanced [S~II] emission and are likely shock heated. (The object is even visible as an extended green nebula in the central panel of Fig. 2, below and right of center.)  The X-ray emission seems to fill the region of both lobes as opposed to being in a shell-like structure, and does not seem to extend into the compact H~II region to the west. \cite{blair88} used an E-W slit at echelle resolution and saw velocities in excess of 100 km $\rm s^{-1}$ for the eastern portion (they called this object M33-20).
The low density limit is indicated by the [S~II] ratio. The extent of the source is only approximated by the oval extraction region.

\item
L10-062  (Fig.\  \ref{fig_atlas15})     %  (fig\_Atlas\_EM70.eps)

This is a large, very faint circular emission region with a hint of limb-brightening especially in the \HA\ image.  It is just 20\arcsec\ west of two other SNRs, G98-51 and 52, which are visible at the edge of the frame.  L10-062 was discovered in the Schmidt survey and is probably just enough fainter than these other SNRs that it was not confidently detected in earlier surveys.  Several stars are seen in projection in and near the shell, but their association is uncertain.  X-rays are not detected, and no optical spectrum is yet available.

\item
L10-063  (Fig.\  \ref{fig_atlas15})   %     (fig\_Atlas\_ EM70A.eps)

This SNR candidate is an exceedingly faint circular region of emission identified in the Schmidt survey.  A single star is seen projected within the region but its association is unknown. 
X-rays are not detected, and no optical spectrum is yet available.

\item
L10-064=G98-49  (Fig.\  \ref{fig_atlas16})    %   (fig\_Atlas\_GKL49.eps)

The optical SNR looks similar in all three optical bands (including \oiii), showing roughly half of an optical shell bright on the west side and open to the east.  An extension to the south in \HA\ does not appear to be part of the SNR.  Although below a 3$\sigma$ detection, the X-rays appear to coincide more with the interior of the SNR as opposed to the shell.  The extent is measured assuming the full shell follows the curvature of the visible portion.  A fairly low density, but above the low density limit, is indicated by the [S~II] ratio in previous spectra.

\item
L10-065=G98-50   (Fig.\  \ref{fig_atlas16})      %  (fig\_Atlas\_GKL50.eps)

This is a moderately bright optical SNR with a well defined partial shell on the south and east sides.  Fainter patchy and diffuse emission fills the interior of the shell and extends somewhat to the  west and north. A bright, compact \hii\ region appears just to the north and another lies 20\arcsec\ to the east.  Both of these \hii\ regions are bright $Spitzer$ 24$\mu$m sources. 
Only a few stars appear in projection and their association is now known.  X-rays are not confidently detected.  Previous optical spectra confirm the enhanced \sii:\HA\ ratio and show densities in the low density limit.

\item
L10-066=G98-52  (Fig.\  \ref{fig_atlas16})    %  (fig\_Atlas\_GKL52.eps)

G98-52 is the northern object of a pair of adjacent SNRs (the other is G98-51).  G98-52 is a faint, diffuse, nearly circular SNR with a hint of limb-brightening in \HA\ and possibly faint X-ray emission (below 2$\sigma$).  Very faint \oiii\ emission is seen from the northern rim.   Despite the optical faintness, an intermediate density is indicated by the [S~II] ratio in existing spectra. To the extent one can tell, the X-ray emission seems to fill the shell.

\item
L10-067=G98-51 (Fig.\  \ref{fig_atlas16})     %     (fig\_Atlas\_GKL51.eps)

G98-51 is comparable in size and surface brightness to G98-52 (just to the north), but is less well-defined, appearing as a faint, diffuse nebula somewhat extended in the northwest-southeast direction.  No X-ray detection is claimed and no optical spectra are available. 

\item
L10-068 (Fig.\  \ref{fig_atlas17})  %   (fig\_Atlas\_EM66.eps)

This is a large, very faint shell of enhanced [S~II] emission (0.42 - 0.55) located on the eastern edge of a bright  H~II complex.  It was located in our Schmidt survey.  The physical size is $\sim$100 $\times$ 120 pc.  Several bright stars and/or tight clusters are projected within the shell.  Although SN shocks are likely to be involved, stellar winds could also be important in producing the observed feature, making a superbubble interpretation likely. A blow-out from the H~II region is another possibility, although a full shell seems to be present in \sii, so this could be a projection effect. Only an X-ray upper limit is available.

\item
L10-069=G98-53 (Fig.\  \ref{fig_atlas17})     %    (fig\_Atlas\_GKL53.eps)

This is a somewhat irregular, patchy shell SNR about 10\arcsec\ west of an extended north-south finger of H~II emission.  No [O~III] is visible.  The X-ray emission appears centrally located in the shell.  Existing spectra indicate densities near the low density limit.

\item
L10-070=G98-54 (Fig.\  \ref{fig_atlas17})    %   (fig\_Atlas\_GKL54.eps)

This is a relatively faint SNR on the western side of a complex region of H~II and related emission.  A partial shell structure is visible in H$\alpha$ with a brighter knot on the north side that also shows in [S~II] and [O~III].  The X-ray emission seems to be concentrated on the eastern side of the SNR, where the shock may be interacting with denser surrounding gas. The neighboring H~II region to the east also seems to have been detected as an extended source.  Existing spectra indicate densities near the low density limit.

\item
L10-071=G98-55     (Fig.\  \ref{fig_atlas17};  also \ref{fig_color_bright2} middle row)

This is a very bright X-ray SNR and is also high surface brightness in all three optical bands. It was discussed in sec. \ref{sec_bright}.  An echelle spectrum by \cite{blair88} shows motions at 270 km $\rm s^{-1}$ associated with this object.  Fairly high densities are indicated by the optical [S~II] ratio.

\item
L10-072    (Fig.\  \ref{fig_atlas18})  %    (fig\_Atlas\_EM12.eps)

Unlike many of the new SNR candidates identified in our Schmidt survey, this one is relatively modest in size and fairly bright.  A well-defined half shell is present, open to the north.  The remainder of the object is invisible in the optical data, but the curvature of the visible portion has been used to define the size.  Despite the appearance of X-ray emission in the region, it is not significant given the fairly high background in the region.  No optical spectrum is yet available to confirm the source.

%\item
%L10-xxx  (FL212)       (fig\_Atlas\_FL212.eps)

\item
L10-073    (Fig.\  \ref{fig_atlas18})   %    (fig\_Atlas\_EM32.eps)

This is a faint, isolated, limb-brightened shell identified in our Schmidt survey, but little is known.  Only an X-ray upper limit is available, and no optical spectra have yet been obtained.

\item
L10-074=G98-57A; L10-076=G98-57B  (Fig.\  \ref{fig_atlas18})     %   (fig\_Atlas\_GKL57AB.eps)

The previous candidate G98-57 has been separated into two candidate objects as follows.
G98-57A is a nearly circular, diffuse, enhanced [S~II] patch in a region of fairly high contamination by other emission.  Interestingly, [O~III] emission forms a partial shell visible only on the southwest side of the object. Two small arcs or fingers of enhanced [S~II] emission are visible just north of (but separate from) the SNR proper, and this region also appears to have X-ray emission, but the relation of this to the SNR is unclear.  We designate this region as G98-57B. Our new MMT spectra of G98-57A confirm the elevated \sii:\HA\ ratio and show a [S~II] ratio near the low density limit.

\item
L10-075=G98-56   (Fig.\  \ref{fig_atlas18})   %    (fig\_Atlas\_GKL56.eps)

This is a faint, diffuse SNR with only a hint of limb-brightening.  It is located in a fairly busy portion of the southern spiral arm, with several nearby compact \hii\ regions and significant extended diffuse emission.  A bright star on the eastern limb of the SNR does not appear to be associated.  Existing optical spectra confirm the elevated \sii:\HA\ ratio and show densities to be low.  Only an X-ray upper limit is available.

\item
L10-077=G98-59 (Fig.\  \ref{fig_atlas19})   %      (fig\_Atlas\_GKL59.eps)

G98-59 is an ill-defined and somewhat marginal candidate in many regards.  Gordon et al. (1998) apparently identified the brighter optical knot at the center of the figure as the SNR, but this knot is just the brightest portion of a more extended region of emission with comparable \sii:\HA\ ratio.  Nonetheless, a previous optical spectra seems to confirm \sii:\HA\ = 0.40, just qualifying it as a SNR candidate under our definition.  No \oiii\ emission is seen, but interestingly, our inspection of data from the $Spitzer$ archive indicates the whole region shown, including the SNR region as defined here, shows significant 24$\mu$m emission.  No X-ray detection is claimed for this object.

\item
L10-078=G98-58  (Fig.\  \ref{fig_atlas19})    %    (fig\_Atlas\_GKL58.eps)

This is a relatively small, bright partial shell SNR at optical wavelengths, brightest in the south.  The SNR appears flattened on the north instead of round.  The X-ray emission appears to correspond more with the northern limb, implying that we are seeing primarily the SE side of the SNR shell at optical wavelengths and the northern rim in X-rays. [O~III] emission is prominent only from the southern partial shell portion. Our new hectospec data show a low density, but above the low density limit.

\item
L10-079     (Fig.\  \ref{fig_atlas19})   %   (fig\_Atlas\_Kip-G.eps)

This object was identified in our diffuse, extended X-ray source search, and inspection of the optical showed a faint nebular structure extended in the east-west direction, in close proximity to two compact \hii\ regions, one just to the north and a smaller one to the south.  There are no clearly associated stars, and an SNR in a very inhomogeneous medium is indicated.  No optical spectra are yet available.

\item
L10-080=G98-60   (Fig.\  \ref{fig_atlas19})     %   (fig\_Atlas\_GKL60.eps)

This is a small diameter, moderately bright optical SNR located in the outer reaches of a bright H~II region complex.  [O~III] is present but faint, and gives an impression of a partial shell open to the north.  A stellar source is located directly adjacent to the south, but does not appear to be related directly.  X-rays are detected but no shape is visible.  Our new hectospec data indicate an intermediate density for this object.

\item
L10-081=G98-62  (Fig.\  \ref{fig_atlas20})    %   (fig\_Atlas\_GKL62.eps)

The optical morphology of this SNR is poorly defined, consisting primarily of a filament or arc of emission at the southwest edge of an H~II complex.  It is unclear whether the fainter arc of emission extending to the south and east are directly related or not. The X-ray emission appears centered southwest of this filament, leaving the impression that the optical emission represents just the northeast edge of the object as the shock encounters the outskirts of the H~II region.  The extent of this poorly defined object is estimated based on both the optical and X-ray emission. Earlier optical spectroscopy confirms the elevated \sii:\HA\ ratio and shows a low density for this object.

\item
L10-082=G98-61  (Fig.\  \ref{fig_atlas20})     %  (fig\_Atlas\_GKL61.eps)

This is a somewhat irregular partial shell SNR at \HA\ and [S~II] wavelengths.  Extended emission visible in \HA\ does not appear to be associated with the SNR directly. Interestingly, in [O~III] the SNR appears diffuse and filled (but faint), showing the full extent.  The X-ray emission seems to fill the object rather than be associated with the bright optical regions.  Low densities are indicated by the [S~II] ratio in existing spectra.

\item
L10-083=G98-65)   (Fig.\  \ref{fig_atlas20})   %   (fig\_Atlas\_GKL65.eps)

This is a faint, ill-defined optical SNR embedded in a complex region of extended optical emission.  In [S~II], there is some indication of a partial shell on the west side, connecting to a possible section on the east side that is confused by other emission.  No [O~III] emission is clearly associated with the shell.  Large groupings of stars are present just to the east and west of the defined region, although no direct association with the SNR is assumed.  Existing spectra confirm elevated \sii:\HA\ ratio and show the low density limit for this object.  X-rays appear to be detected, although the source of the X-ray emission seen to the west-northwest of the SNR is not obvious at optical wavelengths. 

\item
L10-084=G98-64 (Fig.\  \ref{fig_atlas20})    %   (fig\_Atlas\_GKL64.eps)

This is a moderately bright X-ray source coincident with a classic, shell-like \HA\ and [S~II] SNR reminiscent of the Cygnus Loop in our Galaxy.  The [O~III] emission is somewhat fainter and more evenly distributed, with some limb-brightening, but is entirely consistent in extent. New hectospec data give a density just above the low density limit, and confirm a very elevated \sii:\HA\ ratio. The X-ray emission fills the shell, and appear brightest in the south exactly where the \HA\ shell is faintest.  A star located on the southeastern rim does not appear to be associated.

\item
L10-085=G98-63 (Fig.\  \ref{fig_atlas21})     %   (fig\_Atlas\_GKL63.eps)

The optical counterpart of this SNR is peculiar.  It appears almost as a partial shell open to the north, but this partial shell appears as a series of four clumps or small arcs of emission.  The X-ray emission is centered to the north, implying that we are seeing one side (the southern side) of the object at optical wavelengths.  [O~III] emission is visible from the same partial shell, although its distribution is somewhat different in detail from that seen in \HA\ and [S~II].  No spectrum is available for this object, although the image ratio indicates [S~II] is plenty strong to indicate shock heating.  The few faint stars seen in projection do not appear to be related directly to the SNR.

\item
L10-086=FL236  (Fig.\  \ref{fig_atlas21})    %    (fig\_Atlas\_FL236.eps)

This optical counterpart was found via the search for optical emission at the positions of soft X-ray sources. Two faint knots of comparable \HA\ and [S~II] emission are seen, which may represent bright spots on a small, partial shell.  There is no obvious emission in [O~III].  It is unlikely this optical counterpart would have been identified without the X-ray source drawing attention to it. Our hectospec data confirm the elevated \sii:\HA\ ratio and show a fairly high density for the optical knots. The X-ray emission arises on and just north of the optical knots, indicating they may be on the southern edge of a SNR centered slightly north of the optical knots.  Our defined region takes this interpretation into account.

\item
L10-087=G98-66  (Fig.\  \ref{fig_atlas21})   %    (fig\_Atlas\_GKL66.eps)

This is a moderately large, faint diffuse SNR with a hint of limb-brightening or shell-like emission in \HA\ and [S~II], located in a fairly busy nebular field, but still clearly defined.  It stands out best on the [S~II] frame. A faint western extension seen in \HA\ is not seen in [S~II], and is likely not part of the SNR.  No distinct [O~III] is visible.  X-ray emission seems to fill the object, although emission from a point source on the southeast rim of the SNR affects the X-ray source appearance.  Our new MMT-BCS spectra confirm the elevated \sii:\HA\ ratio and indicate a density slightly above the low density limit.

\item
L10-088=G98-67    (Fig.\  \ref{fig_atlas21})     % (fig\_Atlas\_GKL67.eps)

G98-67 is a moderately bright partial shell SNR centrally located between two H~II region (to the north and south).  The shell is brightest on the eastern side.  A moderately bright stellar source is seen in projection inside the shell but is not obviously associated.  [O~III] appears faintly but seems to be somewhat anti-correlated with the H$\alpha$ (thus showing the northwest side of the shell).  The X-ray detection is weak enough so as to make the X-ray morphology uncertain, but no indication of the shell is seen.  Optical spectra from previous work indicate the low density limit for this object.

\item
L10-089  (Fig.\  \ref{fig_atlas22})   %   (fig\_Atlas\_Kip-N.eps)

This large arc of enhanced \sii:\HA\ (0.62 - 0.79) is located on the eastern side of a bright H~II region and appears to be part of a large scale blow-out from the H~II region.  While a number of stars appear in projection, it is not clear whether these stars are associated with the nebular arc or with the general population in the field.    An excess of X-ray emission over the background indicates the source is detected.  The large physical size ($\sim$70 $\times$ 120 pc) makes a multiple SN/stellar wind superbubble origin possible for this intriguing object.

\item
L10-090=G98-68   (Fig.\  \ref{fig_atlas22})   %    (fig\_Atlas\_GKL68.eps)

This is a fairly faint but classic limb-brightened shell-type SNR at all three optical wavelengths.  The distribution in [O~III] appears slightly different, but the extent is the same.   X-ray emission  is detected at just over 3$\sigma$ and seems to be centrally located although it is hard to say with confidence.  Existing optical spectra indicate densities slightly above the low density limit.

\item
L10-091=G98-69 (Fig.\  \ref{fig_atlas22})   %  (fig\_Atlas\_GKL69.eps)

G98-69 is an exceedingly faint, irregular, and patchy optical SNR.  In a more complicated region it probably could not have been identified, but it is located in an unconfused region.  Its size is ill-determined, especially in H$\alpha$ and [S~II].  However, faint [O~III] emission (not shown here) appears to be somewhat more extensive than the other lines.   The X-ray emission is faint, but seems to anti-correlate with the \HA, perhaps indicating a faint shell.  No optical spectra yet exist for this object.

\item
L10-092=G98-70 (Fig.\  \ref{fig_atlas22})       %   (fig\_Atlas\_GKL70.eps)

G98-70 is a large ($\sim$100 pc), patchy emission structure with some indications of shell edges or filaments in conjunction with more diffuse emission.  [O~III] emission is too faint to show structure.  A handful of stellar sources are seen in projection against the nebula, which, along with its size, imply a possible superbubble interpretation is appropriate.  The X-ray emission is faint and not well defined, registering at only 2.3$\sigma$.  Existing optical spectra indicate densities just slightly above the low density limit.

\item
L10-093=FL261  (Fig.\  \ref{fig_atlas23})   %    fig\_Atlas\_FL261.eps)

This optically-faint, small partial shell was found via the search for optical emission at the positions of soft X-ray sources. Emission is comparable in H$\alpha$ and [S~II], and optical [O~III] emission (not shown here) looks identical to H$\alpha$. Our new hectospec data confirm the elevated \sii:\HA\ ratio and show densities significantly above the low density limit.  As with FL236, the X-ray emission is centered just north of the optical partial shell, lending credence to the idea that the optical is showing the southern side of a larger SNR centered to the north.

\item
L10-094=XMM244 (Fig.\  \ref{fig_atlas23})    %   (fig\_Atlas\_XMM244.eps)

This moderately strong X-ray source aligns with a diffuse, poorly defined optical SNR candidate in a complex stellar and nebular field.  However, a region with comparable H$\alpha$ and [S~II] emission does correspond with the X-ray source, and similar diffuse emission is present in [O~III] as well.  A grouping of fairly bright stars surrounds the position.  A longslit MMT-BCS spectrum (this paper) confirms the identification and indicates moderate densities in the optical nebula.

\item
L10-095=G98-71  (Fig.\  \ref{fig_atlas23})     %   (fig\_Atlas\_GKL71.eps)

This is a relatively small, classic shell-like SNR at all three optical wavelengths.  A small break in the southeast is the only part of the shell missing, and almost no emission is seen interior to the shell. Using the optical data to define the size, the associated X-ray emission seems to correlate with the southern half of the SNR, although the detection is quite faint (\EXPU{5.7}{34}{\LUM}).   A longslit MMT-BCS spectrum (this paper) confirms the identification and indicates moderate densities in the optical nebula.

\item
L10-096=G98-73  (Fig.\  \ref{fig_atlas23})    %  (Figure \ref{fig_bright3} right)

This is a bright X-ray source associated with a high surface brightness, small diameter SNR, as discussed in sec. \ref{sec_bright}.  \cite{blair88} found velocities in excess of 180 km $\rm s^{-1}$ for this object.

\item
L10-097=G98-72   (Fig.\  \ref{fig_atlas24})   %    (fig\_Atlas\_GKL72.eps)

All three optical emission line images are consistent, showing an arc or partial shell of emission open to the southwest.  The arc is in the outskirts of a large, bright H~II region to the northwest of the SNR. The X-ray detection is only at the 2.2$\sigma$ level, but is consistent with the partial shell.  The extent of the object is based on an extension of the visible part of the shell.  An intermediate density is indicated by previous optical spectroscopy.

\item
L10-098=G98-74 (Fig.\  \ref{fig_atlas24})    %  (fig\_Atlas\_GKL74.eps)

This is a faint, diffuse emission nebula with marginally enhanced \sii:\HA\ = 0.41-0.49 from image analysis.  Faint \oiii\ emission appears to coincide with the somewhat brighter northeastern quadrant of the nebula as seen in \HA.  A number of bright stars are seen in projection against the nebula (although their association with the nebula is not ironclad) and a compact emission region is also seen in projection on the H$\alpha$ frame, which is a feature seen in some of the other purported superbubbles.  A large, old SNR identification is not outside the realm of possibility. However, a faint superbubble appears to be the preferred classification.  Very low densities are indicated by existing optical spectra.  The upper limit to the X-ray luminosity is \EXPU{2.4}{34}{\LUM}.

\item
L10-099=G98-75  (Fig.\  \ref{fig_atlas24})     %   (fig\_Atlas\_GKL75.eps)

G98-75 is an ill-defined SNR candidate consisting of a few scattered optical filaments with enhanced \sii:\HA\ emission.  The bright compact H~II region to the southeast is a strong $Spitzer$ 24 $\mu$m source, and another SNR, G98-77, is located just to the southeast of that.  Any \oiii\ emission from the SNR is very faint although the H~II regions are detected.  No optical spectrum is available, and the object was not detected in X-rays.

\item
L10-0100=G98-76 (Fig.\  \ref{fig_atlas24})    %   (fig\_Atlas\_GKL76.eps)

This is a faint but fairly complete shell-like SNR at all three optical wavelengths, somewhat brighter in the northeast.  The X-ray detection is only at the 3$\sigma$ level and little can be said about the distribution of the X-ray emission. Intermediate densities are indicated by existing spectroscopy.

\item
L10-101=G98-77    (Fig.\  \ref{fig_atlas25})   %   (fig\_Atlas\_GKL77.eps)

This ill-defined SNR candidate consists of two groupings of faint irregular filaments that are plausibly portions of the same shell, with the brighter filaments aligning with the eastern side and some fainter emission in the north.  An ellipse is fit under these assumptions. No appreciable \oiii\ emission is seen.  Another SNR candidate, G98-75, is located on the top edge of the frame shown.  No optical spectrum is available.  The object does appear to have been detected in X-rays.

\item
L10-102=G98-80   (Fig.\  \ref{fig_atlas25})    %   (fig\_Atlas\_GKL80.eps)

G98-80 is an ill-defined collection of knotty filaments within a fairly confused region of stars and extended emission. Two separate groupings of filaments more or less define the upper half of a shell, which we use to define the region of interest.  Existing optical spectra show \sii:\HA\ = 0.46, which may be a lower limit given the overlying emission, and low densities.  No X-ray detection is claimed for this object.

\item
L10-103=G98-79  (Fig.\  \ref{fig_atlas25})    %   (fig\_Atlas\_GKL79.eps)

G98-79 is a faint, shell-like SNR located just to the southeast of a small H~II region.  It is visible faintly in all three optical bands, looking somewhat like a $\theta$ in \HA\ and [S~II]; the [O~III] emission is very faint and somewhat enhanced on the southern side, but the shell is otherwise similar in extent.  The X-ray emission seems to be center-filled, based on inspection of the X-ray image, but the detection is less than 2$\sigma$ level. Existing optical spectra indicate densities somewhat above the low density limit.

\item
L10-104=G98-81 (Fig.\  \ref{fig_atlas25})    %   (fig\_Atlas\_GKL81.eps)

G98-81 is a fairly diffuse oval SNR with hints of limb-brightening in H$\alpha$ and [S~II], especially on the south side of the SNR.  The [O~III] emission is very faint, but appears brighter in the northwest side. X-ray detection at just under 3$\sigma$ does not allow X-ray morphology to be determined.  The low density limit is indicated by existing optical spectra.

\item
L10-105=G98-78   (Fig.\  \ref{fig_atlas26})   %   (fig\_Atlas\_GKL78.eps)

This is a fairly well-defined shell-like SNR, with a faint protrusion to the east in \HA\ and [S~II] that gives the overall shape a somewhat triangular appearance.  The [O~III] emission comes primarily from the north and west limbs of the shell.  A stellar source is seen projected just inside the northwest limb of the shell, but is not clearly related. We have defined the extent using the main shell, ignoring the faint eastern extension.  X-ray emission seems to fill the upper two-thirds of the shell, but no enhancement corresponding to the shell itself is seen.  Low densities are indicated by the [S~II] ratio in existing spectra.

\item
L10-106   (Fig.\  \ref{fig_atlas26})    %   (fig\_Atlas\_EM50.eps)

This is an exceedingly faint shell at \HA\ and \sii\ detected as part of the Schmidt survey.  A very compact \HA\ emission source coincident with a faint star in the continuum frame is seen in projection against the shell, but is not clearly associated.  With a diameter of $\sim$70 pc and no clearly associated stars, this could be a large, old SNR that has nearly faded from visibility.  The object appears to have been detected in X-rays.  Unfortunately,  no optical spectra are available.  

\item
L10-107=G98-82  (Fig.\  \ref{fig_atlas26})  %   (fig\_Atlas\_GKL82.eps)

G98-82 is a faint, somewhat patchy, diffuse, circular SNR with a probably unrelated stellar source in projection against the northeast limb.  Its extent is well defined despite other nearby diffuse and structured emission.  The [O~III] emission is relatively strong compared with many other SNRs in M33, and the distribution of [O~III] light is fairly uniform compared with H$\alpha$.  X-rays are detected (3.1$\sigma$) and the morphology to some extent seem to be center-filled.  The low density limit is indicated by existing optical spectra.  To the south of the SNR lies a faint extended `banana' of emission with only slightly lower \sii:\HA\ ratio but whose origin is unknown.

\item
L10-108=G98-84  (Fig.\  \ref{fig_atlas26})     %   (fig\_Atlas\_GKL84.eps)

G98-84 is a peculiar, large collection of filaments which we interpret as being a partial shell of emission.  The filaments are part of a much larger network of filamentary and diffuse emission that is only slightly lower in \sii:\HA\ ratio.  Some of the network of filaments shows up faintly in \oiii\ as well. The region defined for the object is 60 $\times$ 100 pc and a handful of stars are seen projected within, raising the possibility of a superbubble interpretation.  Two compact, high surface brightness nebulae are also seen in the \HA\ frame but are not related.  A longslit MMT-BCS spectrum (this paper) confirms the enhanced \sii:\HA\ ratio and indicates densities in the low density limit.  No X-ray detection is claimed for this object.

\item
L10-109=G98-83  (Fig.\  \ref{fig_atlas27})    %    (fig\_Atlas\_GKL83.eps)

This candidate corresponds to a partial shell of raised \sii:\HA\ emission within a confused region of other emission and H~II regions to the north.  From careful inspection of the \sii\ frame, we identify a partial oval shell and define the region based on this.  There is \oiii\ emission in the region, but it is not clearly associated with the \HA-\sii\ shell.  A star is projected onto the northern limb and one interior, but their association is unclear.  A new MMT-BCS spectrum shows a \sii:\HA\ ratio of 0.28, but confusion from overlying emission makes this uncertain.   The density indicated is slightly above the low density limit.  No X-ray detection is claimed for this object.

\item
L10-110  (Fig.\  \ref{fig_atlas27})     %   (fig\_Atlas\_EM17.eps)

This object, which stands out as a ring on the \sii\ frame despite being very confused in the \HA\ frame, was found as part of the Schmidt survey.  Faint X-ray emission appears to have been observed, but  no optical spectra are available.  

\item
L10-111=G98-85    (Fig.\  \ref{fig_atlas27})   %   (fig\_Atlas\_GKL85.eps)

This SNR is very similar in size and morphology to G98-79 (which is $\sim$1\arcmin\ to the northwest) except that it is somewhat brighter.  While a shell is prevalent in \HA\ and [S~II],  some emission is seen in projection toward the center of the shell as well.  The [O~III] emission arises primarily from the outer shell only.  As with many other X-ray detected objects, the X-ray emission seems to be more center-filled.  Low densities are indicated by existing spectroscopy of [S~II].

\item
L10-112  (Fig.\  \ref{fig_atlas27})     %   (fig\_Atlas\_EM62.eps)

This exceedingly faint, nearly circular shell was identified in the Schmidt survey.  Such an object would be very hard to detect in a more confused region.  It is not clear that any of the stars in the region are associated directly with the nebula.  With a diameter near 90 pc, this could be an old SNR that has nearly faded beyond visibility.  No X-ray detection is claimed for this object, and no optical spectra are available. 

\item
L10-113=G98-86   (Fig.\  \ref{fig_atlas28})    %   (fig\_Atlas\_GKL86.eps)

This SNR is the limb-brightened ring that stands out best on the \sii\ image, despite the significant confusion caused by the photoionized regions directly adjacent to the southeast and southwest.  There is an enhancement in \sii\ on the northeast limb of the shell that, in principle, could be a separate partial shell SNR open on the east side, but without independent confirmation we list this as one object.  Another SNR, G98-87, is visible just 25\arcsec\ to the northeast, and much of the general emission visible is only slightly less enhanced in \sii:\HA\ than the SNR candidate(s).  A number of stars are seen projected in and near the SNRs.  Existing spectra confirm the elevated \sii:\HA\ ratio and show densities near the low density limit. While some X-ray emission appears to align with the northeast limb, it is  statistically significant only at the 2.5$\sigma$ level.

\item
L10-114=G98-87  (Fig.\  \ref{fig_atlas28})    %   (fig\_Atlas\_GKL87.eps)

This object is a moderately bright but patchy and ill-defined SNR embedded in a region of extensive diffuse but structured emission.  Two arcs looking somewhat like crab pincers extend eastward from the brightest section and appear to be part of the SNR.  [O~III] emission is present and fairly prominent.  The source is detected in X-rays but the morphology is  ill-defined. Another SNR, G98-86, is adjacent 25\arcsec\ to the southwest and also has a poorly defined morphology (but is apparently not detected at X-ray wavelengths). Despite the moderately high surface brightness, previous optical spectra indicate low densities for this object.

\item
L10-115=G98-88   (Fig.\  \ref{fig_atlas28})    %   (fig\_Atlas\_GKL88.eps)

G98-88 is a faint, fairly diffuse shell SNR, seen mainly in \HA\ and [S~II].  A faint extension from the shell to the southeast gives the impression of a possible break-out in this quadrant.  The X-ray emission is detected at the 2.5$\sigma$ level and seems to fill the shell.  No optical spectrum is available for this SNR.

\item
L10-116=XMM270   (Fig.\  \ref{fig_atlas28})    %  (fig\_Atlas\_XMM270.eps)

This extended but ill-defined X-ray source corresponds with a faint, unresolved knot of H$\alpha$ and [S~II] emission that would not likely have been identified as interesting without the X-ray data pointing the way.  Indeed, the optical emission is directly adjacent to a faint star, making it difficult to confirm from imagery alone.  The optical knot may just be a bright spot on the southeastern edge of the X-ray source centered to the northwest. The extent chosen is based on the X-ray data.  A new MMT-BCS longslit spectrum reported in this paper confirms the shock heated nature of the knot but the densities are in the low density limit.

%\item
%L10-xxx  (EM16)    (fig\_Atlas\_EM16.eps)
%
%This object is a patchy, thick shell filled with diffuse emission, with one moderately bright star or very tight association almost centrally located.  It was located based on our Schmidt survey.  The  [O~III] emission (not shown here) is relatively strong and somewhat more centrally condensed that the H$\alpha$.  From imagery, the \sii:\HA\ is only marginally enhanced, ranging from 0.26-0.43.  With a size of order 110 pc, a superbubble interpretation appears to be indicated.  No X-ray detection is claimed for this object, and no optical spectra are available. 
% Object removed 8/11/09 due to low SII:Ha  WPB.

\item
L10-117=G98-89  (Fig.\  \ref{fig_atlas29})     %  (fig\_Atlas\_GKL89.eps)

This is a large, diffuse but limb-brightened SNR with an H~II region directly to the southwest.  [O~III] emission is nearly non-existent, and a few stars are seen in projection toward the shell, but their relationship to the object is unclear. X-rays are detected and seem to imply a center-filled morphology.  Densities at the low density limit are indicated by [S~II] in the MMT-BCS spectrum reported here.  With a diameter near 70 pc, this is either a very old SNR or a superbubble.

\item
L10-118=G98-90  (Fig.\  \ref{fig_atlas29})     %  (fig\_Atlas\_GKL90.eps)

This is a very faint, oblong, shell-like SNR brightest in the south.  [O~III] is barely visible. The X-ray detection is above 3$\sigma$ and the emission seems to arise in the interior of the shell. The low density limit is indicated by existing optical spectra.

\item
L10-119=FL312   (Fig.\  \ref{fig_atlas29})   %    (fig\_Atlas\_FL312.eps)

This exceedingly faint emission nebula was identified as possibly being associated with the soft, extended X-ray source FL312 as part of our soft-source search. It is too faint optically to determine a reliable \sii:\HA\ ratio from imaging and no spectrum exists.  The X-ray emission is roughly coincident with the optical nebula, but there is no sense that it fills the shell or is necessarily a direct counterpart.  It is not outside the realm of possibility that this source represents a Balmer-dominated SNR in M33, but its true character (and the association of optical and X-ray emission) remains uncertain.

\item
L10-120=G98-91 (Fig.\  \ref{fig_atlas29})    %   (fig\_Atlas\_GKL91.eps)

This patch of enhanced \sii:\HA\ emission is a less extreme version of the phenomenon seen in [G98-41 and G98-44] and described above.  G98-91 is even fainter and more diffuse than these other objects, although the enhanced [S~II] region appears to be connected with more extensive and diffuse emission on the H$\alpha$ image.  No stars appear to be directly associated with the emission, and no fainter shell-like structure appears to be associated.  Existing spectra appear to confirm the candidate as an SNR and show a density very near the low density limit of the \sii\ ratio.  The nature of this nebula remains uncertain.  The \chase\ observations yield an upper limit to the X-ray luminosity of \EXPU{2.9}{34}{\LUM}.

\item
L10-121=G98-92  (Fig.\  \ref{fig_atlas30})    %   (fig\_Atlas\_GKL90.eps)

This moderately large, broken shell SNR is brightest on the west side in \HA\ and \sii.  It is visible in [O~III] as well, although the detailed [O~III] morphology is somewhat different and the shell is more uniform and complete.  The X-ray detection is of low significance (2$\sigma$) but the emission arises within the shell.  Another SNR, G98-93, lies 30\arcsec\ to the southeast, and is not detected in X-rays.  Densities close to the low density limit are indicated by [S~II] in existing spectra.

\item
L10-122 (Fig.\  \ref{fig_atlas30})   %  (fig\_Atlas\_EM34.eps)

This object is only peculiar due to its huge size and oblong shape (~165 x 85 pc).  The observed  \sii:\HA\  = 0.65 - 0.72 easily qualifies the object as being dominated by shocks. This remarkable object was found as part of our Schmidt survey, and is located in the far-northern region of the galaxy. Unlike many of the Schmidt survey objects, this one is fairly bright, appearing as a huge, elongated but complete limb-brightened shell in H$\alpha$.  [O~III] emission is also prominent and looks very similar to H$\alpha$ except for being brighter along the southeastern and southern rim regions. [S~II] is enhanced in selected regions around the shell, especially in the east, north, and upper west sides.  Filaments to the west of the main shell may also have elevated [S~II], but the association of these filaments with the main shell is not entirely clear.  While some stars are seen in projection against this large shell, they are not centrally concentrated like they are in several nearby H~II regions, and their association with the nebula is unclear.  The X-ray detection is marginally significant at 2.3$\sigma$.  Because of the huge size, a fossil superbubble would seem to be indicated.

\item
L10-123=G98-93  (Fig.\  \ref{fig_atlas30})     %  (fig\_Atlas\_GKL93.eps)

This SNR is a diffuse, oval emission nebula just north of a small, compact H~II region.  A hint of limb brightening on the west side is evident, especially in \sii.  Some [O~III] emission is present from the northern third of the object.  No X-ray detection is claimed for this object, and no optical spectra are available. 

\item
L10-124=G98-94 (Fig.\  \ref{fig_atlas30})    % (fig\_Atlas\_GKL94.eps)

G98-94 appears as two bright optical H$\alpha$ and [S~II] knots in the southern extended emission related to the giant H~II region NGC~604.  Only the southeastern of the two knots has associated [O~III] emission.  It is not entirely clear whether this is a single SNR or a double object similar to DEM L316A+B in the outskirts of 30 Doradus in the LMC ((Williams \& Chu 2005).  We have defined a single extraction region,  which appears to be filled with fairly bright X-ray emission, and treat this as a single SNR. Earlier optical spectra show intermediate densities are indicated.  \cite{blair88} found velocities in excess of 220 km $\rm s^{-1}$ for this object.

\item
L10-125    (Fig.\  \ref{fig_atlas31})  %   (fig\_Atlas\_Kip-E.eps)

This X-ray source (3.7$\sigma$ detection) was identified as part of the `extended X-ray source' analysis.  Comparing the position against the optical data revealed a large, very faint, patchy partial shell of emission (brighter on the east side) that was present in H$\alpha$ and [S~II].  Faint [O~III] emission (near the level of detectability in the current optical data) is also present. A small H~II region just south and slightly west of the shell is not obviously related to the SNR candidate.  X-ray morphology appears to be center-filled. No optical spectrum is yet available. 

\item
L10-126=G98-95    (Fig.\  \ref{fig_atlas31})  %   (fig\_Atlas\_GKL95.eps)

G98-95 is a faint, diffuse SNR with a modest brightening on the west side.  In [O~III], the bright spot seems to show a small, half shell that would be indicative of a much smaller object than indicated by the fainter diffuse emission.  (The same structure is hinted at in the \HA\ image.)   No X-ray detection is claimed for this object, and no optical spectra are available. 

\item
L10-127=G98-96  (Fig.\  \ref{fig_atlas31})    %   (fig\_Atlas\_GKL96.eps)

G98-96 is a large, faint, limb-brightened shell that is brightest on its southwest side.  No [O~III] emission is visible.  While a few stars are projected within the shell, their association with the nebula is not obvious. No X-ray detection is claimed for this object, and no optical spectra are available. Another new candidate found in the Schmidt survey, L10-131, is located just to the southeast of the field shown.

\item
L10-129=G98-97A\footnote{Note that although we designate the larger nebula G98-97A and discuss it first, it has a higher right ascension and therefore is L10-129. G98-97B is L10-128.} (Fig.\  \ref{fig_atlas31})    %   (fig\_Atlas\_GKL97A.eps)

The object listed as G98-97 in Gordon et al. (1998) has been broken into two separate SNR listings, based on a careful assessment of the optical imagery.  G98-97A is a large north-south oval of bright H$\alpha$ and [S~II] emission.  The east and west sides of the oval are even more prominent in [O~III]. X-ray emission is well-detected and seems to fill the shell. \cite{blair88} found velocities of nearly 210 km $\rm s^{-1}$ for this object. Existing optical data show densities somewhat above the low density limit.

\item
L10-128=G98-97B  (Fig.\  \ref{fig_atlas31})   %   (fig\_Atlas\_GKL97A.eps)

Directly adjacent to the northwest of G98-97A we have identified G98-97B as a separate complete shell, either overlapping in projection or possibly physically associated.   G98-97B is about half the size of G98-97A, but its [O~III] emission is considerably fainter, and the X-ray detection is clear but of lower significance than for G98-97A. Our new hectospec data on this object confirm elevated [S~II] emission and show densities in the low density limit. 

\item
L10-130   (Fig.\  \ref{fig_atlas32})     %  (fig\_Atlas\_WPB1.eps)

This newly detected object appears as a partial shell in \HA\ and \sii\, open to the north.  In [O~III], the full shell is seen, filling the region shown in the figure.  The object is located about 25\arcsec\ north of the G98-97AB pair.
No X-ray detection is claimed for this object, and no optical spectra are available. 

\item
L10-131  (Fig.\  \ref{fig_atlas32})     %  (fig\_Atlas\_EM36.eps)

This is a huge (120 $\times$ 200 pc), oblong nebula found as part of the Schmidt survey.  Its size and central clustering of stars makes a superbubble origin likely. No X-ray detection is claimed for this object, and no optical spectra are available.  Source L10-127 is in the northwest corner of the frame shown. 

\item
L10-132=G98-98  (Fig.\  \ref{fig_atlas32})  %   (fig\_Atlas\_GKL98.eps)

This is a clumpy but faint and diffuse emission nebula with a clustering of moderately bright stars seen in projection.  While Gordon et al.'s (1998) assessment was that the \sii:\HA\ ratio was above 0.4, our image assessment shows a somewhat lower value of 0.30 - 0.39.  If stellar wind shocks or SNe are involved, this activity has not been successful in organizing the emission into a shell-like structure.  It seems likely to us that this emission is associated with the stars in the region and is not clearly a SNR.  No optical spectra are available and only an X-ray upper limit is available from our survey.

\item
L10-133  (Fig.\  \ref{fig_atlas32})  %  (fig\_Atlas\_EM52.eps)

This is a moderately bright $\sim$80 pc shell found in our Schmidt survey.  There is a larger faint H$\alpha$ halo just visible on the \HA\ image.  There is a bright, compact emission region near center of shell, and several bright stellar condensations within the shell (and likely associated with it).  Upon careful assessment, the \sii:\HA\ ratio is marginally too low (0.35-0.38).  A superbubble is the preferred classification for this nebula.  No optical spectra are available and only an X-ray upper limit is available from our survey.

\item
L10-134  (Fig.\  \ref{fig_atlas33})    %  (fig\_Atlas\_EM48.eps)

This faint, oval complete shell with interior diffuse optical emission was found in our Schmidt survey.  Very faint [O~III] emission is present as well.  Unlike some of the other Schmidt survey objects, this one does not appear to have associated stellar sources.  Although it is fairly large in physical size ($\sim$60 pc), an old, large SNR identification is likely.  The X-ray detection is only marginally significant at  2.5$\sigma$.  No optical spectra are available.

\item
L10-135  (Fig.\  \ref{fig_atlas33})     %  (fig\_Atlas\_EM24.eps)

This is a faint, somewhat irregular shell found in our Schmidt survey, in a region contaminated with other extended emission.  Our assessment of the optical data indicate a shell-like structure with enhanced \sii:\HA\ that we identify as the candidate.  A faint arc on the east side in \HA\ may demarcate a break-out on that side of the shell, but the \sii\ image is too faint to discern clearly.   No optical spectra are available and only an X-ray upper limit is available from our survey.

\item
L10-136 (Fig.\  \ref{fig_atlas33})   %  (fig\_Atlas\_EM26.eps)

This large shell ($\sim$130 pc) was identified in our Schmidt survey as possibly having enhanced \sii:\HA\ emission.  A careful assessment indicates a somewhat low ratio of 0.30-0.37.  A single bright star or tight cluster  is projected within the large shell, and  a smaller interior shell may be indicated in the H$\alpha$ image.  A superbubble interpretation seems likely for this object.     No optical spectra are available and only an X-ray upper limit is available from our survey.

\item
L10-137  (Fig.\  \ref{fig_atlas33})     %  (fig\_Atlas\_EM25.eps)

This large, faint optical shell was identified in our Schmidt survey, and is located on the far eastern edge of the survey region.  
This object is very similar to L10-136.  In this case, the shell is even more uniform and has a sharper limb, and a couple of bright stars or tight associations seem clearly associated.  The \sii:\HA\ ratio seems genuinely enhanced at 0.58 - 0.71, but the physical size is large ($\sim$130 pc).
Spectroscopy with the MMT-BCS reported in this paper confirms the significantly enhanced ratio for the nebula, but the presence of several stars near the center of the shell and the large physical size make a superbubble interpretation likely, similar to the object N70 in the Large Magellanic Cloud (see Danforth \& Blair 2006, and references therein).  The enhanced \sii:\HA\ ratio makes it likely that SNR shocks in addition to stellar winds are responsible for the excitation of the nebula.  While some faint, distributed X-ray emission may be present, the bulk of the X-rays arise in a small, central region adjacent to the brightest stars. A hint of [O~III] emission is present, but only on the southeast side of the shell.

\end{itemize}

\clearpage

%\input{tab_obs.tex}

%\input{table_obs.tex}
%\begin{center}
\begin{deluxetable}{ccccc}
\tablecaption{The \chase\ Observation Log }
\tablehead{\colhead{Field} & 
 \colhead{RA} & 
 \colhead{DEC} & 
 \colhead{Roll~Ang.} & 
 \colhead{Exposure} 
\\
\colhead{~} & 
 \colhead{(J2000)} & 
 \colhead{(J2000)} & 
 \colhead{($\deg$)} & 
 \colhead{(ksec)} 
}
\tabletypesize{\scriptsize}
\tablewidth{0pt}\startdata
F1e1 &  01:33:51.15 &  +30:39:20.5 &  308.5 &  ~93.1 \\ 
F1e2 &  01:33:50.18 &  +30:39:51.4 &  142.0 &  ~92.1 \\ 
F2e1 &  01:34:13.20 &  +30:48:03.0 &  140.2 &  ~94.3 \\ 
F2e2 &  01:34:13.46 &  +30:48:04.3 &  127.9 &  ~98.2 \\ 
F3e1 &  01:33:33.30 &  +30:48:55.6 &  140.2 &  ~89.4 \\ 
F3e2 &  01:33:33.47 &  +30:48:56.5 &  132.2 &  ~98.4 \\ 
F4e1 &  01:33:08.20 &  +30:40:10.5 &  262.2 &  ~96.7 \\ 
F4e2 &  01:33:09.20 &  +30:40:41.1 &  ~97.8 &  ~90.7 \\ 
F5e1 &  01:33:27.20 &  +30:31:39.4 &  145.7 &  100.5 \\ 
F5e2 &  01:33:27.40 &  +30:31:40.7 &  135.7 &  ~89.2 \\ 
F6e1 &  01:34:06.48 &  +30:30:26.6 &  224.2 &  ~86.4 \\ 
F6e2 &  01:34:07.92 &  +30:30:49.3 &  103.2 &  ~98.7 \\ 
F7e1 &  01:34:33.54 &  +30:39:00.4 &  ~94.9 &  ~88.5 \\ 
F7e2 &  01:34:32.55 &  +30:38:29.6 &  265.7 &  ~95.6 \\ 
1730 &  01:33:50.91 &  +30:39:54.3 &  108.6 &  ~46.7 \\ 
2023 &  01:34:34.63 &  +30:47:58.2 &  106.6 &  ~88.3 \\ 
7208 &  01:34:06.99 &  +30:30:18.9 &  259.4 &  ~11.5 \\ 
\enddata 
\label{table_obs}
\end{deluxetable}
%\end{center}

%\input{table_optical.tex}
\center{
\begin{deluxetable}{lccccccccl}
\tabletypesize{\scriptsize}
\tablewidth{0pt}
\tablecaption{Emission-Line Imaging Surveys of M33 from KPNO}
\tablehead{
\colhead{} & \colhead{} & \colhead{} & \colhead{Scale} & \colhead{Line or}& \multicolumn{2}{c}{Filter} &  \multicolumn{2}{c}{Exposure} &
 \colhead{} \\
%\\
%\cline{4-5} \\ 
\colhead{Telescope}   & \colhead{CCD}   &  \colhead{Field} & 
\colhead{($\arcsec\;{\rm pixel}^{-1}$)} &
\colhead{Band}& \colhead{$\lambda_0$\ (\AA)} &\colhead{$\Delta\lambda$\ (\AA)}  & Frames & 
\colhead{Total (s)} & \colhead{Date (U.T.)}
%\colhead{$\Theta$\tablenotemark{b}}
}
\startdata
0.9m Burrell Schmidt &S2KA & $61\arcmin \times 61\arcmin $ & 1.8 & [O~III] & 5010  & 40 & 8 &
12,000  & 1996 Jan 15-16 \\
 & &  &  & Green & 5125  & 92 & 8 &
5,600  & 1996 Jan 15-16 \\
 &  & &  & \HA  & 6567  & 29 & 6 &
5,100  & 1996 Oct 03-11 \\
& &  &  &[S~II] & 6717  & 41 & 6 &
5,700  & 1996 Oct 03-11 \\
 & &  &  & Red & 6852  & 95 & 5 &
2,700  & 1996 Oct 03-11 \\
\\
4m Mayall\tablenotemark{a} &Mosaic & $36\arcmin \times 36\arcmin $ & 0.27 & [O~III] & 5025  & 56 & 5 & 1,500  & 2000 Oct 04-05 \\ 
 & & &  & V & 5387  & 948 & 5 & 300  & 2000 Oct 04-05 \\
& &  &  & \HA  & 6575  & 81 & 5 & 1,500  & 2000 Oct 04-05 \\
& &  &  & [S~II] & 6731  & 81 & 5 & 1,500  & 2000 Oct 04-05 \\
 & & &  & R & 6514  & 1511 & 5 & 250  & 2000 Oct 04-05 \\
\enddata

%% Text for table notes should follow after the \enddata but before
%% the \end{deluxetable}. Make sure there is at least one \tablenotemark
%% in the table for each \tablenotetext.

\tablenotetext{a}{{Local Group Galaxies Survey, see Massey et al. (2006), and http://www.lowell.edu/users/massey/lgsurvey.html for further information.  Exposure is for each of three overlapping LGGS fields.}}
\label{table_optical}
\end{deluxetable}
}
\clearpage

%\input{table_sample.tex}
%\begin{center}
\begin{deluxetable}{rrccrccccccccl}
\tablecaption{M33 SNRs and SNR Candidates -- Basic Data }
\tablehead{\colhead{Name} & 
 \colhead{Other$^{*}$} & 
 \colhead{RA} & 
 \colhead{DEC} & 
 \colhead{Dia.} & 
 \colhead{Morph.} & 
 \colhead{Env.} & 
 \colhead{Radio$^{**}$} & 
 \colhead{Sur.~Bright$_{H\alpha}^{\dag}$} & 
 \colhead{L$_{H\alpha}$} & 
 \colhead{[N~II]:H$\alpha$} & 
 \colhead{[S~II]:H$\alpha$} & 
 \colhead{[S~II]-rat.} & 
 \colhead{Spec.~Ref.} 
\\
\colhead{~} & 
 \colhead{~} & 
 \colhead{(J2000)} & 
 \colhead{(J2000)} & 
 \colhead{(pc)} & 
 \colhead{~} & 
 \colhead{~} & 
 \colhead{~} & 
 \colhead{~} & 
 \colhead{(ergs~s$^{-1})$} & 
 \colhead{~} & 
 \colhead{~} & 
 \colhead{~} & 
 \colhead{~} 
}
\tabletypesize{\scriptsize}
\tablewidth{0pt}\rotate\startdata
L10-001 &  G98-01 &  01:32:30.37 &  +30:27:46.9 &  123 &  A &  2 &  y &  1.6e-16 &  9.6e+36 &  0.18 &  0.77 &  1.30 &  G98 \\ 
L10-002 &  G98-02 &  01:32:31.41 &  +30:35:32.9 &  29 &  C &  3 &  y &  4.6e-16 &  1.6e+36 &  0.26 &  0.44 &  1.34 &  G98 \\ 
L10-003 &  G98-03 &  01:32:42.54 &  +30:20:58.9 &  100 &  A &  2 &  y &  2.0e-16 &  8.0e+36 &  0.22 &  0.55 &  1.47 &  G98 \\ 
L10-004 &  G98-04 &  01:32:44.83 &  +30:22:14.6 &  39 &  A &  1 &  n &  1.5e-16 &  9.3e+35 &  - &  - &  - &  - \\ 
L10-005 &  XMM068 &  01:32:46.73 &  +30:34:37.8 &  45 &  A &  1 &  n &  1.5e-16 &  1.2e+36 &  0.62 &  0.78 &  1.45 &  MMT-BCS$^{ac}$ \\ 
L10-006 &  G98-05 &  01:32:52.76 &  +30:38:12.6 &  56 &  A &  1 &  ? &  1.9e-16 &  2.4e+36 &  0.29 &  0.55 &  1.48 &  S93 \\ 
L10-007 &  (x-ray) &  01:32:53.36 &  +30:48:23.1 &  73 &  A &  1 &  n &  3.6e-17 &  7.6e+35 &  - &  - &  - &  - \\ 
L10-008 &  G98-06 &  01:32:53.40 &  +30:37:56.9 &  51 &  A &  2 &  n &  4.2e-16 &  4.5e+36 &  0.21 &  0.61 &  1.44 &  S93 \\ 
L10-009 &  G98-07 &  01:32:54.10 &  +30:25:31.8 &  39 &  C &  1 &  n &  2.5e-16 &  1.5e+36 &  - &  - &  - &  - \\ 
L10-010 &  G98-08 &  01:32:56.15 &  +30:40:36.4 &  93 &  A &  1 &  y &  1.9e-16 &  6.7e+36 &  0.21 &  0.81 &  1.47 &  S93 \\ 
L10-011 &  G98-09 &  01:32:57.07 &  +30:39:27.1 &  20 &  A$^\prime$  &  2 &  c &  1.8e-15 &  2.9e+36 &  0.28 &  0.84 &  1.11 &  hecto \\ 
L10-012 &  G98-10 &  01:33:00.15 &  +30:30:46.2 &  52 &  B &  3 &  n &  2.5e-15 &  2.7e+37 &  - &  - &  - &  - \\ 
L10-013 &  G98-11 &  01:33:00.42 &  +30:44:08.1 &  33 &  C &  1 &  c &  1.2e-16 &  5.3e+35 &  <0.27 &  0.47 &  1.11 &  G98 \\ 
L10-014 &  G98-12 &  01:33:00.67 &  +30:30:59.3 &  46 &  B &  3 &  n &  1.9e-15 &  1.6e+37 &  - &  - &  - &  - \\ 
L10-015 &  G98-13 &  01:33:01.51 &  +30:30:49.6 &  28 &  B &  3 &  n &  6.7e-16 &  2.1e+36 &  - &  - &  - &  - \\ 
L10-016 &  (opt) &  01:33:02.93 &  +30:32:29.6 &  51 &  A &  2 &  n &  2.4e-16 &  2.5e+36 &  0.37 &  0.97 &  1.39 &  MMT-BCS$^{a}$ \\ 
L10-017 &  G98-14 &  01:33:03.57 &  +30:31:20.9 &  33 &  A &  1 &  y &  3.3e-16 &  1.4e+36 &  0.34 &  1.07 &  1.43 &  G98 \\ 
L10-018 &  G98-15 &  01:33:04.03 &  +30:39:53.7 &  30 &  A &  1 &  y &  1.2e-15 &  4.4e+36 &  0.30 &  0.49 &  1.73 &  S93 \\ 
L10-019 &  G98-16 &  01:33:07.55 &  +30:42:52.5 &  71 &  A &  2 &  n &  3.3e-16 &  6.6e+36 &  - &  - &  - &  - \\ 
L10-020 &  (opt) &  01:33:08.98 &  +30:26:58.9 &  51 &  A &  2 &  n &  1.7e-16 &  1.8e+36 &  - &  - &  - &  - \\ 
L10-021 &  G98-17 &  01:33:09.87 &  +30:39:34.9 &  67 &  A &  3 &  y &  2.1e-16 &  3.8e+36 &  0.38 &  0.66 &  1.38 &  S93 \\ 
L10-022 &  G98-18 &  01:33:10.18 &  +30:42:22.0 &  27 &  B &  1 &  n &  2.0e-16 &  6.0e+35 &  0.39 &  0.86 &  1.36 &  hecto \\ 
L10-023 &  G98-20 &  01:33:11.10 &  +30:39:43.7 &  25 &  A$^\prime$  &  2 &  c &  3.3e-16 &  8.5e+35 &  0.36 &  0.78 &  1.11 &  hecto \\ 
L10-024 &  G98-19 &  01:33:11.28 &  +30:34:23.5 &  99 &  A &  2 &  y &  3.0e-16 &  1.2e+37 &  0.32 &  0.49 &  1.53 &  S93 \\ 
L10-025 &  G98-21 &  01:33:11.76 &  +30:38:41.5 &  25 &  A &  3 &  c &  8.2e-16 &  2.0e+36 &  0.21 &  0.55 &  1.39 &  MMT-BCS \\ 
L10-026 &  (x-ray) &  01:33:16.73 &  +30:46:10.3 &  73 &  B &  2 &  n &  2.5e-16 &  5.2e+36 &  - &  - &  - &  - \\ 
L10-027 &  G98-22 &  01:33:17.44 &  +30:31:28.5 &  44 &  C &  2 &  y &  1.3e-16 &  1.0e+36 &  0.50 &  0.91 &  1.46 &  G98 \\ 
L10-028 &  (opt) &  01:33:18.80 &  +30:27:04.4 &  179 &  A &  1 &  n &  1.2e-16 &  1.5e+37 &  - &  - &  - &  - \\ 
L10-029 &  (opt) &  01:33:18.94 &  +30:46:51.9 &  66 &  A &  1 &  n &  9.6e-17 &  1.6e+36 &  - &  - &  - &  - \\ 
L10-030 &  G98-23 &  01:33:21.64 &  +30:31:31.1 &  76 &  A &  1 &  n &  2.0e-16 &  4.7e+36 &  - &  - &  - &  - \\ 
L10-031 &  G98-24 &  01:33:22.67 &  +30:27:04.0 &  20 &  A &  1 &  c &  4.4e-16 &  6.9e+35 &  0.41 &  0.95 &  1.39 &  hecto \\ 
L10-032 &  G98-25 &  01:33:23.85 &  +30:26:13.5 &  21 &  A &  1 &  c &  9.6e-16 &  1.8e+36 &  0.51 &  1.06 &  1.31 &  BK85 \\ 
L10-033 &  G98-26 &  01:33:27.07 &  +30:47:48.6 &  67 &  C &  2 &  y &  1.3e-16 &  2.4e+36 &  0.27 &  0.57 &  1.46 &  MMT-BCS \\ 
L10-034 &  G98-27 &  01:33:28.08 &  +30:31:35.0 &  32 &  B &  2 &  c &  5.7e-16 &  2.4e+36 &  0.41 &  0.55 &  1.41 &  G98 \\ 
L10-035 &  XMM156 &  01:33:28.96 &  +30:47:43.5 &  19 &  C &  2 &  n &  4.1e-16 &  5.8e+35 &  0.20 &  0.36 &  1.52 &  MMT-BCS \\ 
L10-036 &  G98-28 &  01:33:29.05 &  +30:42:17.0 &  18 &  A &  1 &  c &  5.0e-15 &  6.5e+36 &  0.59 &  1.13 &  1.11 &  S93 \\ 
L10-037 &  G98-29 &  01:33:29.45 &  +30:49:10.8 &  32 &  B &  1 &  y &  1.8e-16 &  7.3e+35 &  0.39 &  0.75 &  1.27 &  G98 \\ 
L10-038 &  G98-30 &  01:33:30.21 &  +30:47:43.8 &  51 &  C &  3 &  ? &  2.6e-16 &  2.7e+36 &  0.25 &  0.31 &  1.58 &  MMT-BCS \\ 
L10-039 &  G98-31 &  01:33:31.25 &  +30:33:33.4 &  13 &  A &  1 &  c &  1.5e-14 &  9.6e+36 &  0.86 &  0.95 &  0.78 &  MMT-BCS \\ 
L10-040 &  G98-32 &  01:33:31.34 &  +30:42:18.3 &  55 &  A &  2 &  n &  2.4e-16 &  2.8e+36 &  0.26 &  0.65 &  1.61 &  S93 \\ 
L10-041 &  (opt) &  01:33:31.80 &  +30:31:01.1 &  97 &  A &  1 &  n &  4.5e-17 &  1.7e+36 &  - &  - &  - &  - \\ 
L10-042 &  G98-33 &  01:33:35.14 &  +30:23:07.5 &  43 &  A &  1 &  n &  1.3e-16 &  9.6e+35 &  - &  - &  - &  - \\ 
L10-043 &  (x-ray) &  01:33:35.39 &  +30:42:32.1 &  85 &  C &  3 &  n &  1.6e-16 &  4.6e+36 &  - &  - &  - &  - \\ 
L10-044 &  G98-34 &  01:33:35.61 &  +30:49:23.0 &  29 &  A &  1 &  n &  3.3e-16 &  1.1e+36 &  0.37 &  1.01 &  1.43 &  hecto$^a$ \\ 
L10-045 &  G98-35 &  01:33:35.90 &  +30:36:27.4 &  30 &  A$^\prime$  &  2 &  c &  5.8e-15 &  2.0e+37 &  0.57 &  0.82 &  1.10 &  MMT-BCS \\ 
L10-046 &  G98-36 &  01:33:37.09 &  +30:32:53.5 &  39 &  A &  1 &  n &  2.2e-16 &  1.3e+36 &  0.48 &  1.01 &  1.39 &  MMT-BCS \\ 
L10-047 &  G98-37 &  01:33:37.75 &  +30:40:09.2 &  50 &  A &  1 &  n &  1.1e-16 &  1.1e+36 &  0.62 &  1.11 &  1.43 &  S93 \\ 
L10-048 &  G98-38 &  01:33:38.01 &  +30:42:18.2 &  21 &  A &  2 &  n &  5.3e-16 &  9.7e+35 &  0.40 &  0.66 &  1.40 &  S93 \\ 
L10-049 &  G98-39 &  01:33:40.66 &  +30:39:40.8 &  43 &  A &  2 &  n &  1.9e-16 &  1.4e+36 &  - &  - &  - &  - \\ 
L10-050 &  G98-41 &  01:33:40.73 &  +30:42:35.7 &  71 &  C &  2 &  n &  2.8e-16 &  5.6e+36 &  0.44 &  0.61 &  1.44 &  S93 \\ 
L10-051 &  G98-40 &  01:33:40.87 &  +30:52:13.7 &  60 &  A &  2 &  y &  1.0e-16 &  1.5e+36 &  0.31 &  0.69 &  1.37 &  G98 \\ 
L10-052 &  G98-42 &  01:33:41.30 &  +30:32:28.4 &  33 &  C &  2 &  y &  2.0e-16 &  8.9e+35 &  - &  - &  - &  - \\ 
L10-053 &  G98-43A &  01:33:41.71 &  +30:21:04.1 &  34 &  A &  2 &  y &  1.1e-15 &  5.1e+36 &  0.23 &  0.44 &  1.47 &  G98 \\ 
L10-054 &  G98-43B &  01:33:42.24 &  +30:20:57.8 &  45 &  A &  2 &  n &  4.1e-16 &  3.3e+36 &  - &  - &  - &  - \\ 
L10-055 &  G98-44 &  01:33:42.91 &  +30:41:49.5 &  44 &  C &  2 &  y &  1.6e-16 &  1.2e+36 &  - &  - &  - &  - \\ 
L10-056 &  G98-45 &  01:33:43.49 &  +30:41:03.8 &  23 &  B &  2 &  y &  3.8e-16 &  8.3e+35 &  0.54 &  0.89 &  1.45 &  G98 \\ 
L10-057 &  G98-46 &  01:33:43.70 &  +30:36:11.5 &  36 &  C &  3 &  n &  2.4e-16 &  1.2e+36 &  0.37 &  0.54 &  1.40 &  S93 \\ 
L10-058 &  (opt) &  01:33:45.26 &  +30:32:20.1 &  67 &  A &  2 &  n &  4.0e-16 &  7.2e+36 &  - &  - &  - &  - \\ 
L10-059 &  (opt) &  01:33:47.46 &  +30:39:44.7 &  42 &  A &  1 &  n &  1.7e-16 &  1.2e+36 &  - &  - &  - &  - \\ 
L10-060 &  G98-48 &  01:33:48.35 &  +30:39:28.4 &  14 &  C &  2 &  ? &  1.4e-15 &  1.0e+36 &  0.51 &  0.74 &  1.49 &  G98 \\ 
L10-061 &  G98-47 &  01:33:48.50 &  +30:33:07.9 &  60 &  B &  3 &  y &  1.1e-15 &  1.6e+37 &  0.40 &  0.74 &  1.43 &  S93 \\ 
L10-062 &  (opt) &  01:33:49.75 &  +30:30:49.7 &  73 &  A &  1 &  n &  8.9e-17 &  1.9e+36 &  - &  - &  - &  - \\ 
L10-063 &  (opt) &  01:33:49.90 &  +30:30:16.7 &  54 &  A &  1 &  n &  4.8e-17 &  5.5e+35 &  - &  - &  - &  - \\ 
L10-064 &  G98-49 &  01:33:50.12 &  +30:35:28.6 &  48 &  B &  1 &  y &  2.2e-16 &  2.0e+36 &  0.53 &  0.83 &  1.35 &  S93 \\ 
L10-065 &  G98-50 &  01:33:51.06 &  +30:43:56.2 &  50 &  B &  2 &  y &  5.4e-16 &  5.4e+36 &  0.46 &  0.63 &  1.44 &  S93 \\ 
L10-066 &  G98-52 &  01:33:51.67 &  +30:30:59.6 &  59 &  A &  1 &  y &  9.0e-17 &  1.3e+36 &  0.43 &  1.65 &  1.21 &  G98 \\ 
L10-067 &  G98-51 &  01:33:51.71 &  +30:30:43.4 &  45 &  C &  1 &  n &  1.2e-16 &  9.5e+35 &  - &  - &  - &  - \\ 
L10-068 &  (opt) &  01:33:52.15 &  +30:56:33.4 &  109 &  A &  2 &  n &  5.6e-17 &  2.7e+36 &  - &  - &  - &  - \\ 
L10-069 &  G98-53 &  01:33:54.28 &  +30:33:47.9 &  47 &  A &  1 &  y &  3.6e-16 &  3.2e+36 &  0.36 &  0.82 &  1.40 &  S93 \\ 
L10-070 &  G98-54 &  01:33:54.51 &  +30:45:18.7 &  21 &  B &  3 &  c &  2.2e-15 &  3.7e+36 &  0.47 &  0.83 &  1.41 &  G98 \\ 
L10-071 &  G98-55 &  01:33:54.91 &  +30:33:11.0 &  20 &  A &  2 &  c &  5.0e-15 &  8.1e+36 &  0.48 &  0.83 &  1.12 &  S93 \\ 
L10-072 &  (opt) &  01:33:55.01 &  +30:39:57.3 &  32 &  B &  1 &  n &  1.1e-16 &  4.5e+35 &  - &  - &  - &  - \\ 
L10-073 &  (opt) &  01:33:56.49 &  +30:21:27.0 &  57 &  A &  1 &  n &  1.7e-16 &  2.2e+36 &  - &  - &  - &  - \\ 
L10-074 &  G98-57A &  01:33:56.97 &  +30:34:58.7 &  31 &  B &  2 &  y &  3.2e-16 &  1.2e+36 &  0.35 &  0.67 &  1.39 &  MMT-BCS \\ 
L10-075 &  G98-56 &  01:33:57.13 &  +30:40:48.5 &  44 &  A &  1 &  ? &  1.3e-16 &  1.0e+36 &  0.70 &  1.16 &  1.46 &  G98 \\ 
L10-076 &  G98-57B &  01:33:57.13 &  +30:35:06.1 &  21 &  B &  2 &  n &  7.5e-16 &  1.3e+36 &  - &  - &  - &  - \\ 
L10-077 &  G98-59 &  01:33:58.06 &  +30:32:09.6 &  24 &  C &  3 &  ? &  8.6e-16 &  1.9e+36 &  0.30 &  0.40 &  1.45 &  S93 \\ 
L10-078 &  G98-58 &  01:33:58.07 &  +30:37:54.6 &  16 &  A &  1 &  n &  1.0e-15 &  1.1e+36 &  0.69 &  1.19 &  1.35 &  hecto \\ 
L10-079 &  (x-ray) &  01:33:58.15 &  +30:48:36.4 &  58 &  A &  2 &  n &  3.2e-16 &  4.3e+36 &  - &  - &  - &  - \\ 
L10-080 &  G98-60 &  01:33:58.42 &  +30:36:24.3 &  8 &  A &  1 &  n &  1.2e-15 &  2.9e+35 &  0.64 &  0.96 &  1.22 &  hecto \\ 
L10-081 &  G98-62 &  01:33:58.51 &  +30:33:32.3 &  34 &  C &  3 &  c &  3.6e-16 &  1.7e+36 &  0.44 &  0.73 &  1.44 &  S93 \\ 
L10-082 &  G98-61 &  01:33:58.52 &  +30:51:54.3 &  56 &  A &  1 &  y &  1.2e-16 &  1.5e+36 &  0.52 &  1.07 &  1.42 &  G98 \\ 
L10-083 &  G98-65 &  01:33:59.93 &  +30:34:21.2 &  31 &  B &  3 &  n &  1.4e-15 &  5.2e+36 &  0.32 &  0.63 &  1.64 &  G98 \\ 
L10-084 &  G98-64 &  01:34:00.31 &  +30:42:19.4 &  32 &  A &  1 &  c &  4.4e-16 &  1.9e+36 &  0.81 &  1.23 &  1.39 &  hecto \\ 
L10-085 &  G98-63 &  01:34:00.32 &  +30:47:24.1 &  43 &  B &  1 &  n &  1.7e-16 &  1.2e+36 &  - &  - &  - &  - \\ 
L10-086 &  FL236 &  01:34:00.60 &  +30:49:04.2 &  13 &  C &  1 &  n &  1.4e-16 &  9.5e+34 &  0.51 &  0.80 &  1.03 &  hecto$^{a}$ \\ 
L10-087 &  G98-66 &  01:34:01.34 &  +30:35:20.2 &  48 &  A &  2 &  n &  1.2e-16 &  1.1e+36 &  0.41 &  0.90 &  1.38 &  MMT-BCS$^{ab}$ \\ 
L10-088 &  G98-67 &  01:34:02.24 &  +30:31:06.8 &  60 &  A &  1 &  y &  1.2e-16 &  1.7e+36 &  0.43 &  0.96 &  1.49 &  S93 \\ 
L10-089 &  (x-ray) &  01:34:03.31 &  +30:36:22.9 &  92 &  A &  2 &  n &  3.7e-16 &  1.3e+37 &  - &  - &  - &  - \\ 
L10-090 &  G98-68 &  01:34:03.48 &  +30:44:43.8 &  42 &  A &  1 &  y &  1.6e-16 &  1.1e+36 &  0.61 &  1.03 &  1.37 &  S93 \\ 
L10-091 &  G98-69 &  01:34:04.26 &  +30:32:57.1 &  36 &  B &  1 &  y &  4.3e-17 &  2.3e+35 &  - &  - &  - &  - \\ 
L10-092 &  G98-70 &  01:34:07.23 &  +30:36:22.0 &  101 &  A &  1 &  n &  1.9e-16 &  7.7e+36 &  0.38 &  0.71 &  1.35 &  S93 \\ 
L10-093 &  FL261 &  01:34:07.50 &  +30:37:08.0 &  20 &  B &  1 &  n &  1.4e-16 &  2.1e+35 &  0.57 &  0.95 &  1.32 &  hecto$^{ab}$ \\ 
L10-094 &  XMM244 &  01:34:08.37 &  +30:46:33.2 &  20 &  C &  2 &  n &  3.3e-16 &  5.1e+35 &  0.49 &  0.75 &  1.21 &  MMT-BCS$^{ad}$ \\ 
L10-095 &  G98-71 &  01:34:10.02 &  +30:47:14.9 &  23 &  A &  1 &  n &  1.9e-16 &  4.0e+35 &  0.55 &  0.83 &  1.26 &  MMT-BCS$^{b}$ \\ 
L10-096 &  G98-73 &  01:34:10.70 &  +30:42:24.0 &  18 &  A &  1 &  c &  3.7e-15 &  5.0e+36 &  0.59 &  1.25 &  1.15 &  S93 \\ 
L10-097 &  G98-72 &  01:34:11.04 &  +30:38:59.9 &  14 &  B &  2 &  n &  3.1e-16 &  2.5e+35 &  0.65 &  1.15 &  1.27 &  G98 \\ 
L10-098 &  G98-74 &  01:34:12.69 &  +30:35:12.0 &  67 &  A &  2 &  n &  2.2e-16 &  3.8e+36 &  0.36 &  0.46 &  1.95 &  S93 \\ 
L10-099 &  G98-75 &  01:34:13.02 &  +30:48:36.1 &  51 &  C &  3 &  n &  3.0e-16 &  3.1e+36 &  - &  - &  - &  - \\ 
L10-100 &  G98-76 &  01:34:13.65 &  +30:43:27.0 &  26 &  A &  1 &  n &  1.6e-16 &  4.2e+35 &  0.51 &  0.87 &  1.25 &  S93 \\ 
L10-101 &  G98-77 &  01:34:13.71 &  +30:48:17.5 &  58 &  B &  2 &  n &  2.4e-16 &  3.2e+36 &  - &  - &  - &  - \\ 
L10-102 &  G98-80 &  01:34:14.10 &  +30:34:30.9 &  39 &  B &  3 &  n &  8.5e-16 &  5.1e+36 &  0.26 &  0.46 &  1.42 &  S93 \\ 
L10-103 &  G98-79 &  01:34:14.35 &  +30:41:53.6 &  48 &  A &  1 &  n &  1.1e-16 &  1.0e+36 &  0.61 &  1.08 &  1.34 &  S93 \\ 
L10-104 &  G98-81 &  01:34:14.38 &  +30:39:41.6 &  39 &  A &  1 &  n &  1.7e-16 &  1.0e+36 &  0.52 &  1.05 &  1.50 &  S93 \\ 
L10-105 &  G98-78 &  01:34:14.41 &  +30:53:51.9 &  50 &  A &  1 &  n &  2.8e-16 &  2.8e+36 &  0.39 &  1.02 &  1.40 &  G98 \\ 
L10-106 &  (opt) &  01:34:14.67 &  +30:31:50.9 &  66 &  A &  1 &  n &  4.9e-17 &  8.6e+35 &  - &  - &  - &  - \\ 
L10-107 &  G98-82 &  01:34:15.57 &  +30:32:59.9 &  33 &  A &  2 &  y &  2.1e-16 &  9.3e+35 &  0.44 &  0.74 &  1.50 &  S93 \\ 
L10-108 &  G98-84 &  01:34:16.31 &  +30:52:32.7 &  77 &  A &  2 &  y &  1.5e-16 &  3.6e+36 &  0.23 &  0.80 &  1.59 &  MMT-BCS$^{ab}$ \\ 
L10-109 &  G98-83 &  01:34:17.00 &  +30:51:47.1 &  25 &  B &  3 &  ? &  2.0e-15 &  5.2e+36 &  0.25 &  0.28 &  1.36 &  MMT-BCS$^{c}$ \\ 
L10-110 &  (opt) &  01:34:17.03 &  +30:33:57.7 &  54 &  B &  3 &  n &  4.4e-16 &  5.1e+36 &  - &  - &  - &  - \\ 
L10-111 &  G98-85 &  01:34:17.61 &  +30:41:23.3 &  51 &  A &  1 &  y &  2.3e-16 &  2.4e+36 &  0.58 &  1.04 &  1.47 &  S93 \\ 
L10-112 &  (opt) &  01:34:18.32 &  +30:54:05.8 &  84 &  A &  1 &  n &  4.9e-17 &  1.4e+36 &  - &  - &  - &  - \\ 
L10-113 &  G98-86 &  01:34:19.28 &  +30:33:45.9 &  38 &  A &  3 &  n &  1.1e-15 &  6.4e+36 &  0.29 &  0.62 &  1.42 &  S93 \\ 
L10-114 &  G98-87 &  01:34:19.87 &  +30:33:56.0 &  22 &  A &  2 &  n &  1.6e-15 &  3.1e+36 &  0.33 &  0.74 &  1.43 &  S93 \\ 
L10-115 &  G98-88 &  01:34:23.23 &  +30:25:24.9 &  47 &  A &  1 &  n &  9.5e-17 &  8.4e+35 &  - &  - &  - &  - \\ 
L10-116 &  XMM270 &  01:34:23.27 &  +30:54:23.9 &  42 &  A &  1 &  n &  1.4e-17 &  9.8e+34 &  0.52 &  0.86 &  1.48 &  MMT-BCS$^{a}$ \\ 
L10-117 &  G98-89 &  01:34:25.09 &  +30:54:58.1 &  66 &  A &  2 &  n &  3.5e-16 &  6.0e+36 &  0.31 &  0.78 &  1.45 &  MMT-BCS$^{ab}$ \\ 
L10-118 &  G98-90 &  01:34:25.41 &  +30:48:30.9 &  53 &  A &  1 &  n &  8.0e-17 &  8.9e+35 &  0.44 &  1.06 &  1.70 &  G98 \\ 
L10-119 &  FL312 &  01:34:25.87 &  +30:33:16.8 &  33 &  A &  1 &  n &  5.4e-17 &  2.4e+35 &  - &  - &  - &  - \\ 
L10-120 &  G98-91 &  01:34:29.61 &  +30:41:33.4 &  43 &  C &  2 &  ? &  1.6e-16 &  1.2e+36 &  0.60 &  0.82 &  1.37 &  G98 \\ 
L10-121 &  G98-92 &  01:34:30.29 &  +30:35:44.8 &  54 &  A &  1 &  y &  1.3e-16 &  1.5e+36 &  0.36 &  1.04 &  1.39 &  G98 \\ 
L10-122 &  (opt) &  01:34:31.85 &  +30:56:41.5 &  111 &  A &  1 &  n &  1.4e-16 &  6.9e+36 &  - &  - &  - &  - \\ 
L10-123 &  G98-93 &  01:34:32.41 &  +30:35:32.6 &  41 &  A &  1 &  n &  1.6e-16 &  1.0e+36 &  - &  - &  - &  - \\ 
L10-124 &  G98-94 &  01:34:33.02 &  +30:46:39.2 &  11 &  A$^\prime$  &  3 &  n &  9.4e-15 &  4.5e+36 &  0.29 &  0.66 &  0.97 &  BK85 \\ 
L10-125 &  (x-ray) &  01:34:35.41 &  +30:52:12.7 &  39 &  B &  1 &  n &  8.6e-17 &  5.1e+35 &  - &  - &  - &  - \\ 
L10-126 &  G98-95 &  01:34:36.22 &  +30:36:23.6 &  46 &  A &  1 &  n &  1.2e-16 &  1.1e+36 &  - &  - &  - &  - \\ 
L10-127 &  G98-96 &  01:34:39.00 &  +30:37:59.8 &  86 &  A &  1 &  n &  1.5e-16 &  4.3e+36 &  - &  - &  - &  - \\ 
L10-128 &  G98-97B &  01:34:40.73 &  +30:43:36.4 &  22 &  A &  2 &  n &  1.3e-15 &  2.6e+36 &  0.33 &  0.77 &  1.43 &  hecto$^{a}$ \\ 
L10-129 &  G98-97A &  01:34:41.10 &  +30:43:28.3 &  39 &  A &  2 &  y &  1.2e-15 &  7.2e+36 &  0.41 &  1.07 &  1.36 &  BK85 \\ 
L10-130 &  (opt) &  01:34:41.23 &  +30:43:55.4 &  35 &  A &  1 &  n &  7.6e-17 &  3.7e+35 &  - &  - &  - &  - \\ 
L10-131 &  (opt) &  01:34:41.89 &  +30:37:35.2 &  156 &  A &  1 &  n &  8.7e-17 &  8.4e+36 &  - &  - &  - &  - \\ 
L10-132 &  G98-98 &  01:34:44.62 &  +30:42:38.8 &  55 &  C &  1 &  n &  2.4e-16 &  2.9e+36 &  - &  - &  - &  - \\ 
L10-133 &  (opt) &  01:34:54.88 &  +30:41:17.0 &  75 &  A &  1 &  n &  3.0e-16 &  6.9e+36 &  - &  - &  - &  - \\ 
L10-134 &  (opt) &  01:34:56.44 &  +30:36:23.2 &  58 &  A &  1 &  n &  5.0e-17 &  6.8e+35 &  - &  - &  - &  - \\ 
L10-135 &  (opt) &  01:35:00.36 &  +30:40:05.0 &  65 &  A &  2 &  n &  1.3e-16 &  2.1e+36 &  - &  - &  - &  - \\ 
L10-136 &  (opt) &  01:35:01.22 &  +30:38:17.1 &  128 &  A &  1 &  n &  8.1e-17 &  5.3e+36 &  - &  - &  - &  - \\ 
L10-137 &  (opt) &  01:35:03.18 &  +30:37:09.6 &  127 &  A &  1 &  n &  7.4e-17 &  4.7e+36 &  - &  - &  - &  - \\ 
\tablenotetext{*}{ Other names or for new objects whether the object was initially suspected to be SNR based on optical or X-ray data.}
\tablenotetext{**}{ The letter ``y" implies reported by Gordon et al.\ (1999) as a radio-detected SNR; ``c" detected by Gordon et al.\ and confirmed by our new radio observations; ``?" claimed as a radio detected SNR by Gordon et al.\, but seen in our higher-resolution data as radio emission arising from a nearby HII region or point source, not the SNR; ```n" not yet detected at radio wavelengths. As discussed in the text, because our radio observations were designed for studies of small diameter SNRs, we did not expect to detect all of the SNRs detected by Gordon et al.\ (1999).}
\tablenotetext{\dag}{ H$\alpha$ surface brightness in units of ergs~cm$^{-2}$s$^{-1}$arcsec$^{-2}$; both surface brightness and $L_{H\alpha}$ include some contribution from [N~II] $\lambda\lambda$ 6548, 6583.}
\tablenotetext{a}{ First spectroscopic confirmation that [S~II]:H$\alpha$ ratio supports a classification as a SNR.}
\tablenotetext{b}{ [S~II]:H$\alpha$ varies for two spectra of separate sections; average shown.}
\tablenotetext{c}{ G98-83 has a low SII:H$\alpha$ ratio for a SNR and this makes its identification as a SNR suspect.}
\tablenotetext{d}{ [S~II]:H$\alpha$ varies for two spectra of separate sections; average shown.}
\enddata 
\label{tab_sample}
\end{deluxetable}
%\end{center}

%\input{table_results.tex}
%\begin{center}
\begin{deluxetable}{rcrccccc}
\tablecaption{M33 SNRs and SNR Candidates -- X-ray Results }
\tablehead{\colhead{Name} & 
 \colhead{Other~name$^{*}$} & 
 \colhead{Exp.} & 
 \colhead{Angle$^{a}$} & 
 \colhead{Counts} & 
 \colhead{L$_{x}$$^{b}$} & 
 \colhead{Hardness$^{c}$} 
\\
\colhead{~} & 
 \colhead{~} & 
 \colhead{(ksec)} & 
 \colhead{($\arcmin$)} & 
 \colhead{(0.35-2.0~keV)} & 
 \colhead{(0.35-2.0~keV)} & 
 \colhead{~} 
}
\tabletypesize{\scriptsize}
\tablewidth{0pt}\startdata
L10-002 &  G98-02 &  96.0 &  9.2 &  ~~~0.6$\pm$~3.8 &  $<$7.1e+34 &  - \\ 
L10-005 &  FL016 &  276.0 &  7.9 &  ~~43.8$\pm$~9.5 &  ~1.4e+35 &  -0.5 \\ 
L10-006 &  G98-05 &  187.0 &  4.1 &  ~~22.8$\pm$~7.2 &  ~9.5e+34 &  0.1 \\ 
L10-007 &  (x-ray) &  375.0 &  8.6 &  ~~28.2$\pm$10.8 &  ~6.2e+34 &  -1.0 \\ 
L10-008 &  G98-06 &  187.0 &  4.1 &  ~~-0.3$\pm$~4.6 &  $<$3.9e+34 &  - \\ 
L10-010 &  G98-08 &  187.0 &  2.7 &  ~~12.5$\pm$~6.8 &  $<$1.3e+35 &  - \\ 
L10-011 &  G98-09 &  187.0 &  2.7 &  ~~80.6$\pm$10.2 &  ~3.0e+35 &  -0.6 \\ 
L10-012 &  G98-10 &  189.0 &  5.9 &  ~~26.1$\pm$~7.6 &  ~1.2e+35 &  -0.7 \\ 
L10-013 &  G98-11 &  187.0 &  4.1 &  ~~60.6$\pm$~9.2 &  ~2.3e+35 &  -0.1 \\ 
L10-014 &  G98-12 &  189.0 &  5.8 &  ~~16.1$\pm$~6.7 &  ~6.6e+34 &  0.1 \\ 
L10-015 &  G98-13 &  189.0 &  5.6 &  ~~18.5$\pm$~6.4 &  ~7.7e+34 &  -0.1 \\ 
L10-016 &  (opt) &  377.0 &  6.6 &  ~~44.4$\pm$10.3 &  ~9.1e+34 &  -0.9 \\ 
L10-017 &  G98-14 &  189.0 &  5.1 &  ~~~5.3$\pm$~5.0 &  $<$6.4e+34 &  - \\ 
L10-018 &  G98-15 &  187.0 &  1.2 &  ~~39.9$\pm$~7.7 &  ~1.9e+35 &  -0.4 \\ 
L10-019 &  G98-16 &  187.0 &  2.4 &  ~~-2.2$\pm$~4.3 &  $<$4.4e+34 &  - \\ 
L10-020 &  (opt) &  189.0 &  6.1 &  ~~18.5$\pm$~7.0 &  ~8.0e+34 &  -0.8 \\ 
L10-021 &  G98-17 &  562.0 &  5.9 &  ~~-4.1$\pm$10.3 &  $<$2.9e+34 &  - \\ 
L10-022 &  G98-18 &  187.0 &  2.0 &  ~~44.7$\pm$~8.0 &  ~1.9e+35 &  -0.7 \\ 
L10-023 &  G98-20 &  562.0 &  6.1 &  ~427.7$\pm$22.5 &  ~6.7e+35 &  -0.6 \\ 
L10-024 &  G98-19 &  377.0 &  5.3 &  ~~-4.9$\pm$~9.4 &  $<$3.9e+34 &  - \\ 
L10-025 &  G98-21 &  562.0 &  5.7 &  9146.7$\pm$96.8 &  ~1.3e+37 &  -0.5 \\ 
L10-026 &  (x-ray) &  421.0 &  5.8 &  ~~25.4$\pm$11.0 &  ~4.9e+34 &  -1.0 \\ 
L10-027 &  G98-22 &  372.0 &  5.3 &  ~~~8.6$\pm$~8.0 &  $<$5.2e+34 &  - \\ 
L10-028 &  (opt) &  111.0 &  4.9 &  ~-25.3$\pm$11.2 &  $<$1.5e+35 &  - \\ 
L10-029 &  (opt) &  375.0 &  5.3 &  ~~24.3$\pm$~9.2 &  ~5.1e+34 &  -0.3 \\ 
L10-030 &  G98-23 &  372.0 &  4.5 &  ~-19.5$\pm$~7.4 &  $<$3.3e+34 &  - \\ 
L10-031 &  G98-24 &  189.0 &  4.7 &  ~~~7.9$\pm$~4.6 &  $<$6.7e+34 &  - \\ 
L10-032 &  G98-25 &  189.0 &  5.5 &  ~~45.5$\pm$~8.1 &  ~1.8e+35 &  -0.2 \\ 
L10-033 &  G98-26 &  469.0 &  6.5 &  ~~13.9$\pm$~9.5 &  $<$7.5e+34 &  - \\ 
L10-034 &  G98-27 &  524.0 &  5.3 &  ~163.3$\pm$15.6 &  ~2.4e+35 &  -0.7 \\ 
L10-035 &  XMM156 &  561.0 &  6.1 &  ~286.1$\pm$18.7 &  ~4.2e+35 &  -0.4 \\ 
L10-036 &  G98-28 &  607.0 &  5.6 &  1745.3$\pm$43.0 &  ~2.3e+36 &  -0.5 \\ 
L10-037 &  G98-29 &  470.0 &  5.6 &  ~759.4$\pm$29.2 &  ~1.3e+36 &  -0.5 \\ 
L10-038 &  G98-30 &  563.0 &  6.0 &  ~~~7.3$\pm$~9.0 &  $<$3.6e+34 &  - \\ 
L10-039 &  G98-31 &  805.0 &  6.7 &  2967.5$\pm$55.9 &  ~3.1e+36 &  -0.3 \\ 
L10-040 &  G98-32 &  696.0 &  6.1 &  ~~32.5$\pm$12.6 &  ~3.7e+34 &  0.5 \\ 
L10-041 &  (opt) &  622.0 &  6.0 &  ~~10.6$\pm$15.2 &  $<$5.6e+34 &  - \\ 
L10-042 &  G98-33 &  201.0 &  8.8 &  ~~-6.8$\pm$~5.0 &  $<$4.1e+34 &  - \\ 
L10-043 &  (x-ray) &  607.0 &  5.6 &  ~~88.1$\pm$16.3 &  ~1.2e+35 &  -0.7 \\ 
L10-044 &  G98-34 &  472.0 &  6.8 &  ~~12.8$\pm$~5.5 &  ~4.9e+34 &  -1.0 \\ 
L10-045 &  G98-35 &  733.0 &  6.4 &  ~998.2$\pm$33.6 &  ~1.2e+36 &  -0.2 \\ 
L10-046 &  G98-36 &  714.0 &  6.0 &  ~131.7$\pm$16.3 &  ~1.4e+35 &  -0.4 \\ 
L10-047 &  G98-37 &  796.0 &  6.2 &  ~~99.7$\pm$15.3 &  ~1.0e+35 &  -0.4 \\ 
L10-048 &  G98-38 &  607.0 &  5.5 &  ~~-0.6$\pm$~7.2 &  $<$1.8e+34 &  - \\ 
L10-049 &  G98-39 &  796.0 &  6.2 &  ~~68.4$\pm$14.6 &  ~6.9e+34 &  -0.4 \\ 
L10-050 &  G98-41 &  607.0 &  5.6 &  ~~~8.0$\pm$12.5 &  $<$4.2e+34 &  - \\ 
L10-051 &  G98-40 &  380.0 &  5.7 &  ~~~3.9$\pm$~7.5 &  $<$3.8e+34 &  - \\ 
L10-052 &  G98-42 &  618.0 &  5.5 &  ~~-3.6$\pm$~7.1 &  $<$1.9e+34 &  - \\ 
L10-055 &  G98-44 &  513.0 &  6.3 &  ~~13.3$\pm$10.0 &  $<$6.3e+34 &  - \\ 
L10-056 &  G98-45 &  606.0 &  5.3 &  ~~70.5$\pm$12.0 &  ~8.9e+34 &  0.0 \\ 
L10-057 &  G98-46 &  709.0 &  5.9 &  ~~33.6$\pm$11.1 &  ~3.7e+34 &  -0.5 \\ 
L10-058 &  (opt) &  618.0 &  5.6 &  ~~19.1$\pm$12.2 &  $<$5.3e+34 &  - \\ 
L10-059 &  (opt) &  231.0 &  0.8 &  ~~30.5$\pm$~8.5 &  ~1.1e+35 &  0.2 \\ 
L10-060 &  G98-48 &  231.0 &  0.6 &  ~~15.4$\pm$~5.8 &  ~5.6e+34 &  -0.6 \\ 
L10-061 &  G98-47 &  618.0 &  5.4 &  ~456.5$\pm$24.3 &  ~5.5e+35 &  -0.5 \\ 
L10-062 &  (opt) &  571.0 &  5.7 &  ~-16.7$\pm$~9.9 &  $<$2.7e+34 &  - \\ 
L10-063 &  (opt) &  571.0 &  5.9 &  ~~-6.5$\pm$~8.7 &  $<$2.6e+34 &  - \\ 
L10-064 &  G98-49 &  804.0 &  6.3 &  ~~31.3$\pm$13.0 &  ~2.9e+34 &  -0.3 \\ 
L10-065 &  G98-50 &  612.0 &  5.5 &  ~~17.6$\pm$~9.8 &  $<$4.5e+34 &  - \\ 
L10-066 &  G98-52 &  571.0 &  5.5 &  ~~17.8$\pm$10.3 &  $<$5.2e+34 &  - \\ 
L10-067 &  G98-51 &  386.0 &  4.3 &  ~~~1.8$\pm$~6.3 &  $<$2.7e+34 &  - \\ 
L10-068 &  (opt) &  187.0 &  8.6 &  ~~-3.3$\pm$~8.0 &  $<$7.5e+34 &  - \\ 
L10-069 &  G98-53 &  713.0 &  5.9 &  ~~99.7$\pm$14.7 &  ~1.1e+35 &  -0.6 \\ 
L10-070 &  G98-54 &  612.0 &  5.5 &  ~~46.6$\pm$~9.1 &  ~6.7e+34 &  -1.0 \\ 
L10-071 &  G98-55 &  713.0 &  5.9 &  2094.2$\pm$47.2 &  ~2.2e+36 &  -0.5 \\ 
L10-072 &  (opt) &  327.0 &  3.4 &  ~~~4.8$\pm$~7.4 &  $<$5.6e+34 &  - \\ 
L10-073 &  (opt) &  86.0 &  9.2 &  ~~-1.1$\pm$~4.5 &  $<$9.4e+34 &  - \\ 
L10-074 &  G98-57 &  428.0 &  4.9 &  ~~33.5$\pm$~8.6 &  ~5.8e+34 &  -0.5 \\ 
L10-075 &  G98-56 &  608.0 &  5.4 &  ~~~7.6$\pm$~9.6 &  $<$3.7e+34 &  - \\ 
L10-076 &  G98-57A &  428.0 &  4.9 &  ~~19.8$\pm$~7.1 &  ~3.4e+34 &  -0.7 \\ 
L10-077 &  G98-59 &  386.0 &  4.6 &  ~~~7.9$\pm$~5.1 &  $<$4.6e+34 &  - \\ 
L10-078 &  G98-58 &  524.0 &  4.9 &  ~~34.3$\pm$~9.1 &  ~5.0e+34 &  -1.0 \\ 
L10-079 &  (x-ray) &  380.0 &  4.3 &  ~~42.4$\pm$~9.5 &  ~8.2e+34 &  -0.7 \\ 
L10-080 &  G98-60 &  713.0 &  6.3 &  ~~35.9$\pm$~9.0 &  ~4.5e+34 &  -0.4 \\ 
L10-081 &  G98-62 &  713.0 &  6.0 &  ~333.7$\pm$21.0 &  ~3.6e+35 &  -0.4 \\ 
L10-082 &  G98-61 &  468.0 &  6.1 &  ~~27.2$\pm$10.0 &  ~4.4e+34 &  -0.5 \\ 
L10-083 &  G98-65 &  618.0 &  5.7 &  ~~23.2$\pm$~9.5 &  ~2.8e+34 &  -0.4 \\ 
L10-084 &  G98-64 &  608.0 &  5.6 &  ~216.5$\pm$17.3 &  ~2.8e+35 &  -0.5 \\ 
L10-085 &  G98-63 &  700.0 &  5.9 &  ~123.6$\pm$14.6 &  ~1.3e+35 &  0.0 \\ 
L10-086 &  FL236 &  700.0 &  6.2 &  ~~67.7$\pm$11.1 &  ~7.5e+34 &  -0.2 \\ 
L10-087 &  G98-66 &  428.0 &  4.9 &  ~~33.1$\pm$~8.8 &  ~7.9e+34 &  -0.3 \\ 
L10-088 &  G98-67 &  571.0 &  6.0 &  ~~41.8$\pm$11.3 &  ~6.6e+34 &  -0.2 \\ 
L10-089 &  (x-ray) &  802.0 &  6.2 &  ~114.2$\pm$20.0 &  ~1.1e+35 &  -0.6 \\ 
L10-090 &  G98-68 &  884.0 &  6.5 &  ~~37.6$\pm$12.0 &  ~3.3e+34 &  -0.4 \\ 
L10-091 &  G98-69 &  802.0 &  6.3 &  ~~73.5$\pm$13.0 &  ~7.4e+34 &  -0.6 \\ 
L10-092 &  G98-70 &  713.0 &  6.0 &  ~~38.2$\pm$16.9 &  ~4.2e+34 &  -0.3 \\ 
L10-093 &  FL261 &  612.0 &  5.4 &  ~~48.9$\pm$~9.6 &  ~6.0e+34 &  -0.9 \\ 
L10-094 &  XMM244 &  884.0 &  6.5 &  ~294.7$\pm$19.7 &  ~2.8e+35 &  -0.6 \\ 
L10-095 &  G98-71 &  607.0 &  5.4 &  ~~43.0$\pm$~9.8 &  ~5.7e+34 &  -0.6 \\ 
L10-096 &  G98-73 &  696.0 &  5.9 &  ~856.9$\pm$30.7 &  ~1.1e+36 &  -0.6 \\ 
L10-097 &  G98-72 &  881.0 &  6.8 &  ~~21.0$\pm$~9.6 &  ~2.0e+34 &  -0.1 \\ 
L10-098 &  G98-74 &  713.0 &  6.2 &  ~-15.0$\pm$11.3 &  $<$2.4e+34 &  - \\ 
L10-099 &  G98-75 &  515.0 &  5.3 &  ~~~7.9$\pm$~8.5 &  $<$4.2e+34 &  - \\ 
L10-100 &  G98-76 &  696.0 &  5.8 &  ~~26.6$\pm$~8.8 &  ~2.9e+34 &  -1.0 \\ 
L10-101 &  G98-77 &  515.0 &  5.7 &  ~~26.0$\pm$~9.8 &  ~5.2e+34 &  0.3 \\ 
L10-102 &  G98-80 &  713.0 &  6.3 &  ~~10.1$\pm$~9.6 &  $<$3.2e+34 &  - \\ 
L10-103 &  G98-79 &  696.0 &  5.9 &  ~~17.8$\pm$~9.9 &  $<$4.5e+34 &  - \\ 
L10-104 &  G98-81 &  881.0 &  6.9 &  ~~36.4$\pm$12.2 &  ~3.5e+34 &  -0.9 \\ 
L10-105 &  G98-78 &  280.0 &  6.3 &  ~~63.5$\pm$10.6 &  ~1.7e+35 &  -0.3 \\ 
L10-106 &  (opt) &  469.0 &  5.5 &  ~~28.8$\pm$11.1 &  ~4.6e+34 &  -0.2 \\ 
L10-107 &  G98-82 &  663.0 &  6.8 &  ~~30.9$\pm$~9.9 &  ~4.1e+34 &  -1.0 \\ 
L10-108 &  G98-84 &  370.0 &  6.0 &  ~~16.4$\pm$10.0 &  $<$7.3e+34 &  - \\ 
L10-109 &  G98-83 &  280.0 &  4.3 &  ~~~5.1$\pm$~5.3 &  $<$4.1e+34 &  - \\ 
L10-110 &  (opt) &  621.0 &  6.4 &  ~~22.9$\pm$11.3 &  ~2.8e+34 &  0.2 \\ 
L10-111 &  G98-85 &  696.0 &  5.9 &  ~137.8$\pm$15.5 &  ~1.5e+35 &  -0.7 \\ 
L10-112 &  (opt) &  280.0 &  6.4 &  ~~~0.1$\pm$~8.2 &  $<$4.4e+34 &  - \\ 
L10-113 &  G98-86 &  520.0 &  5.8 &  ~~21.4$\pm$~8.5 &  ~3.5e+34 &  -0.5 \\ 
L10-114 &  G98-87 &  520.0 &  5.8 &  ~~34.4$\pm$~8.6 &  ~5.8e+34 &  -0.1 \\ 
L10-115 &  G98-88 &  196.0 &  6.3 &  ~~16.7$\pm$~6.6 &  ~6.6e+34 &  -0.8 \\ 
L10-116 &  XMM270 &  280.0 &  6.7 &  ~~51.9$\pm$~9.7 &  ~1.5e+35 &  -0.5 \\ 
L10-117 &  G98-89 &  280.0 &  7.3 &  ~~68.0$\pm$11.8 &  ~1.9e+35 &  0.1 \\ 
L10-118 &  G98-90 &  379.0 &  4.3 &  ~~30.7$\pm$~9.3 &  ~6.0e+34 &  -0.0 \\ 
L10-119 &  FL312 &  380.0 &  5.2 &  ~153.5$\pm$14.1 &  ~3.0e+35 &  0.4 \\ 
L10-120 &  G98-91 &  603.0 &  6.2 &  ~~~5.0$\pm$~8.7 &  $<$2.9e+34 &  - \\ 
L10-121 &  G98-92 &  520.0 &  6.1 &  ~~20.6$\pm$10.0 &  ~3.0e+34 &  -0.7 \\ 
L10-122 &  (opt) &  186.0 &  9.1 &  ~~25.4$\pm$10.7 &  ~1.3e+35 &  -0.5 \\ 
L10-123 &  G98-93 &  427.0 &  5.7 &  ~~10.9$\pm$~7.6 &  $<$4.5e+34 &  - \\ 
L10-124 &  G98-94 &  464.0 &  4.5 &  ~127.3$\pm$13.4 &  ~2.4e+35 &  -0.3 \\ 
L10-125 &  (x-ray) &  280.0 &  5.5 &  ~~27.8$\pm$~7.6 &  ~8.1e+34 &  -0.9 \\ 
L10-126 &  G98-95 &  434.0 &  6.0 &  ~~-8.5$\pm$~7.2 &  $<$2.6e+34 &  - \\ 
L10-127 &  G98-96 &  282.0 &  4.1 &  ~~10.0$\pm$~8.6 &  $<$7.7e+34 &  - \\ 
L10-128 &  G98-97B &  464.0 &  5.8 &  ~~37.0$\pm$~8.3 &  ~6.3e+34 &  -0.6 \\ 
L10-129 &  G98-97A &  464.0 &  5.9 &  ~150.7$\pm$14.3 &  ~2.6e+35 &  -0.7 \\ 
L10-130 &  (opt) &  464.0 &  5.8 &  ~~~7.4$\pm$~6.7 &  $<$3.7e+34 &  - \\ 
L10-131 &  (opt) &  375.0 &  5.5 &  ~-14.9$\pm$13.8 &  $<$5.9e+34 &  - \\ 
L10-132 &  G98-98 &  272.0 &  4.8 &  ~~~2.7$\pm$~5.5 &  $<$4.8e+34 &  - \\ 
L10-133 &  (opt) &  272.0 &  6.2 &  ~~~8.7$\pm$~8.4 &  $<$7.3e+34 &  - \\ 
L10-134 &  (opt) &  184.0 &  5.6 &  ~~17.1$\pm$~6.9 &  ~7.2e+34 &  0.0 \\ 
L10-135 &  (opt) &  184.0 &  6.0 &  ~~~7.5$\pm$~6.4 &  $<$8.6e+34 &  - \\ 
L10-136 &  (opt) &  184.0 &  6.1 &  ~~~1.3$\pm$~8.8 &  $<$7.8e+34 &  - \\ 
L10-137 &  (opt) &  184.0 &  6.7 &  ~~18.5$\pm$11.9 &  $<$1.8e+35 &  - \\ 
\tablenotetext{*}{ Other names or for new objects whether the object was initially suspected to be SNR based on optical or X-ray data.}
\tablenotetext{a}{ Characteristic off-axis angle at which this object was observed.}
\tablenotetext{b}{ 2 $\sigma$ upper limits are shown for objects that were not detected.}
\tablenotetext{c}{ Hardness ratio constructed from the ratio of (M-S)/Total counts, where S corresponds to 0.35-1.2 keV counts, M to 1.2-2.6 keV counts, and Total to 0.35-8 keV counts.}
\enddata 
\label{table_results}
\end{deluxetable}
%\end{center}

% Table values updated 090723
%\input{table_morph.tex}
%\begin{center}
\begin{deluxetable}{lcrrrrcrrrrcrrrr}
\tablecaption{Effects of SNR Morphology and Environment on X-ray Detection }
\tablehead{\colhead{} & \colhead{~} & 
\multicolumn{4}{c}{Sample Number$^{a}$} & 
\phm{nn} & \multicolumn{4}{c}{X-ray Detected} & 
\phm{nn} &\multicolumn{4}{c}{Percentage Detected}  
\\
\cline{3-6} \cline{8-11} \cline{13-16}
\colhead{} & \colhead{Environ.} & 
 \colhead{1} & 
 \colhead{2} & 
 \colhead{3} & 
 \colhead{Sum} & &
 \colhead{1} & 
 \colhead{2} & 
 \colhead{3} & 
 \colhead{Sum} & &
 \colhead{1} & 
 \colhead{2} & 
 \colhead{3} & 
 \colhead{Sum} 
}
\tabletypesize{\scriptsize}
\tablewidth{0pt}\startdata
& Morph. &   &   &   &   &   & &   &   &   &   &  &   &   &   \\ 
%All & &  ~ &  ~ &  ~ &  ~ &  ~ & ~ &  ~ &  ~ &  ~ & ~ &  ~ &  ~ &  ~ &  ~ \\ 
& A, A$^{\prime}$ &  56 &  26 &  4 &  86 & &  34 &  16 &  3 &  53 & &  61  &  62  &  75  &  62  \\ 
All & B &  8 &  8 &  9 &  25 & &  7 &  7 &  7 &  21 & &  88  &  88  &  78  &  84  \\ 
& C &  4 &  9 &  7 &  20 & &  2 &  3 &  3 &  8 & &  50  &  33  &  43  &  40  \\ 
%D &  0 &  3 &  1 &  4 & &  0 &  3 &  1 &  4 & &  x &  1.0 &  1.0 &  1.0 \\ 
& Sum &  68 &  43 &  20 & 131 & &  43 &  26 &  13 &  82  & &  63  &  60  &  65  &  63  \\ 
\hline
%D$<$50~pc &  ~ &  ~ &  ~ &  ~ &  ~ & ~ &  ~ &  ~ & ~ &  ~ &  ~ &  ~ &  ~ &  ~ \\ 
& A, A$^{\prime}$ &  30 &  11 &  3 &  44 & & 22 &  10 &  3 &  35 & &  73  &  91  &  100 &  80  \\ 
$D<50\,$pc & B &  8 &  6 &  6 &  20 & &  7 &  5 &  4 &  16 & &  88  &  83  &  67  &  80  \\ 
& C &  3 &  7 &  4 &  14 & &  2 &  3 &  2 &  7 & &  67  &  43  &  50  &  50  \\ 
%D &  0 &  3 &  1 &  4 & &  0 &  3 &  1 &  4 & & -- &  1.0 &  1.0 &  1.0 \\ 
& Sum &  41 &  24 &  13 &  78 & &  31 &  18 &  9 &  58 & &  76  &  75  &  69  &  74  \\ 
\hline
%D$>$50~pc &  ~ &  ~ &  ~ &  ~ &  ~ &  ~ &  ~ & ~ &  ~ &  ~ &  ~ &  ~ &  ~ \\ 
& A, A$^{\prime}$ &  26 &  15 &  1 &  42 & &  12 &  6 &  0 &  18 & &  46  &  40  &    0 &  43  \\ 
$D>50\,$pc & B &  0 &  2 &  3 &  5 & &  0 &  2 &  3 &  5 & &  x &  100 &  100 & 100 \\ 
& C &  1 &  2 &  3 &  6 & &  0 &  0 &  1 &  1 & &  0   &  0   &  33  &  17  \\ 
%D &  0 &  0 &  0 &  0 & &  0 &  0 &  0 &  0 & &  -- &  -- &  -- &  -- \\ 
& Sum &  27 &  19 &  7 &  53 & &  12 &  8 &  4 &  24 & &  44  &  42  &  57  &  45  \\ 
\enddata 
\tablenotetext{a}{ The total number of SNRs sums to 131 rather than 137 because the 6 objects not observed in X-rays were not counted.}
\label{tab_morph}
\end{deluxetable}
%\end{center}

% Next table is only for comparison to the real table_morph above, which has better formatting, but which must be hand edited to agree with ksl's less prettily formatted version
%\input{table_morph_ksl.tex}

%\input{table_nonSNR.tex}
%\begin{center}
\begin{deluxetable}{lcl}
\tablecaption{Possible Superbubbles and Peculiar SNR Candidates\tablenotemark{a} }
\tablehead{\colhead{Object} & 
 \colhead{Alternate Name} &  
 \colhead{Comments} 
}
\tabletypesize{\scriptsize}
\tablewidth{0pt}
\startdata
L10-001 & G98-01 & Very large SNR or fossil superbubble with few stars remaining. Not in \chase\ fields\\
L10-003 & G98-03 & Likely superbubble. Not in \chase\ fields. \\
L10-010 & G98-08 & Very large SNR or fossil superbubble with few stars remaining. \\
L10-012 & G98-19 & Large shell with bright central star(s); likely superbubble. \\
L10-026 & (x-ray) & Likely stars associated and only slightly enhanced [S~II]; likely superbubble. \\ 
L10-028 & (opt) & [S~II] enhanced, but stars seen in projection; likely superbubble. \\
L10-030 & G98-23 & Associated stars likely; old SNR or superbubble both possible. \\ 
L10-035 & XMM156 & Bright, extended X-ray emission but compact [O~III] optical source; peculiar SNR. \\
L10-043 & (x-ray) & Associated stars likely; complex region; old SNR or superbubble both possible. \\
L10-050 & G98-41 & Long filament of enhanced [S~II] associated with larger region of emission. \\
L10-055 & G98-44 & Long linear filament of enhanced [S~II] in much larger region of emission. \\
L10-068 & (opt) & Possible blow-out on east side of H~II region, but projected stars not clearly associated.\\
L10-080 & G98-60 & Single optical filament on edge of H~II region but X-rays associated to one side; peculiar SNR. \\
L10-089 &(x-ray) & Possible blow-out from bright H~II region to west; budding superbubble? \\
L10-092 & G98-70 & Cluster of centrally located stars; likely superbubble. \\
L10-098 & G98-74 & Associated stars likely; old SNR or superbubble both possible. \\
L10-108 & G98-84 & Enhanced [S~II] filaments in much larger region of emission; old SNR or fossil superbubble both possible. \\ 
L10-117 & G98-89 & Possible associated stars; old SNR or superbubble both possible. \\
L10-119 & FL312 & Possible but uncertain association of X-ray source with faint emission nebula seen in \HA. \\ 
L10-120 & G98-91 & Modest sized linear filament in more diffuse gas.  [S~II] enhanced, but no shell or associated structure. \\
L10-122 & (opt) & Exceedingly large size (165$\times$85 pc) makes a single SNR origin unlikely; no clearly associated stars. \\
L10-131 & (opt) & Large size and star cluster associated; likely superbubble. \\ 
L10-132 & G98-98 & Clumpy, faint H$\alpha$ with marginally too low [S~II]:\HA; stars seen in projection; likely superbubble.  \\
L10-133 & (opt) & Large shell with central stars; also central bright, compact emission region; likely superbubble. \\
L10-136 & (opt) & Large shell with marginally enhanced [S~II]:\HA; bright star(s) centrally located; likely superbubble. \\
L10-137 & (opt) & Large shell with stars associated; likely superbubble. \\
\enddata
\tablenotetext{a}{See individual object descriptions in Appendix A for more details.} 
\label{table_nonsnr}
\end{deluxetable}
%\end{center}

%\input{pshock.tex}
%\input{table_pshock.tex}
%\begin{center}
%
% 12 Dec 2009 - PPP updates the analysis after the new spectral extractions from Knox
% 
%
\begin{deluxetable}{cccccccc}
\tablecaption{Plane Parallel Shock Model$^a$  Fits}
\tablehead{\colhead{Parameter} & 
 \colhead{G98-21} & 
 \colhead{G98-28} & 
 \colhead{G98-29} & 
 \colhead{G98-31} & 
 \colhead{G98-35} & 
 \colhead{G98-55} & 
 \colhead{G98-73} 
}
\tabletypesize{\scriptsize}
\tablewidth{0pt}\startdata
%% &                           & G98-21           &  G98-28           &	G98-29		&  G98-31           & G98-35            & G98-55            & G98-73  %%  
N$_H$($10^{21}cm^{-2}$)         & 0.0[0.0,0.3]$^b$    &  0.0[0.0,0.3]    & 0.0[0.0,2.0] 	& 5.8[4.3,7.3]    &  1.0[0.0,3.2]    &  0.0[0.0,1.1]    &  0.0[0.0,1.0] \\ 
kT(keV)                      & 0.57[0.55,0.59] &  0.61[0.60,0.66] & 0.84[0.63,1.40] 	& 0.43[0.38,0.48] &  0.78[0.65,0.99] &  0.60[0.53,0.66] &  1.05[0.66,2.47] \\ 
Abundance  & 0.36[0.32,0.42] &  0.46[0.35,0.69] & 0.46[0.24,1.17]	& 0.19[0.14,0.28] &  0.18[0.10,0.35] &  0.33[0.23,0.45] &  0.66[0.33,1.81] \\ 
$\tau$($\POW{11}cm^{-3}{s}$) & 5.23[4.44,6.14] &  4.61[3.01,5.87] & 0.88[0.32,1.68] 	& 0.56[0.34,0.98] &  3.40[1.55,7.98] &  2.29[1.58,3.06] &  0.43[0.25,0.62] \\ 
Norm($10^{-4}$)              & 1.95[1.71,2.33] &  0.27[0.19,0.35] & 0.10[0.03,0.21] 	& 2.09[1.30,3.30] &  0.24[0.13,0.43] &  0.31[0.23,0.46] &  0.05[0.02,0.14] \\ 
$\CHINU$    & 1.39            &  1.33            & 0.83		& 3.13            &  1.39            &  0.95            &  1.10 \\ 
\enddata 
\tablenotetext{a}{XSPEC model: tbabs(tbvarabs*pshock).  See text for explanation.}
\tablenotetext{b}{90\% ($\chi^{2}_{min}+2.706$) confidence intervals are indicated in brackets.}
\label{table_pshock}
\end{deluxetable}
%\end{center}

% \input{vpshock.tex}
%\input{table_vpshock.tex}
%\begin{center}
\begin{deluxetable}{ccc}
\tablecaption{Variable Abundance Plane Parallel Shock Model$^a$ Fits}
\tablehead{\colhead{Parameter} & 
 \colhead{G98-31} & 
 \colhead{G98-35} 
}
\tabletypesize{\scriptsize}
\tablewidth{0pt}\startdata
N$_H$($10^{21}cm^{-2}$)        &  3.5[1.2,5.6]$^b$     &  0.0[0.0,1.9] \\ 
kT(keV)                     &  0.48[0.40,0.66]  &  0.60[0.49,0.67] \\ 
O                           &  0.21[0.15,0.31]  &  1.06[0.32,4.00] \\ 
Ne                          &  0.23[0.17,0.34]  &  0.99[0.51,2.64] \\ 
Mg                          &  0.61[0.48,0.85]  &  0.93[0.37,2.26] \\
Fe                          &  0.10[0.09,0.19]  &  0.15[0.07,0.41] \\
$\tau$($\POW{11}cm^{-3}{s}$) &  1.55[0.84,2.76] &  $>24.8$ \\ 
Norm($10^{-4}$)             &  1.42[0.53,2.82]  &  0.24[0.10,0.59] \\ 
$\CHINU$                    &  1.57             &  0.64  \\ 
\enddata 
\tablenotetext{a}{XSPEC model:  tbabs(tbvarabs*vpshock).  See text for explanation.}
\tablenotetext{b}{90\% ($\chi^{2}_{min}+2.706$) confidence intervals are indicated in brackets.}
\label{table_vpshock}
\end{deluxetable}
%\end{center}

% Parviz' table
%\input{table_bright_compare.tex}
\clearpage

%temporary end for quicker processing
%\end{document}
 
\newpage
\pagestyle{empty}

\clearpage

\begin{figure}
%\plotone{m33snrs_f1_bw.ps}
\plotone{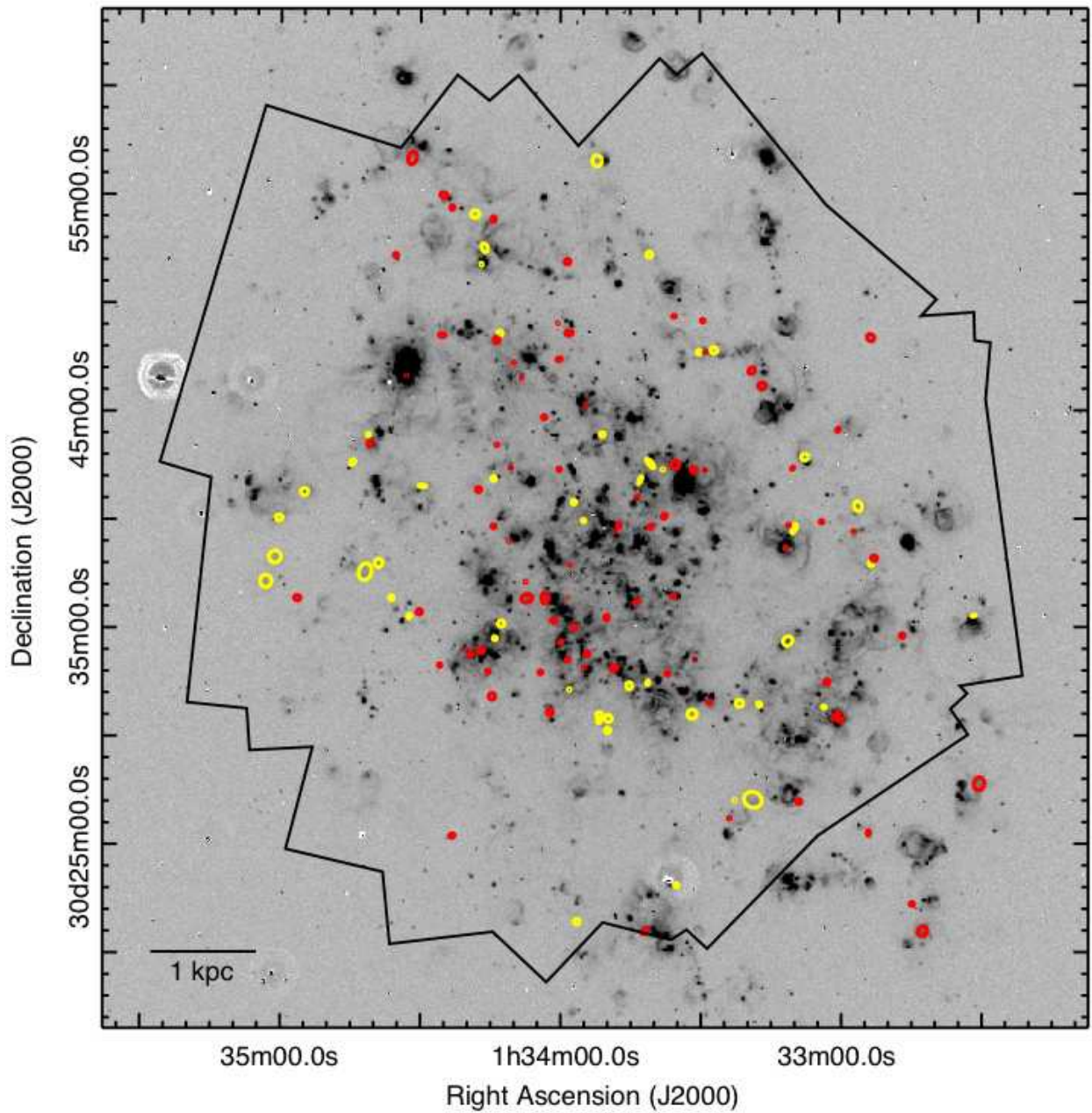}
\figcaption[snr_overview]{Continuum-subtracted \HA\  image of M33 with the positions of SNRs and SNR candidates indicated.  Objects in yellow are objects with high \sii:\HA\  ratios that  were detected in X-rays at greater than 2$\sigma$. Objects in red were also included in our target list of SNRs, but were not detected at this level.  The portion of M33 that was surveyed with \chandra\ is indicated.  Note that a few of the catalog objects were outside of the survey field.     \label{fig_snr_overview}}
\end{figure}

\begin{figure}
%\plotone{fig_multiwave.eps}
\centering
\includegraphics[height=7in]{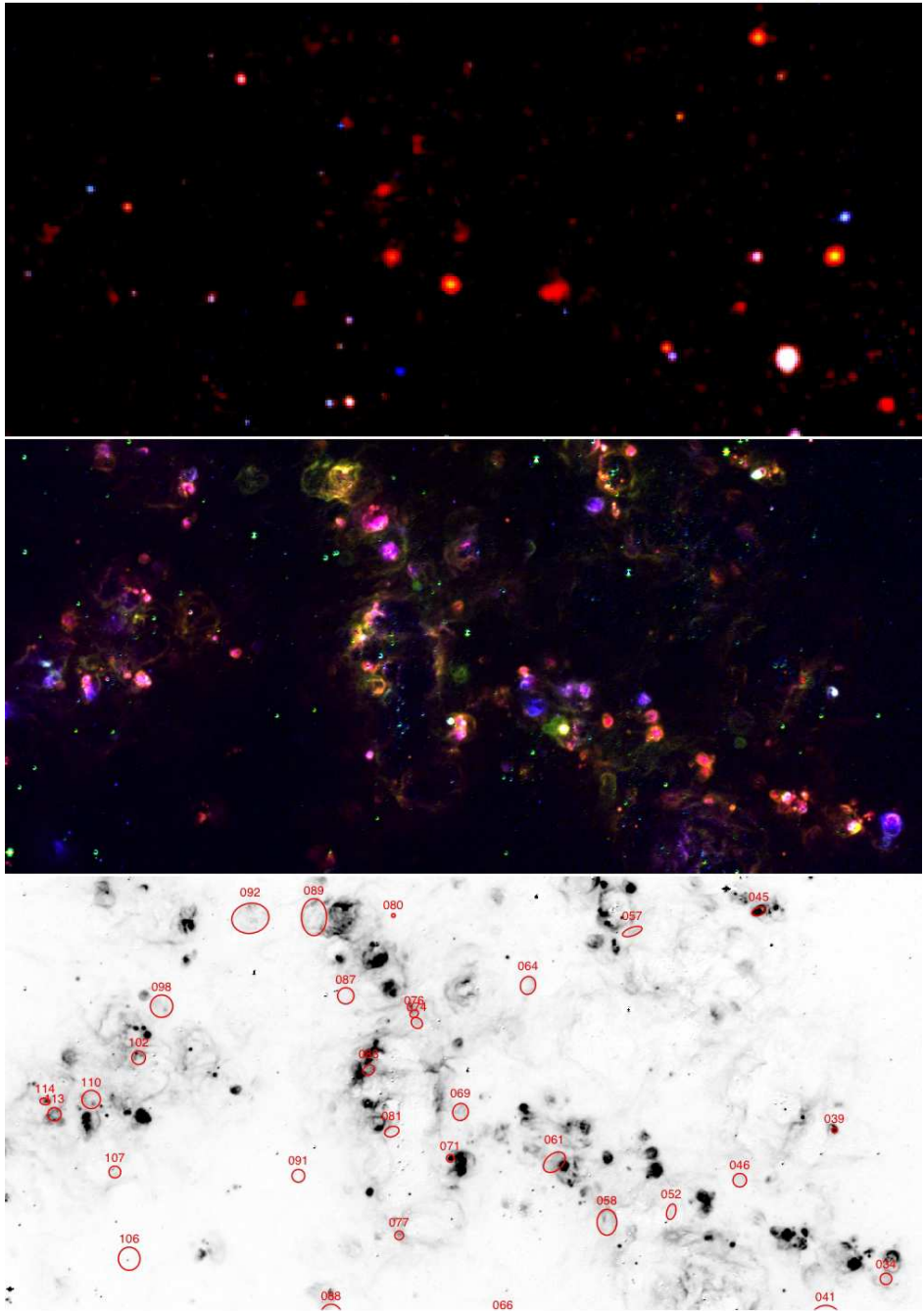}
\figcaption[multiwave]{ A portion of the southern spiral arm of M33.  The top panel is a three-color rendition of the \chandra\ X-ray mosaic image ({\it red}: 0.35--1.1 keV; {\it green}: 1.1--2.6 keV; {\it blue}: 2.6--8.0 keV).   The center panel is a three-color optical image from the continuum-subracted LGGS data ({\it red}: \HA; {\it green}: \sii; {\it blue}: \oiii).  The bottom panel shows the continuum-subtracted \HA\ image with the SNRs marked.   In the center panel, SNRs generally appear as green or yellow, while \hii\ regions are generally magenta or red.  The field measures $12.2\arcmin \times 5.8\arcmin$ and is oriented N up, E left. \label{fig_multiwave}}
\end{figure}

% Replaced 080812
\begin{figure}
%\plotone{fig_lum.eps}
\plotone{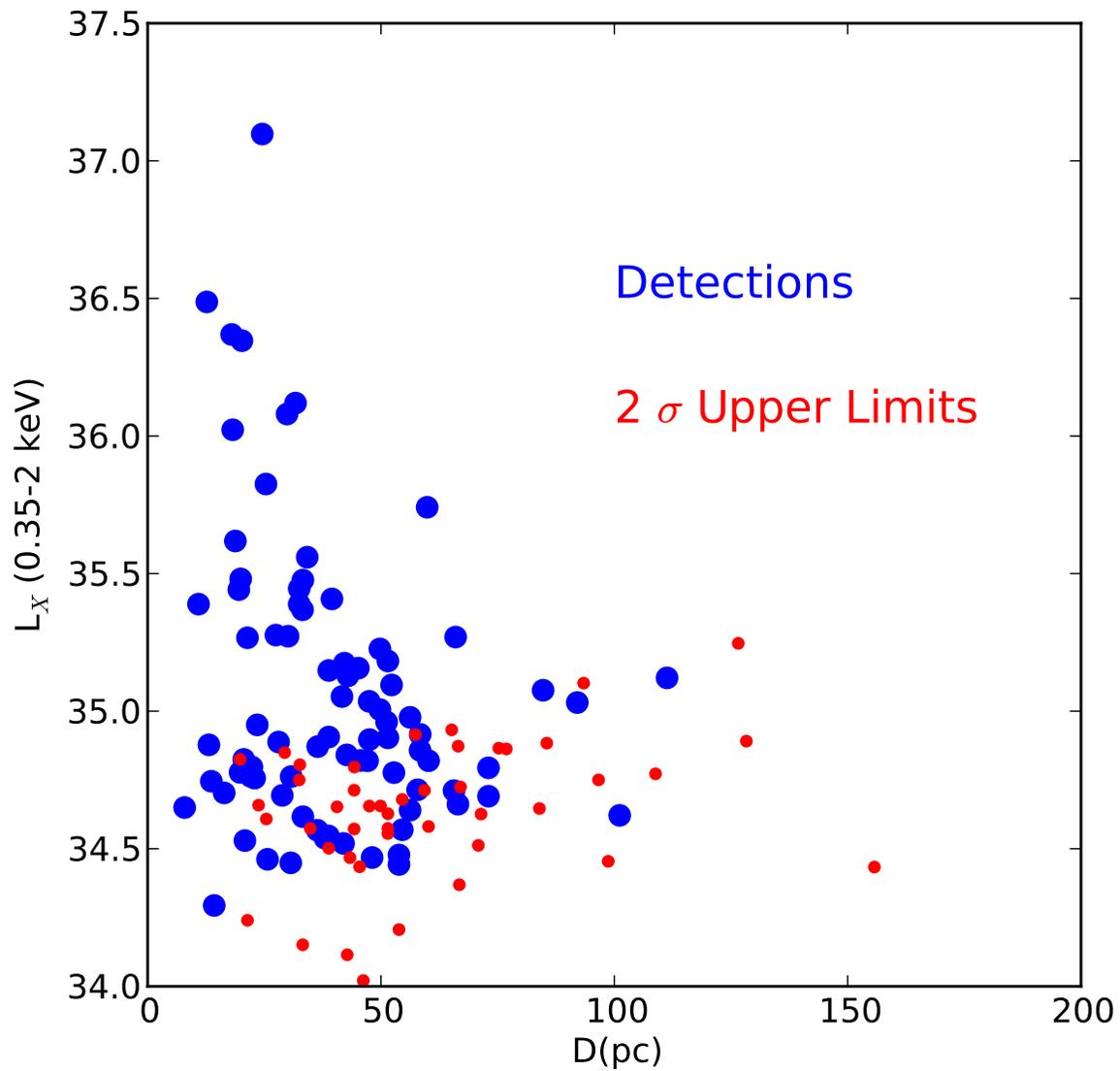}
\figcaption[lum]{The 0.35--2 keV luminosities of   SNRs in M33 as a function of diameter in parsecs.      Objects detected in X-rays at $>2\sigma$  are shown as larger blue dots.  The red dots represent 2$\sigma$ upper limits for those SNRs that were not detected in X-rays.  \label{fig_lum}}
\end{figure}

% This is in galcen. Replace 080812
\begin{figure}
%\plottwo{fig_gal.eps}{fig_gal_lum.eps}
%\plotone{fig_gal.eps}
\plotone{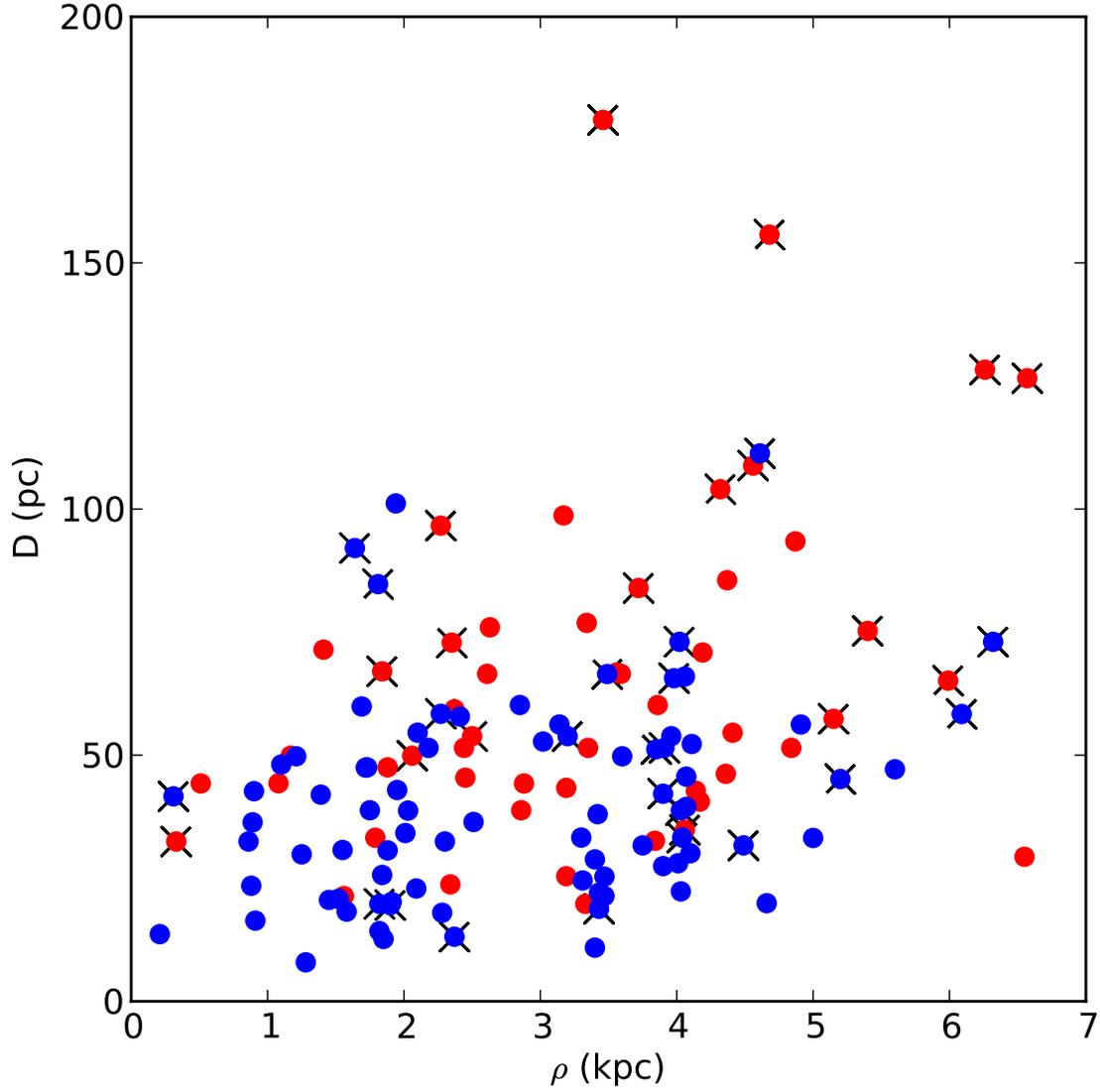}

\figcaption[galacto]{The diameter $D$ of M33 SNRs and candidates as a function of galactocentric radius $\rho$ in M33.  Points in blue are objects detected in X-rays at 2$\sigma$ or greater; other objects from our list are shown in red.  The pure circles are objects listed by \cite{gordon98}; the ones with superimposed x's have been added subsequent to their list. The new objects are not concentrated in one area of the figure.  There are no obvious trends of detection with galactocentric radius.  There are no very large ($>$100 pc) objects within 3 kpc of the nucleus. \label{fig_galacto}
}
\end{figure}

% These are the original numbers, before some additional objects were added
%KS (-0.48043478260869565, 1.3724149820543426e-06)
%Med 32.2393731588 55.44

% in fig_rates   Replace 080812
\begin{figure}
%\plotone{fig_histo_dia.eps}
\plotone{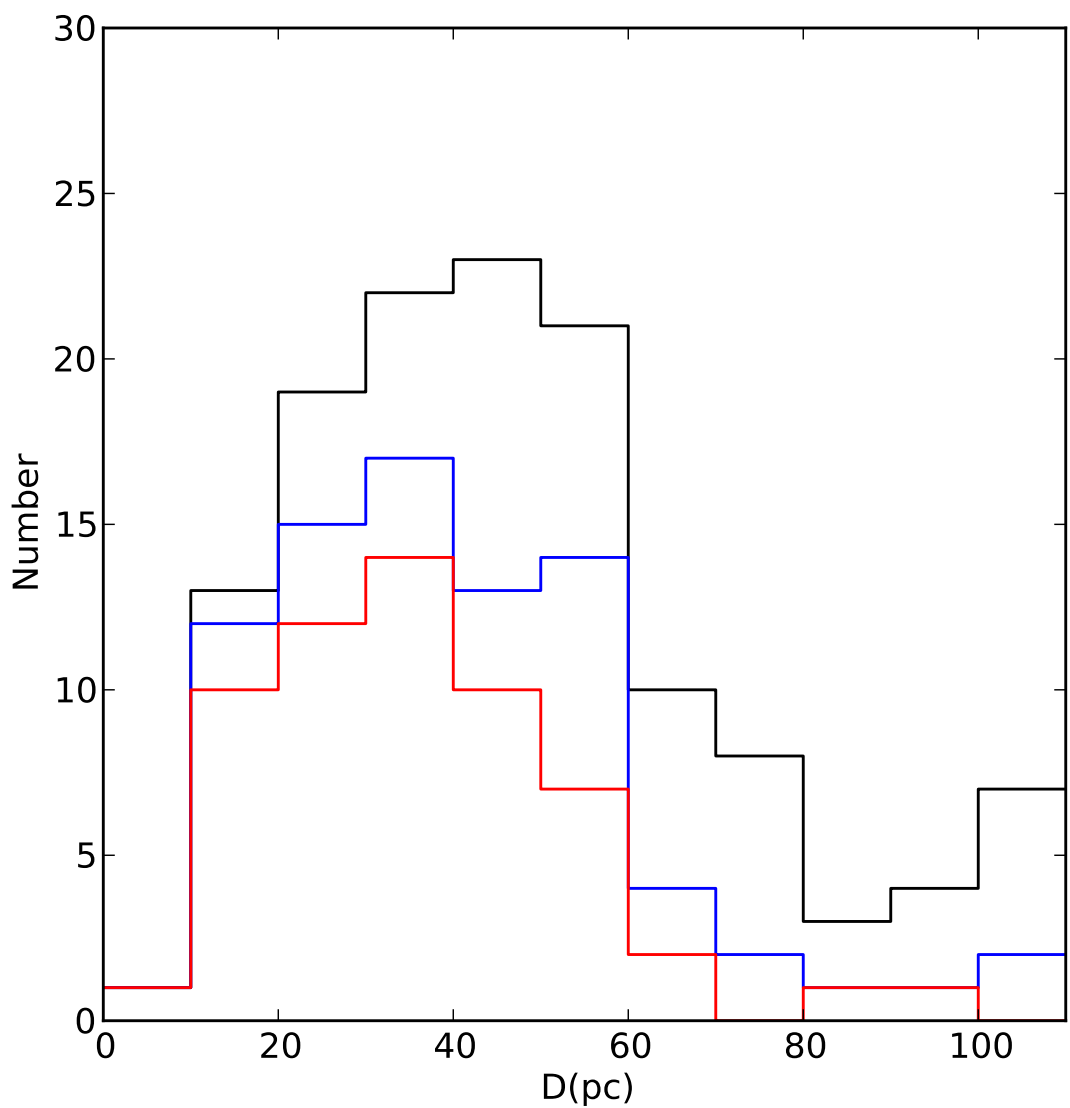}
\figcaption[radial_dist]{A histogram of the SNRs and SNR candidates as a function of diameter in bins of 10 pc.  The entire sample is shown in black.  The portion of the sample that was detected at 2$\sigma$ or higher is shown in blue, and at 3$\sigma$ or higher in red.  All of the objects with diameters greater than 100 pc are plotted in the last bin. The median diameter of objects which were detected in X-rays is 38 pc, compared to 55 pc for the objects which were not detected.  
%A K-S test comparing the two populations that the possibility they are drawn from the same parent population is about 1 in a million. {\bf These numbers need to updated assuming we use them as they are for 3 $\sigma$} 
\label{fig_dia}}
\end{figure}

% In fig hard.  Replaced 080812
\begin{figure}
%\plotone{fig_histo_hard.eps}
\plotone{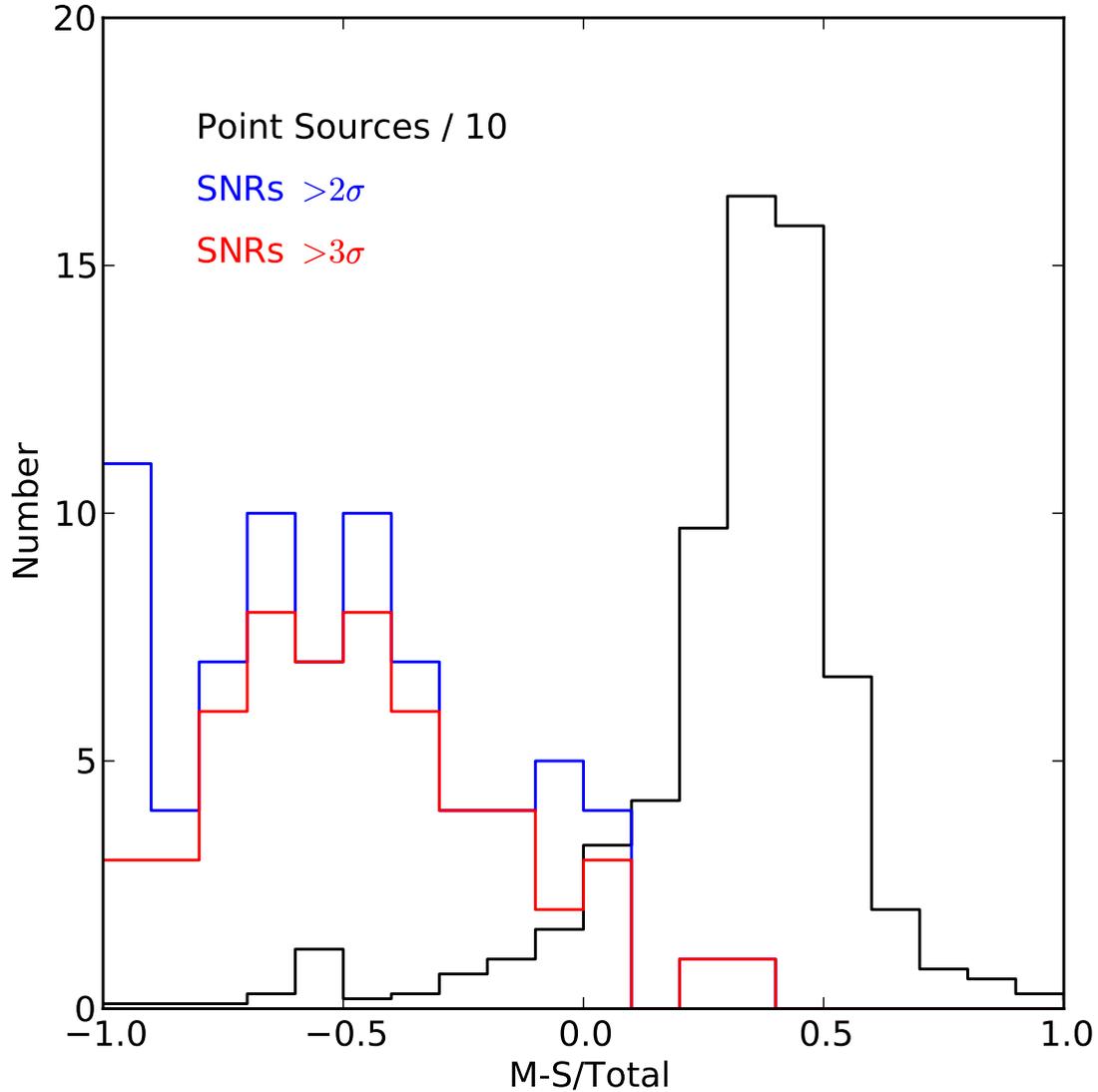}
\figcaption[hardness]{A comparison of the distribution of hardness ratios for point sources (shown in black) and SNRs seen in M33 with Chandra.  The hardness ratio M--S/Total is defined so that  S corresponds to net source counts detected between 0.35 and 1.2 keV, M, to counts beween 1.2 and 2.6 keV, and Total, to counts between 0.35 and 8 keV. The SNRs detected at more than 2$\sigma$ and 3$\sigma$ are plotted in blue and red, respectively.   
%As one would expect, the distribution is broader when the sources between 2$\sigma$ and 3$\sigma$ are included.{\bf It is not obvious to me it is that much broader -- same statement is in the text. DJH}  
The number of point sources has been scaled down by a factor of 10 so that the difference between the hardness-ratio distributions would be more apparent. SNRs have been excluded from the point source list; the small peak near $-0.5$ in the point-source distribution is due, at least in part, to foreground stars.
\label{fig_hardness}}
\end{figure}

 \clearpage
 
 % fig_ha   Replaced 080812
 \begin{figure}
%\plotone{fig_dia_ha_lum.eps}
\plotone{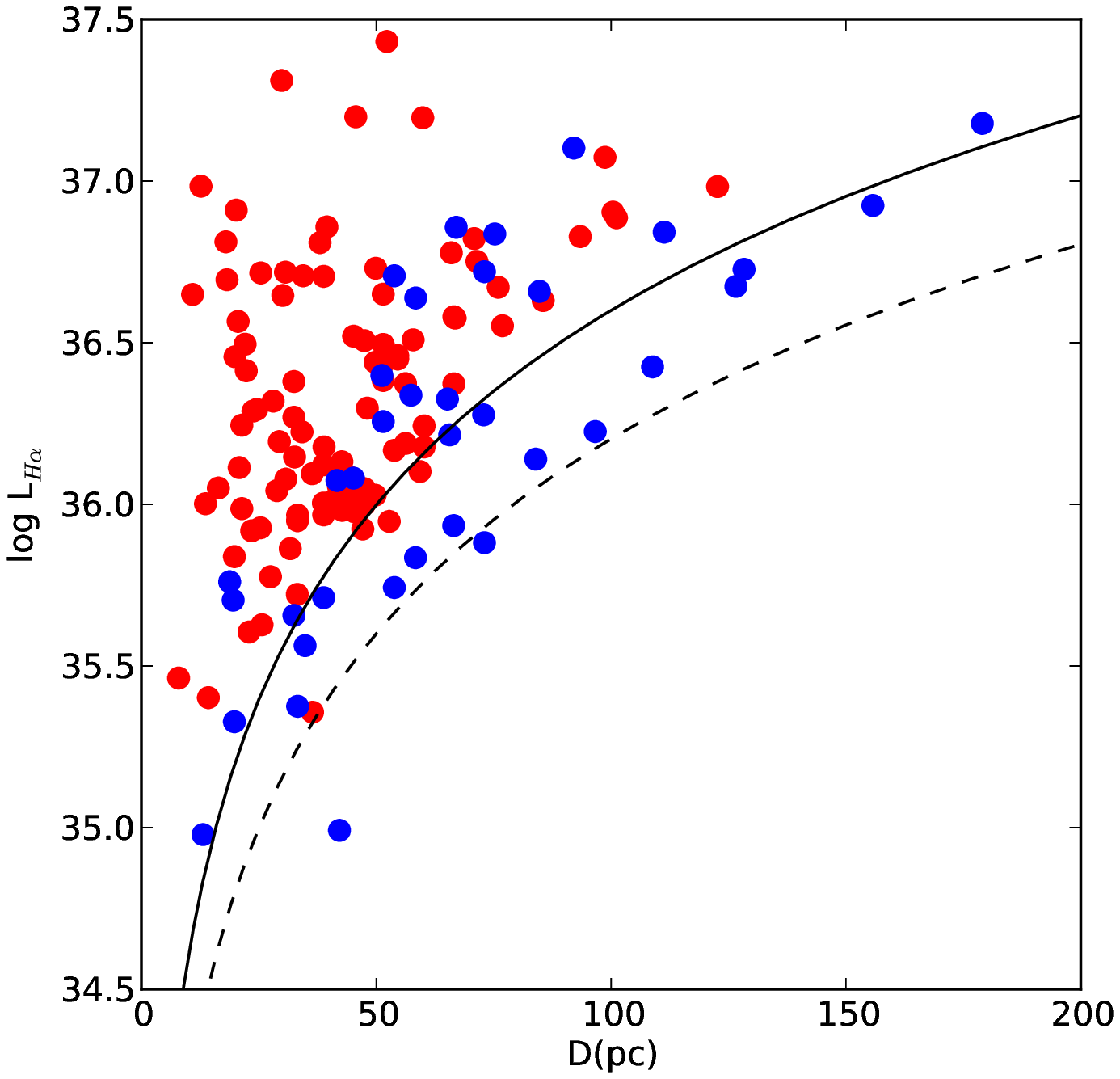}
\figcaption[dia_ha_lum]{\HA\  luminosities of the objects in our sample as a function of diameter in parsec.  Objects from the list of \cite{gordon98} are plotted as 
%the lighter 
red dots while the new objects that have been identified as SNRs or candidates since then are plotted in blue.  The solid black line corresponds to a surface brightness of \EXPU{1}{-16}{ergs~cm^{-2}s^{-1}arcsec^{-2}}, which is our estimate of the surface-brightness limit above which SNRs can be identified in the LGGS.  The dashed curve corresponds to a surface brightness of  \EXPU{4}{-17}{ergs~cm^{-2}s^{-1}arcsec^{-2}}, the approximate limit of the Schmidt images (but only in very uncrowded  regions).  Many of the new objects are large, low surface-brightness ones.
\label{fig_ha_lum}
}
\end{figure}

\begin{figure}
%\plotone{fig_FL281.eps}
\plotone{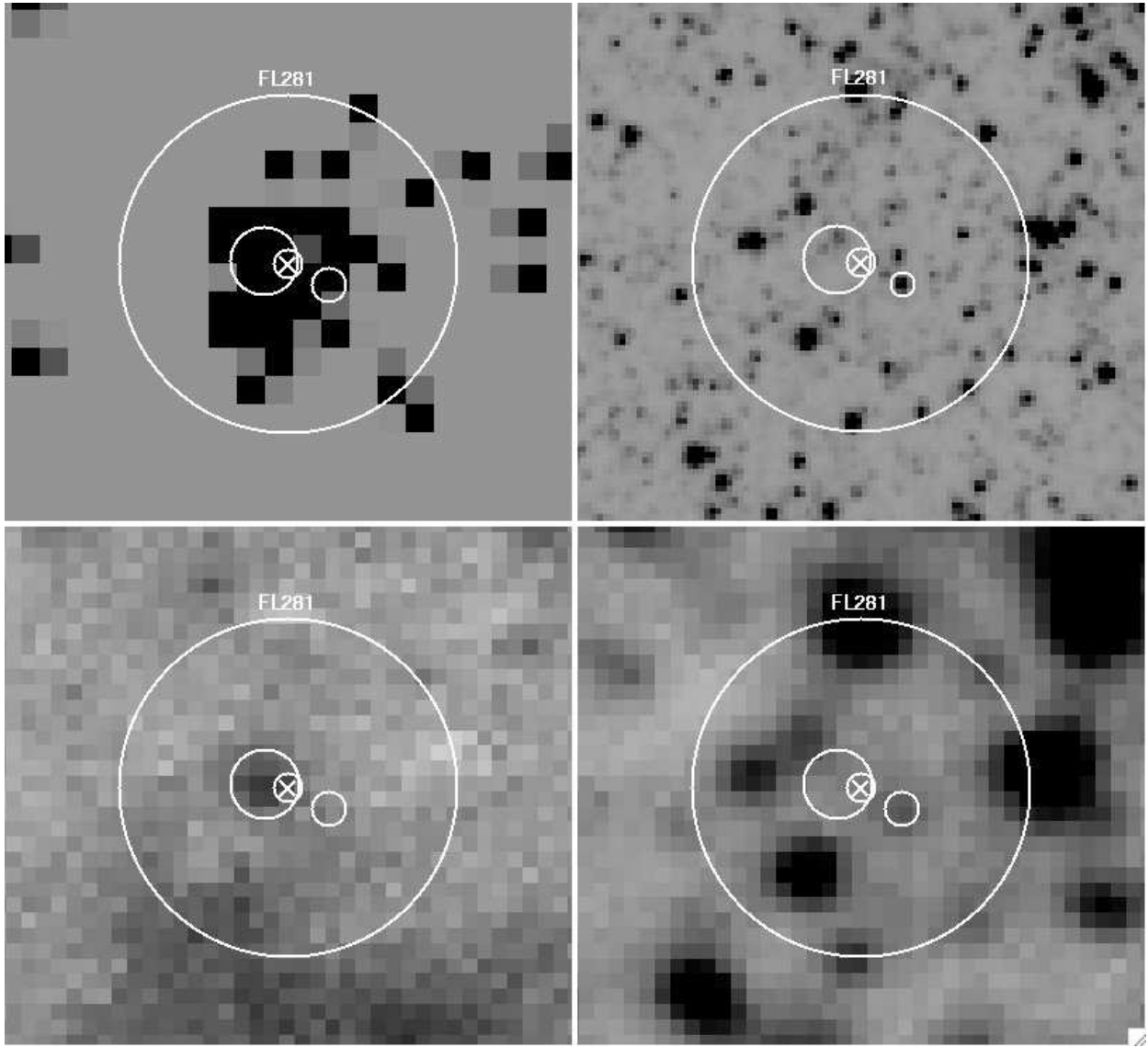}
\figcaption{X-ray and optical images of FL281, the candidate PWN in M33.  The  upper two panels show ({\it left}) the 0.35-2 keV X-ray image of the region and ({\it right}) the {\it HST} F606 W image.  The lower two panels show ({\it left}) a continuum-subtracted \HA\  image and ({\it right}) a V band image, both from LGGS.  The region files show a large 3\arcsec\ radius circle centered on the X-ray source, the radio position as an $\times$, a small circle showing the \HA\  source, and a smaller circle centered on a bluish star.  Comparison of the positions of 23 point-like X-ray and radio matches suggests the X-ray$-$radio astrometric offsets are $+0.10\arcsec\pm 0.16\arcsec$ in RA and $-0.19\arcsec\pm 0.17\arcsec$ in declination. \label{fig_pwn}
}
\end{figure}

%\begin{figure}
%\plottwo{fig_color.eps}{fig_color_rate.eps}
%\figcaption[colors]{left.  The hardness ratio as a function of size, where the soft band is 0.35-1 keV and the medium band is 1-2 keV.  right.   The hardness ratio as a function of broad band count rate.   The big outlier is EM25 again.  There is no obvious trend of hardness with radius, as one might expect if kT is lower for large diameter SNRs.  It's not clear this is the best way to explore this in a more systematic fashion, but it would seem that if the temperature of a SNR declines with age, that a systematic sample like this would be the best way to find out.}
%\end{figure}

%fig_ha Replaced 080812
\begin{figure}
%\plotone{fig_ha_surbri.eps}
\plotone{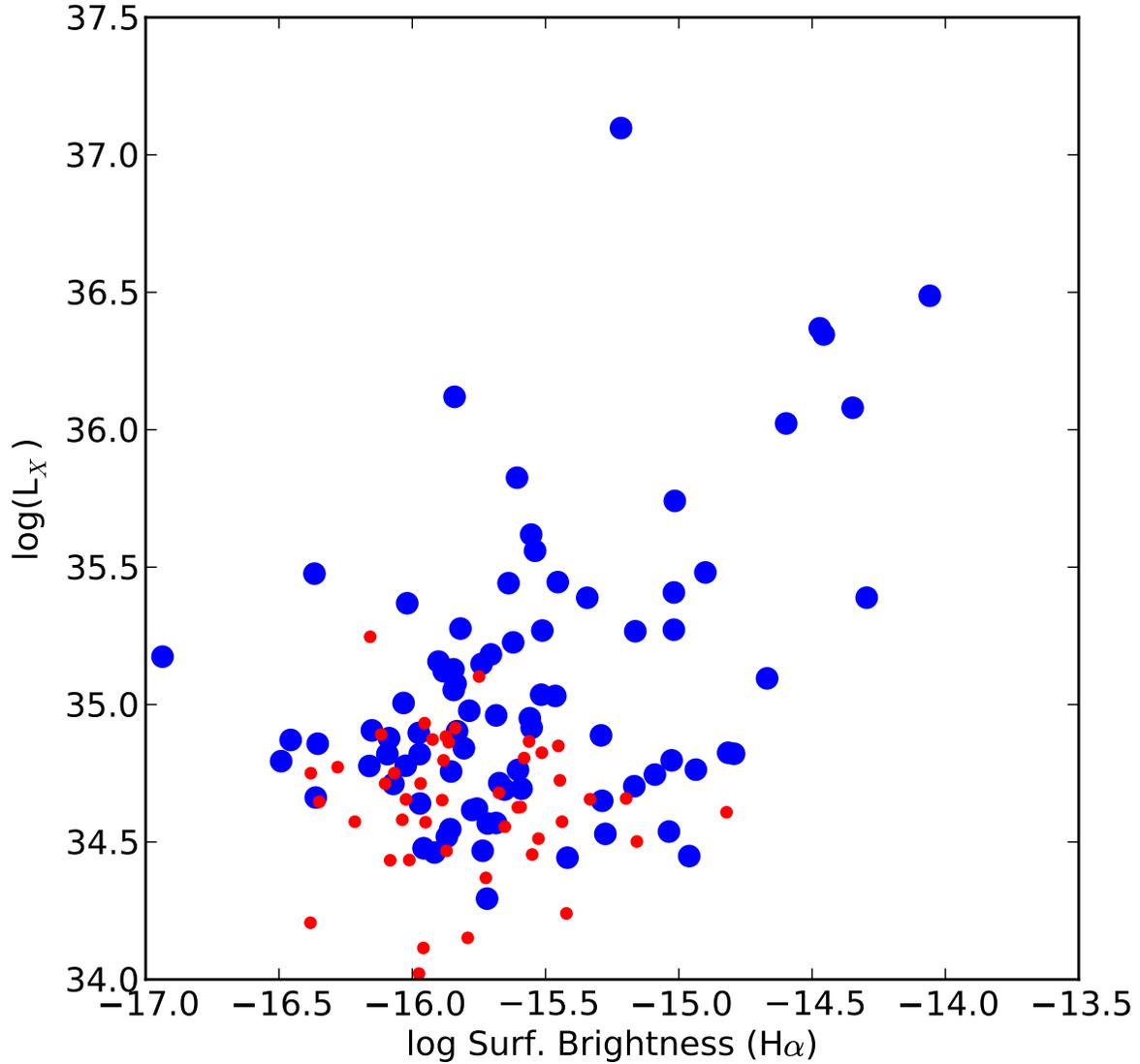}

\figcaption[ha]{The  0.35--2 keV luminosity from M33 SNRs as a function of average H$\alpha$ surface brightness.  As in earlier figures, the objects shown as blue dots are objects in our list of SNRs that were detected, while the red dots represent 2$\sigma$ upper limits for objects that were not detected in X-rays.  \label{fig_ha}}
\end{figure}

% fig_ha Replaced 080812
\begin{figure}
%\plotone{fig_ha_lum.eps}
\plotone{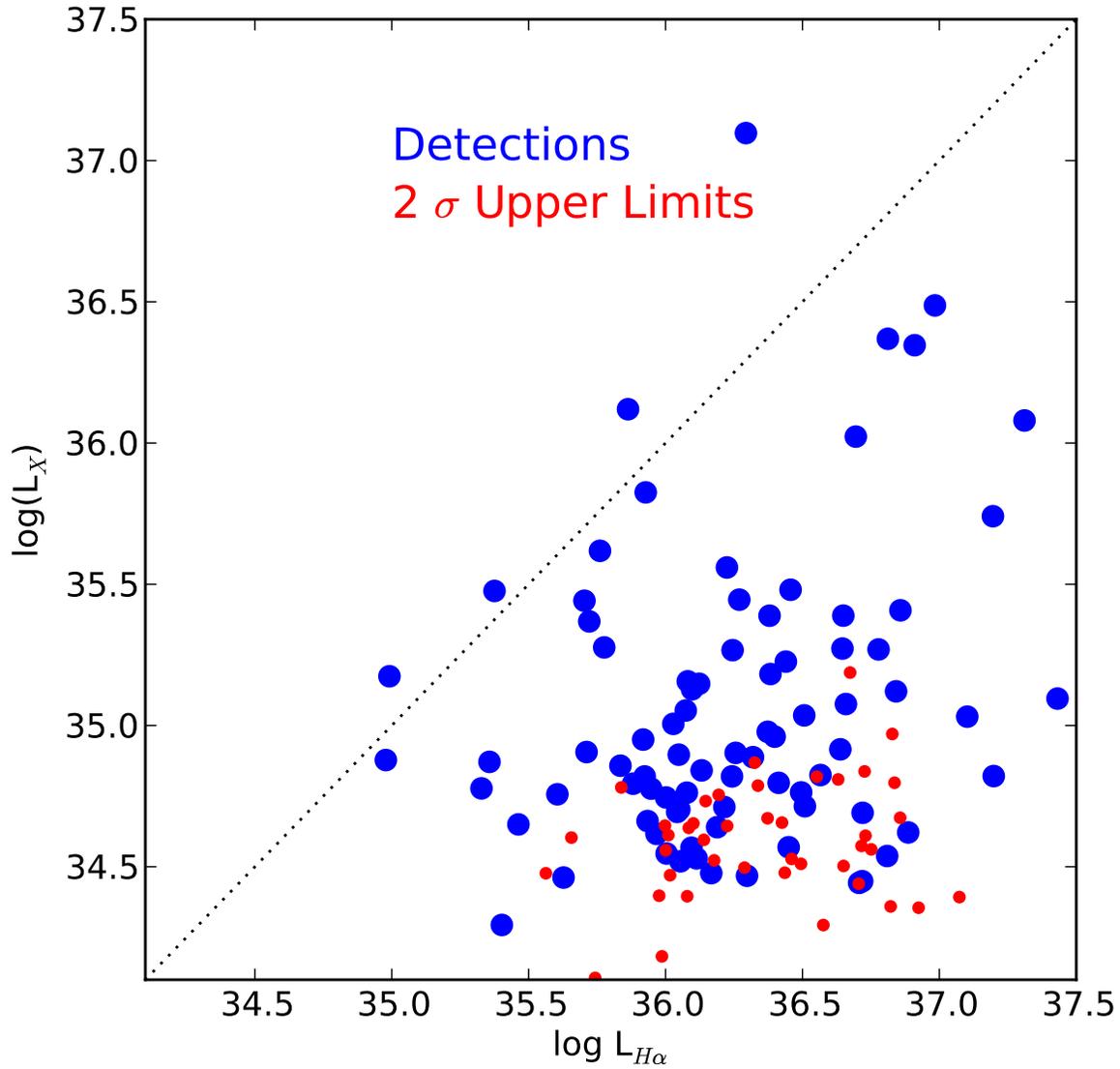}
\figcaption[ha_lum]{The X-ray luminosity as a function of H$\alpha$ luminosity.  The way the points are plotted is identical to Fig.\ \ref{fig_ha}.  Most of the objects---those to the lower right of the dotted line---have H$\alpha$ luminosities that exceed their X-ray luminosities.   G98-21, the brightest of the X-ray SNRs is the only SNR with an X-ray luminosity significantly higher than its \HA\  luminosity. \label{fig_ha_xray}}
\end{figure}

% Replaced 080812
\begin{figure}
%\plottwo{fig_ratio_dia.eps}{fig_ratio.eps}
\plottwo{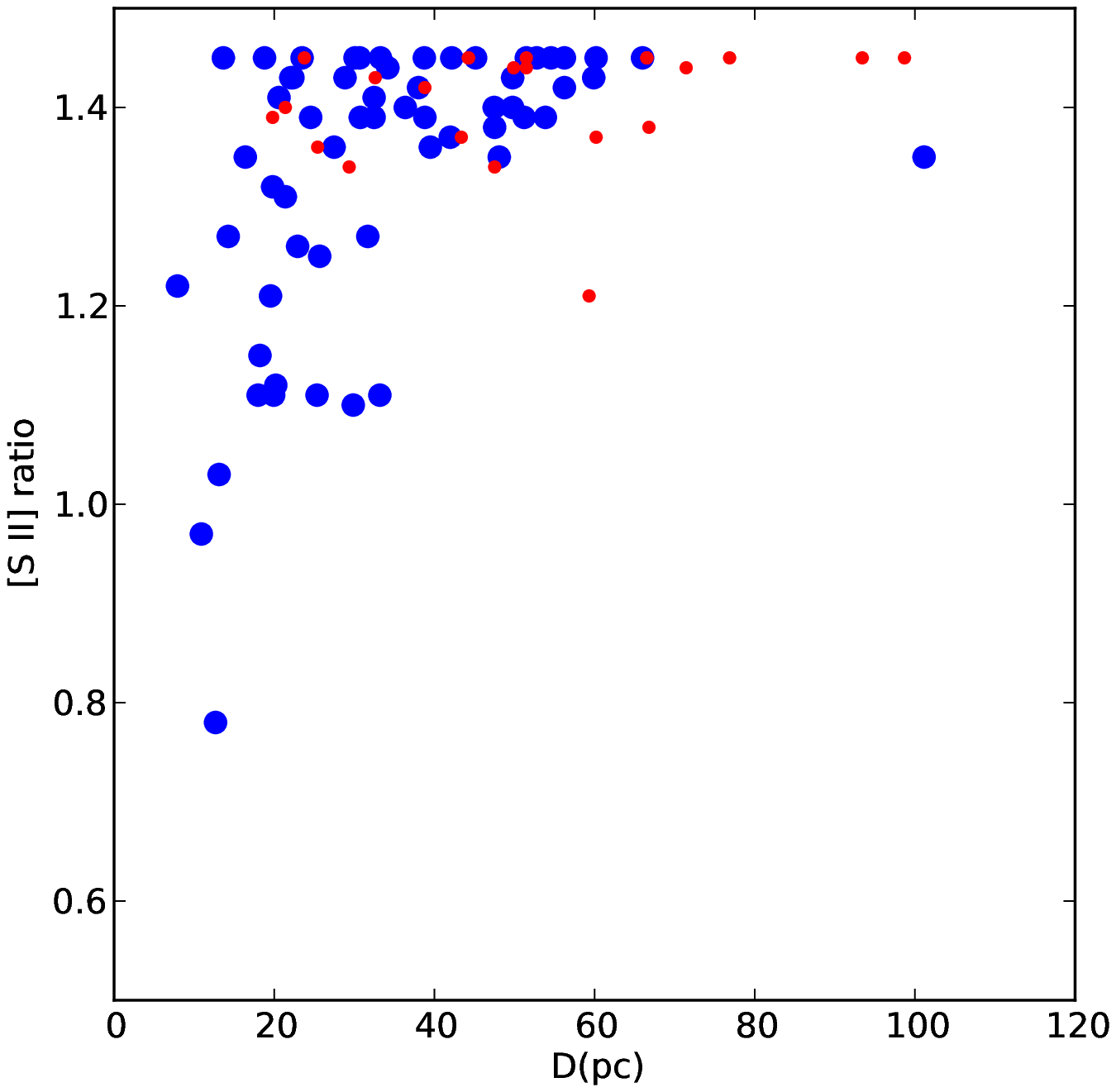}{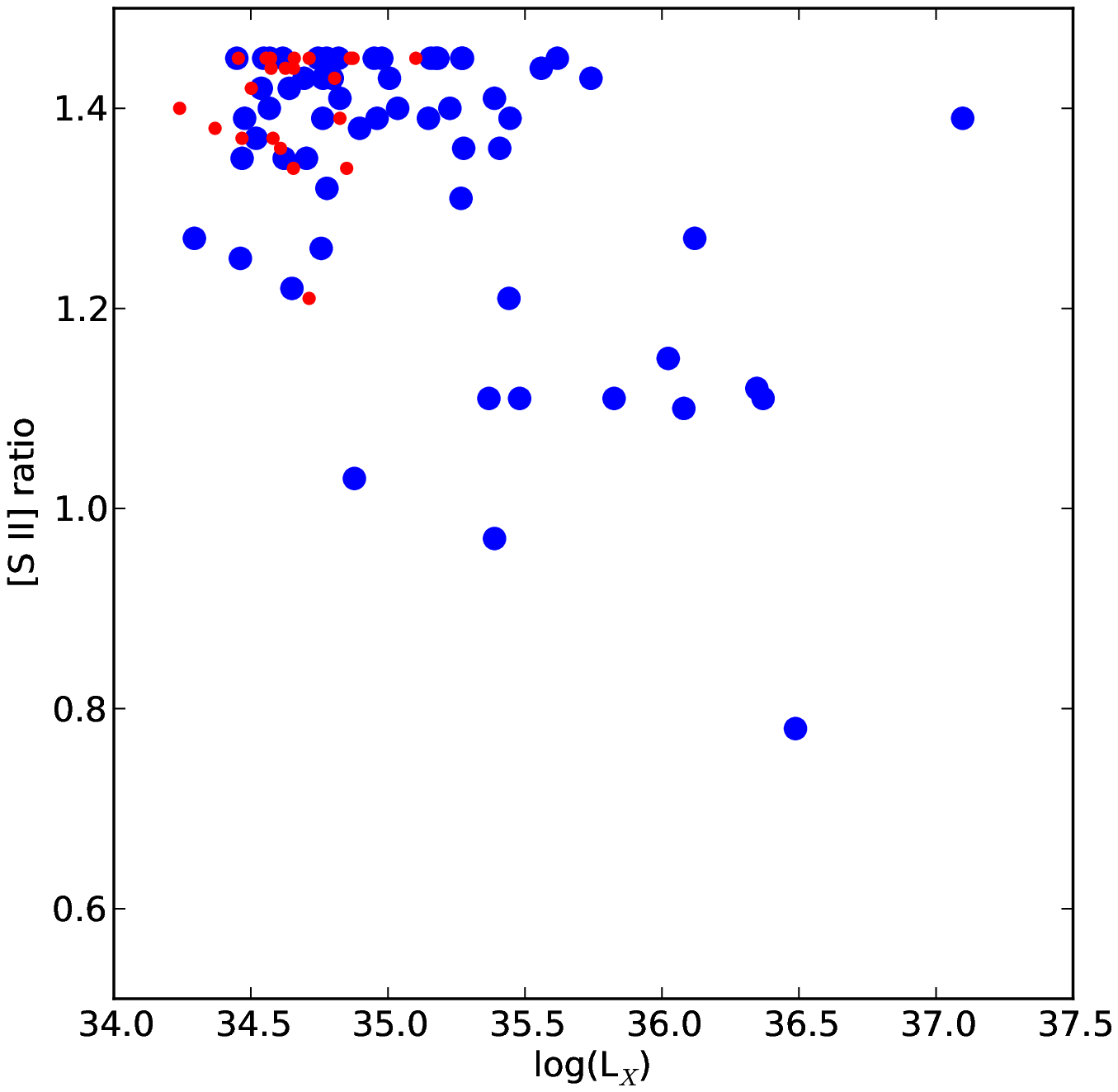}

\figcaption[opt_spec]{  
({\it left}) The density-sensitive line ratio of \sii\ $\lambda\lambda 6717,6731$ as a function of SNR diameter. Unphysical \sii\  line ratios greater than the low-density limit have been plotted at 1.45.    Large-diameter SNRs have \sii\ ratios that are nearly always near the low-density limit.  Ratios of 1.4, 1.2, 1.0, and 0.8 correspond to densities in the zone where \sii\ is produced of 50, 250, 700, and 1400 cm$^{-3}$, respectively \citep{cai93}.
({\it right})
The density-sensitive \sii\ line ratio  as a function of the 0.35--2 keV X-ray luminosity.  
Objects with line ratios $\lesssim 1.2$ (density $\gtrsim 250\;{\rm cm}^{-3}$) are nearly always detected in X-rays.  
\label{fig_s2}
}
\end{figure}

\begin{figure}
\plotone{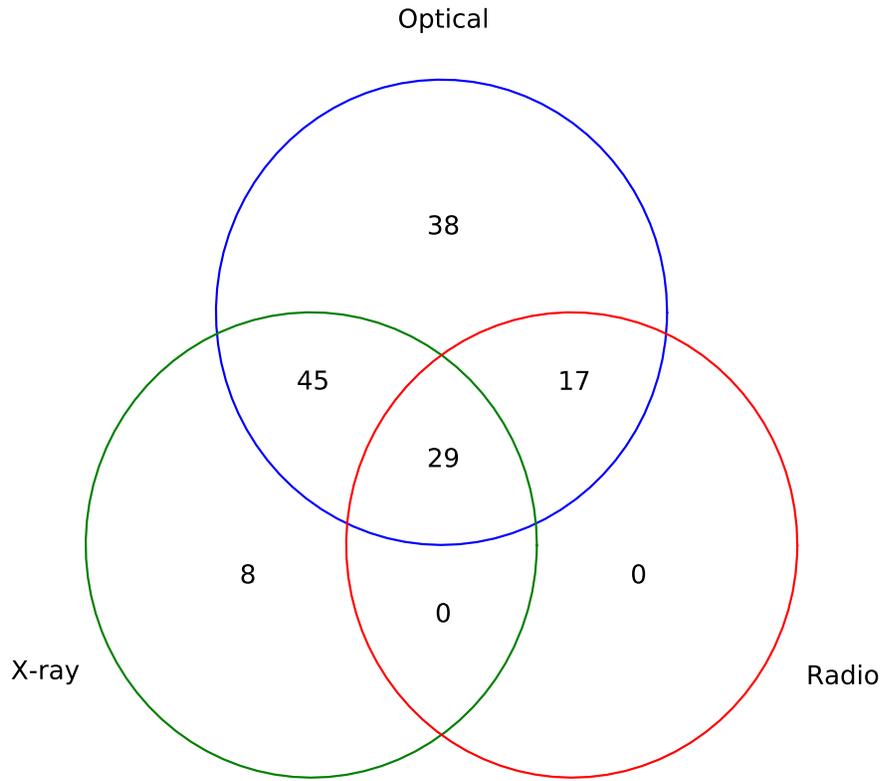}
\figcaption[venn]{A Venn diagram providing a snapshot of the current demographics of wavelength detections for  SNRs in M33.  In this figure, the optical candidates refer only to those objects for which there is evidence of shock heating, based on elevated \sii\ compared to \HA.  
%The 8 X-ray only objects are associated with some optical emission, either \HA\ or \oiii, but do not appear to show elevated \sii.  
As discussed in \S \ref{sec_final}, the diagram reflects to a substantial degree the sensitivity of searches for SNRs in M33, which is especially apparent in the absence or radio-only detections.  
%In particular, while \cite{gordon99} did detect  a substantial number of the SNRs that had been identified optically at radio wavelengths, their survey sensitivity and angular resolution were not sufficient for them to identify a radio-only sample of SNRs.  Indeed our higher angular resolution radio observations of M33 showed that seven of the 53 objects that they thought were coincident with optically-identified SNRs were nearby H II regions or background objects.  
The object FL281 (which we suggest may be a PWN) is not included in the diagram. \label{fig_venn}
}
\end{figure}

\begin{figure}
%\plotone{fig_Color_GKL21.eps}
\plotone{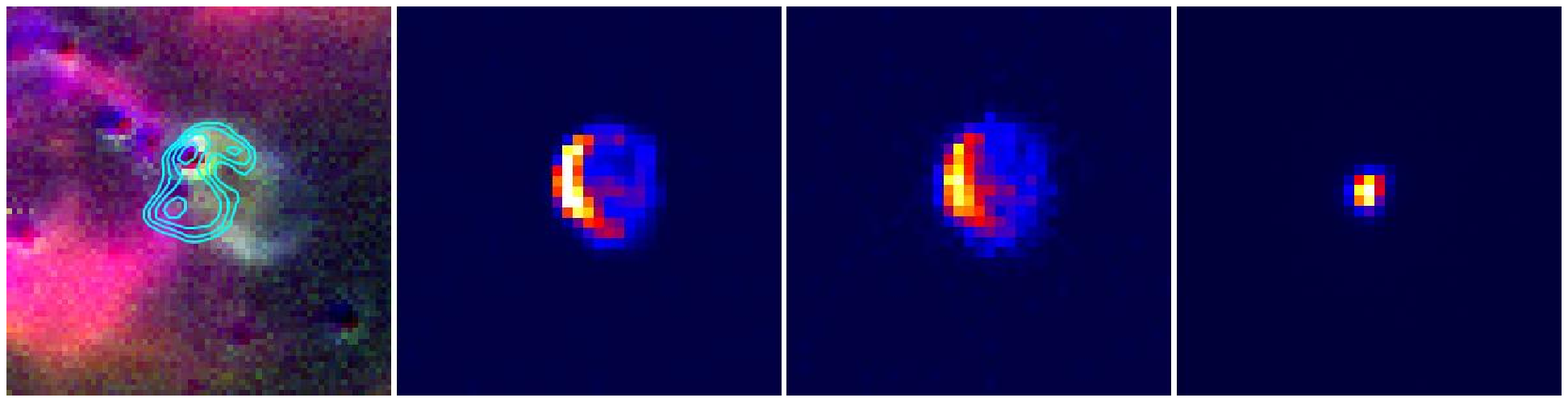}
\vspace{0.05in}
%\plotone{fig_Color_GKL28.eps}
\plotone{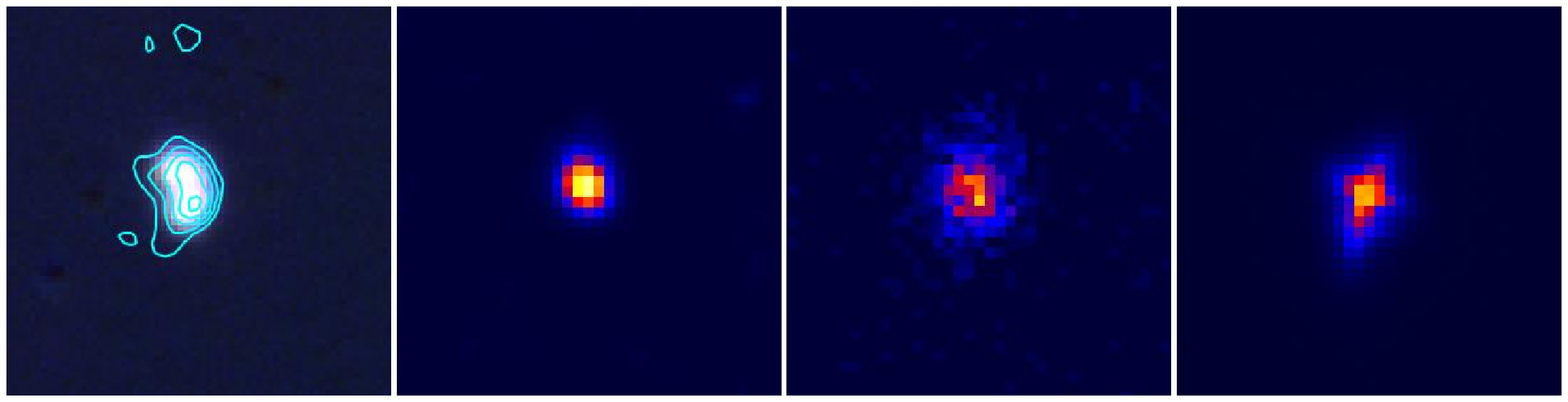}
\vspace{0.05in}
%\plotone{fig_Color_GKL29.eps}
\plotone{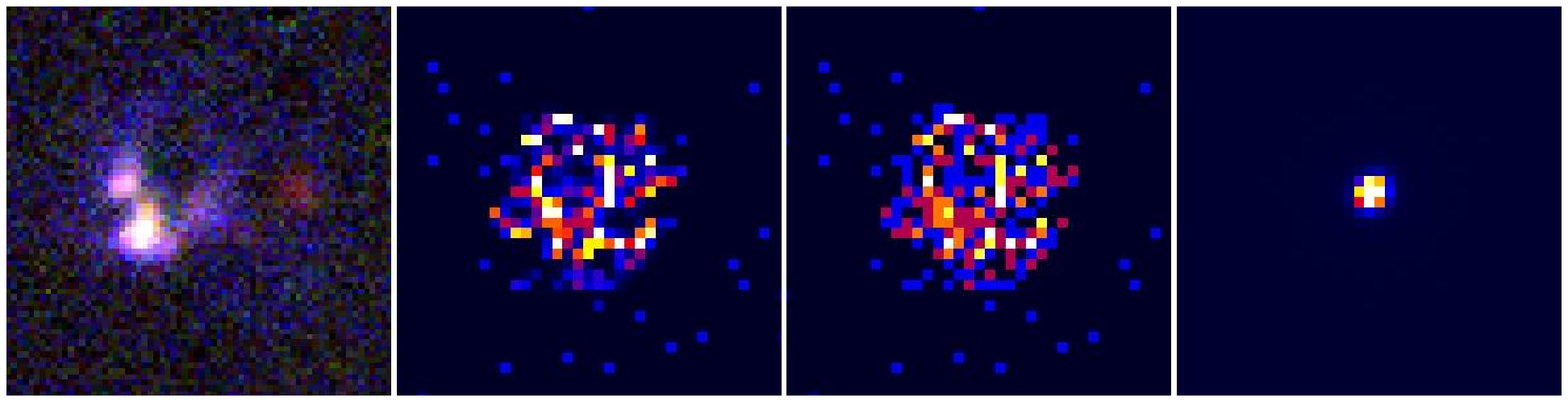}
\vspace{0.05in}
%\plotone{fig_Color_GKL31.eps}
\plotone{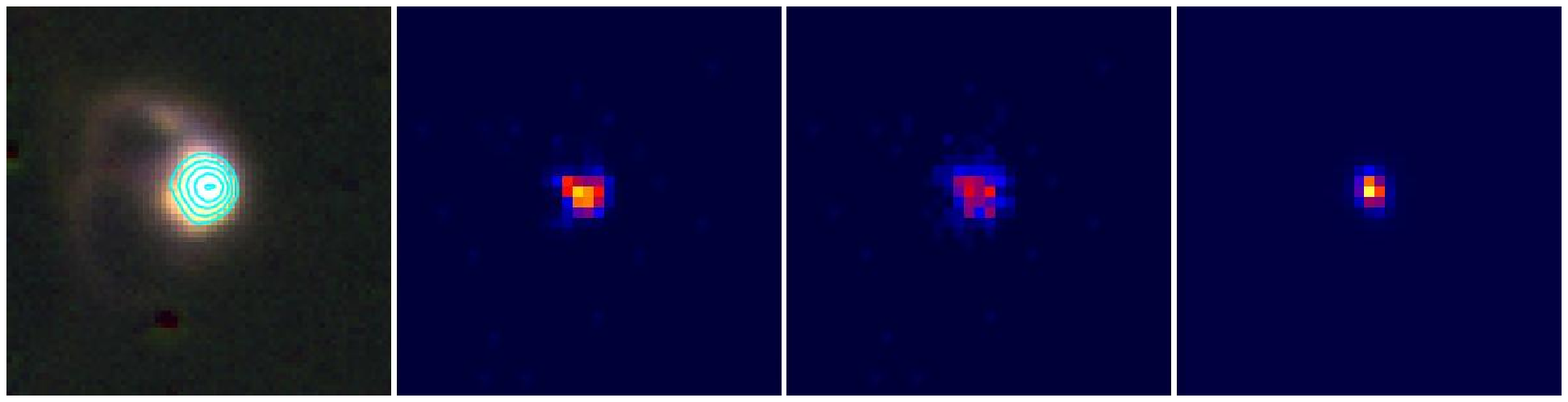}
\figcaption{Images, from top to bottom of  G98-21,  G98-28, G98-29 and  G98-31.  The left most panel in each row is a color composite made from continuum-subracted \HA\  (red),  \sii\  ( green), and  \oiii\ (blue) LGGS images.  Radio contours from our VLA observations of M33 are overlaid on the color composite, except in the case of GKL-29, which we did not detect.  The remaining panels in each row are (left to right) the deconvolved X-ray image, the raw X-ray image, and the PSF for the observations in the fields used.  \label{fig_color_bright1}}
\end{figure}

\begin{figure}
%\plotone{fig_Color_GKL35.eps}
\plotone{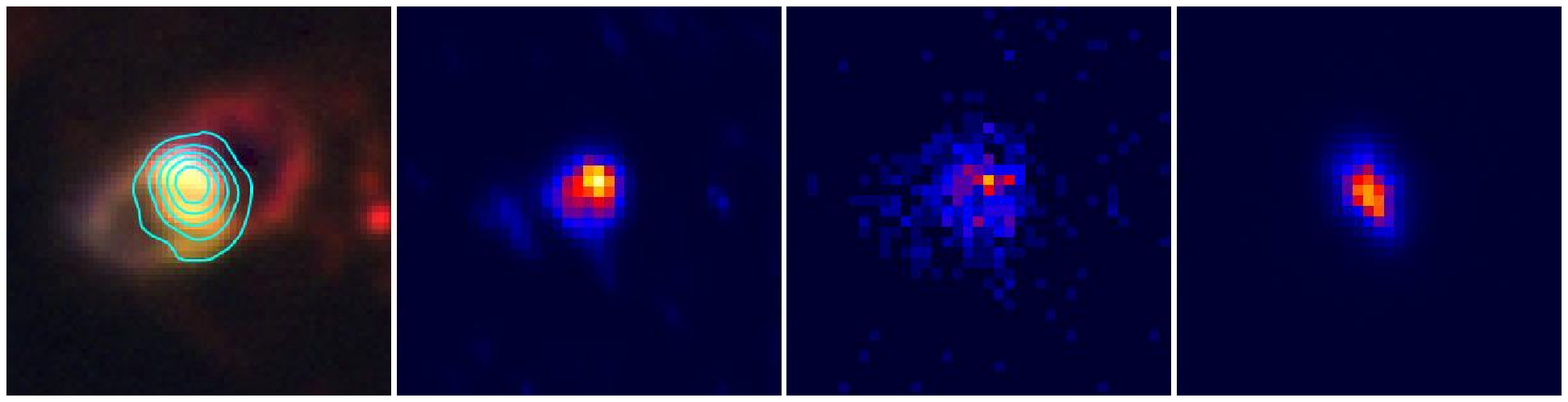}
\vspace{0.05in}
%\plotone{fig_Color_GKL55.eps}
\plotone{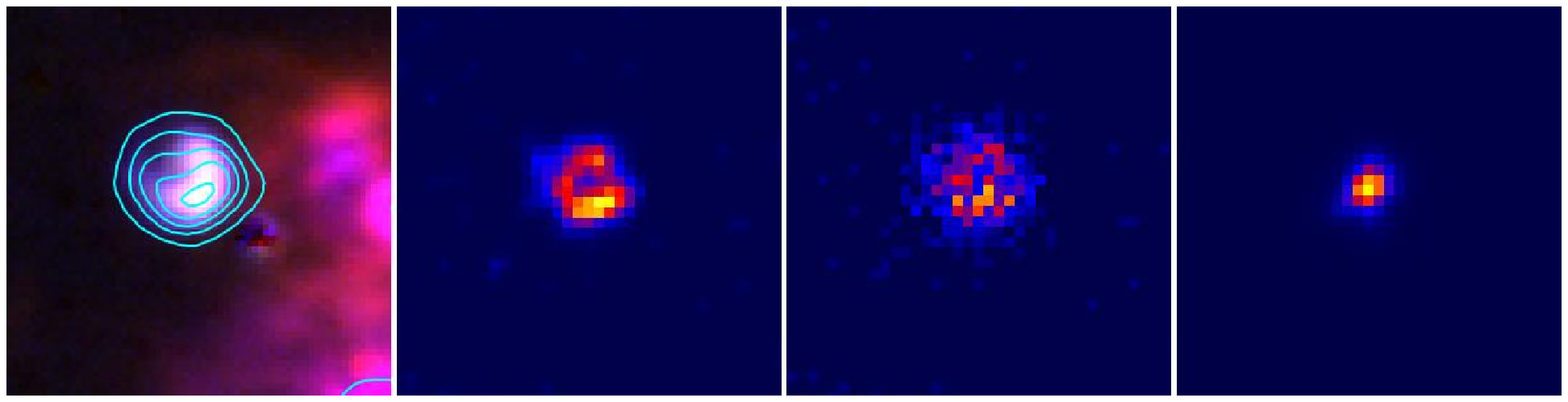}
\vspace{0.05in}
%\plotone{fig_Color_GKL73.eps}
\plotone{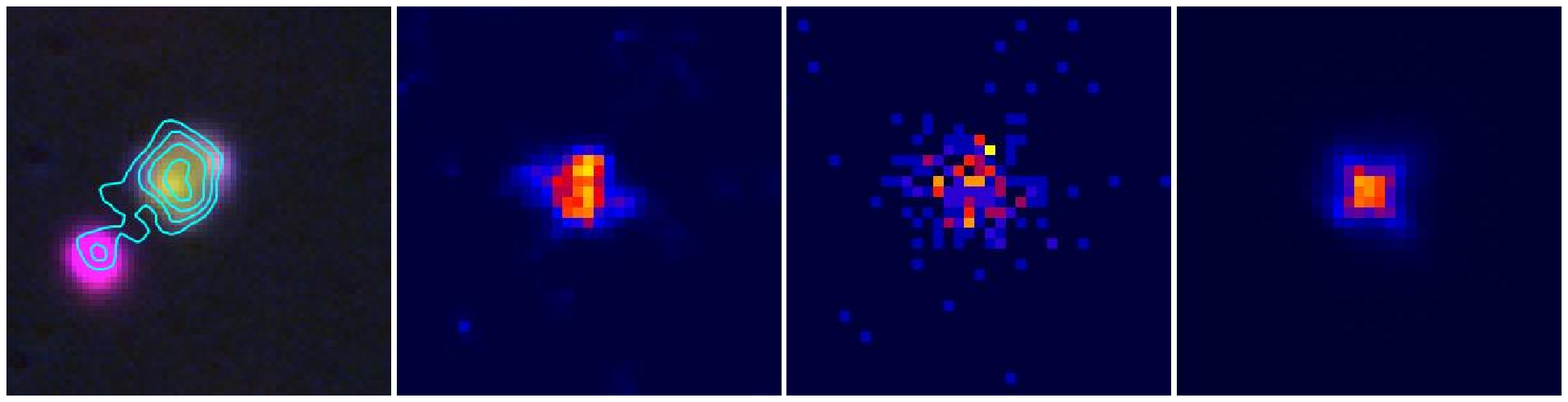}
\figcaption{From top to bottom, images of G98-35,  G98-55,  G98-73.  The format is identical to Fig.\ \ref{fig_color_bright1}.  \label{fig_color_bright2} }
\end{figure}

\begin{figure}[htbp] %  figure placement: here, top, bottom, or page
   \centering
\includegraphics[width=2in,angle=-90]{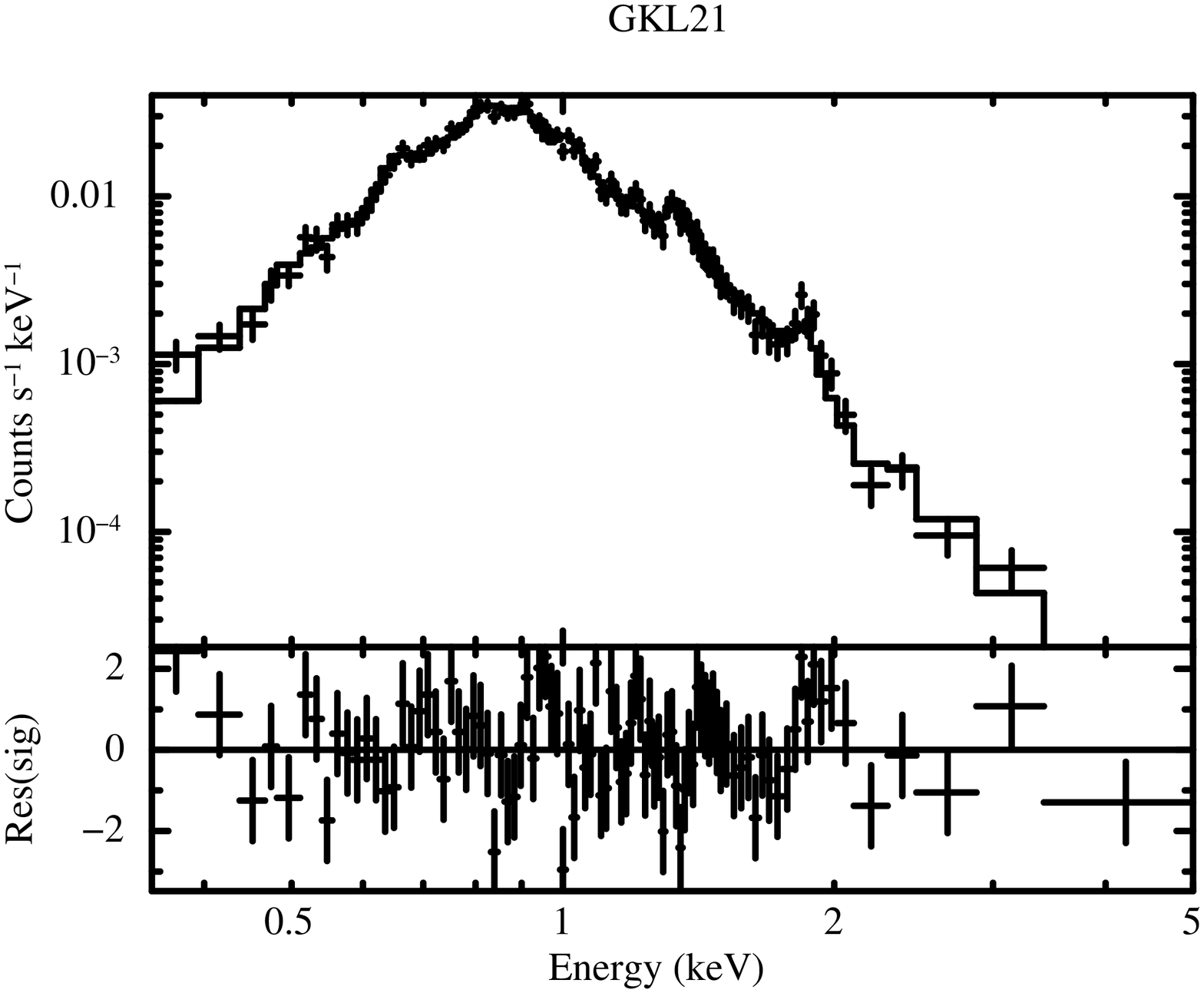}
\includegraphics[width=2in,angle=-90]{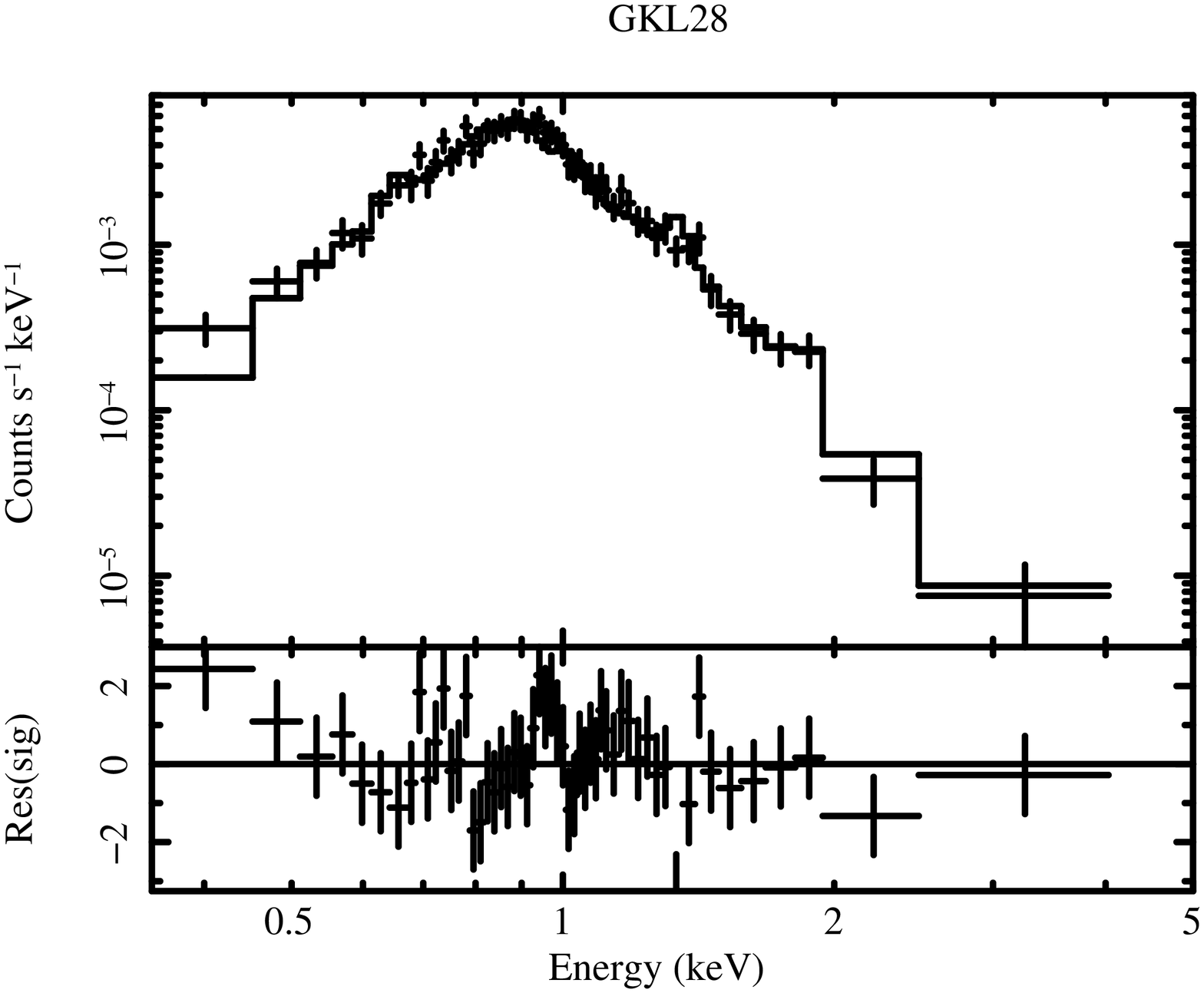}
\includegraphics[width=2in,angle=-90]{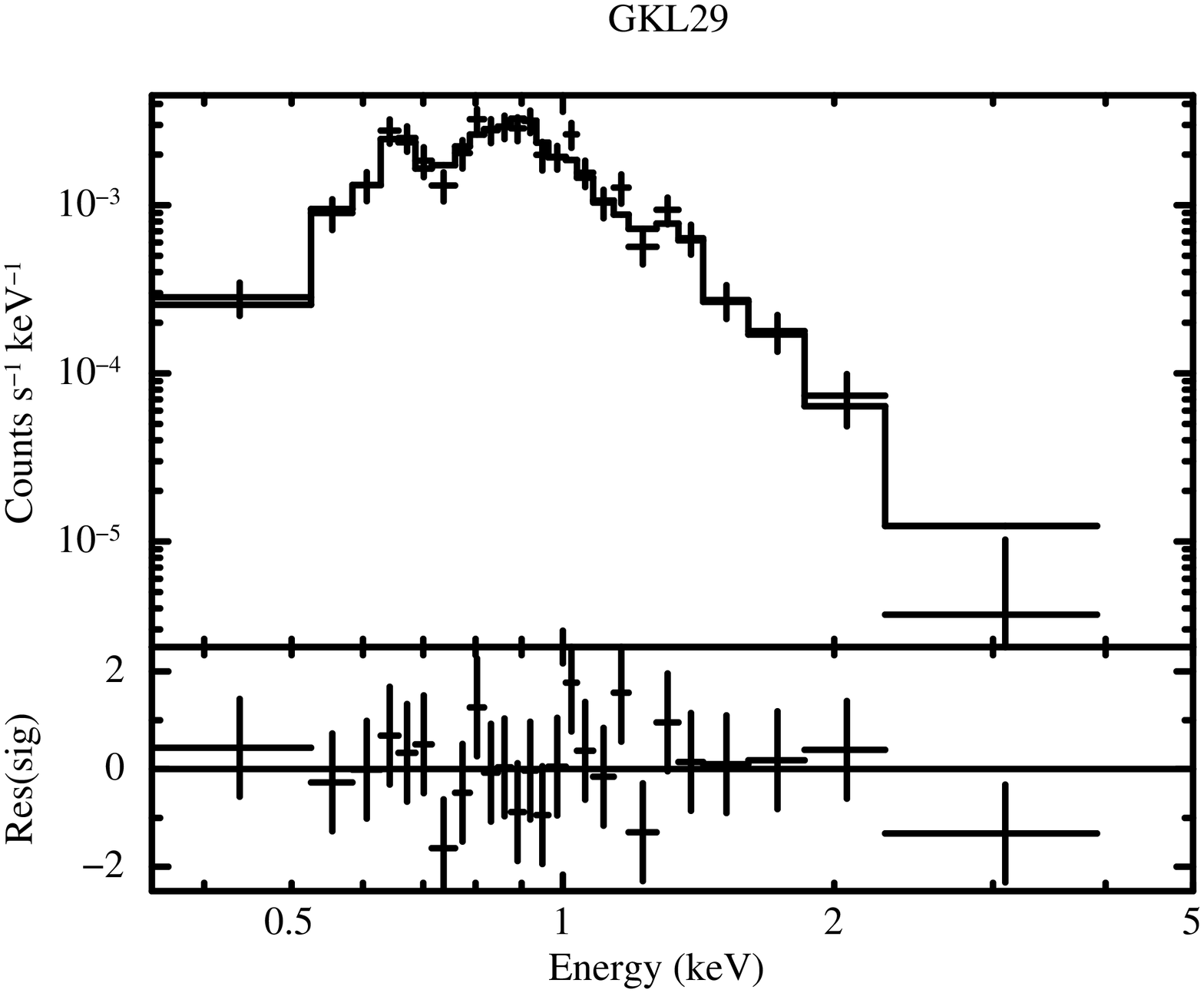}
\includegraphics[width=2in,angle=-90]{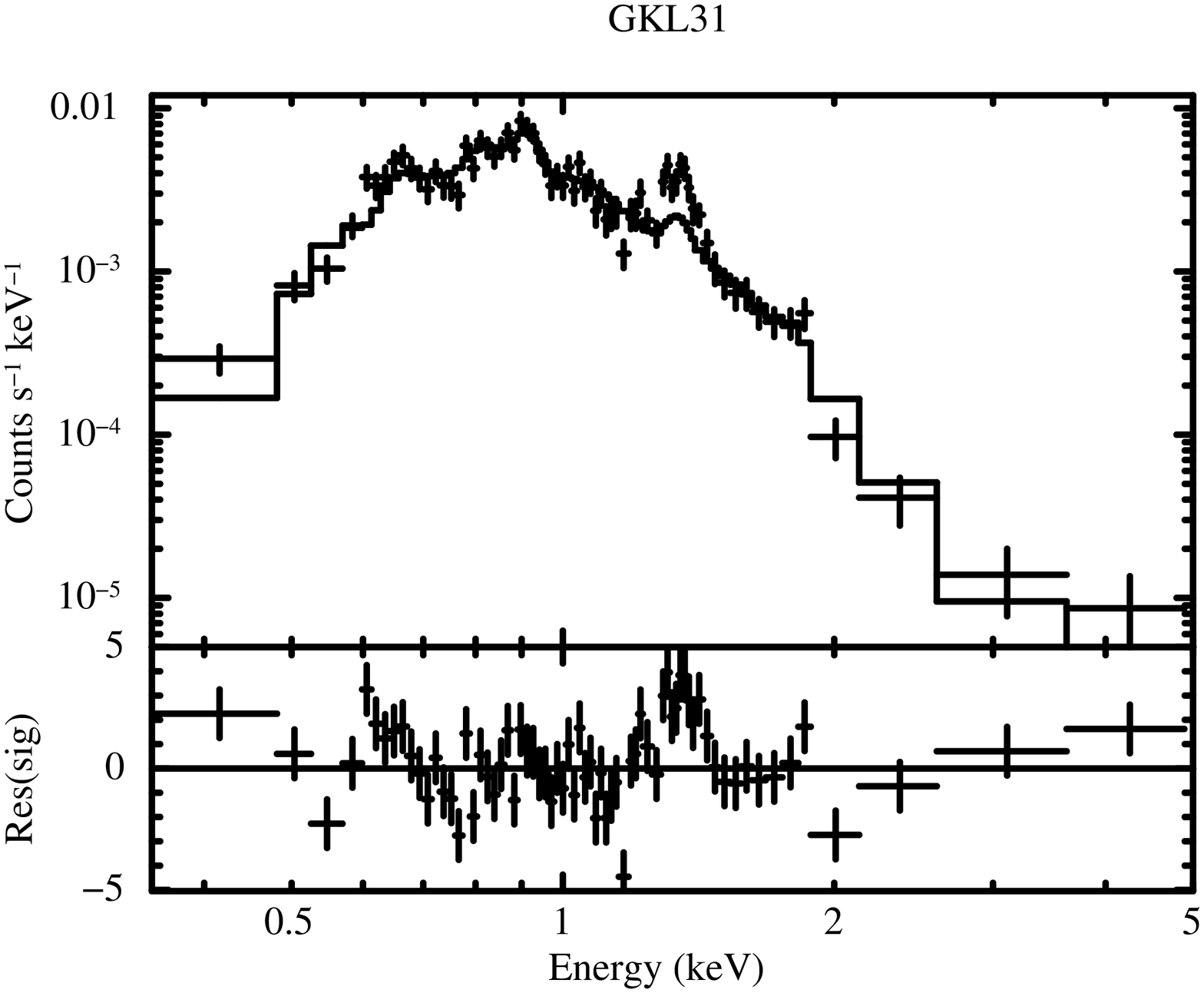}
\includegraphics[width=2in,angle=-90]{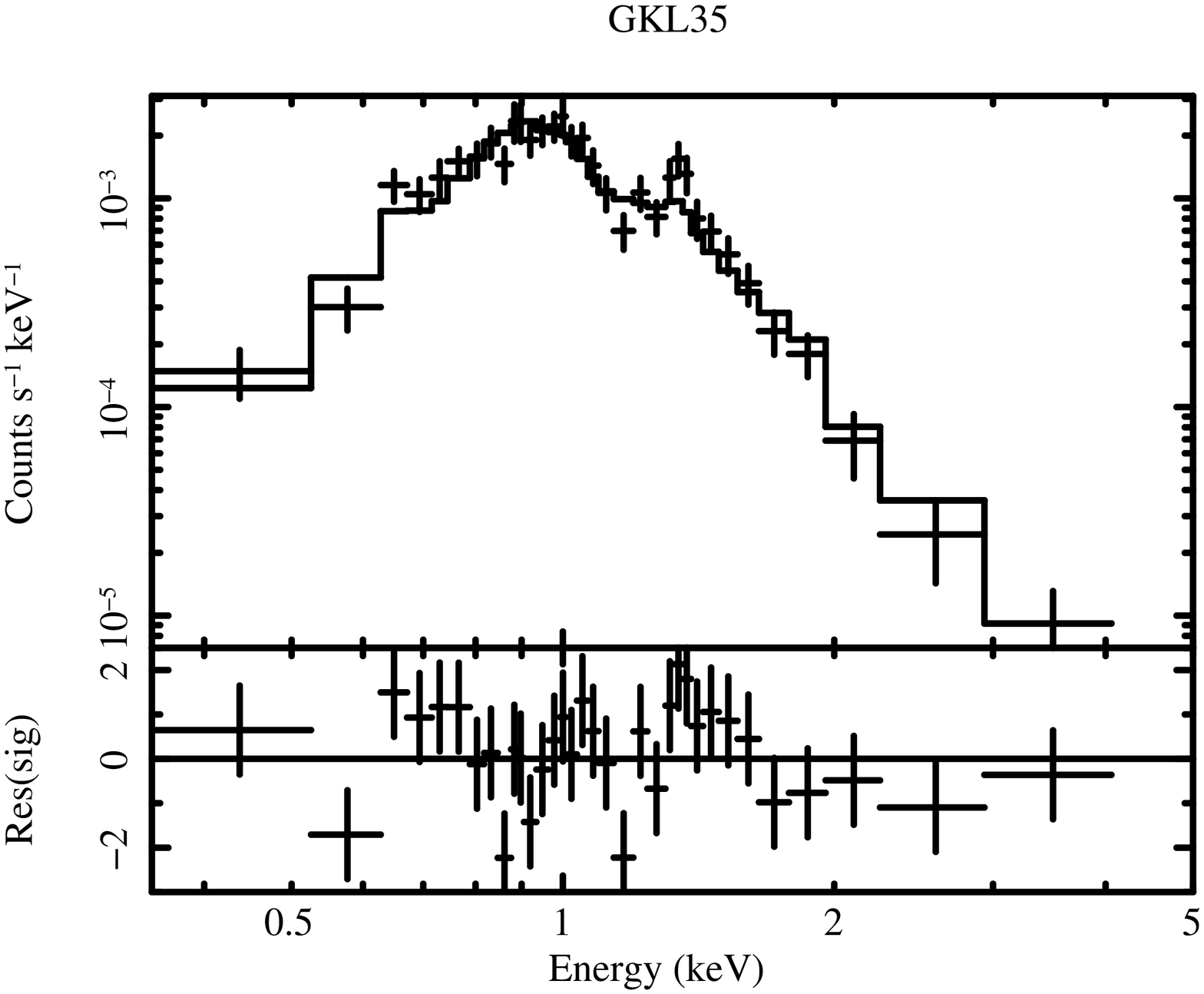}
\includegraphics[width=2in,angle=-90]{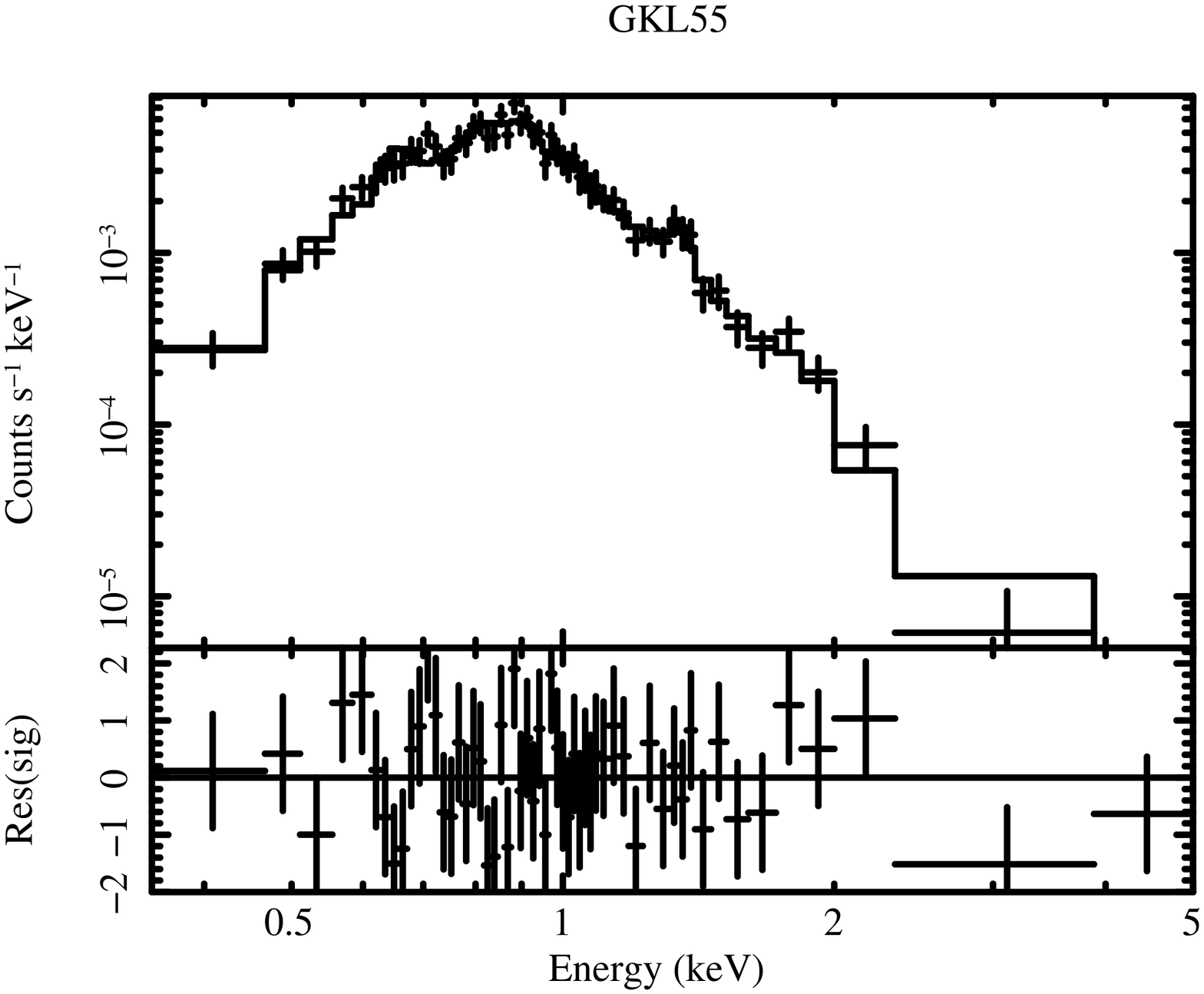}
\includegraphics[width=2in,angle=-90]{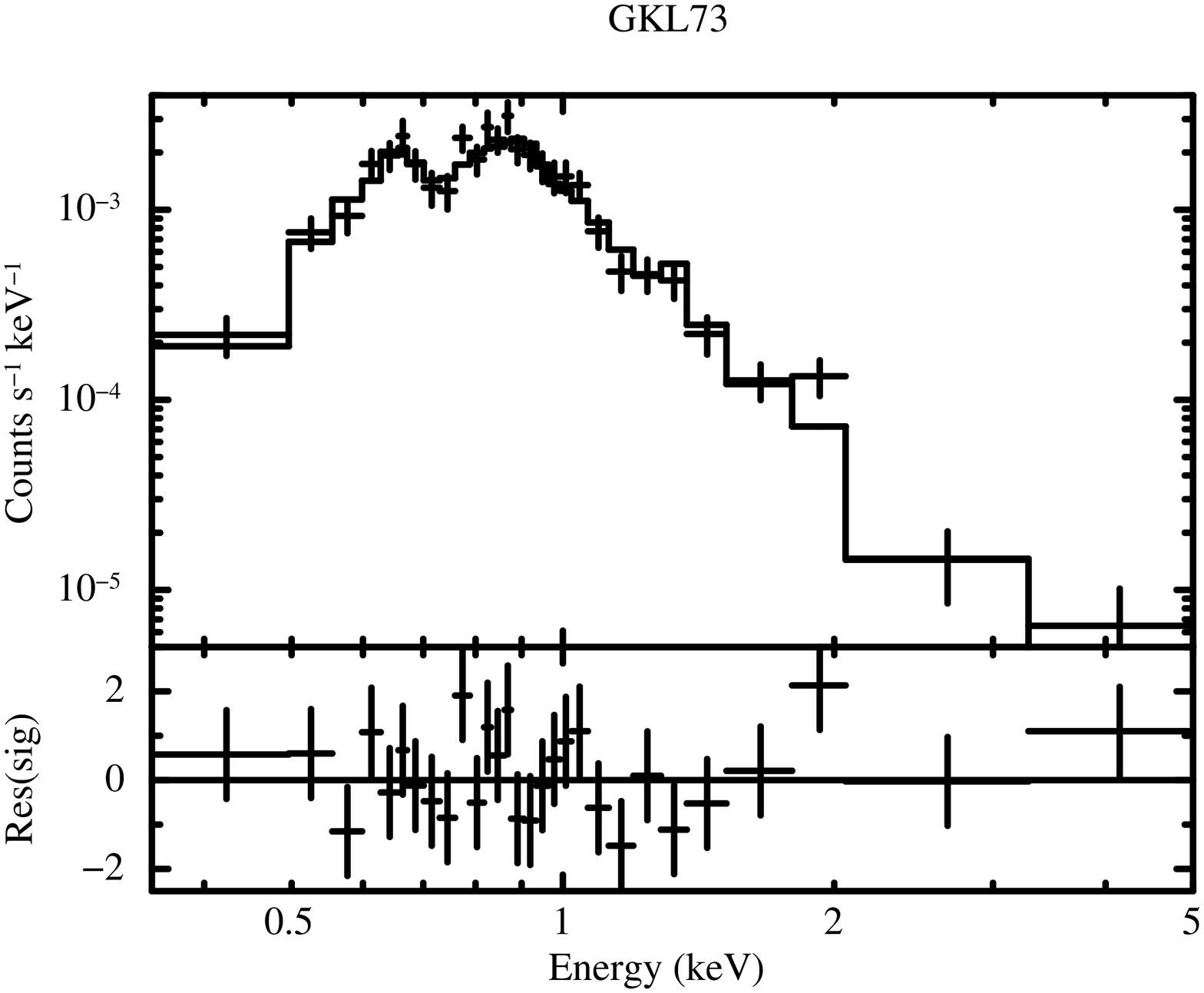}
     
   \figcaption[pshock]{Spectral fits using a pshock model for the seven brightest SNRs in M33.  Qualitatively, the model fits reproduce the data for all of the SNRs, except for G98-31, and to a lessor degree G98-35.  Both of these objects show strong emission lines especially for \ion{Mg}{11} at 1.4 keV. \label{fig_pshock}}
\end{figure}

\begin{figure}
%\plottwo{GKL31_vpshock.eps}{GKL35_vpshock.eps}

%     \includegraphics[width=2in,angle=-90]{GKL31_vpshock.eps} 
\includegraphics[width=2in,angle=-90]{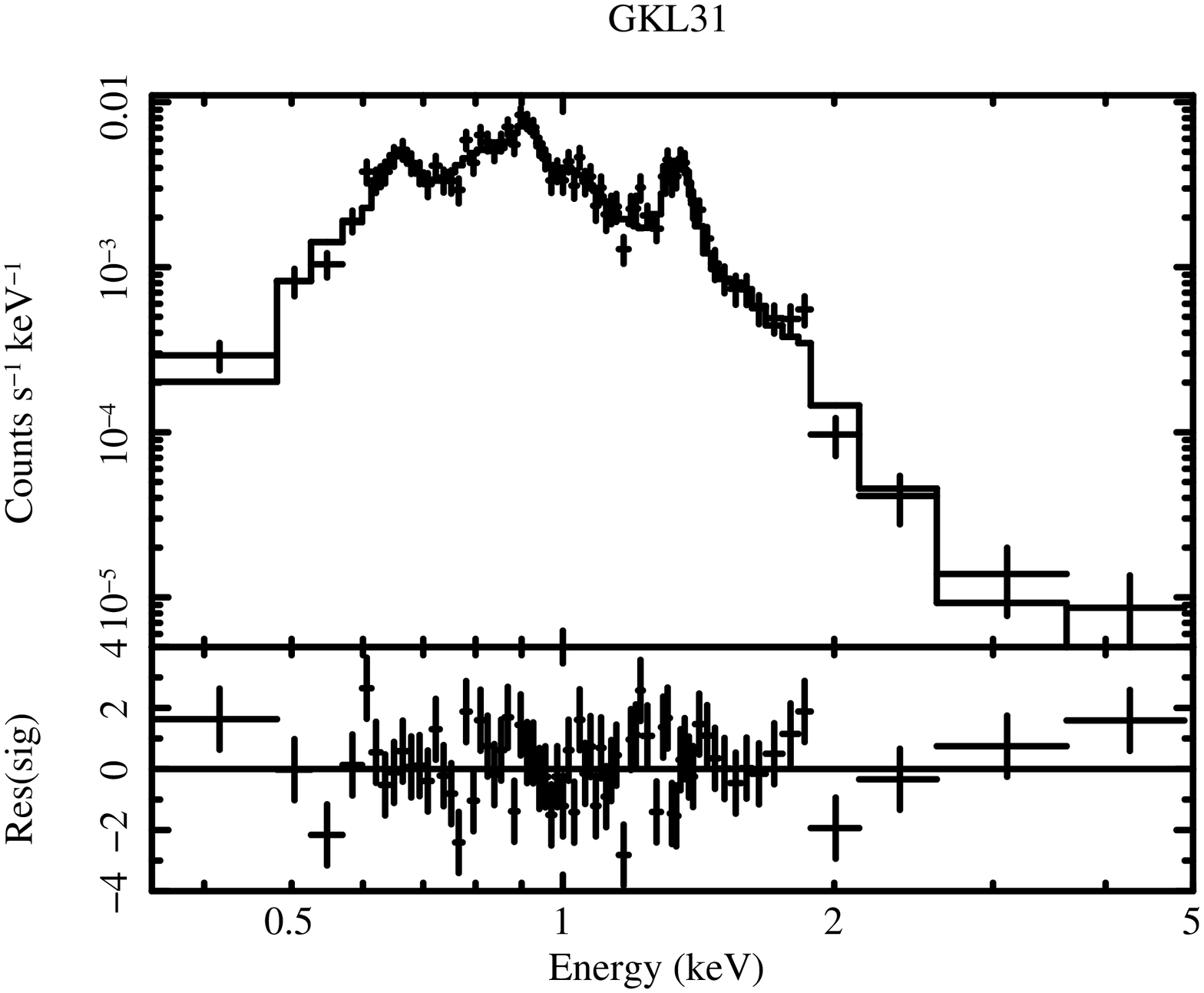}
\includegraphics[width=2in,angle=-90]{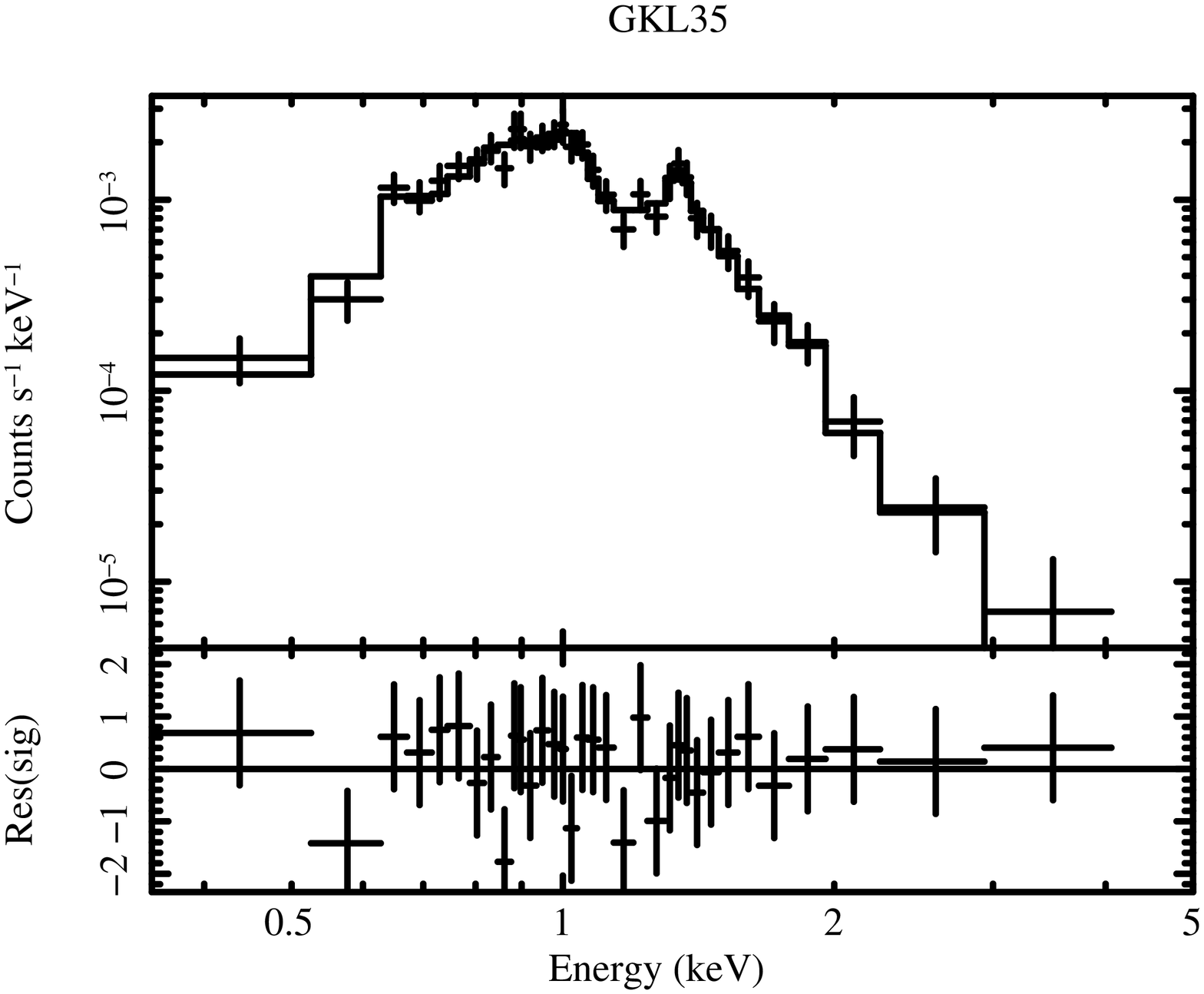}

\figcaption[vpshock]{Spectral fits using a vpshock model for G98-31 and G98-35, where in both cases O, Ne, Mg and Fe abundances have been allowed to vary.  The abundances for remainder of the elements were fixed at the metallicity values derived from the pshock fits. \label{fig_vpshock}}
\end{figure}

%Replaced 080812
\begin{figure}
%\plotone{fig_lumfunc.eps}
\plotone{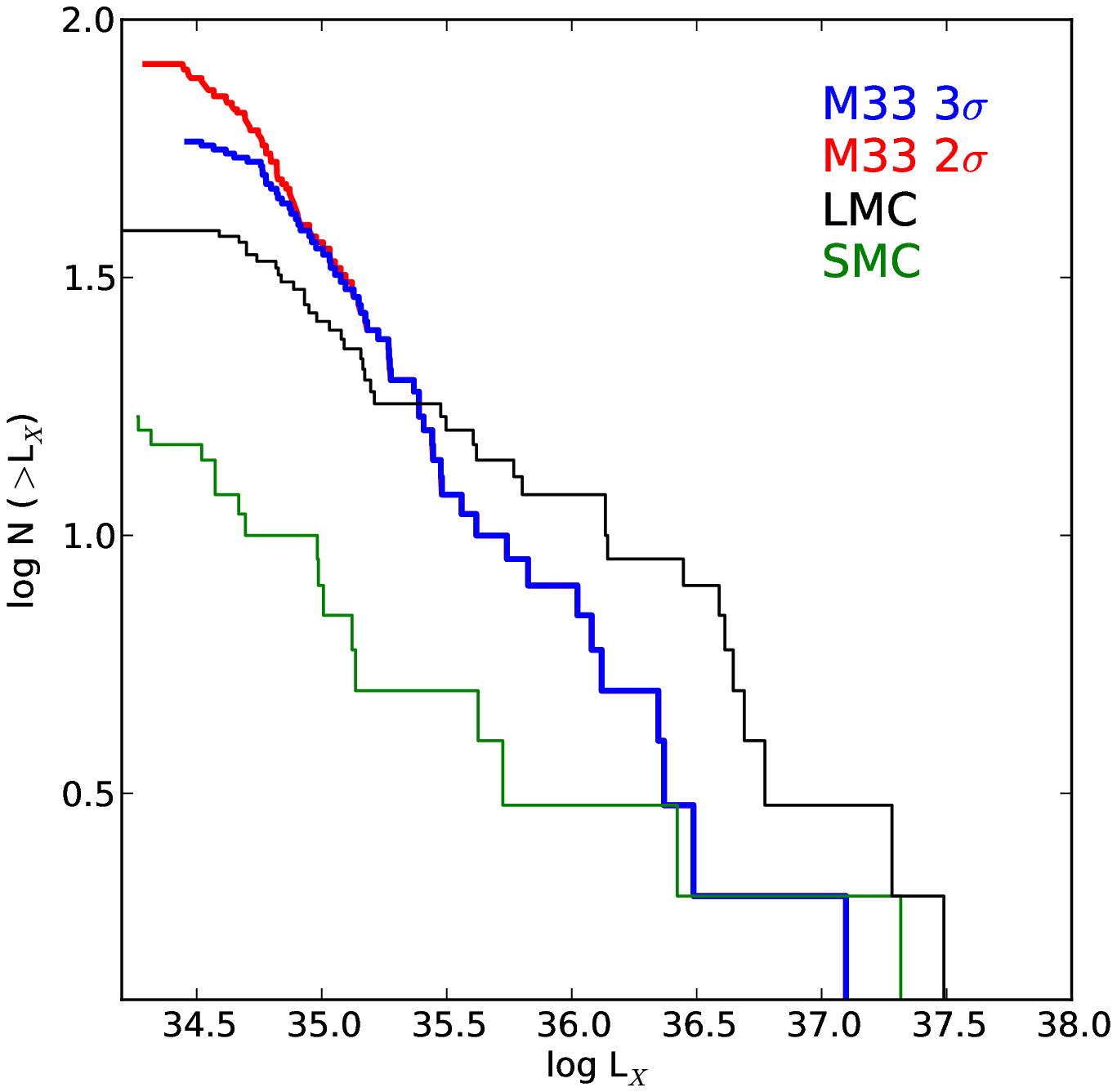}
\figcaption[lumfunc]{
The 0.35--2 keV  luminosity function for SNRs in M33.  The sources detected at 3$\sigma$ or greater are shown in blue; those only detected at 2$\sigma$ are shown red. 
For comparison, the luminosity functions for SNRs in the Large and Small Magellanic Clouds are also shown in black and green, respectively.  For the LMC, we used data from  \cite{williams99, williams04}, \cite{haberl99} and \cite{bamba06}; for the SMC we used \cite{filipovic08} and \cite{vanderHeyden04}. In all cases we converted count rates to luminosities assuming simple thermal plasma models with kT of 0.6 kev and an absorbing N(H) of \EXPU{5}{20}{cm^{-2}}.  We assumed a distances of 50 and 60 kpc for the Large and Small Clouds, respectively. \label{fig_lumfunc}
}
\end{figure}

\clearpage

%APPENDIXFIGS
%\begin{appendix}
\appendix

\begin{figure}
%\plotone{fig_atlas_GKL01.eps}
\plotone{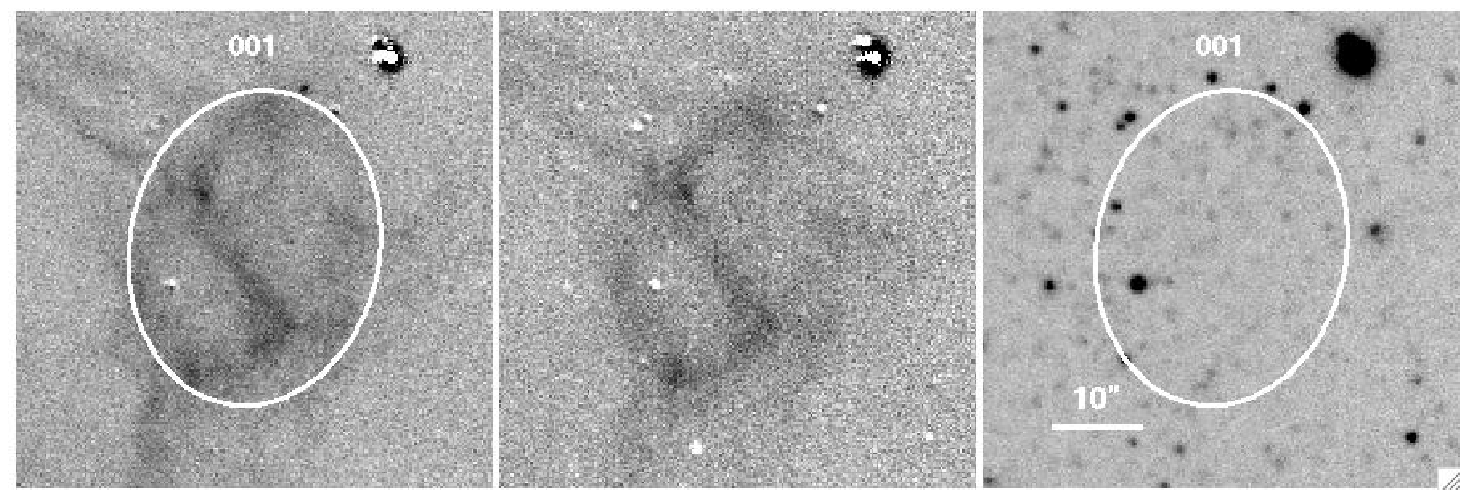}
\vspace{0.05in}
%\plotone{fig_atlas_GKL02.eps}
\plotone{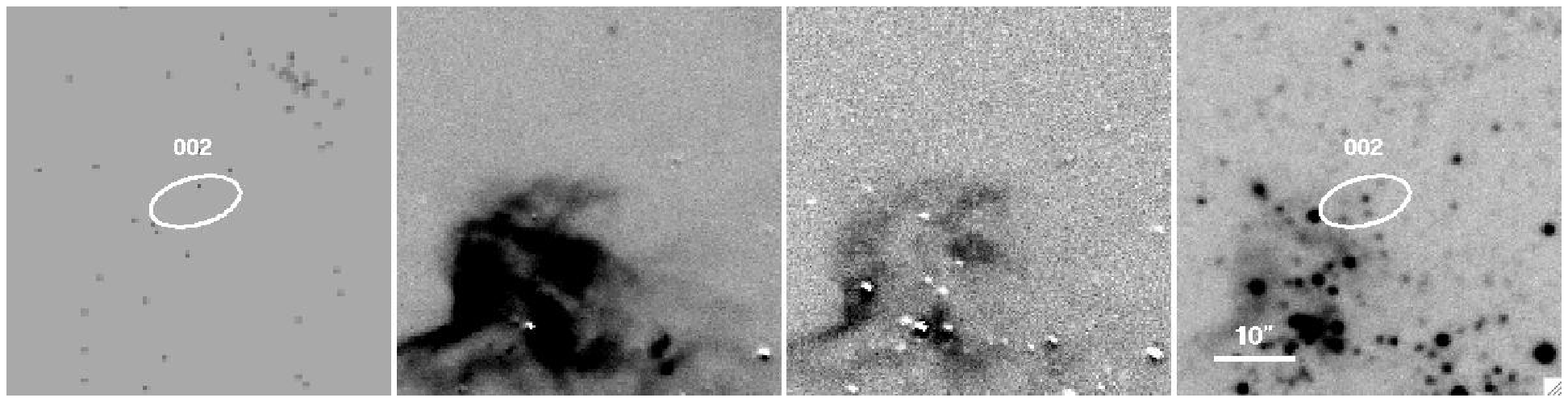}
\vspace{0.05in}
%\plotone{fig_atlas_GKL03.eps}
\plotone{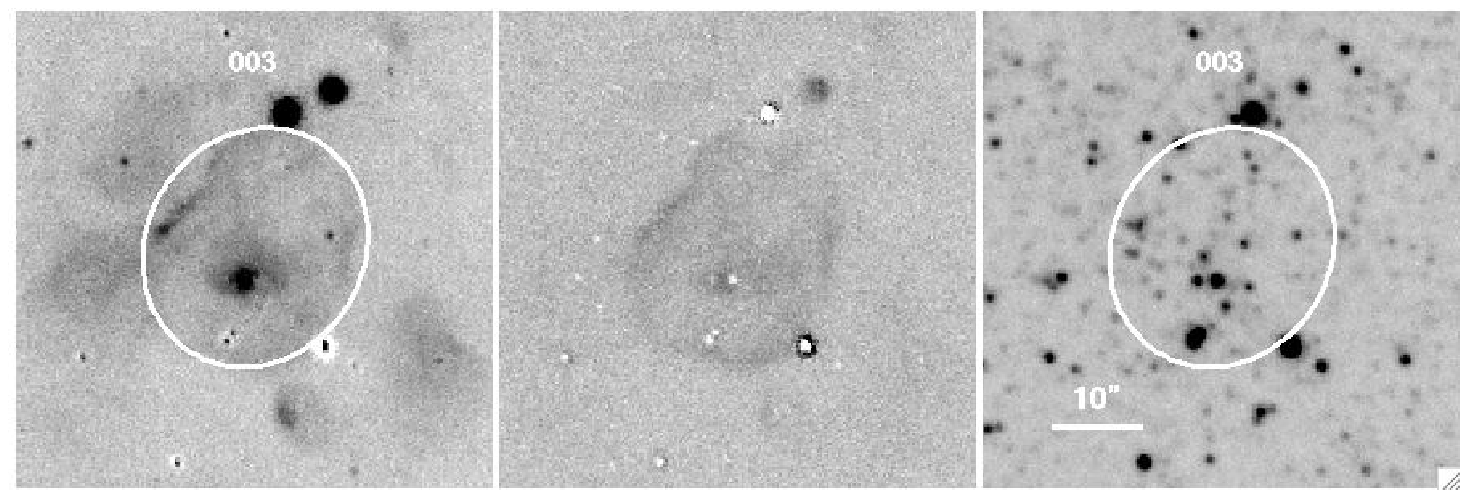}
\vspace{0.05in}
%\plotone{fig_atlas_GKL04.eps}
\plotone{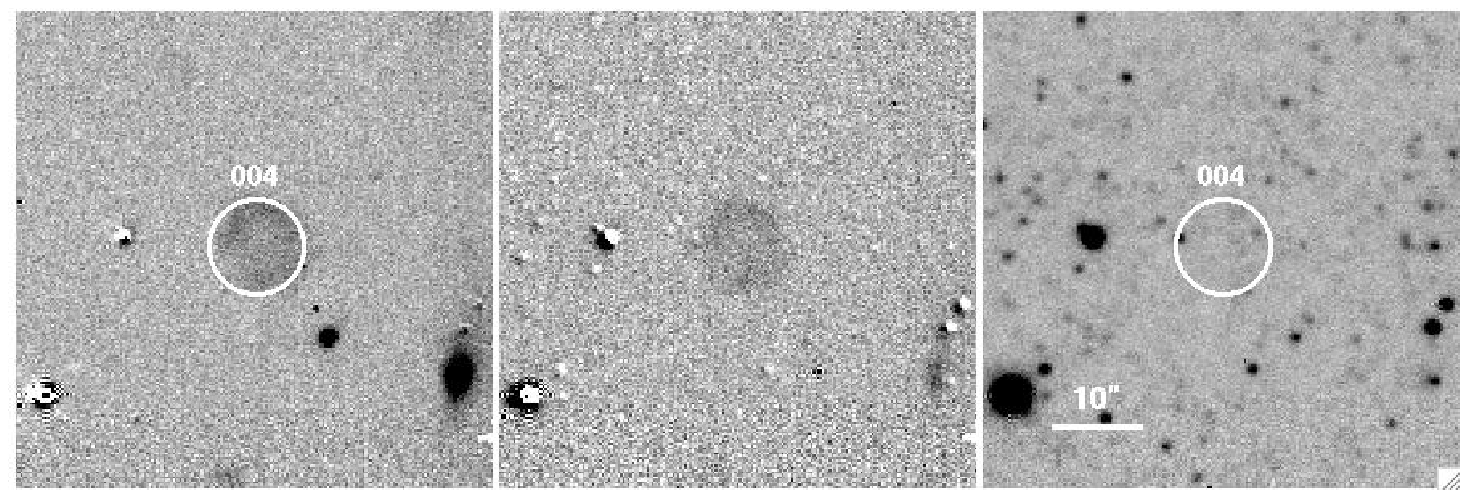}
\figcaption{Images from top to bottom of  G98-01,  G98-02,  G98-03,  G98-04.  Each row contains panels showing from left to right the X-ray, \HA, \sii, and V-band images. The region used to extract the X-ray count rates and optical fluxes are shown.   For the X-ray data this region file was convolved with the point-spread function to produce an appropriate region file at each off-axis angle.  The labels for the regions correspond to the the names listed in Table 3, in this case L10-001, L10-002, L10-003, and L10-004.  Objects for which the X-ray panel is missing are objects that were outside the \chase\ survey region. \label{fig_atlas01}  }
\end{figure}

\begin{figure}
%\plotone{fig_atlas_FL016.eps}
\plotone{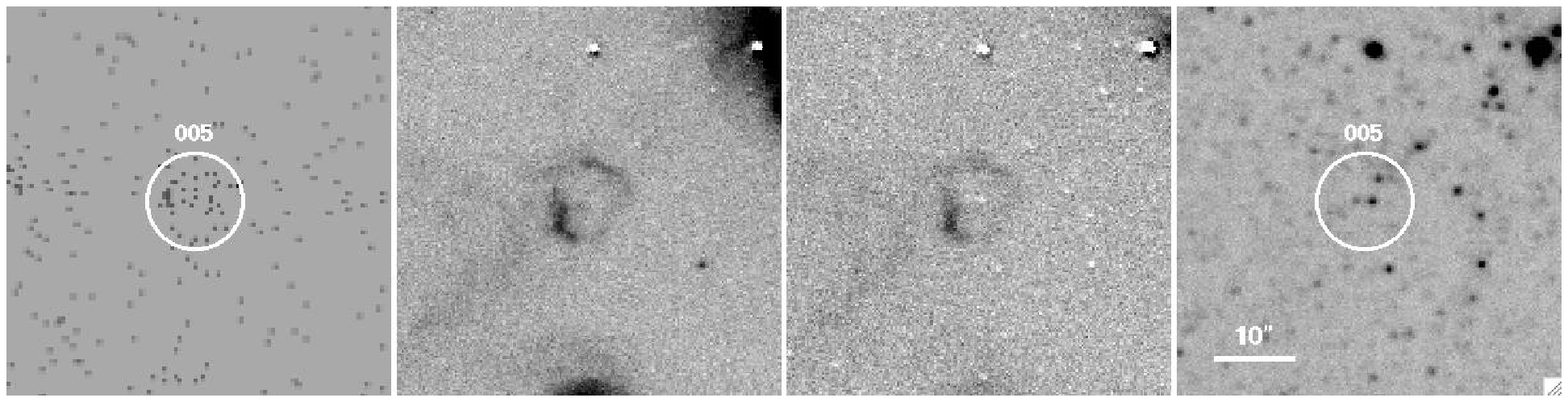}
\vspace{0.05in}
%\plotone{fig_atlas_GKL05.eps}
\plotone{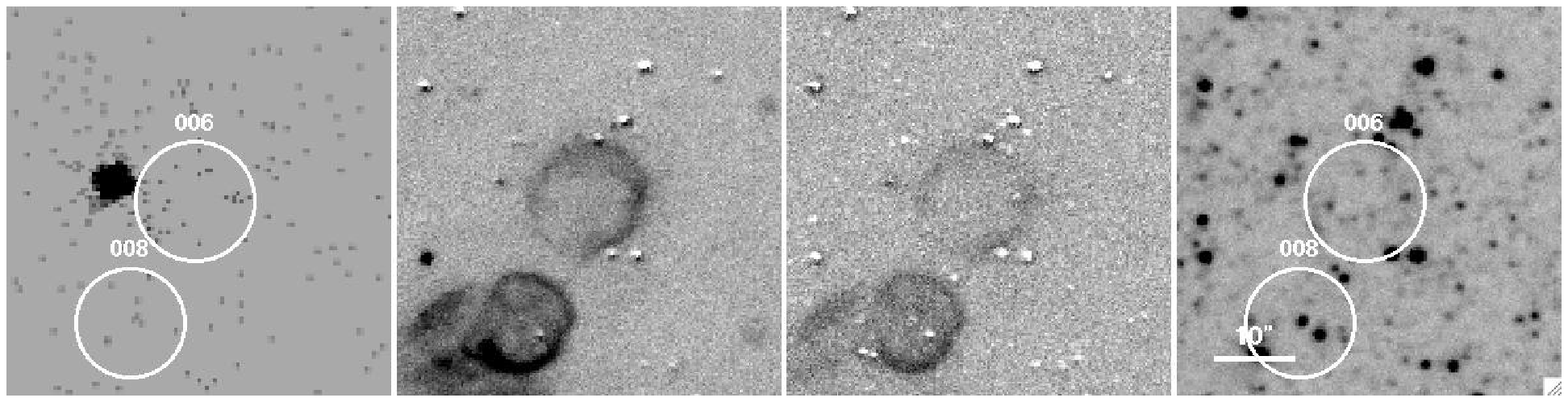}
\vspace{0.05in}
%\plotone{fig_atlas_kip-S.eps}
\plotone{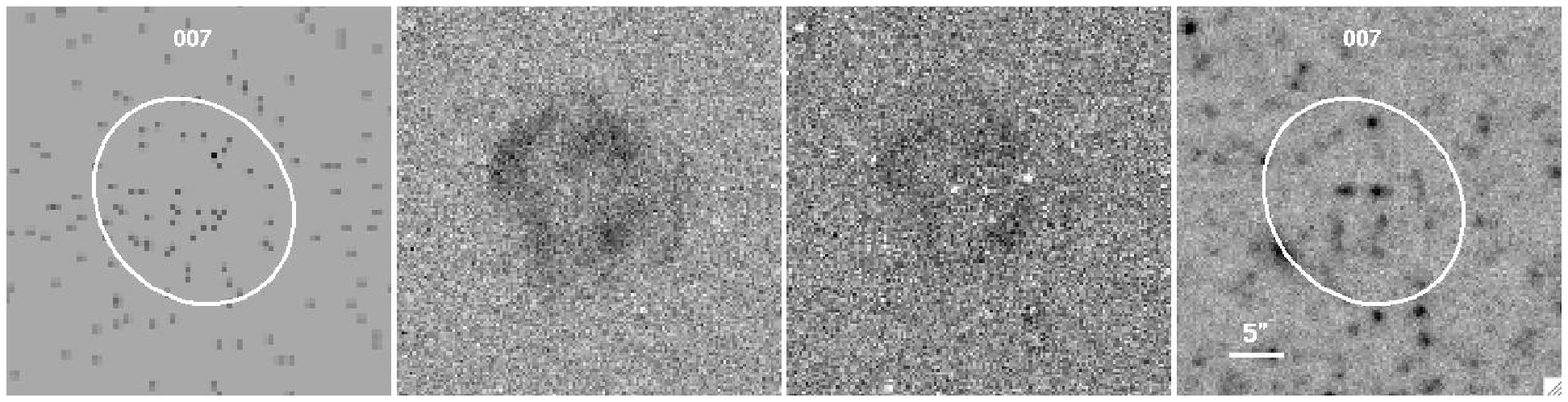}
\vspace{0.05in}
%\plotone{fig_atlas_GKL06.eps}
\plotone{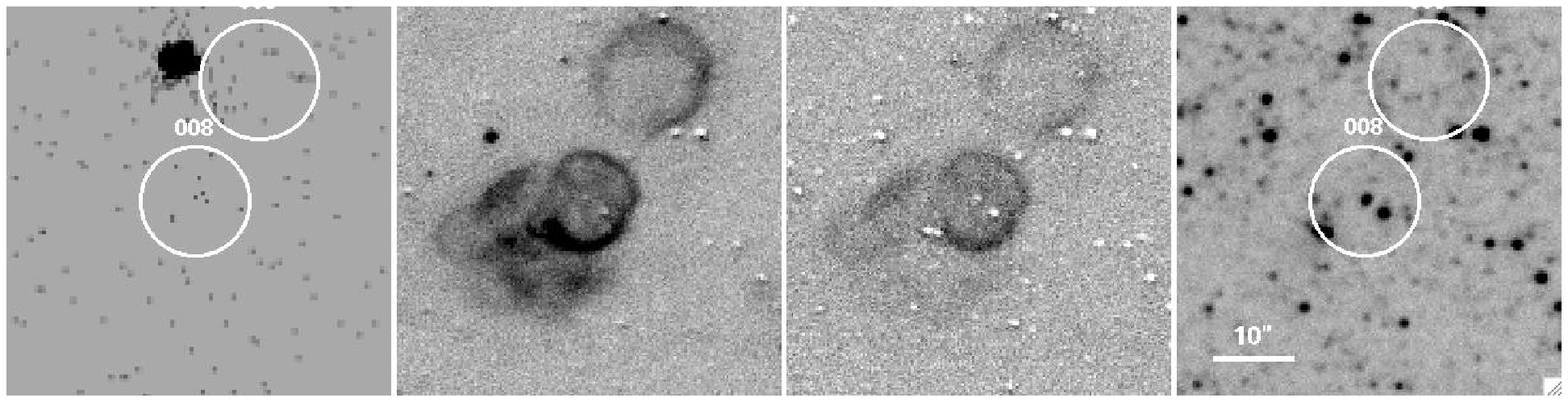}
\figcaption{Images from top to bottom of  XMM068,  G98-05,  L10-007,  G98-06.  The format is identical to Fig.\ \ref{fig_atlas01}. \label{fig_atlas02}  }
\end{figure}

\begin{figure}
%\plotone{fig_atlas_GKL07.eps}
\plotone{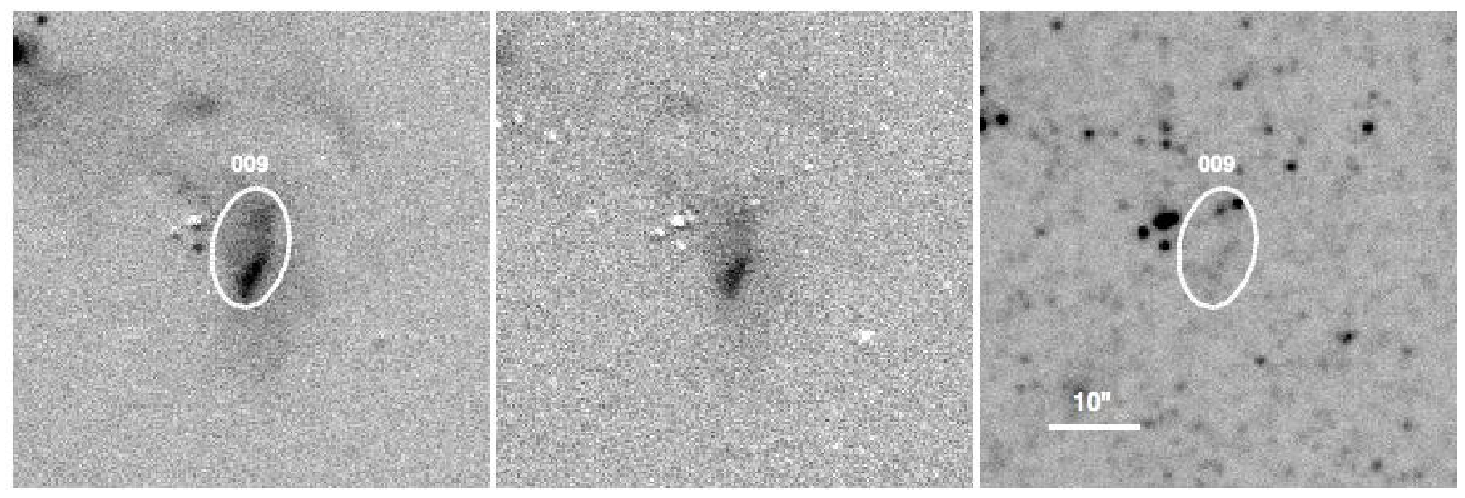}
\vspace{0.05in}
%\plotone{fig_atlas_GKL08.eps}
\plotone{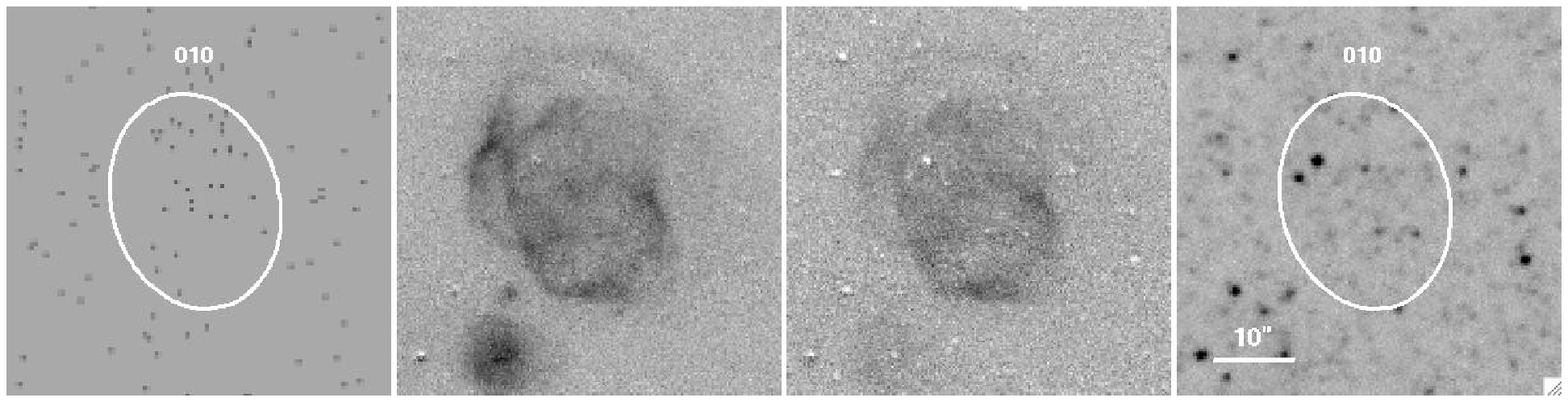}
\vspace{0.05in}
%\plotone{fig_atlas_GKL09.eps}
\plotone{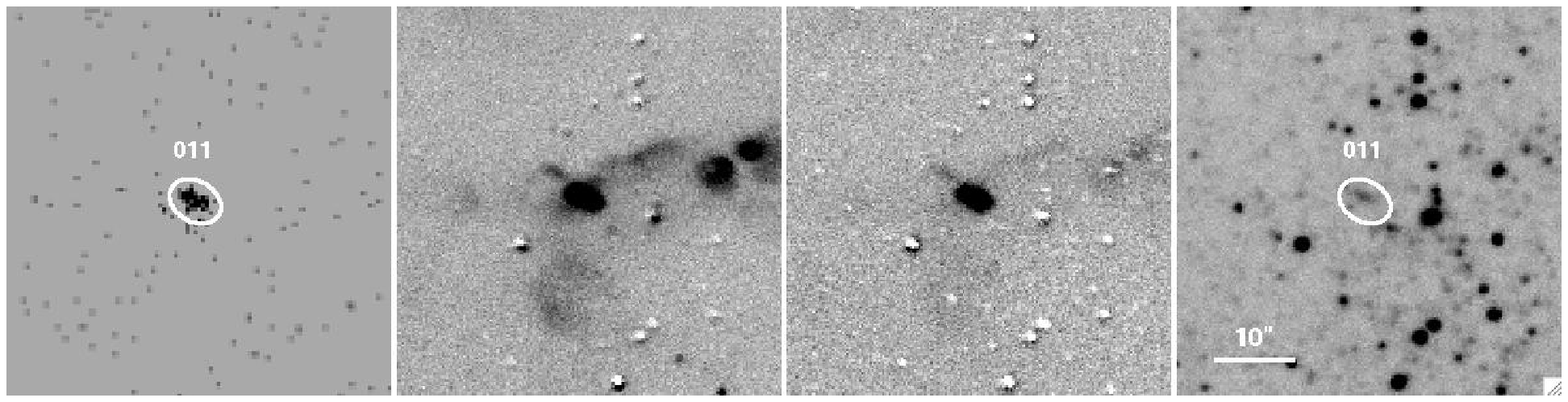}
\vspace{0.05in}
%\plotone{fig_atlas_GKL_10_12_13.eps}
\plotone{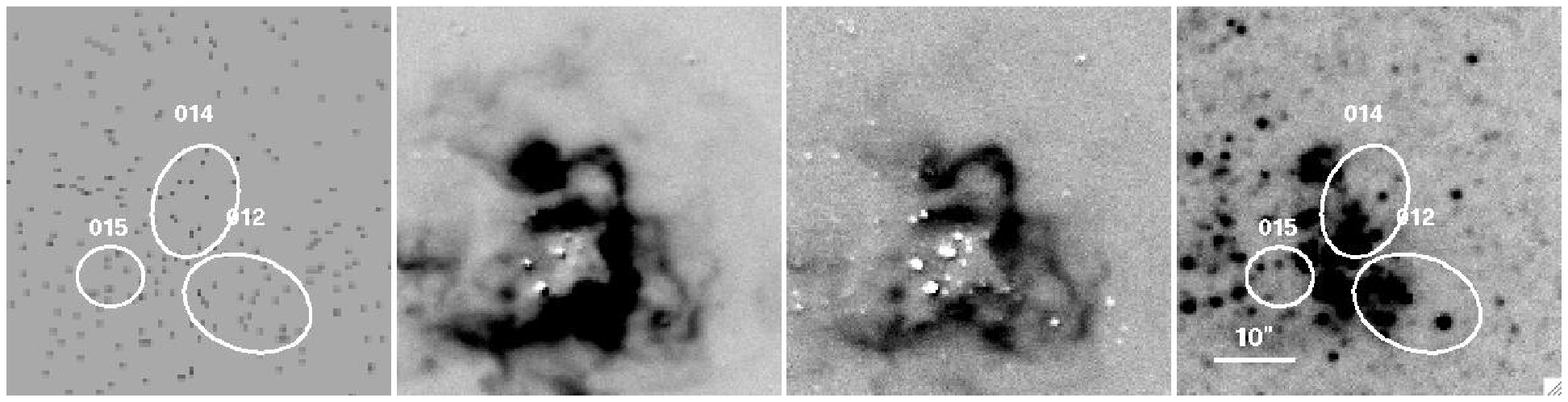}
\figcaption{Images from top to bottom of  G98-07,  G98-08,  G98-09, and (a single row) for G98-10, G98-12 and G98-13.  The format is identical to Fig.\ \ref{fig_atlas01}. \label{fig_atlas03}  Note that in the bottom panel, G98-10 corresponds to L10-012, G98-12 to L10-014, and G98-13 to L10-015.  }
\end{figure}

\begin{figure}
%\plotone{fig_atlas_GKL11.eps}
\plotone{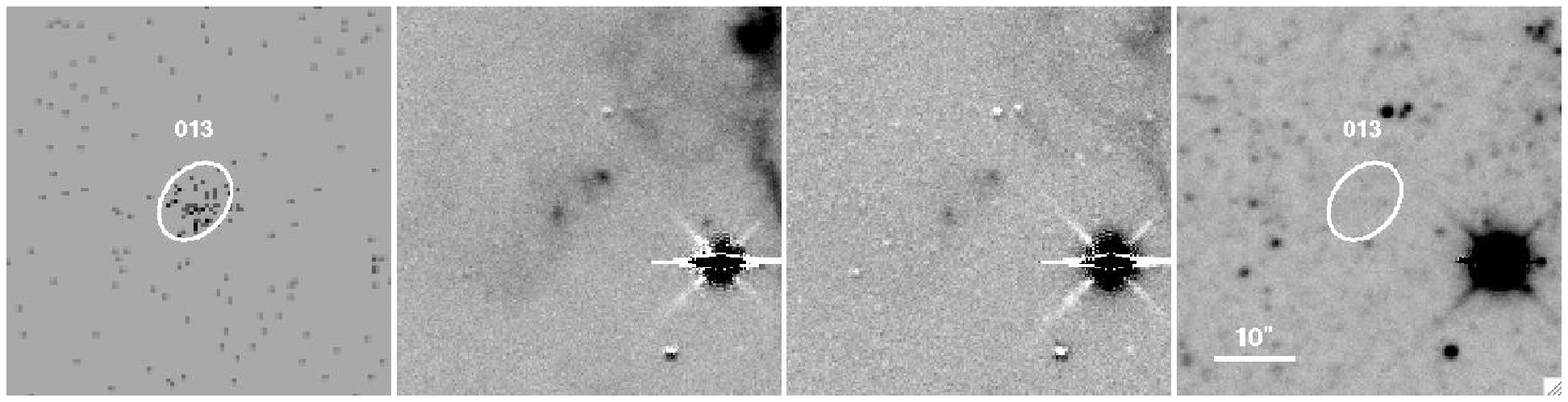}
\vspace{0.05in}
%\plotone{fig_atlas_EM20.eps}
\plotone{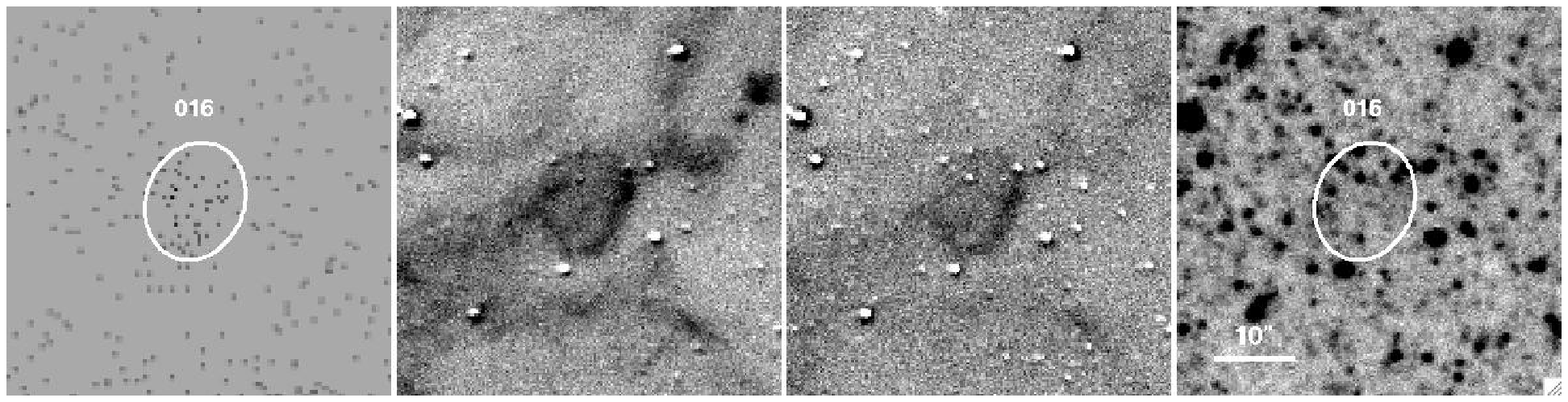}
\vspace{0.05in}
%\plotone{fig_atlas_GKL14.eps}
\plotone{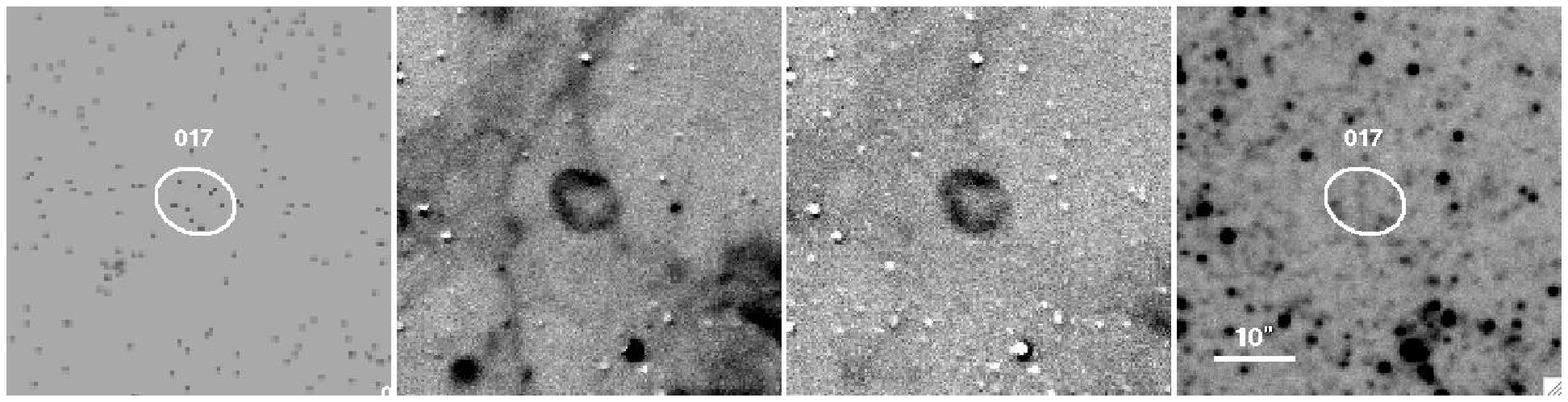}
\vspace{0.05in}
%\plotone{fig_atlas_GKL15.eps}
\plotone{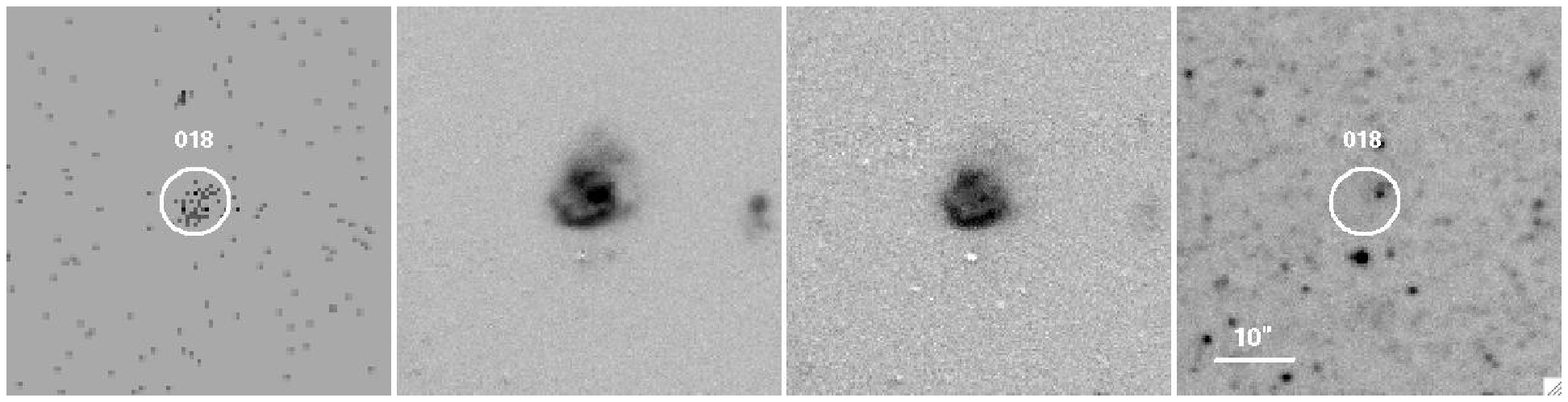}
\figcaption{Images from top to bottom of   G98-11,  L10-016,  G98-14,  G98-15.  The format is identical to Fig.\ \ref{fig_atlas01}. \label{fig_atlas04}    }
\end{figure}

\begin{figure}
%\plotone{fig_atlas_GKL16.eps}
\plotone{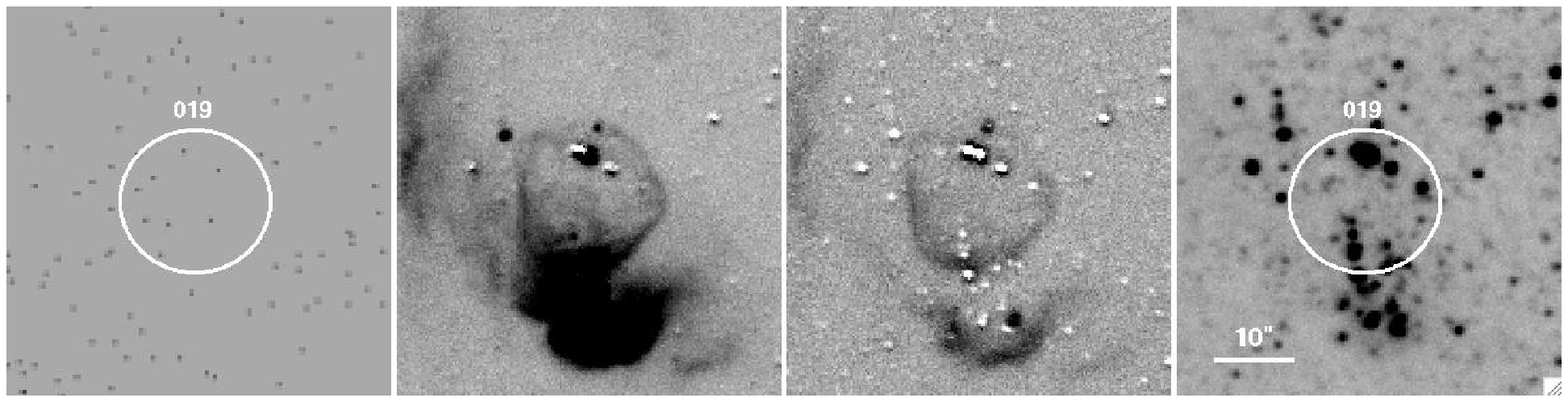}
\vspace{0.05in}
%\plotone{fig_atlas_EM80.eps}
\plotone{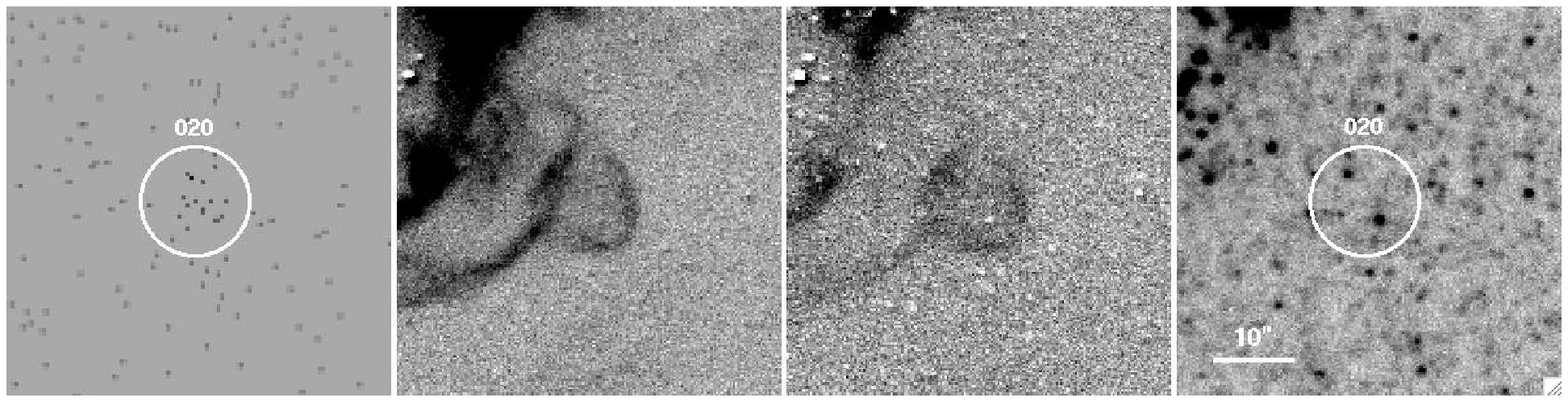}
\vspace{0.05in}
%\plotone{fig_atlas_GKL17.eps}
\plotone{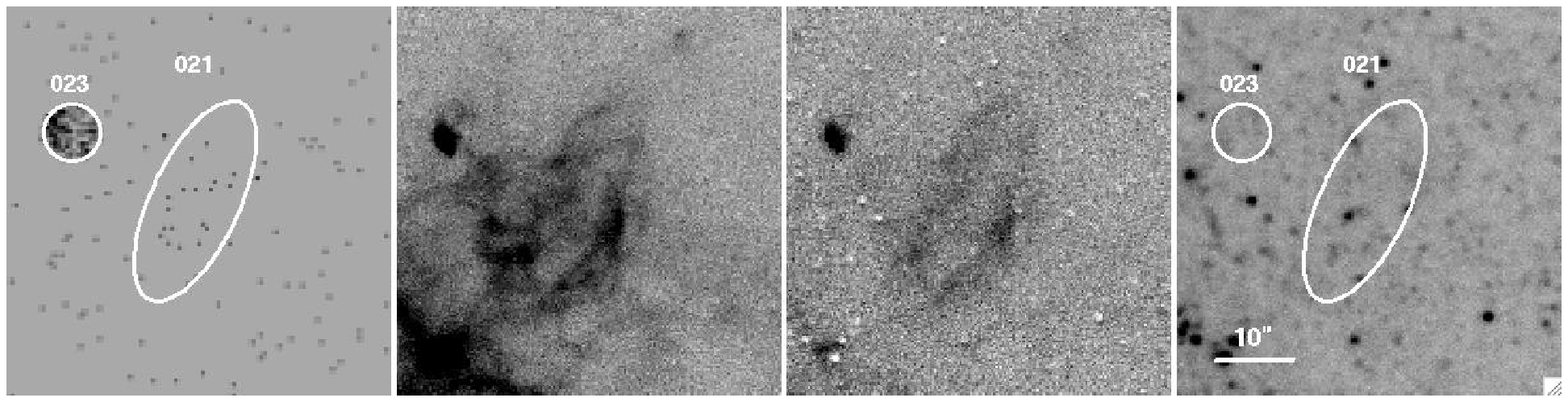}
\vspace{0.05in}
%\plotone{fig_atlas_GKL18.eps}
\plotone{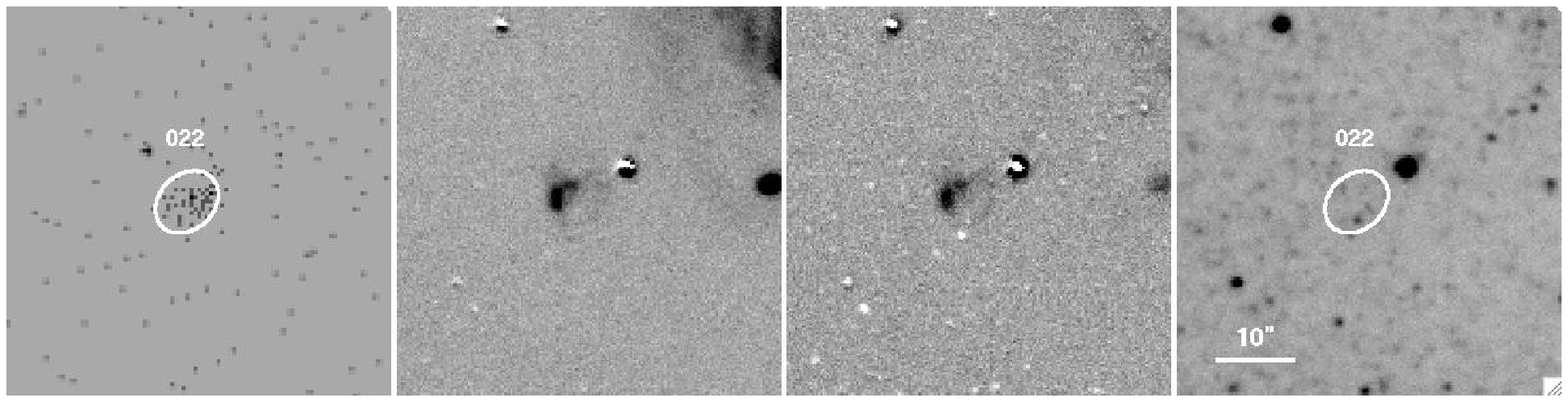}
\figcaption{Images from top to bottom of   G98-16,  L10-020,  G98-17,  G98-18.  The format is identical to Fig.\ \ref{fig_atlas01}.  \label{fig_atlas05}  }
\end{figure}

\begin{figure}
%\plotone{fig_atlas_GKL20.eps}
\plotone{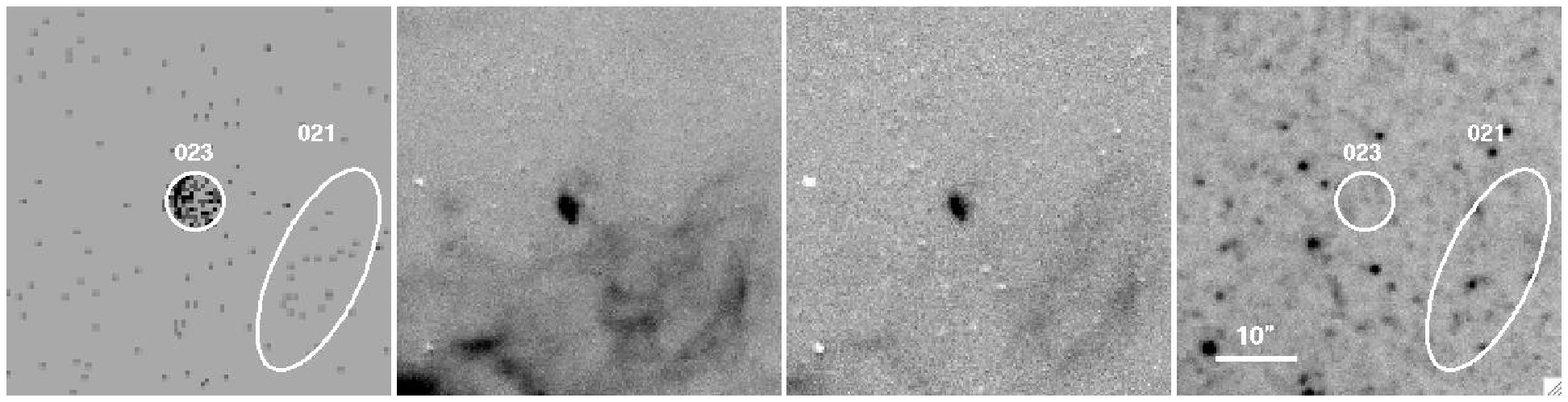}
\vspace{0.05in}
%\plotone{fig_atlas_GKL19.eps}
\plotone{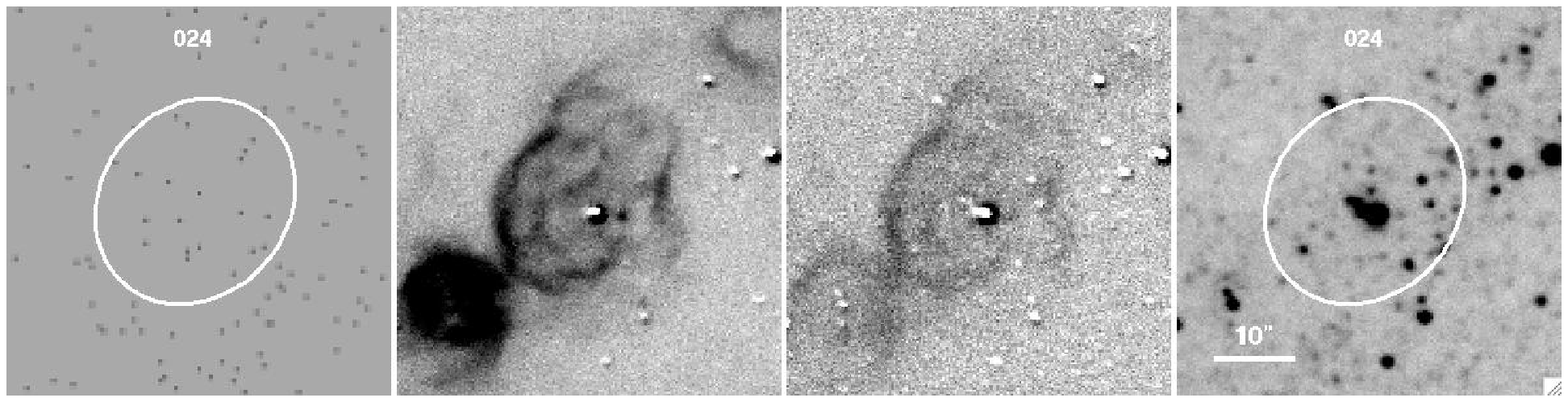}
\vspace{0.05in}
%\plotone{fig_atlas_GKL21.eps}
\plotone{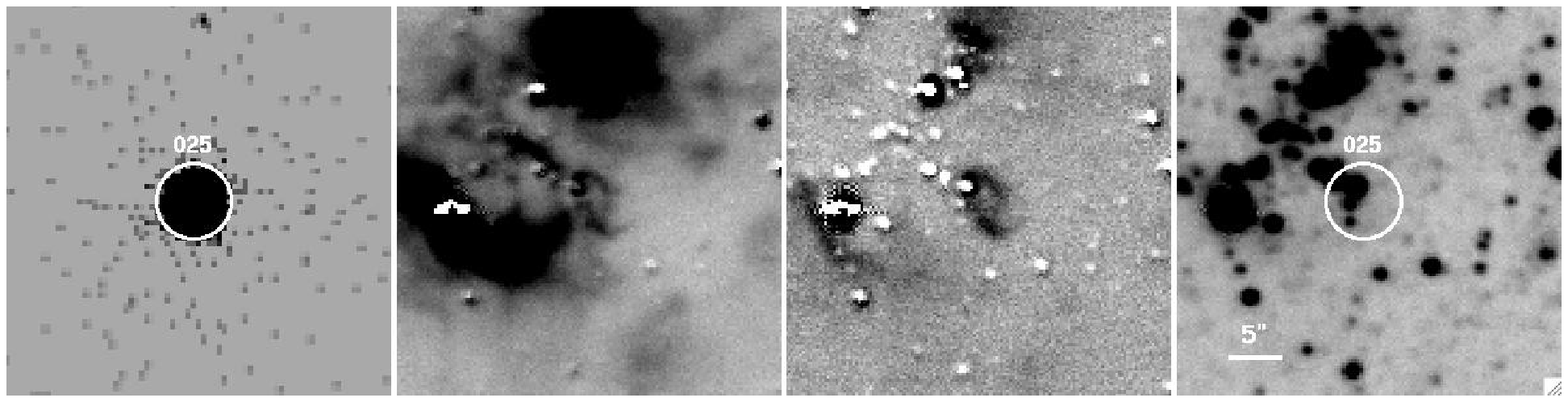}
\vspace{0.05in}
%\plotone{fig_atlas_kip-I.eps}
\plotone{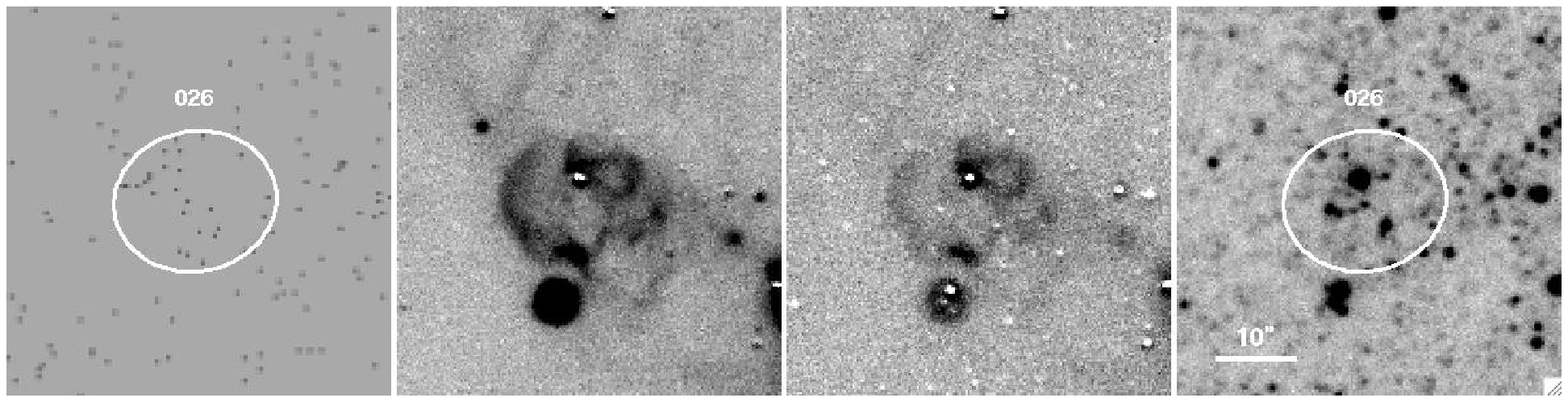}
\figcaption{Images from top to bottom of   G98-20,  G98-19,  G98-21,  L10-026.  The format is identical to Fig.\ \ref{fig_atlas01}. \label{fig_atlas06}   }
\end{figure}

\begin{figure}
%\plotone{fig_atlas_GKL22.eps}
\plotone{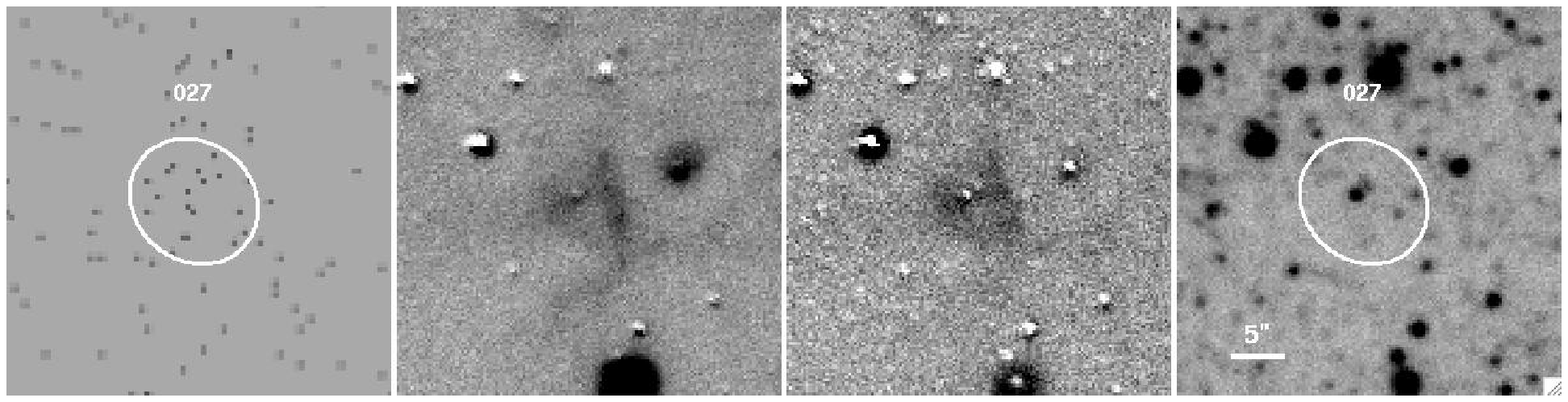}
\vspace{0.05in}
%\plotone{fig_atlas_EM22.eps}
\plotone{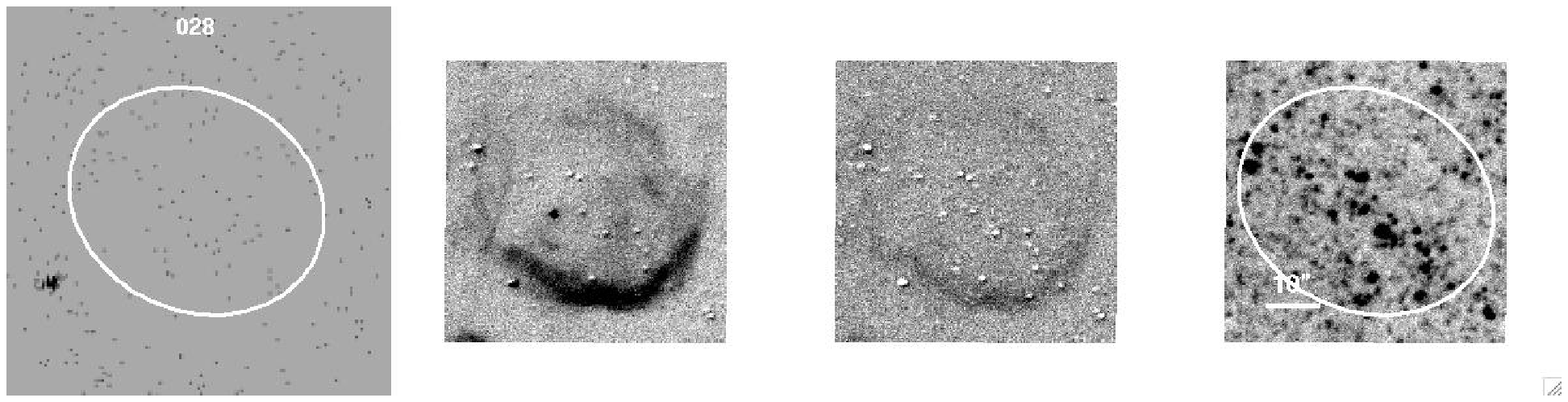}
\vspace{0.05in}
%\plotone{fig_atlas_EM04.eps}
\plotone{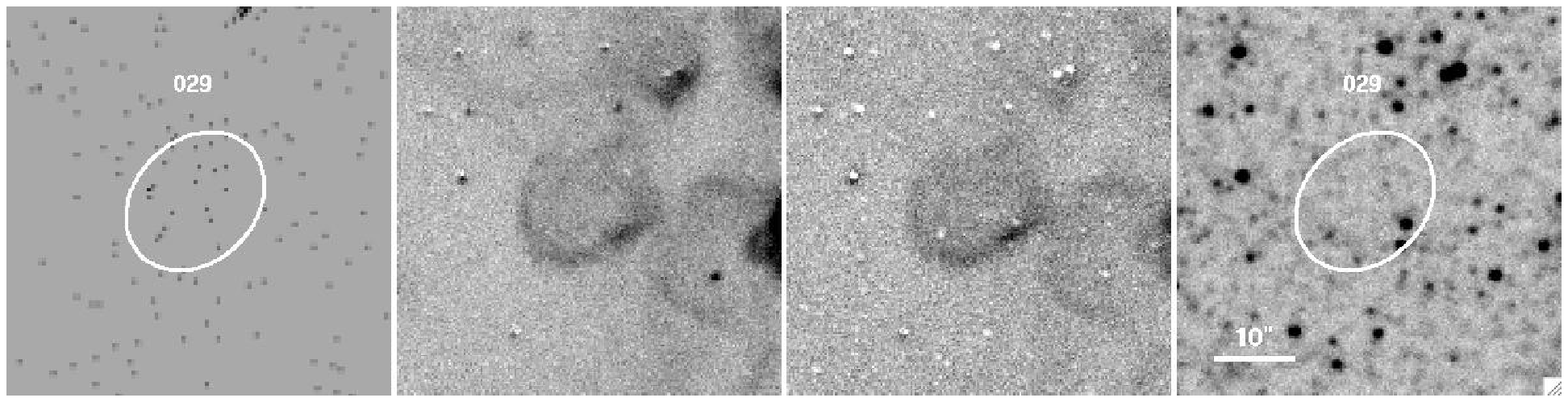}
\vspace{0.05in}
%\plotone{fig_atlas_GKL23.eps}
\plotone{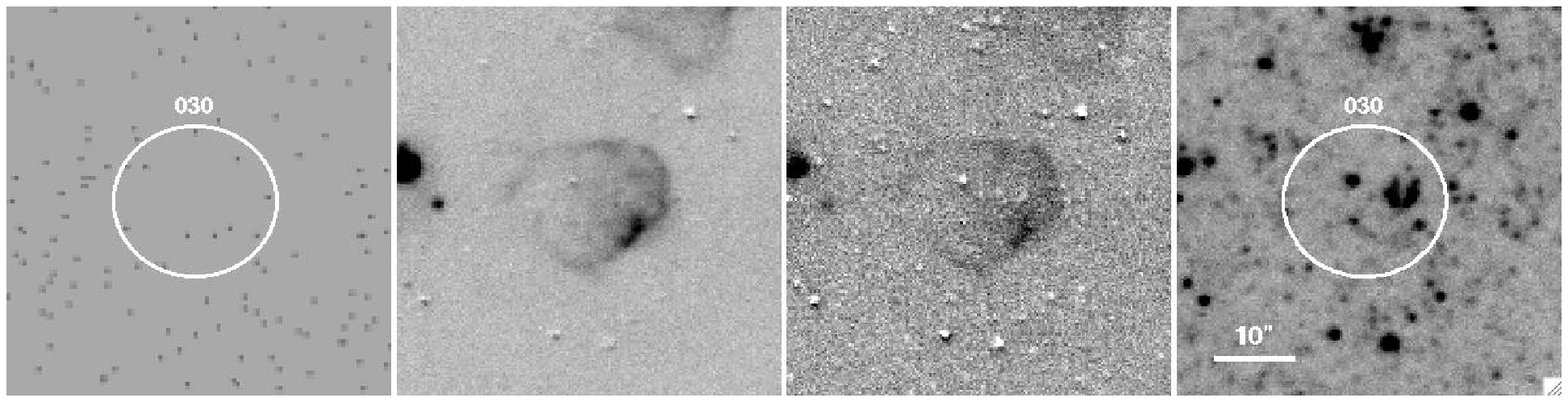}
\figcaption{Images from top to bottom of   G98-22,  L10-028,  L10-029,  G98-23.  The format is identical to Fig.\ \ref{fig_atlas01}.  \label{fig_atlas07}  }
\end{figure}

\begin{figure}
%\plotone{fig_atlas_GKL24.eps}
\plotone{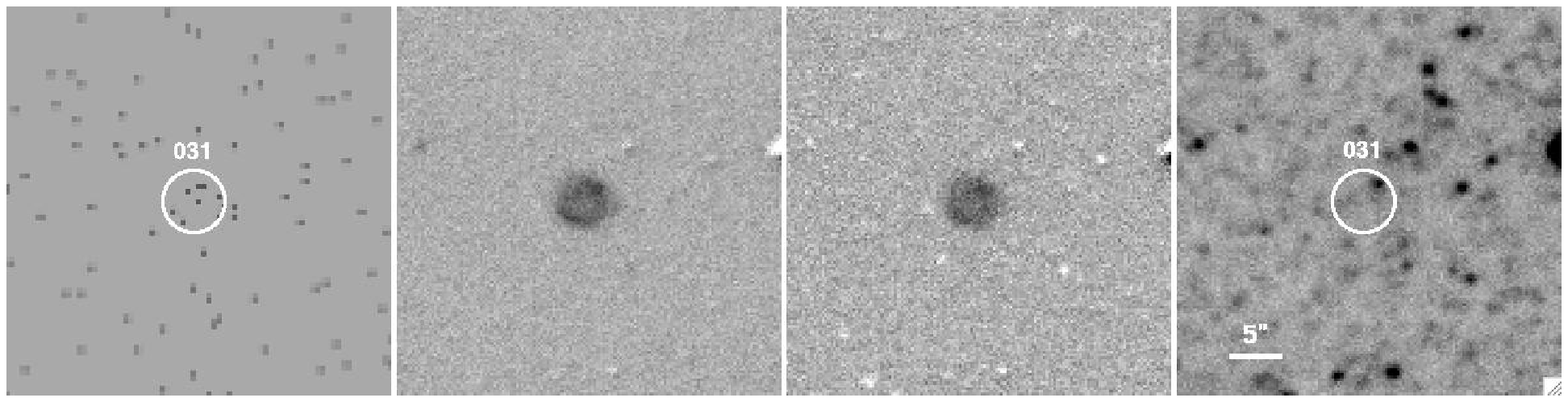}
\vspace{0.05in}
%\plotone{fig_atlas_GKL25.eps}
\plotone{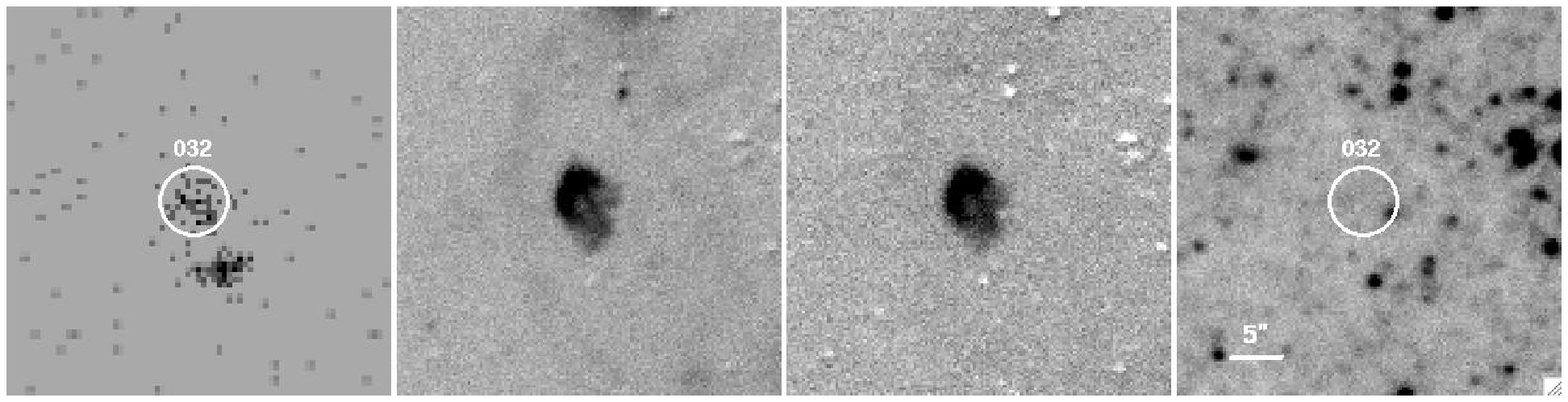}
\vspace{0.05in}
%\plotone{fig_atlas_GKL26.eps}
\plotone{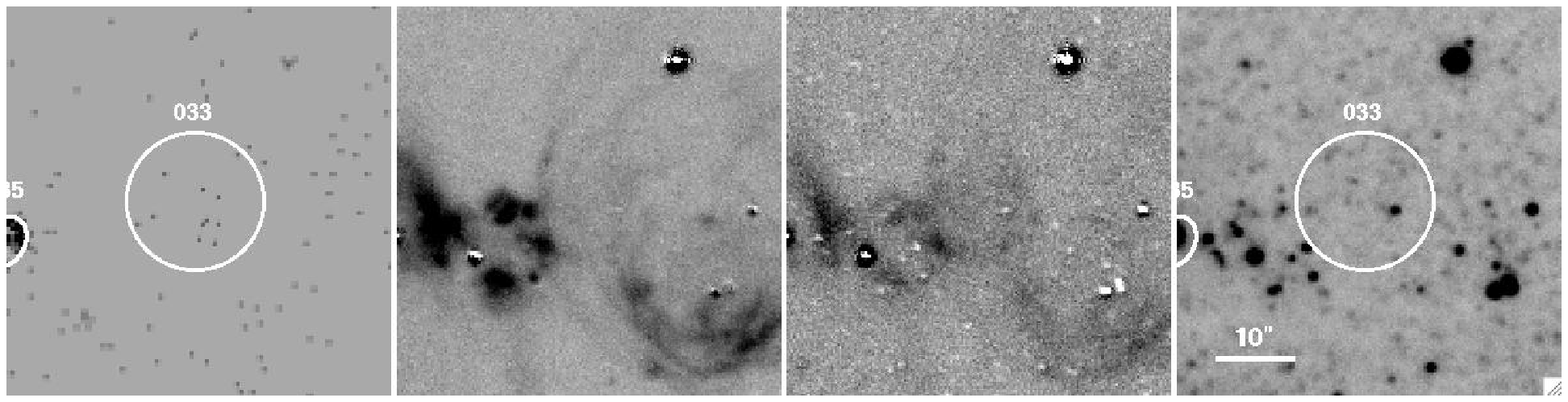}
\vspace{0.05in}
%\plotone{fig_atlas_GKL27.eps}
\plotone{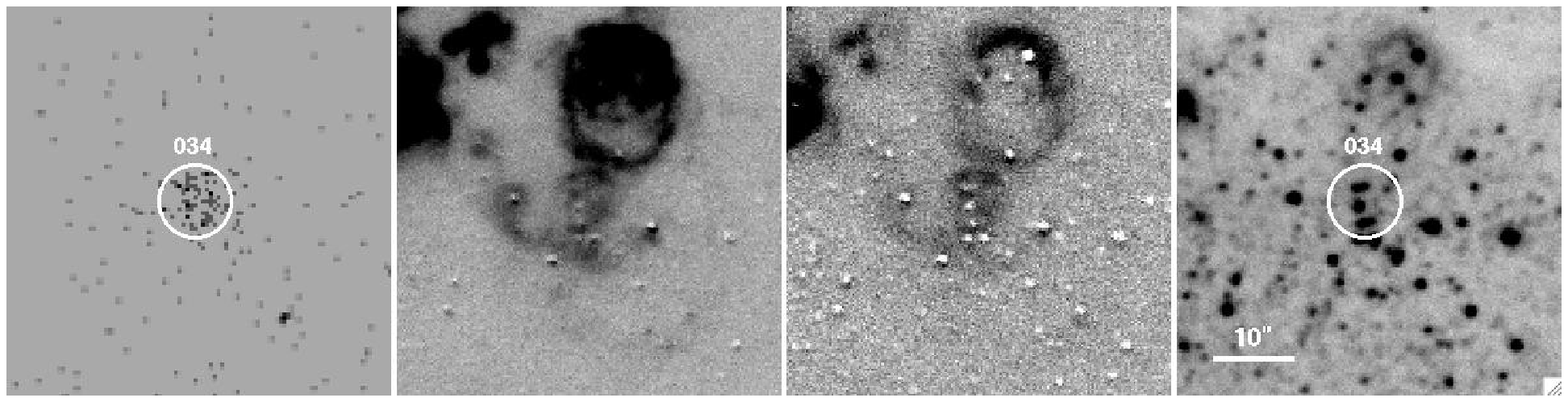}
\figcaption{Images from top to bottom of  G98-24,  G98-25,  G98-26,  G98-27.  The format is identical to Fig.\ \ref{fig_atlas01}. \label{fig_atlas08}   }
\end{figure}

\begin{figure}
%\plotone{fig_atlas_XMM156.eps}
\plotone{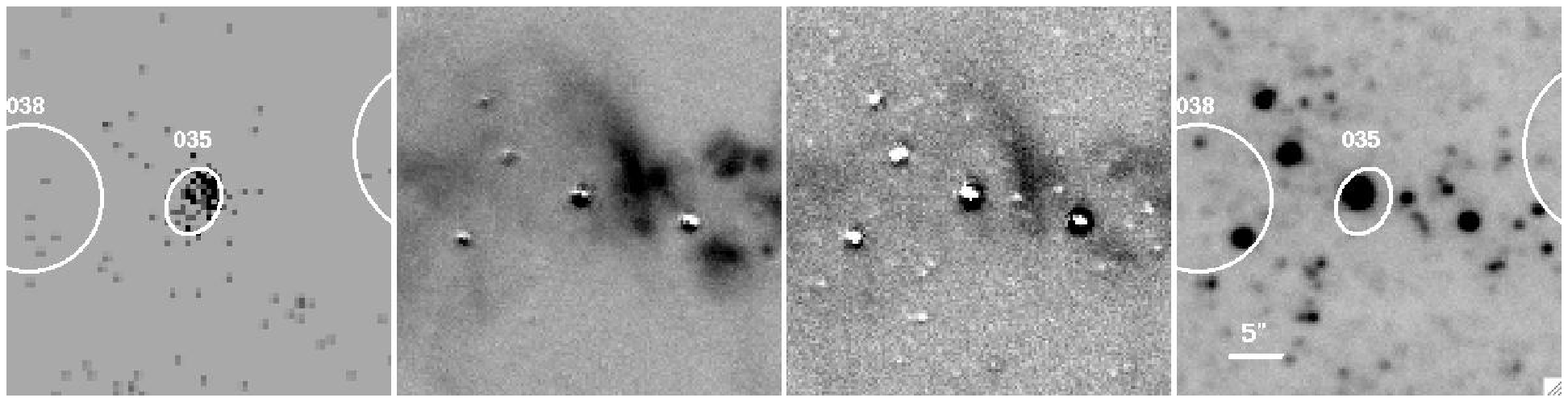}
\vspace{0.05in}
%\plotone{fig_atlas_GKL28.eps}
\plotone{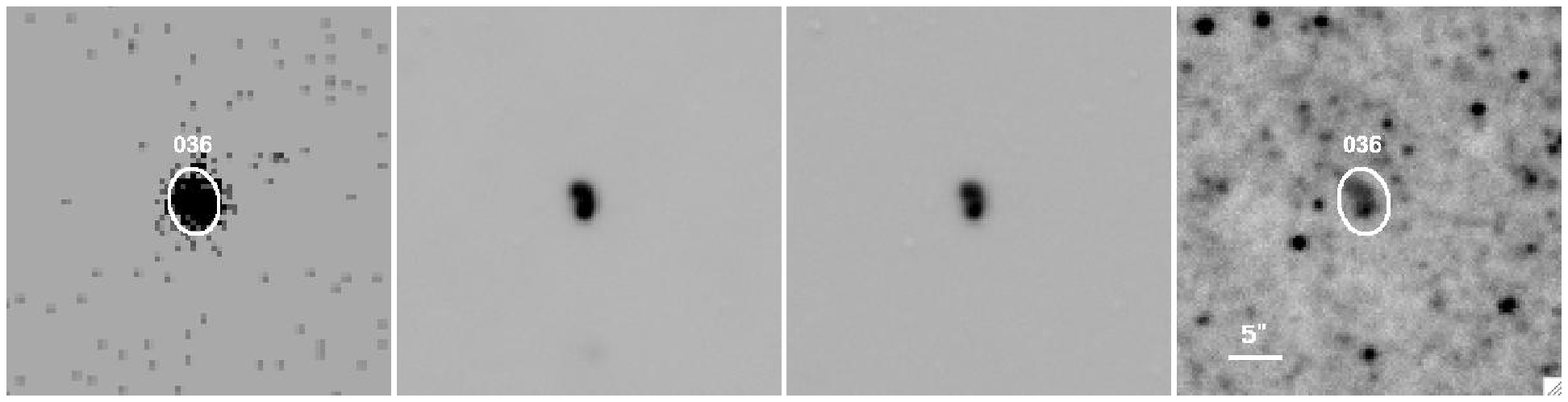}
\vspace{0.05in}
%\plotone{fig_atlas_GKL29.eps}
\plotone{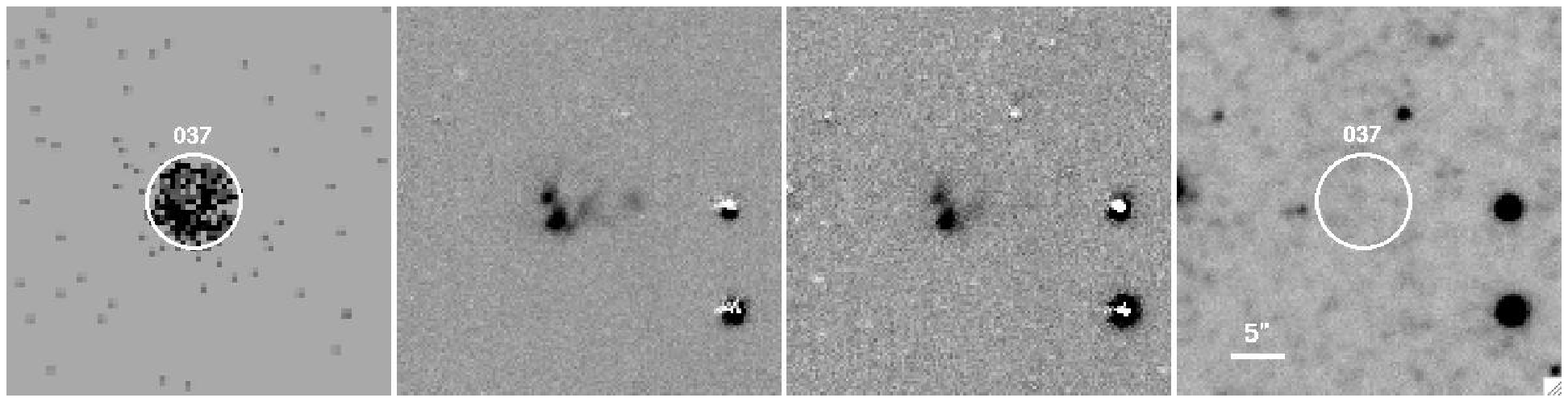}
\vspace{0.05in}
%\plotone{fig_atlas_GKL30.eps}
\plotone{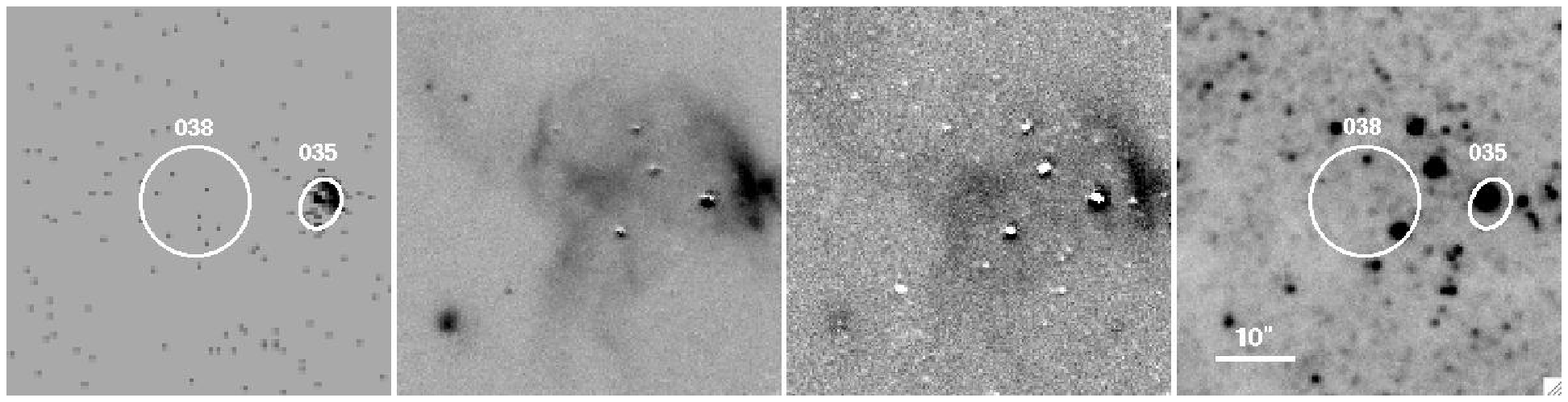}
\figcaption{Images from top to bottom of  XMM156,  G98-28,  G98-29,  G98-30.  The format is identical to Fig.\ \ref{fig_atlas01}. \label{fig_atlas09}   }
\end{figure}

\begin{figure}
%\plotone{fig_atlas_GKL31.eps}
\plotone{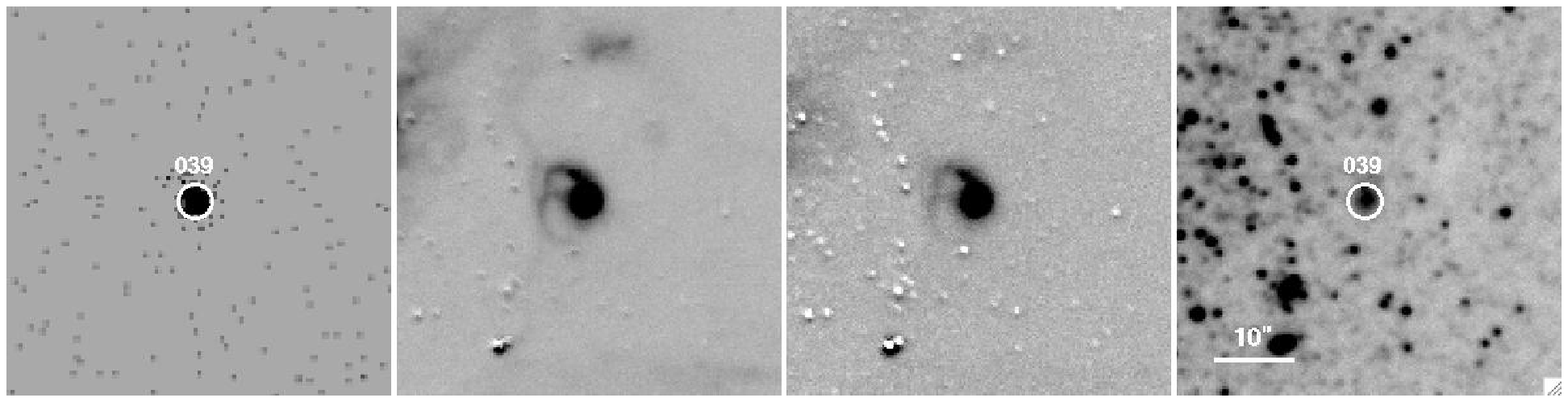}
\vspace{0.05in}
%\plotone{fig_atlas_GKL32.eps}
\plotone{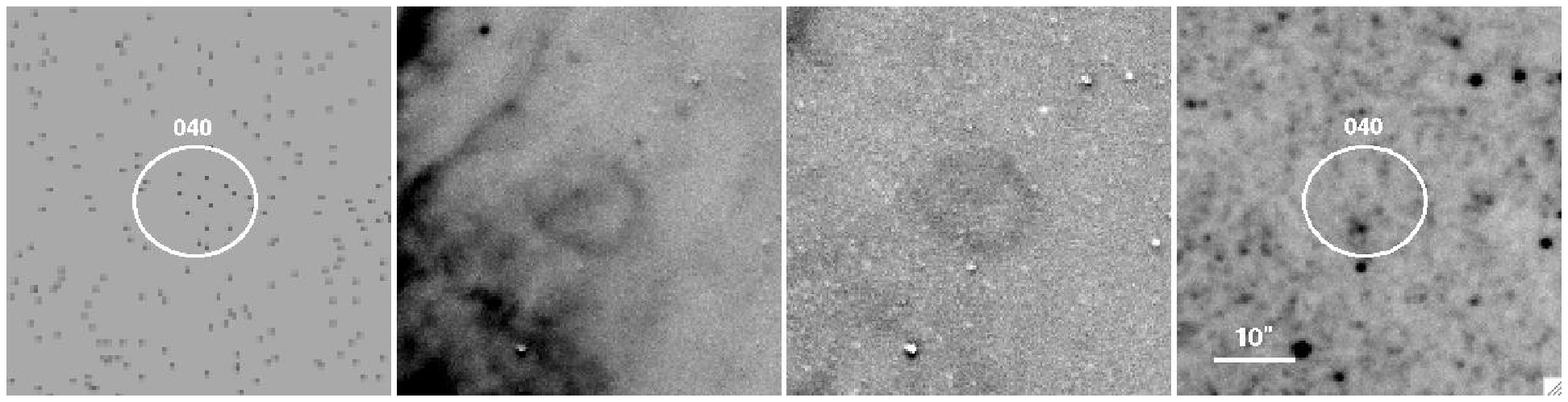}
\vspace{0.05in}
%\plotone{fig_atlas_EM18.eps}
\plotone{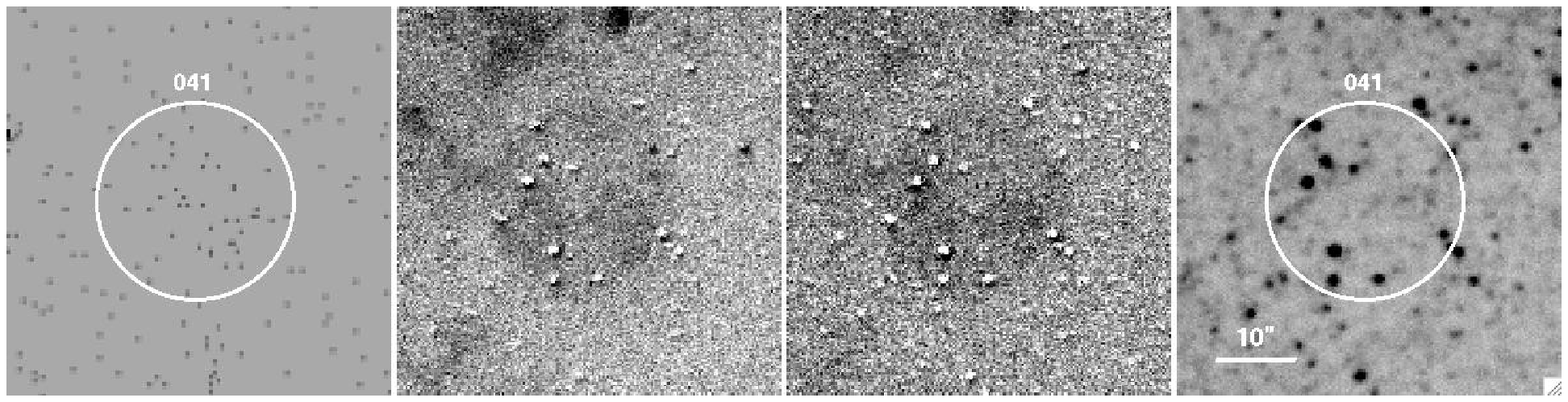}
\vspace{0.05in}
%\plotone{fig_atlas_GKL33.eps}
\plotone{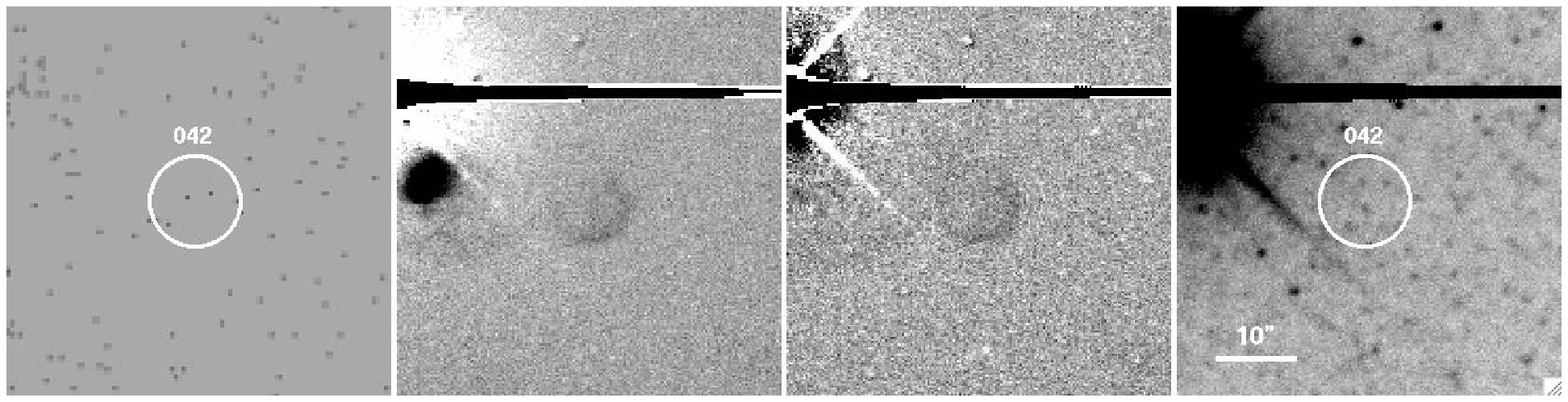}
\figcaption{Images from top to bottom of G98-31,  G98-32,  L10-041,  G98-33.  The format is identical to Fig.\ \ref{fig_atlas01}. \label{fig_atlas10}    }
\end{figure}

\begin{figure}
%\plotone{fig_atlas_kip-H.eps}
\plotone{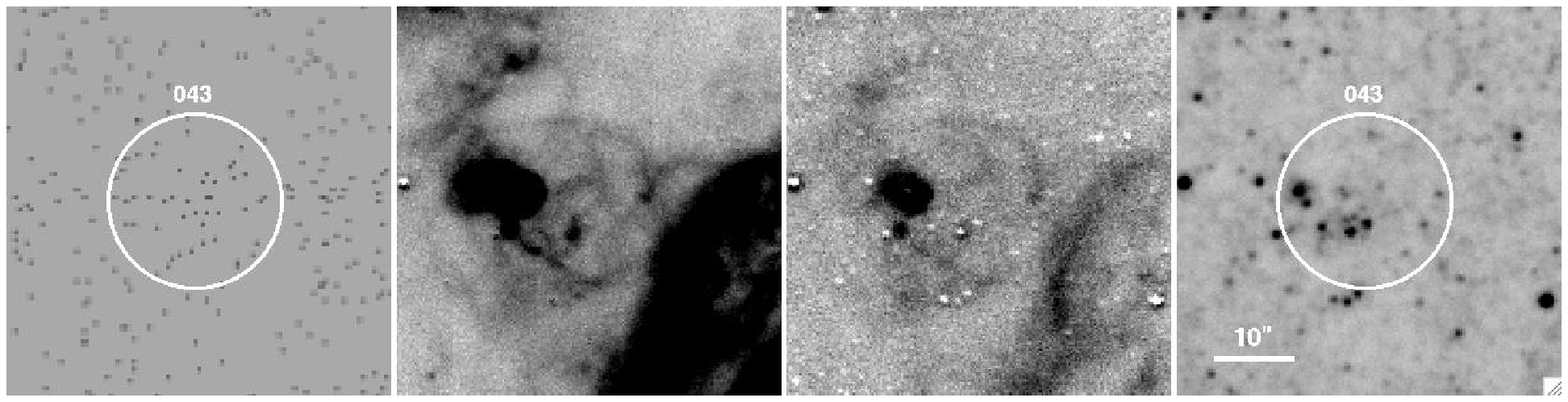}
\vspace{0.05in}
%\plotone{fig_atlas_GKL34.eps}
\plotone{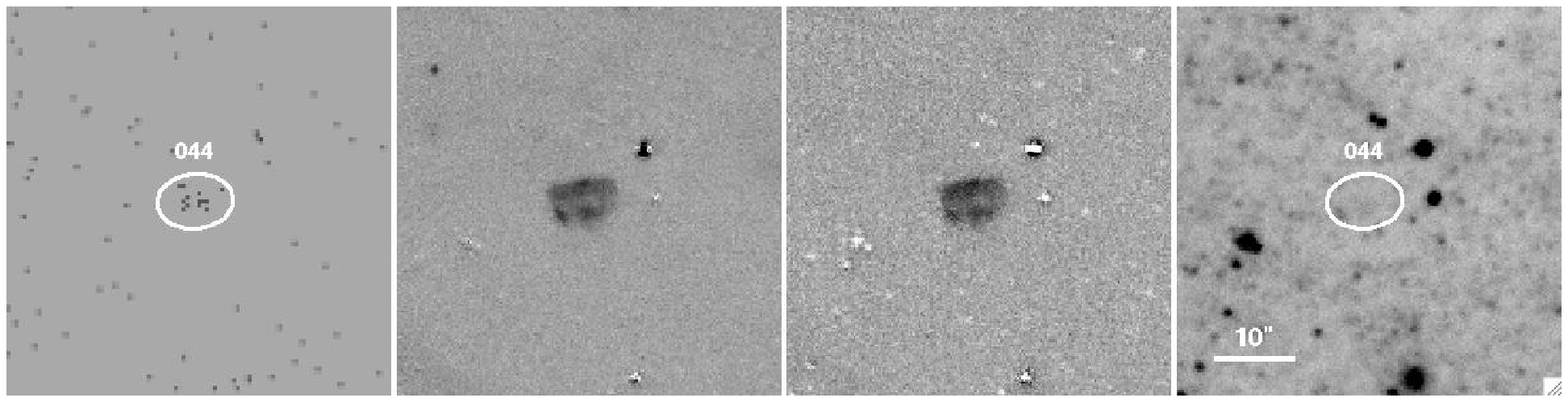}
\vspace{0.05in}
%\plotone{fig_atlas_GKL35.eps}
\plotone{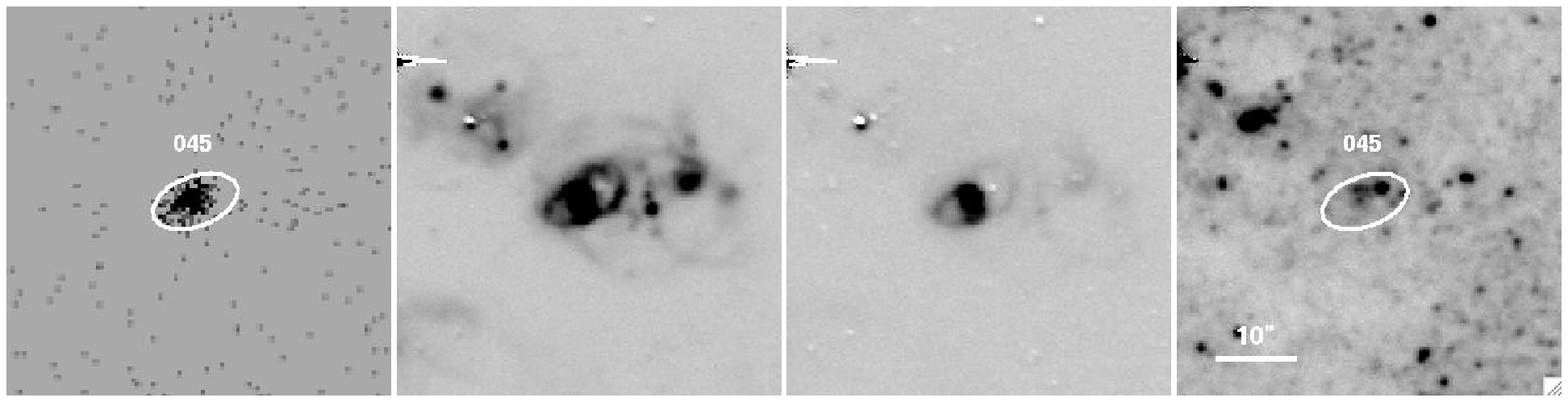}
\vspace{0.05in}
%\plotone{fig_atlas_GKL36.eps}
\plotone{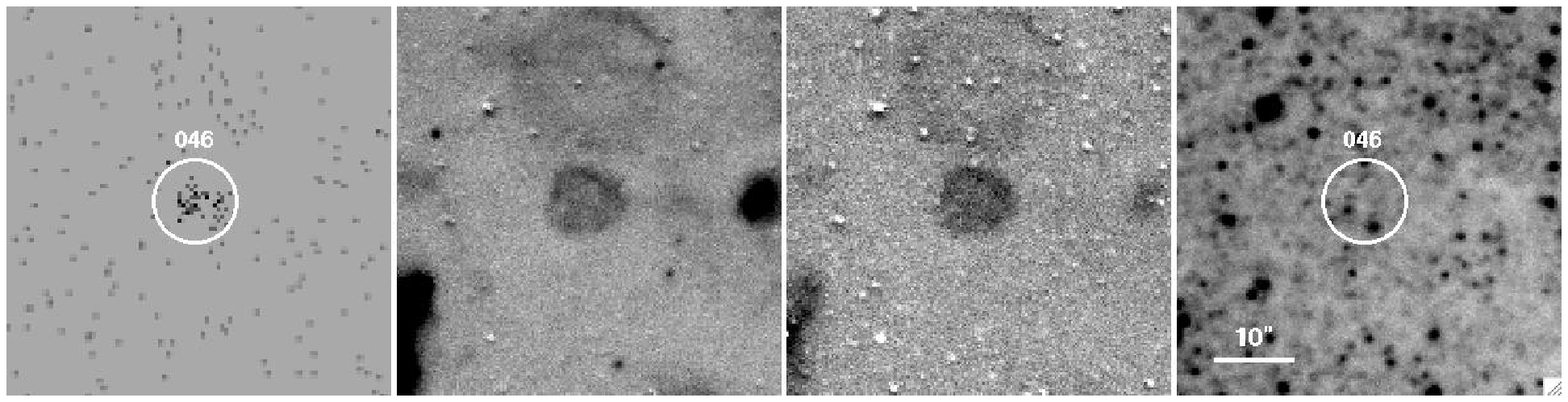}
\figcaption{Images from top to bottom of L10-043,  G98-34,  G98-35,  G98-36.  The format is identical to Fig.\ \ref{fig_atlas01}.  \label{fig_atlas11}  }
\end{figure}

\begin{figure}
%\plotone{fig_atlas_GKL37.eps}
\plotone{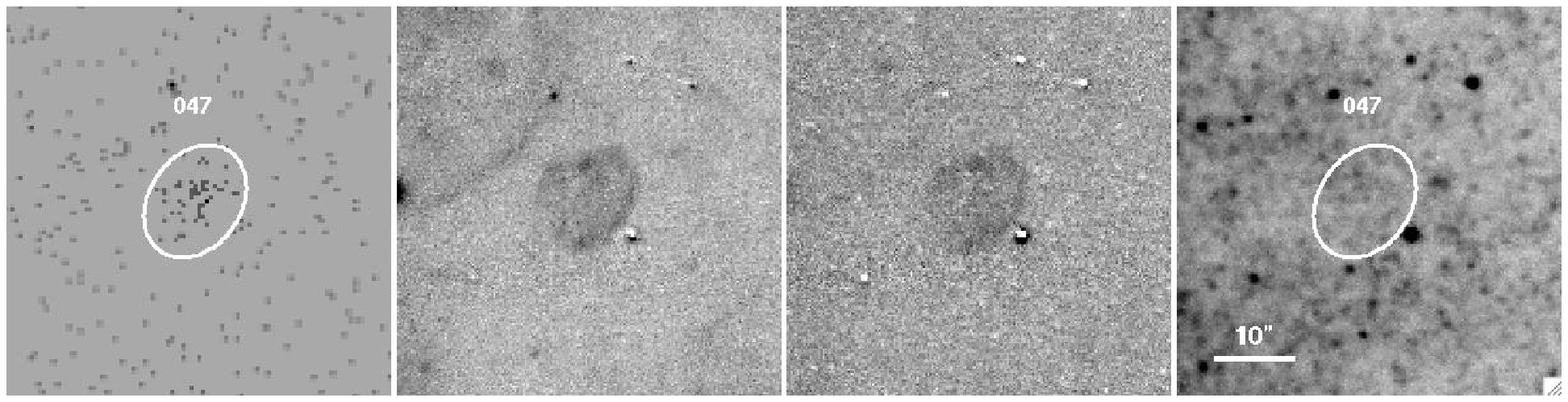}
\vspace{0.05in}
%\plotone{fig_atlas_GKL38.eps}
\plotone{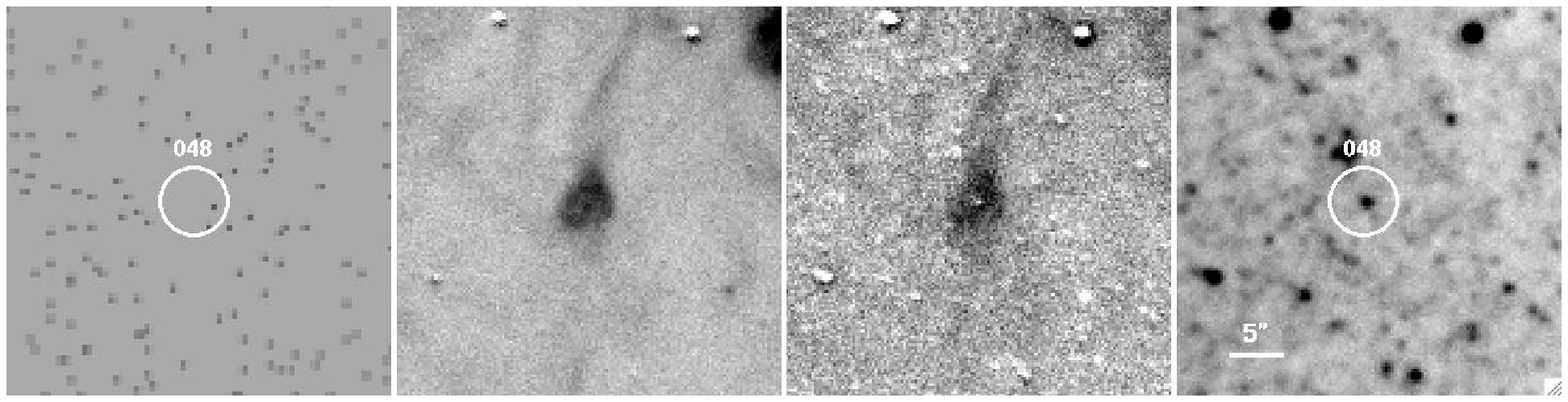}
\vspace{0.05in}
%\plotone{fig_atlas_GKL39.eps}
\plotone{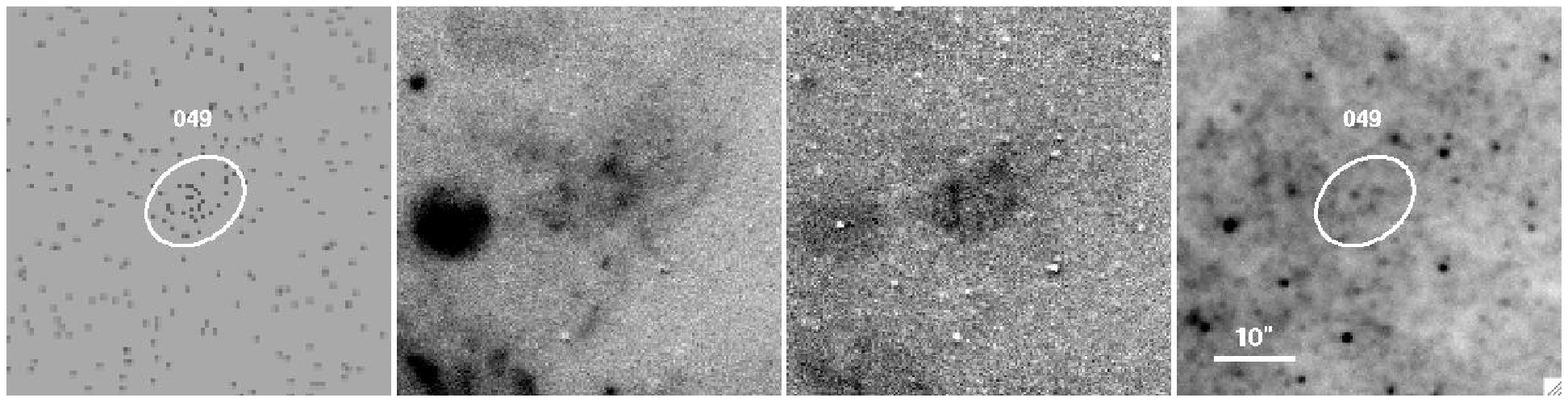}
\vspace{0.05in}
%\plotone{fig_atlas_GKL41.eps}
\plotone{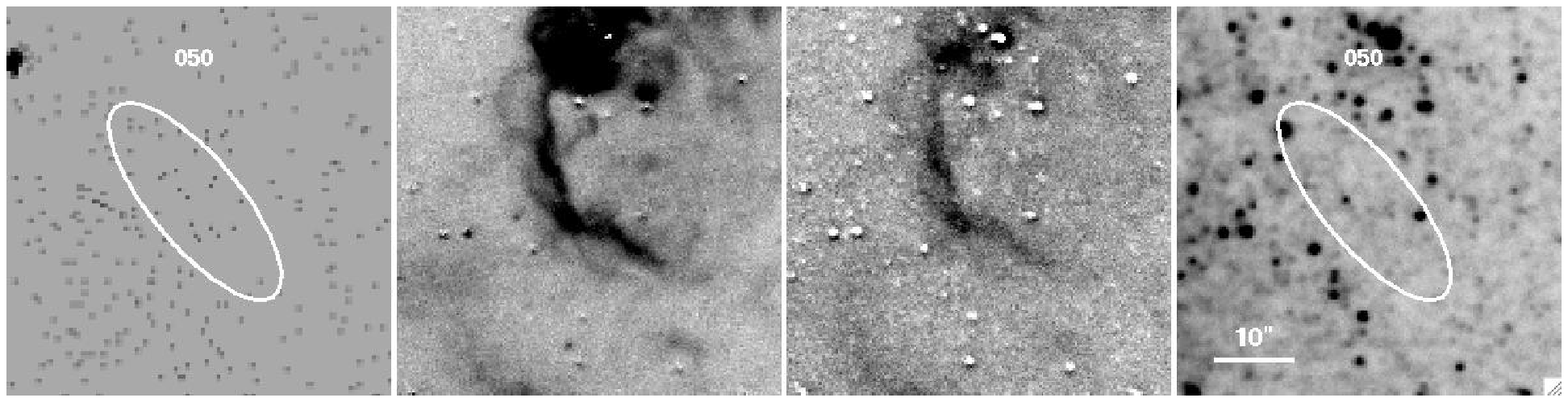}
\figcaption{Images from top to bottom of G98-37,  G98-38,  G98-39,  G98-41.  The format is identical to Fig.\ \ref{fig_atlas01}. \label{fig_atlas12}   }
\end{figure}

\begin{figure}
%\plotone{fig_atlas_GKL40.eps}
\plotone{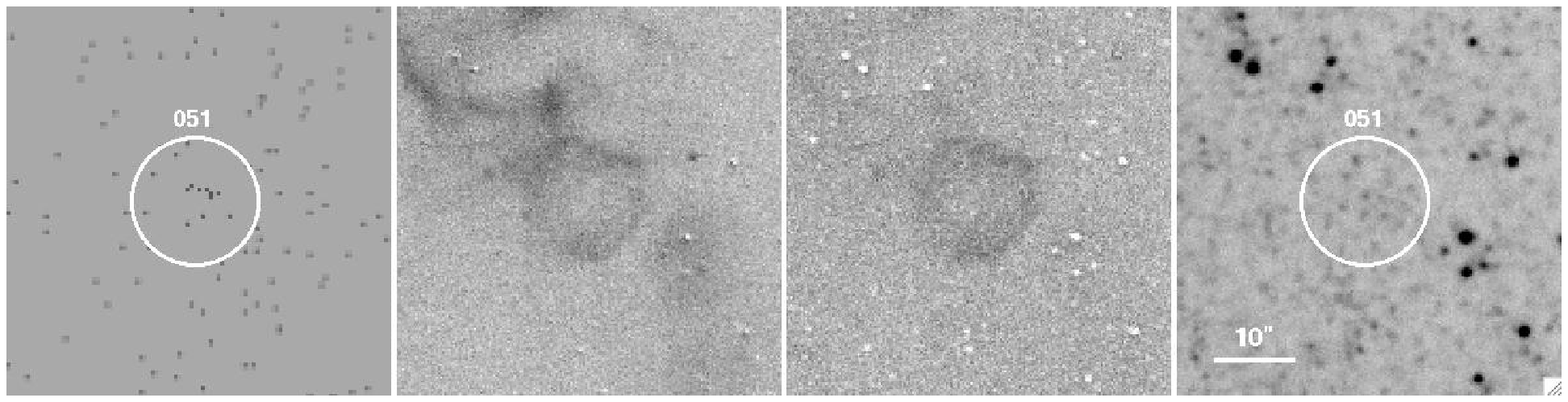}
\vspace{0.05in}
%\plotone{fig_atlas_GKL42.eps}
\plotone{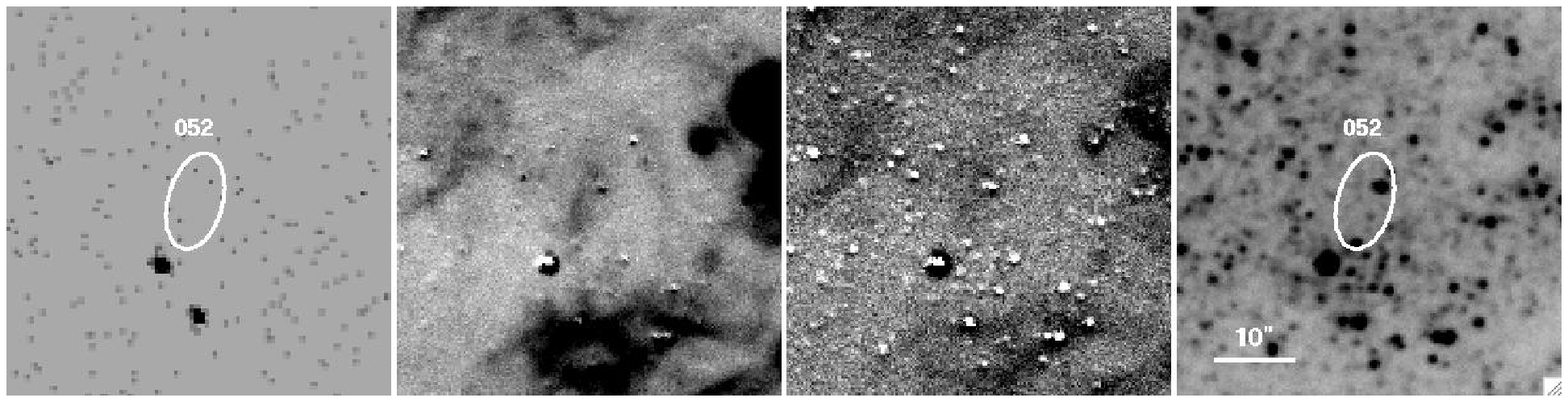}
\vspace{0.05in}
%\plotone{fig_atlas_GKL43.eps}
\plotone{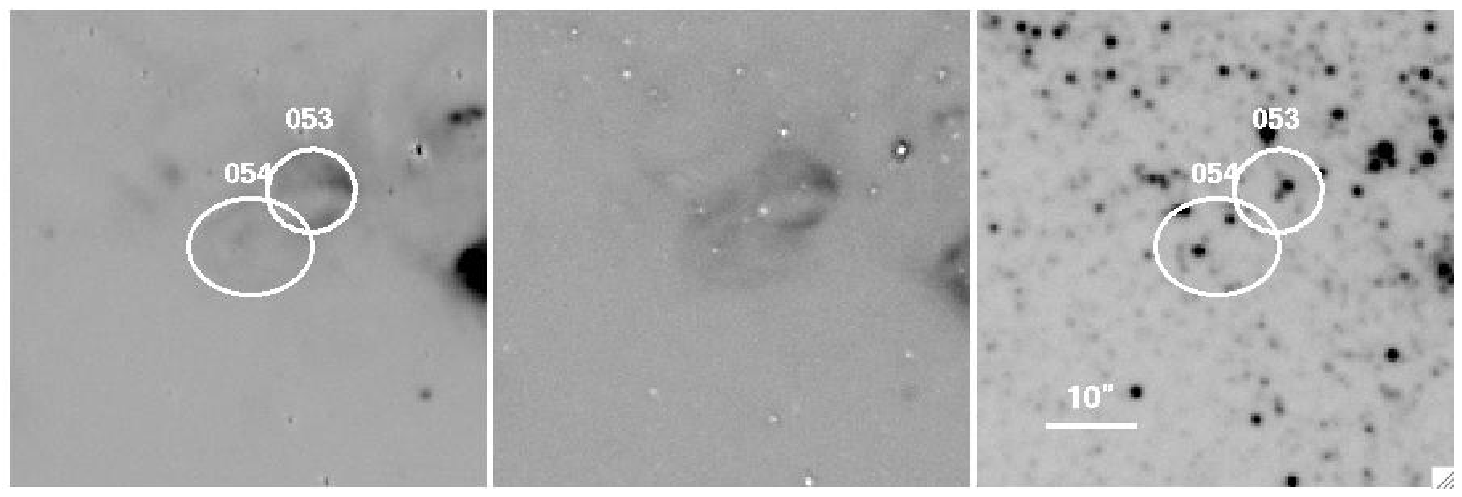}
\vspace{0.05in}
%\plotone{fig_atlas_GKL44.eps}
\plotone{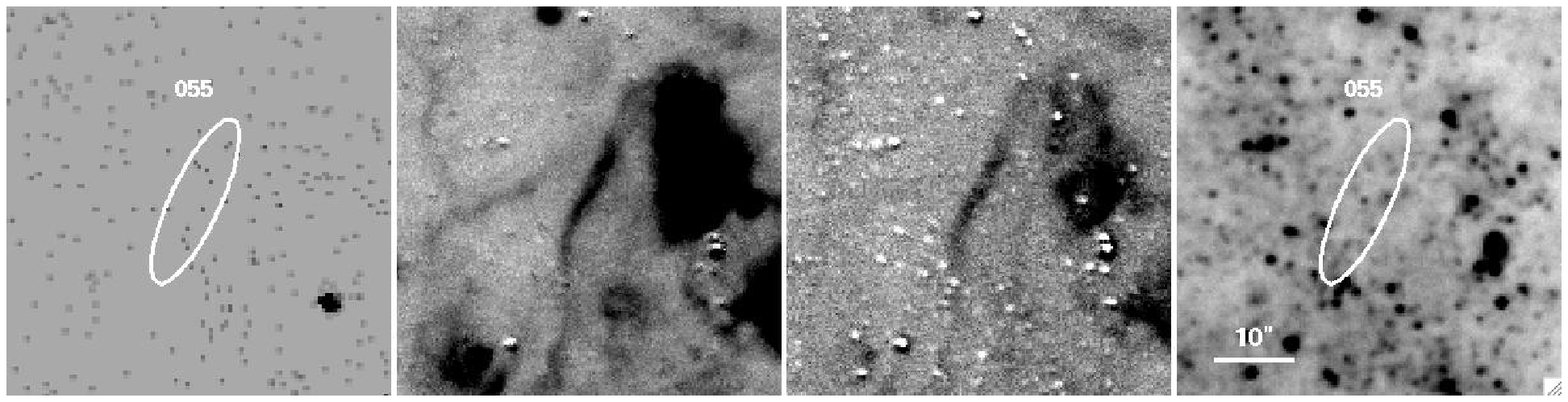}
\figcaption{Images from top to bottom of G98-40,  G98-42,  G98-43,  G98-44.  The format is identical to Fig.\ \ref{fig_atlas01}.  \label{fig_atlas13}  }
\end{figure}

\begin{figure}
%\plotone{fig_atlas_GKL45.eps}
\plotone{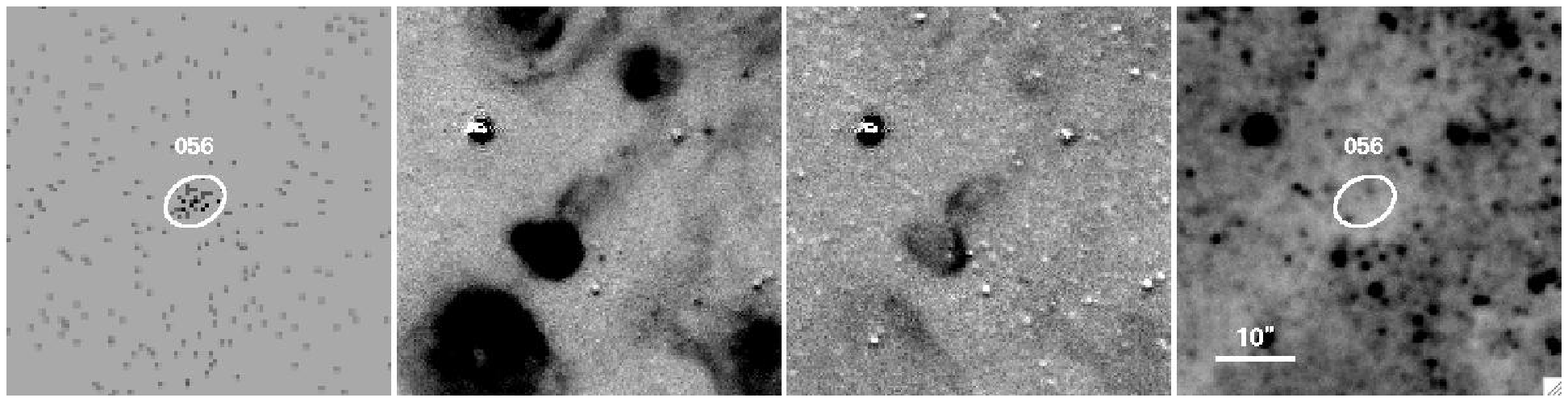}
\vspace{0.05in}
%\plotone{fig_atlas_GKL46.eps}
\plotone{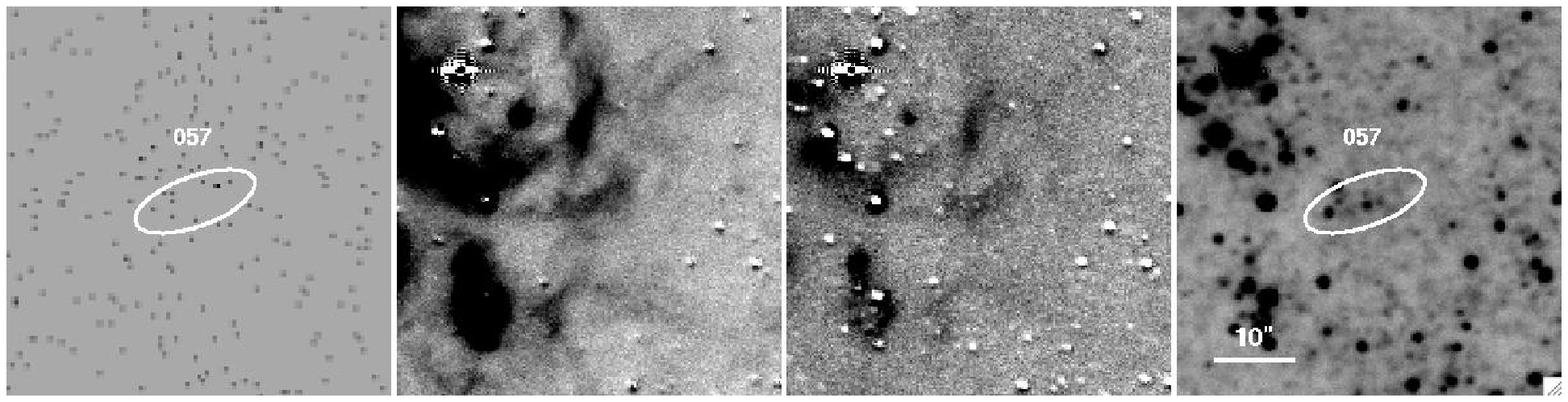}
\vspace{0.05in}
%\plotone{fig_atlas_PFW1.eps}
\plotone{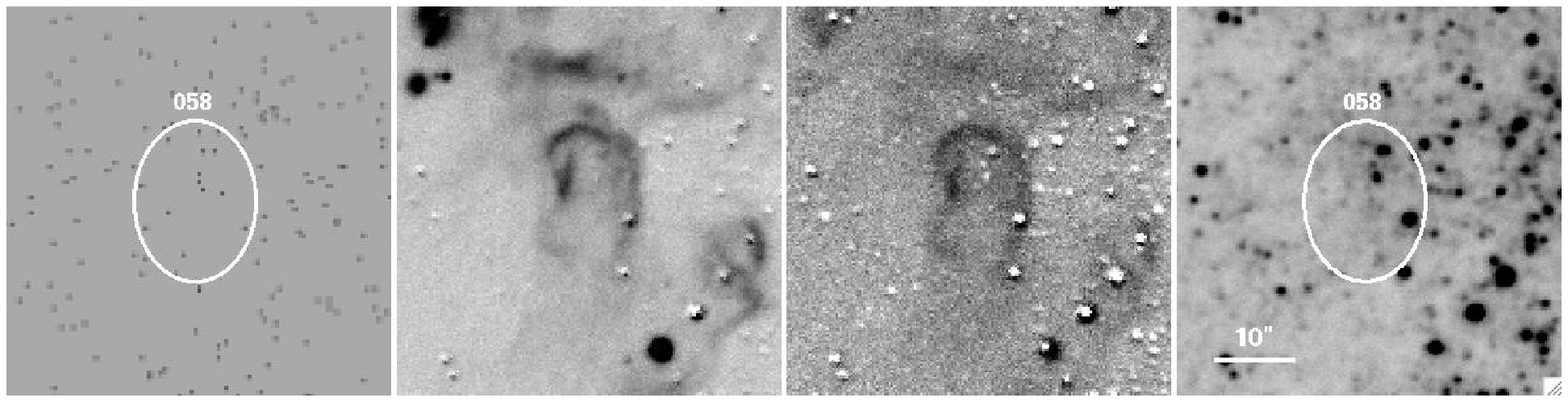}
\vspace{0.05in}
%\plotone{fig_atlas_WPB2.eps}
\plotone{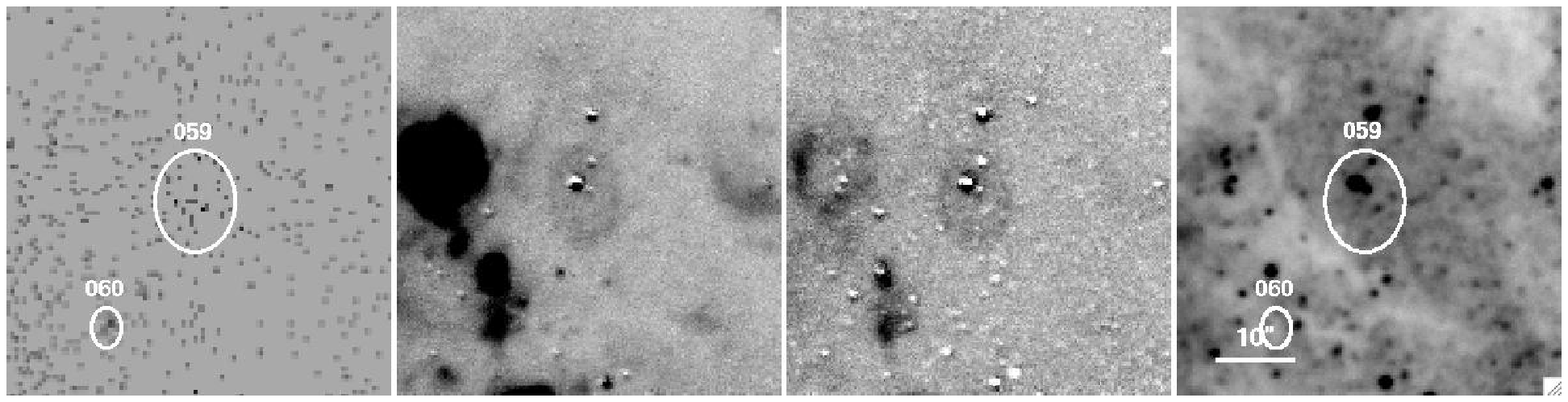}
\figcaption{Images from top to bottom of G98-45,  G98-46,  L10-058,  L10-059.  The format is identical to Fig.\ \ref{fig_atlas01}. \label{fig_atlas14}   }
\end{figure}

\begin{figure}
%\plotone{fig_atlas_GKL48.eps}
\plotone{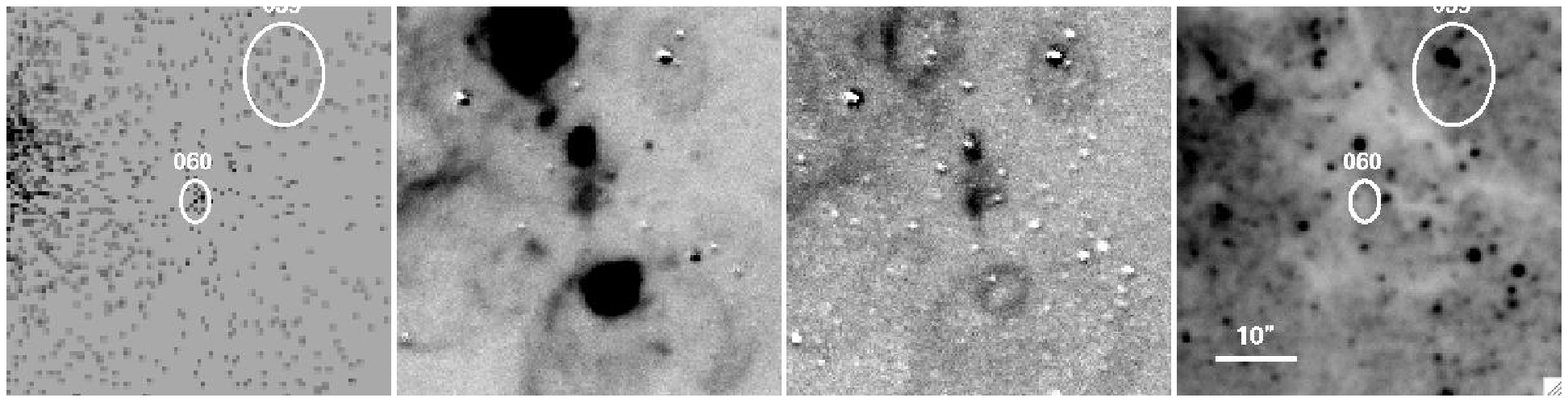}
\vspace{0.05in}
%\plotone{fig_atlas_GKL47.eps}
\plotone{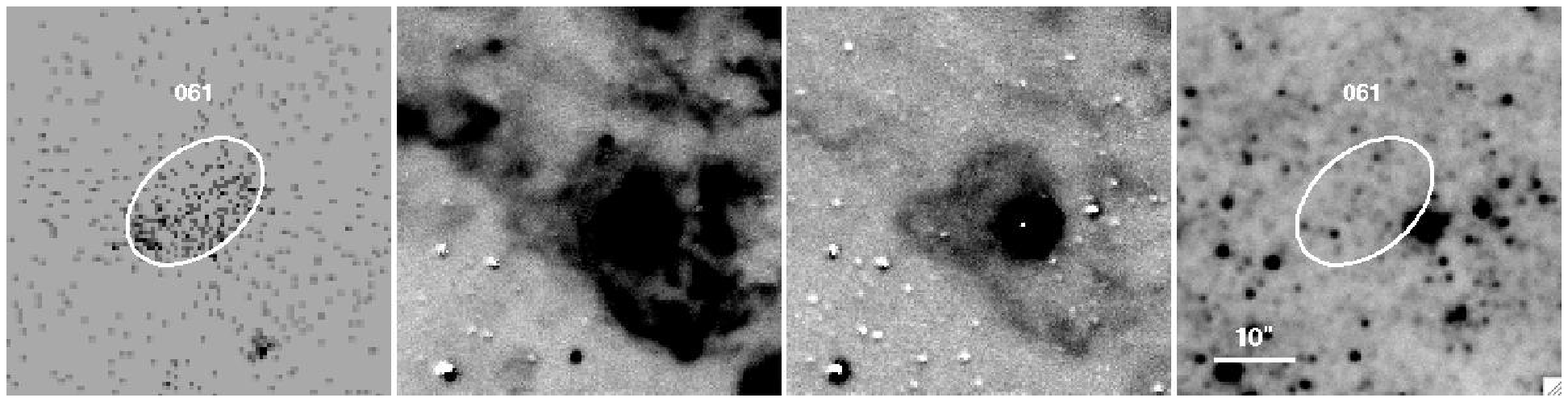}
\vspace{0.05in}
%\plotone{fig_atlas_EM70.eps}
\plotone{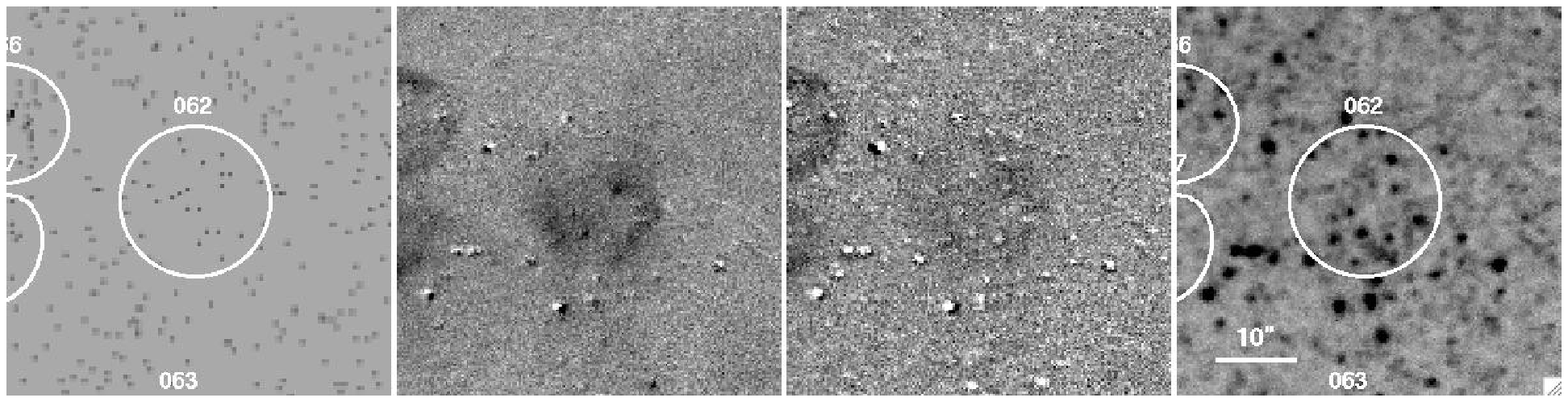}
\vspace{0.05in}
%\plotone{fig_atlas_EM70A.eps}
\plotone{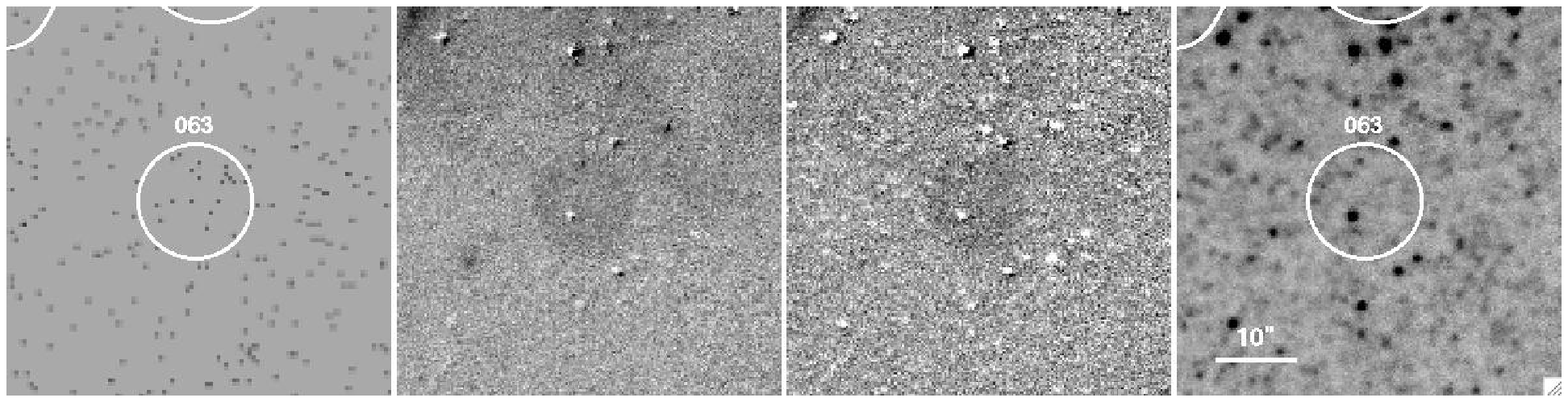}
\figcaption{Images from top to bottom of G98-48,  G98-47,  L10-062,  L10-063.  The format is identical to Fig.\ \ref{fig_atlas01}.  \label{fig_atlas15}  }
\end{figure}

\begin{figure}
%\plotone{fig_atlas_GKL49.eps}
\plotone{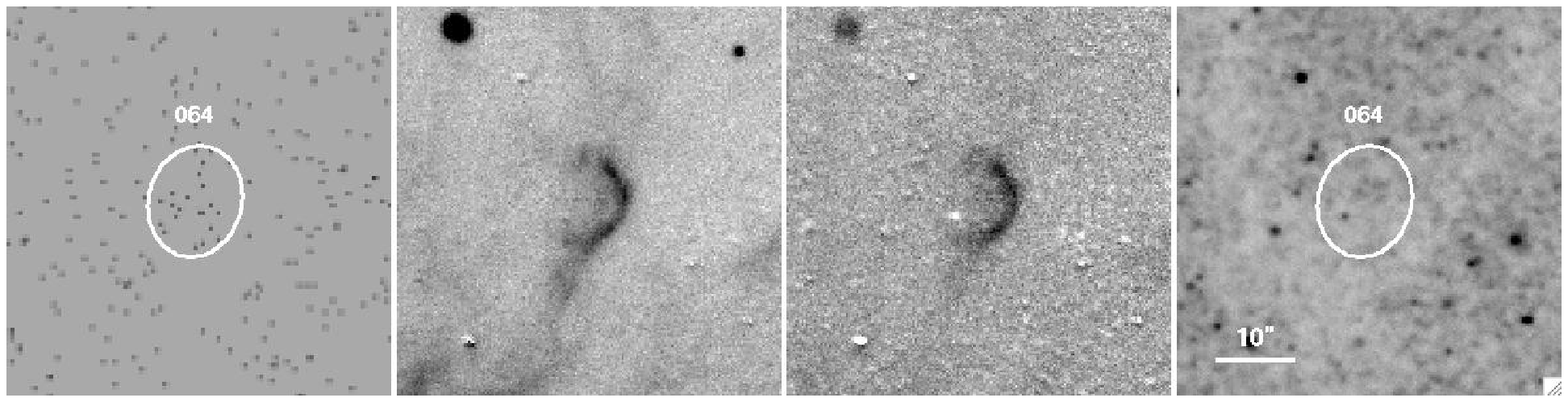}
\vspace{0.05in}
%\plotone{fig_atlas_GKL50.eps}
\plotone{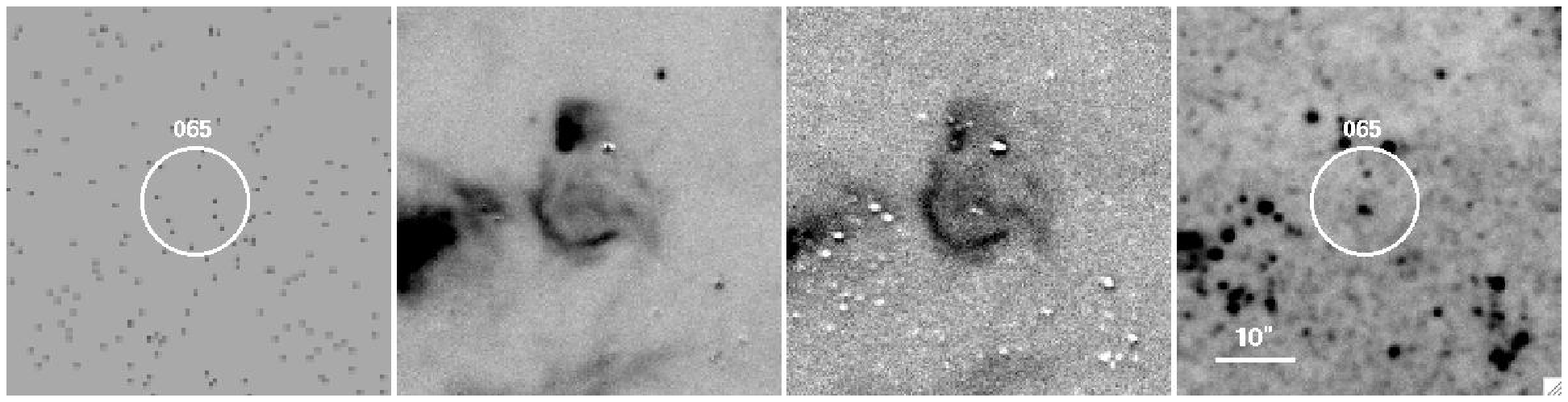}
\vspace{0.05in}
%\plotone{fig_atlas_GKL52.eps}
\plotone{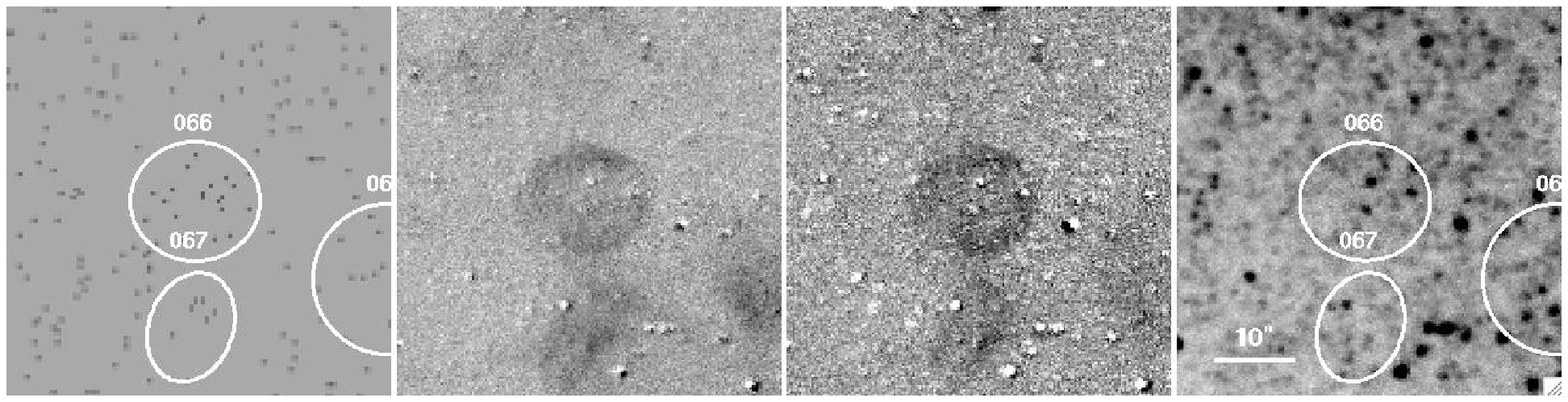}
\vspace{0.05in}
%\plotone{fig_atlas_GKL51.eps}
\plotone{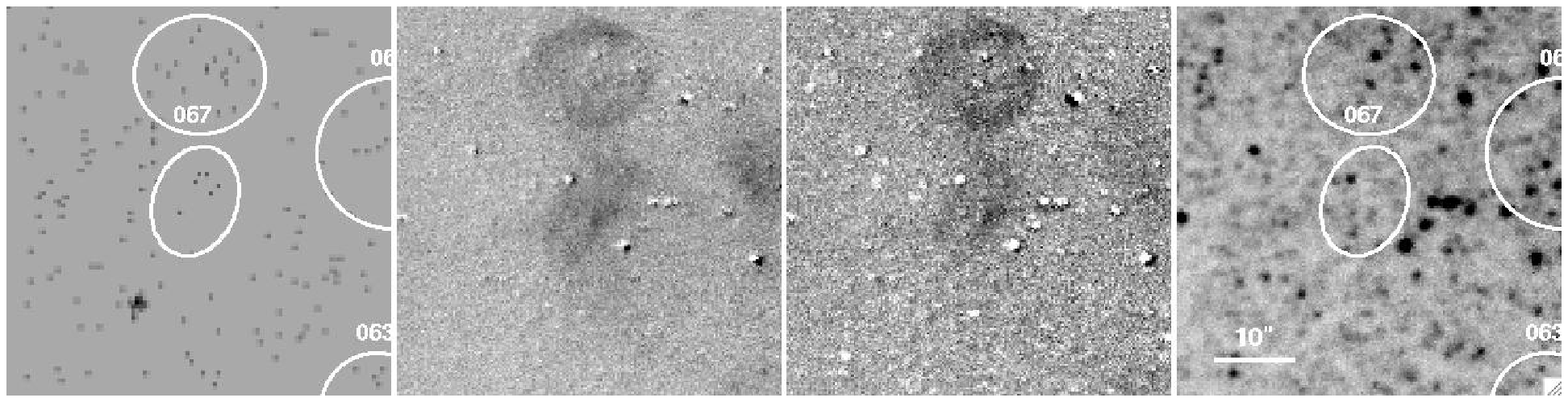}
\figcaption{Images from top to bottom of G98-49,  G98-50,  G98-52,  G98-51.  The format is identical to Fig.\ \ref{fig_atlas01}. \label{fig_atlas16}   }
\end{figure}

\begin{figure}
%\plotone{fig_atlas_EM66.eps}
\plotone{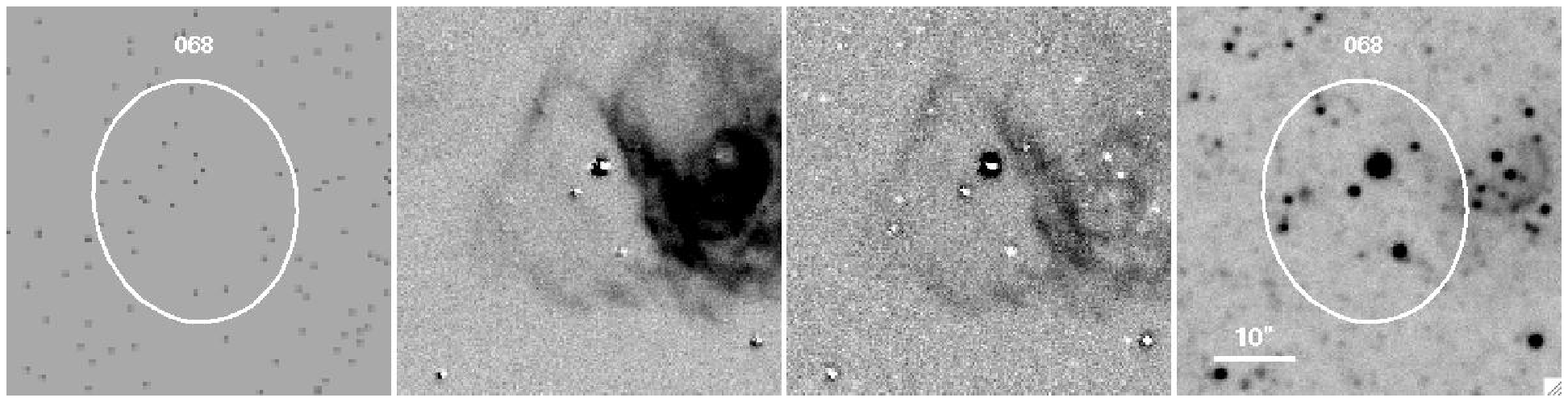}
\vspace{0.05in}
%\plotone{fig_atlas_GKL53.eps}
\plotone{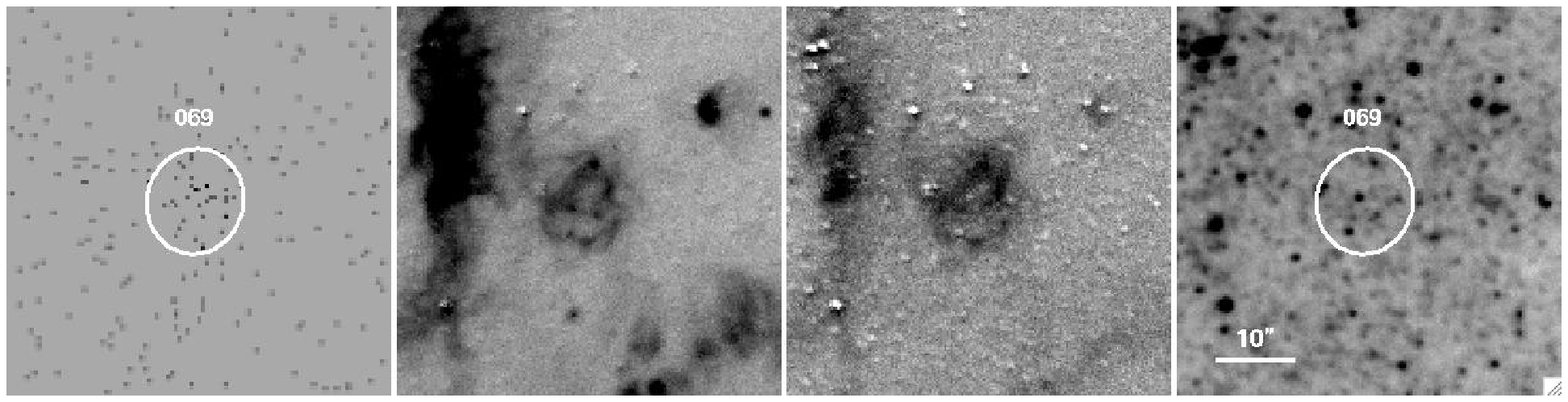}
\vspace{0.05in}
%\plotone{fig_atlas_GKL54.eps}
\plotone{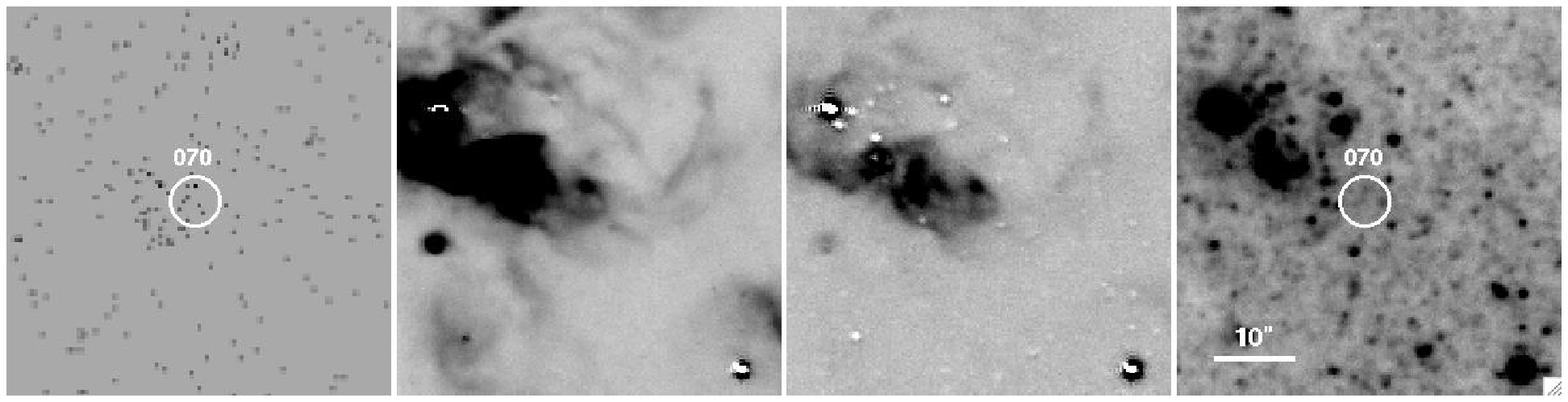}
\vspace{0.05in}
%\plotone{fig_atlas_GKL55.eps}
\plotone{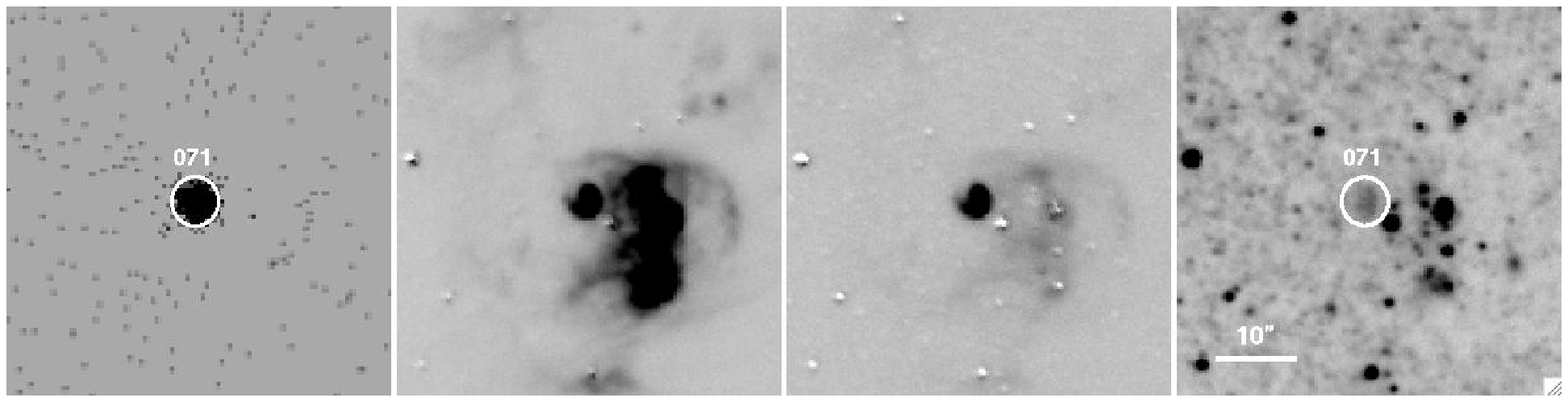}
\figcaption{Images from top to bottom of L10-068,  G98-53,  G98-54,  G98-55.  The format is identical to Fig.\ \ref{fig_atlas01}.  \label{fig_atlas17}  }
\end{figure}

\clearpage

\begin{figure}
%\plotone{fig_atlas_EM12.eps}
\plotone{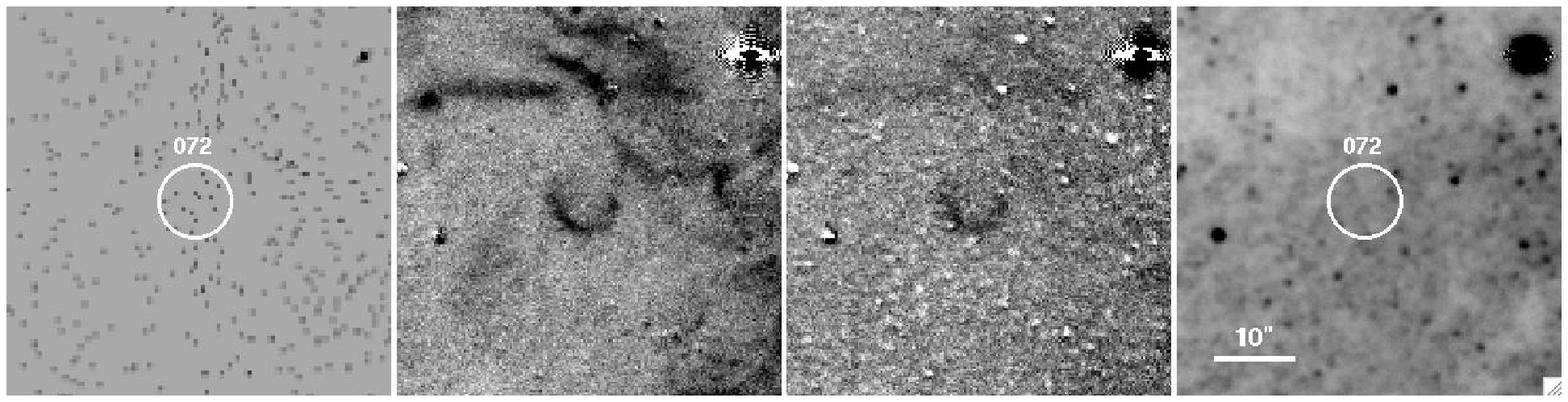}
\vspace{0.05in}
%\plotone{fig_atlas_EM32.eps}
\plotone{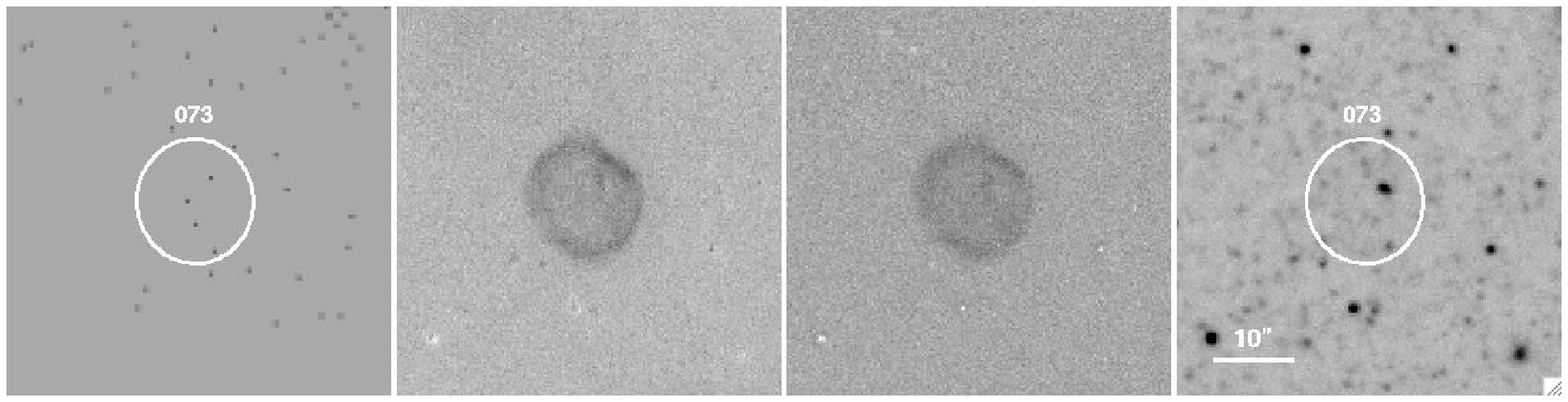}
\vspace{0.05in}
%\plotone{fig_atlas_GKL57AB.eps}
\plotone{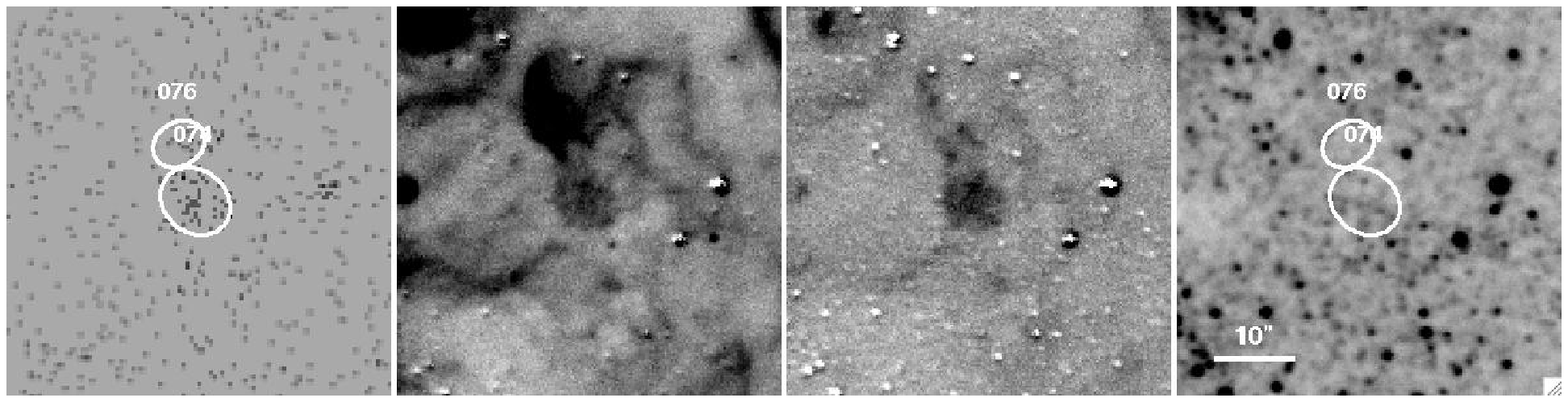}
\vspace{0.05in}
%\plotone{fig_atlas_GKL56.eps}
\plotone{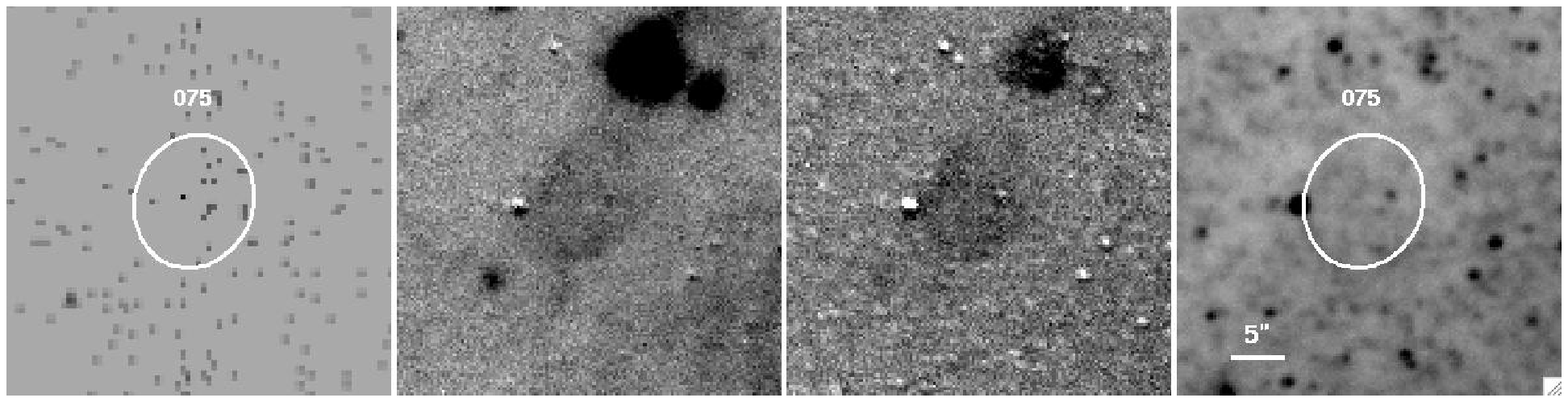}
\figcaption{Images from top to bottom of L10-072,  L10-073,  G98-57A and B,  G98-56.  The format is identical to Fig.\ \ref{fig_atlas01}.  \label{fig_atlas18}  }
\end{figure}

\begin{figure}
%\plotone{fig_atlas_GKL59.eps}
\plotone{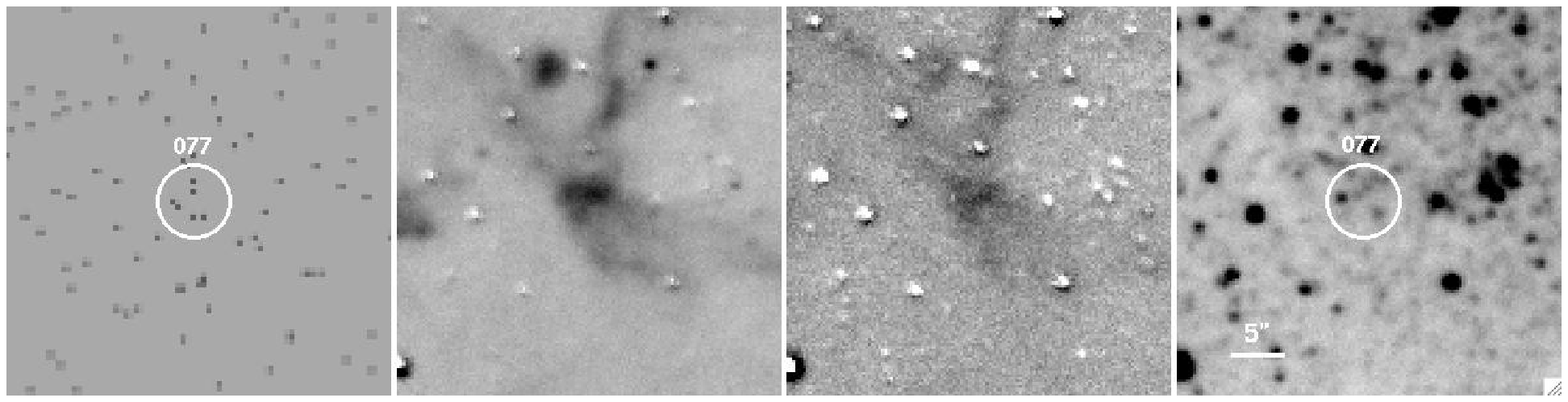}
\vspace{0.05in}
%\plotone{fig_atlas_GKL58.eps}
\plotone{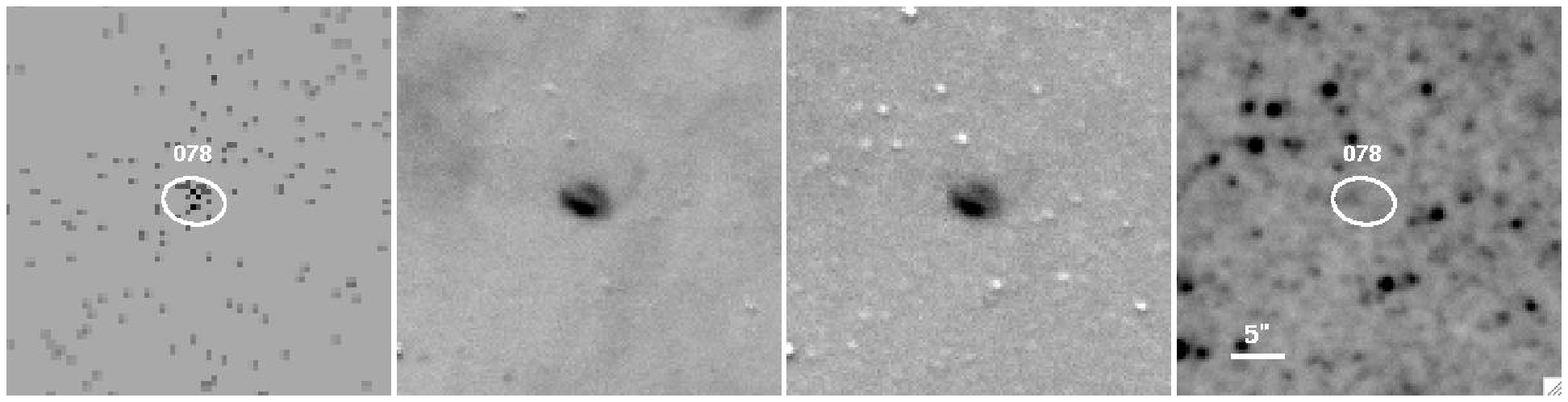}
\vspace{0.05in}
%\plotone{fig_atlas_kip-G.eps}
\plotone{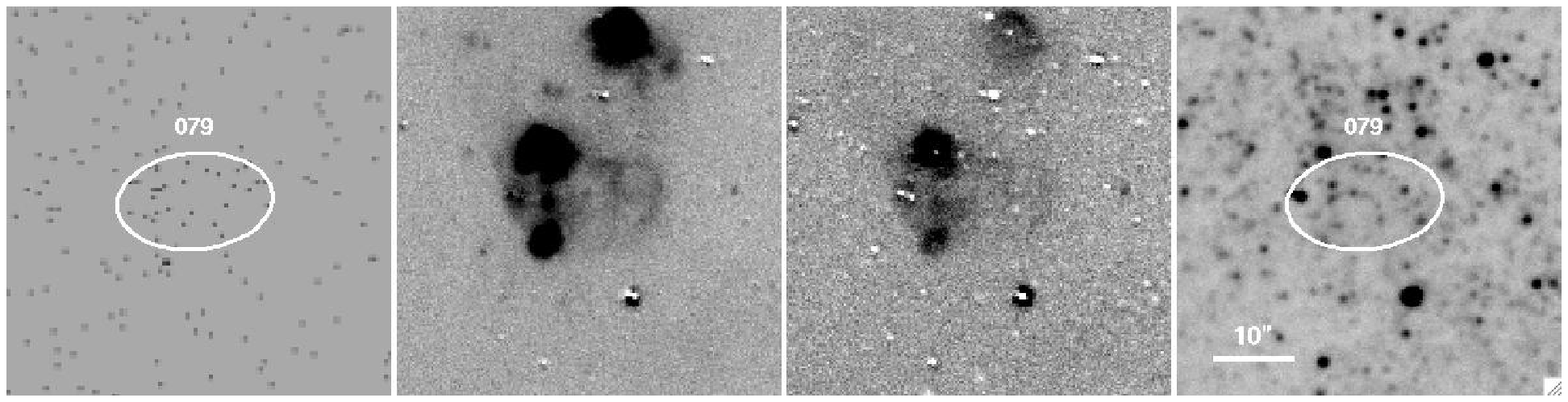}
\vspace{0.05in}
%\plotone{fig_atlas_GKL60.eps}
\plotone{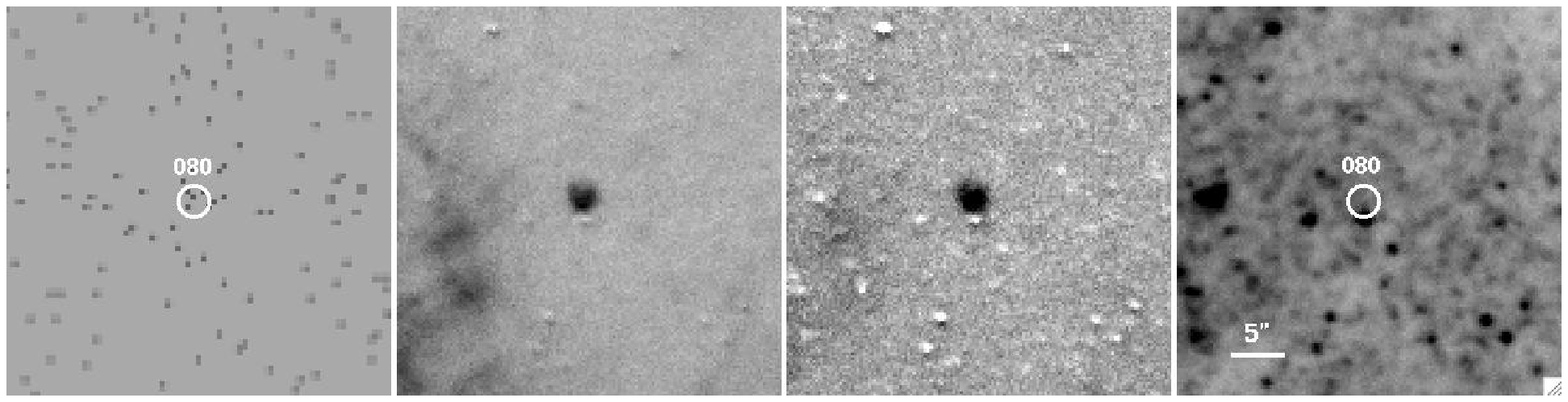}
\figcaption{Images from top to bottom of G98-59,  G98-58,  L10-079,  G98-60.  The format is identical to Fig.\ \ref{fig_atlas01}. \label{fig_atlas19}   }
\end{figure}

\begin{figure}
%\plotone{fig_atlas_GKL62.eps}
\plotone{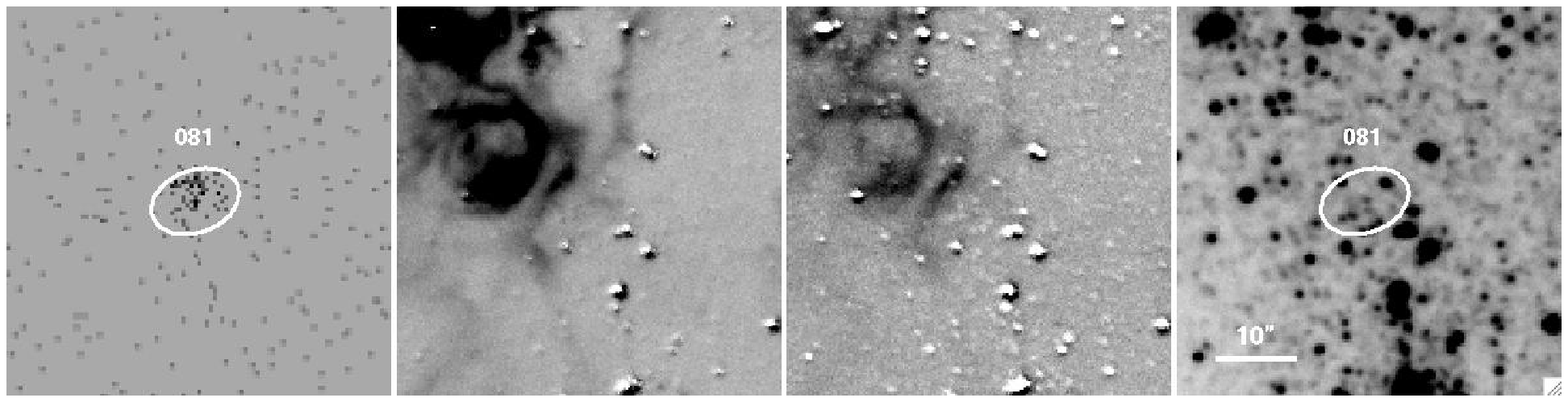}
\vspace{0.05in}
%\plotone{fig_atlas_GKL61.eps}
\plotone{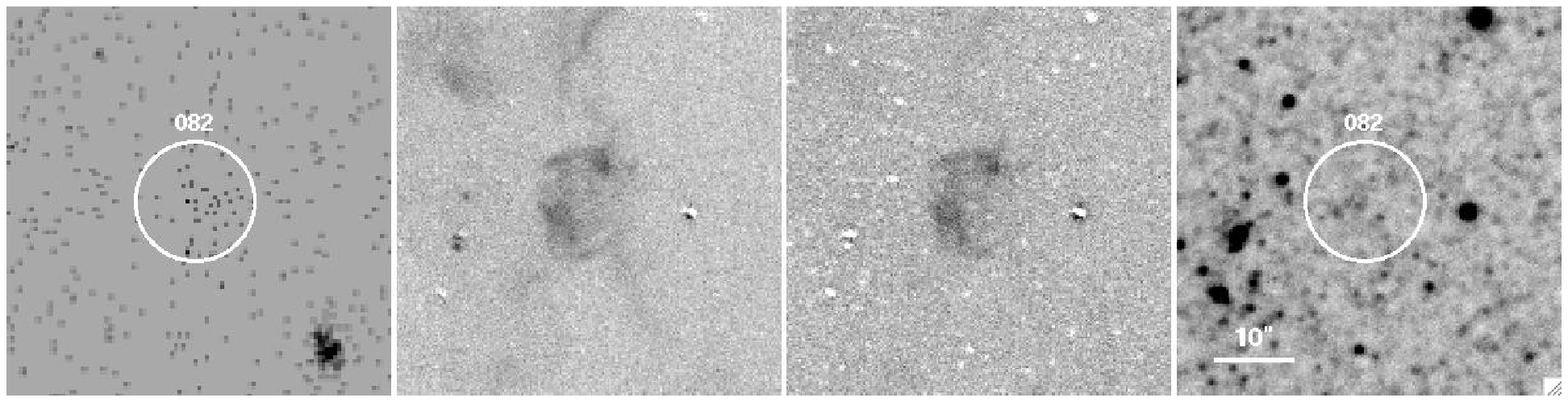}
\vspace{0.05in}
%\plotone{fig_atlas_GKL65.eps}
\plotone{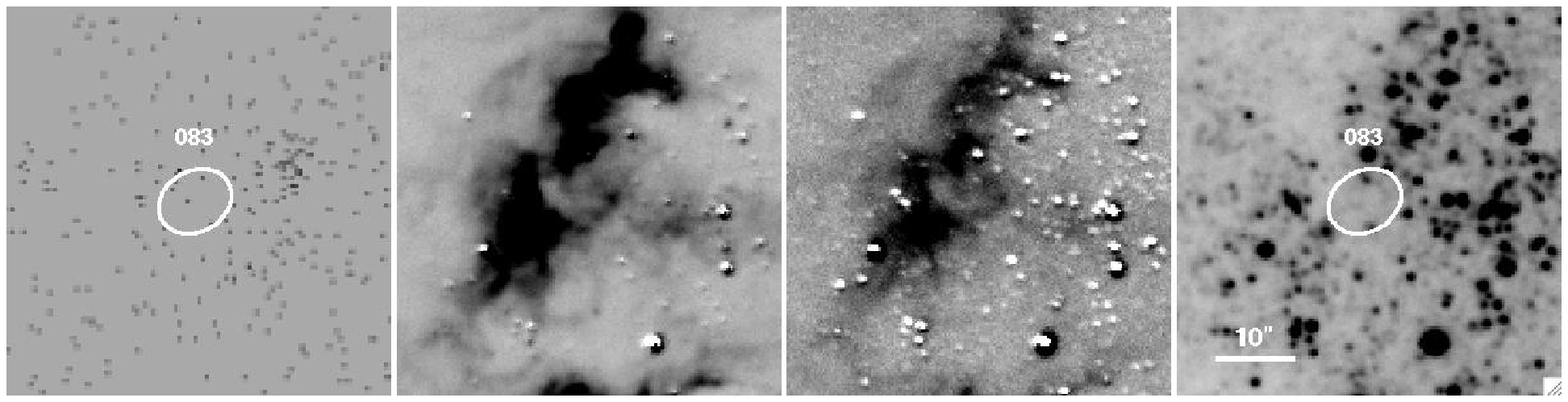}
\vspace{0.05in}
%\plotone{fig_atlas_GKL64.eps}
\plotone{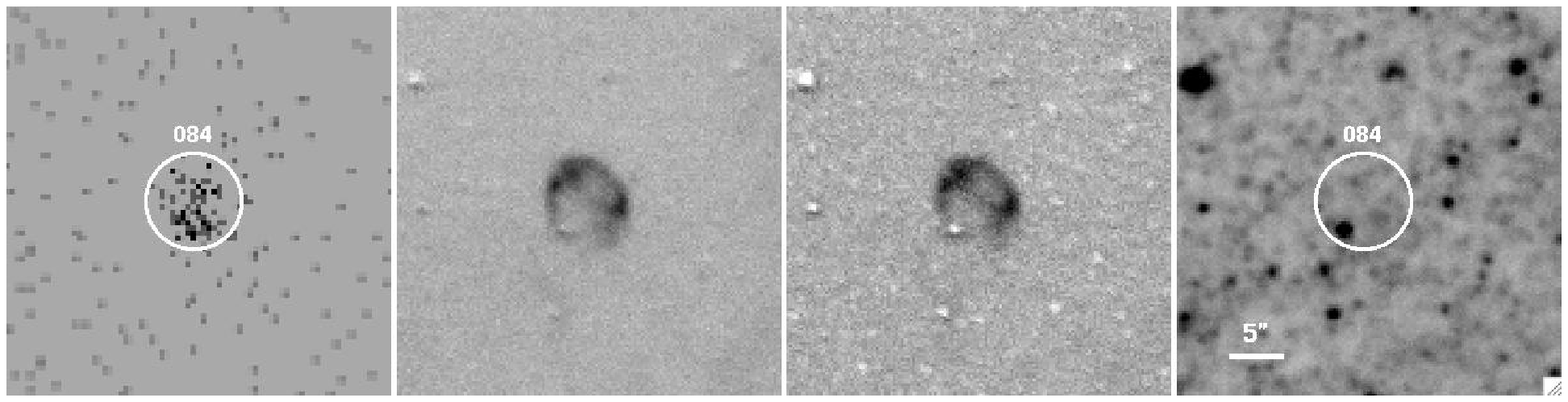}
\figcaption{Images from top to bottom of G98-62,  G98-61,  G98-65,  G98-64.  The format is identical to Fig.\ \ref{fig_atlas01}.  \label{fig_atlas20}  }
\end{figure}

\begin{figure}
%\plotone{fig_atlas_GKL63.eps}
\plotone{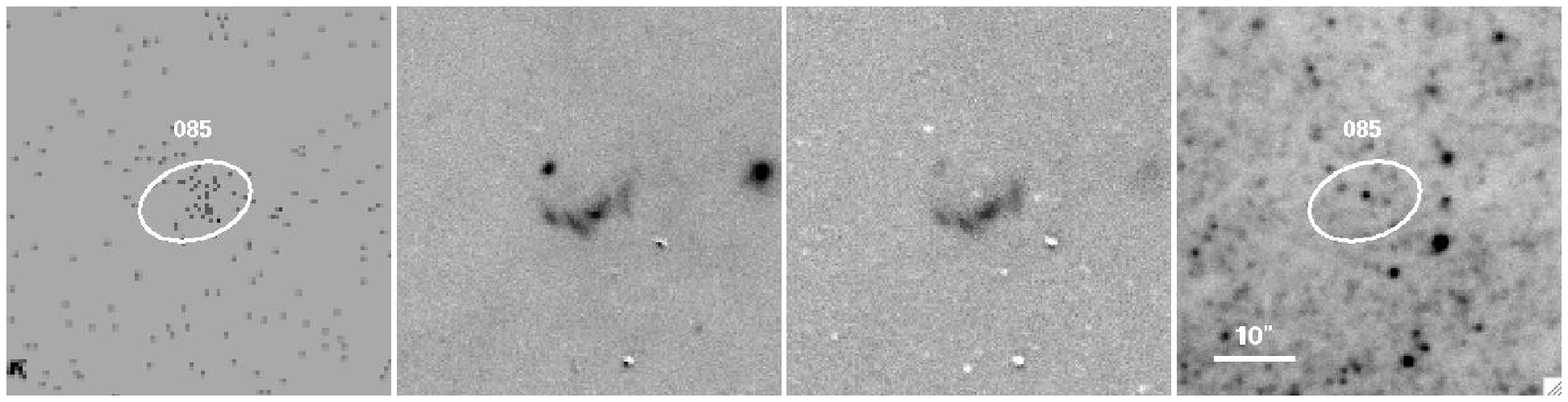}
\vspace{0.05in}
%\plotone{fig_atlas_FL236.eps}
\plotone{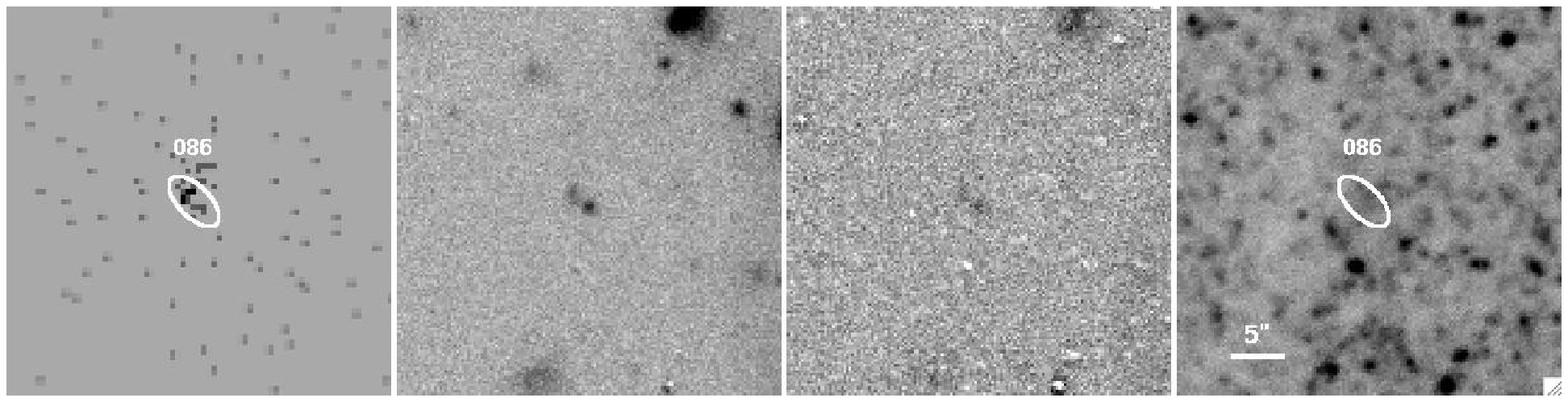}
\vspace{0.05in}
%\plotone{fig_atlas_GKL66.eps}
\plotone{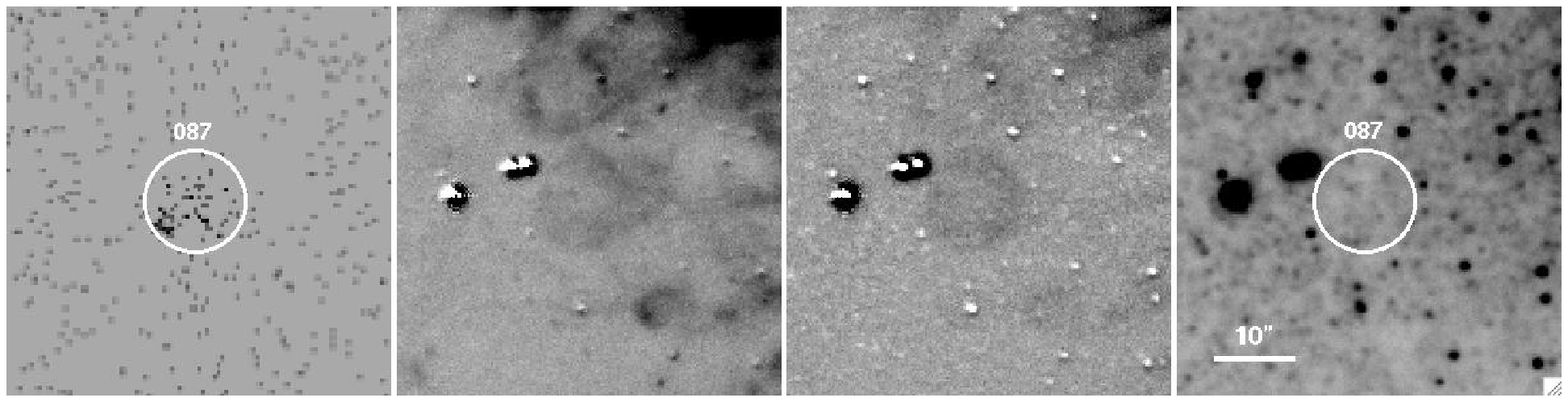}
\vspace{0.05in}
%\plotone{fig_atlas_GKL67.eps}
\plotone{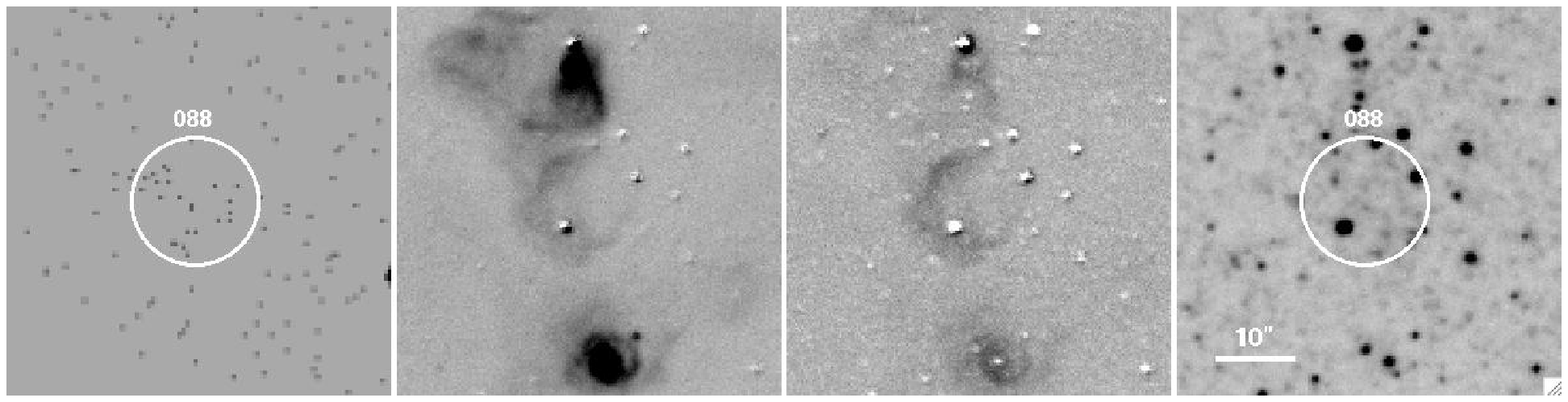}
\figcaption{Images from top to bottom of G98-63,  FL236,  G98-66,  G98-67.  The format is identical to Fig.\ \ref{fig_atlas01}. \label{fig_atlas21}   }
\end{figure}

\begin{figure}
%\plotone{fig_atlas_kip-N.eps}
\plotone{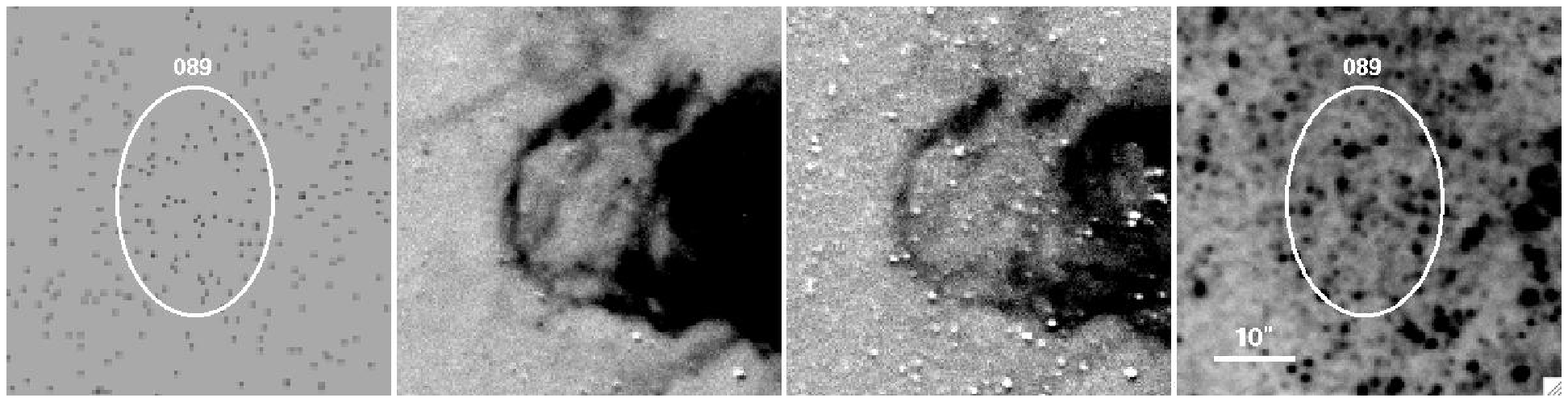}
\vspace{0.05in}
%\plotone{fig_atlas_GKL68.eps}
\plotone{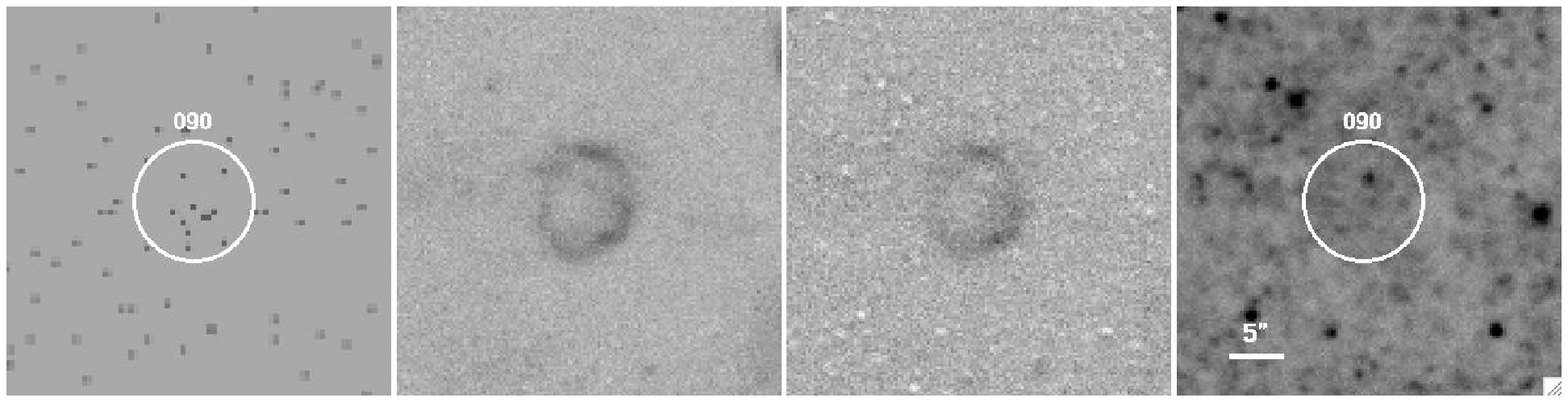}
\vspace{0.05in}
%\plotone{fig_atlas_GKL69.eps}
\plotone{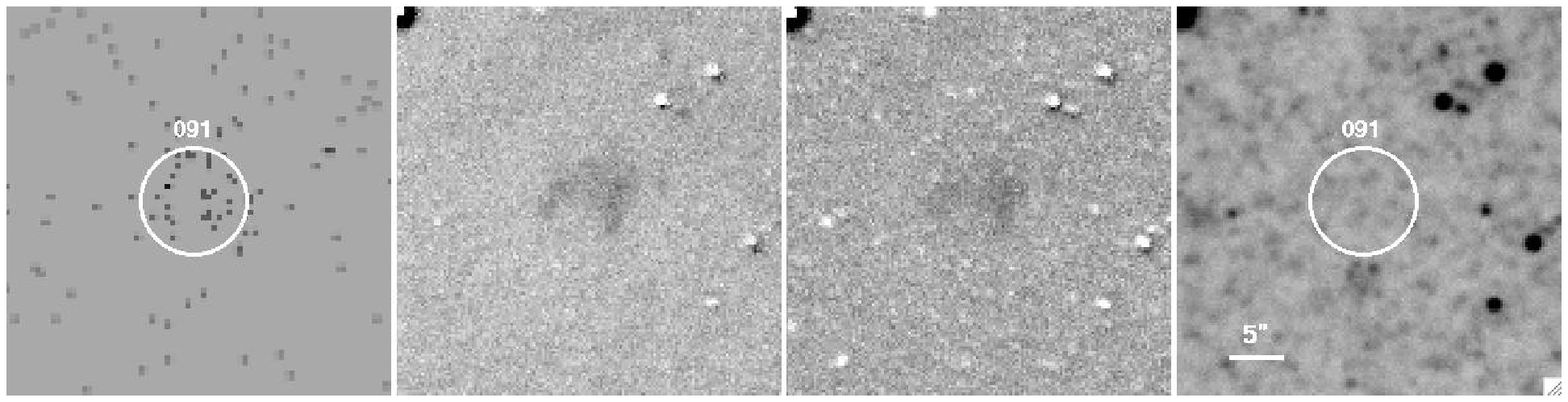}
\vspace{0.05in}
%\plotone{fig_atlas_GKL70.eps}
\plotone{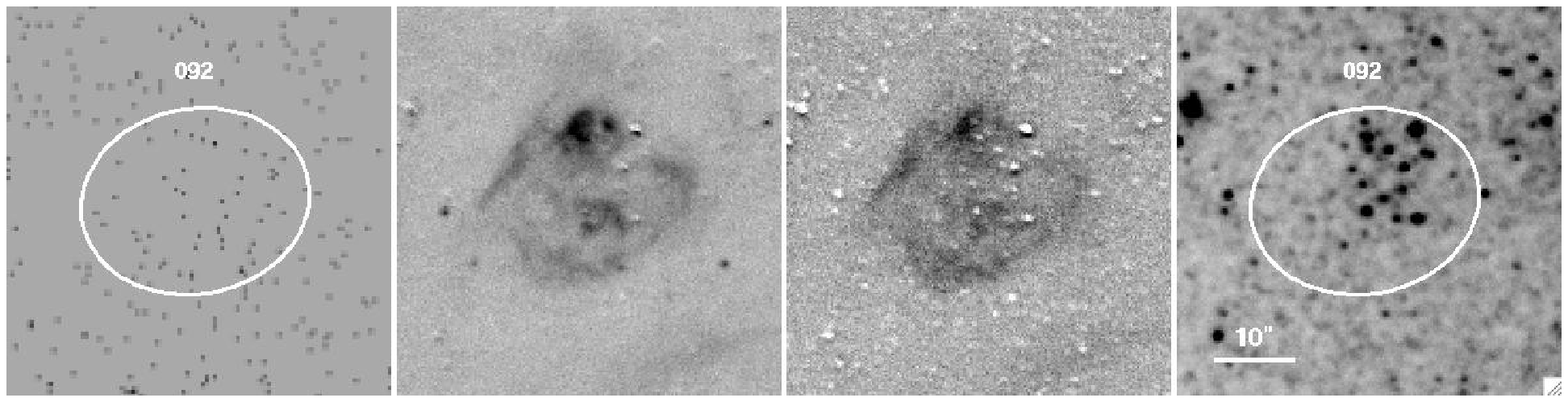}
\figcaption{Images from top to bottom of L10-089,  G98-68,  G98-69,  G98-70.  The format is identical to Fig.\ \ref{fig_atlas01}. \label{fig_atlas22}   }
\end{figure}

\begin{figure}
%\plotone{fig_atlas_FL261.eps}
\plotone{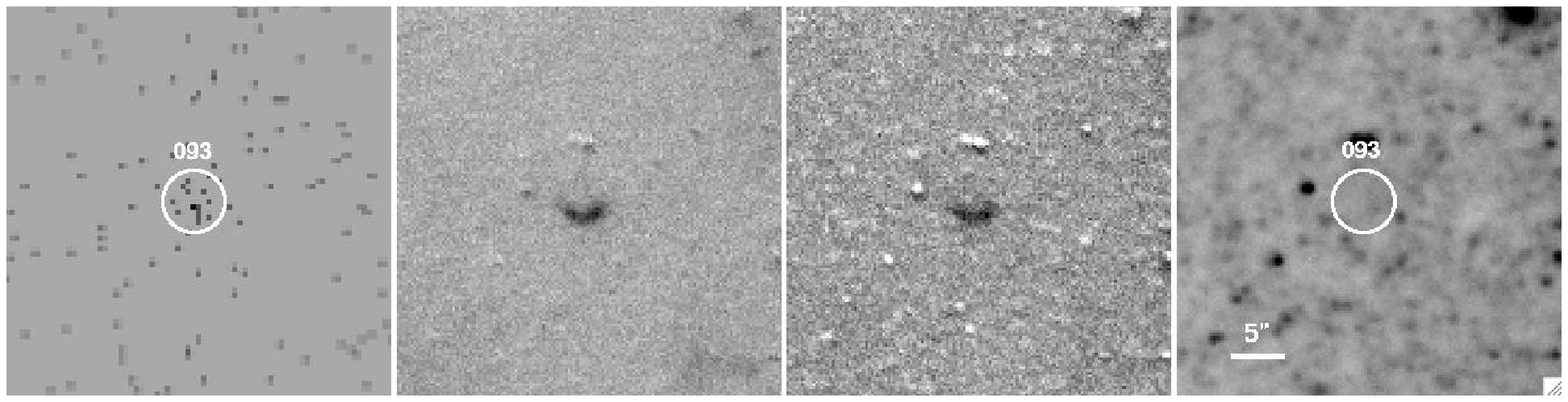}
\vspace{0.05in}
%\plotone{fig_atlas_XMM244.eps}
\plotone{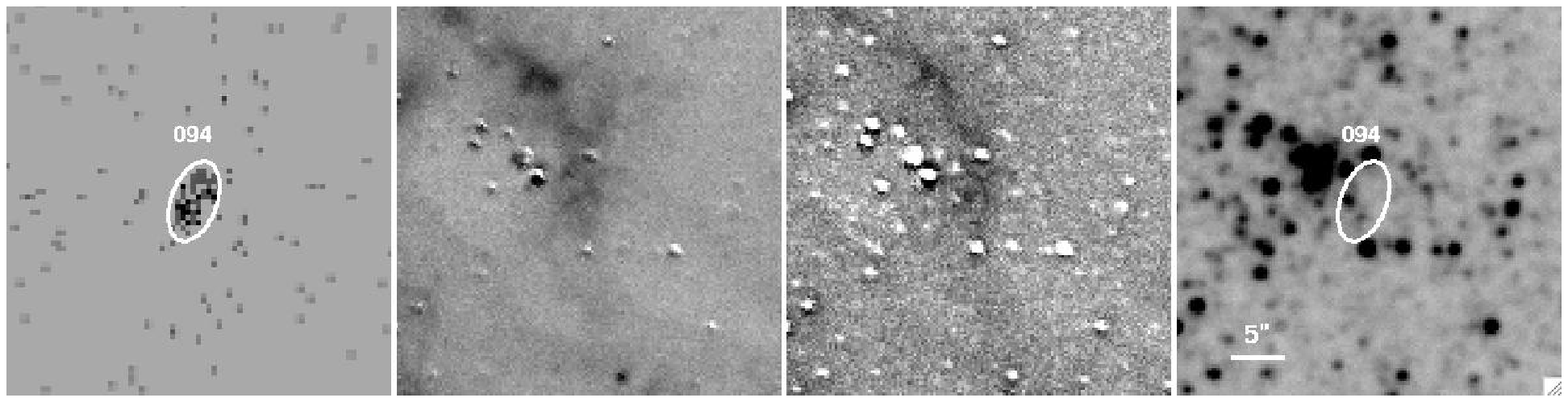}
\vspace{0.05in}
%\plotone{fig_atlas_GKL71.eps}
\plotone{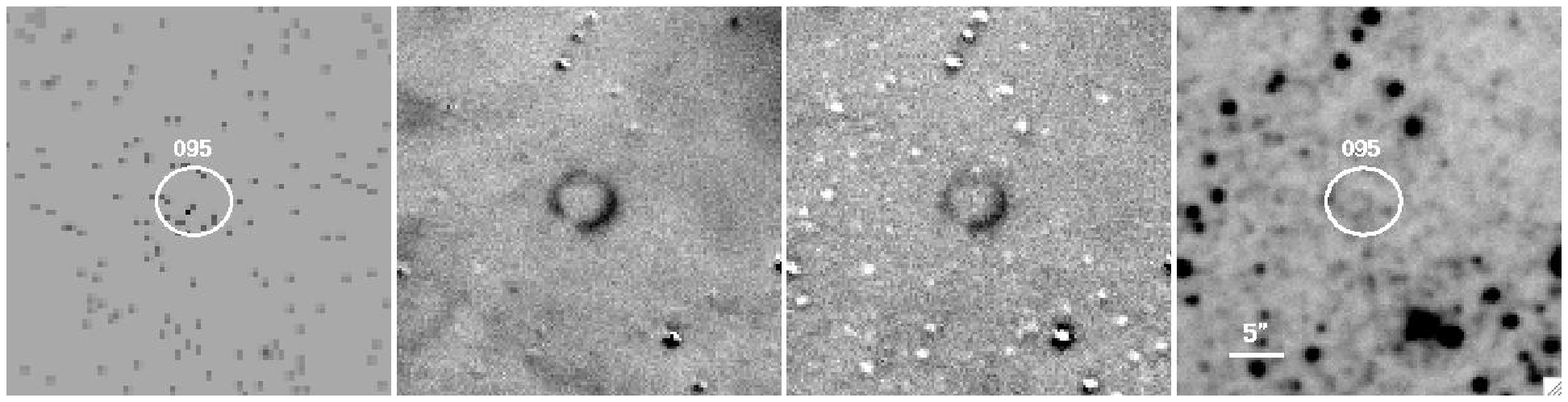}
\vspace{0.05in}
%\plotone{fig_atlas_GKL73.eps}
\plotone{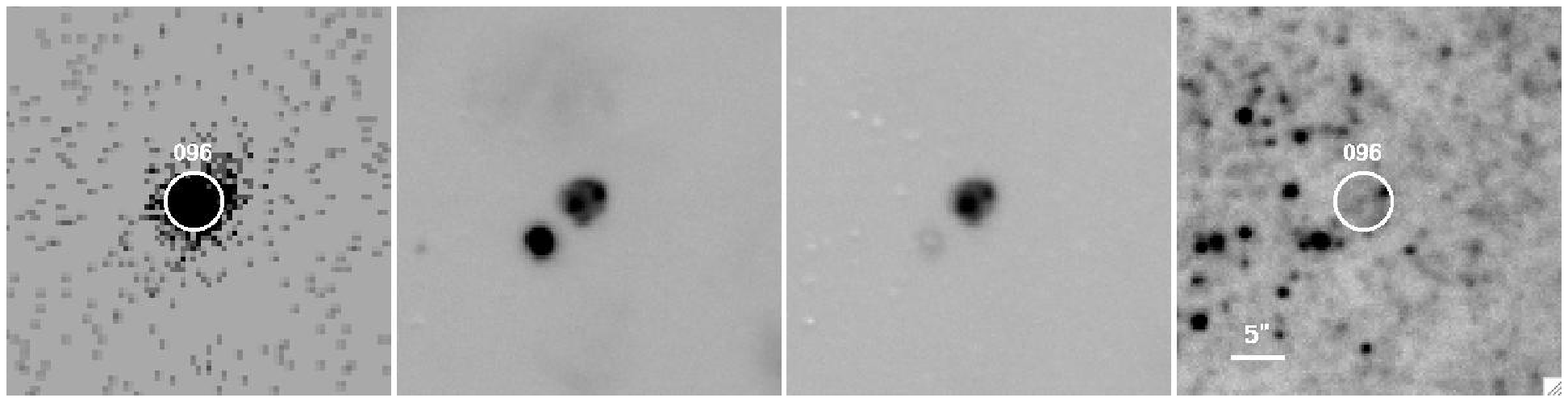}
\figcaption{Images from top to bottom of FL261,  XMM244,  G98-71,  G98-73.  The format is identical to Fig.\ \ref{fig_atlas01}. \label{fig_atlas23}   }
\end{figure}

\begin{figure}
%\plotone{fig_atlas_GKL72.eps}
\plotone{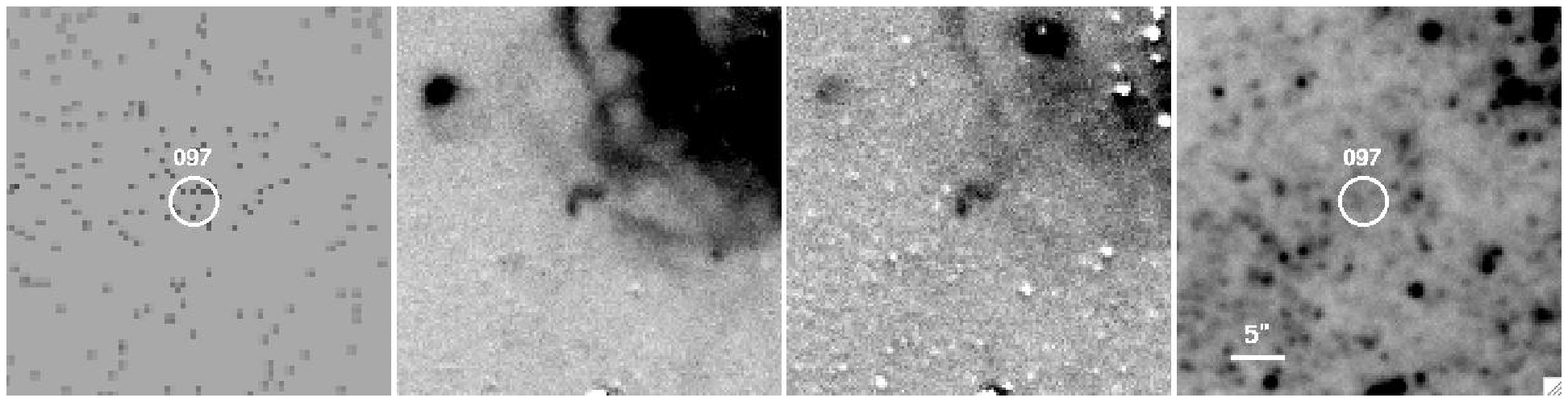}
\vspace{0.05in}
%\plotone{fig_atlas_GKL74.eps}
\plotone{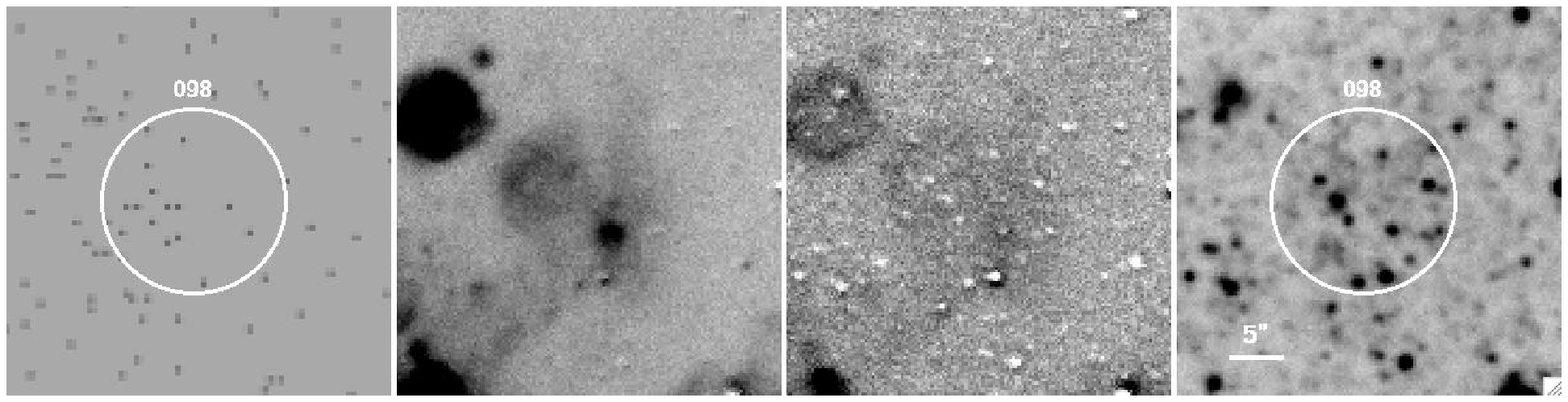}
\vspace{0.05in}
%\plotone{fig_atlas_GKL75.eps}
\plotone{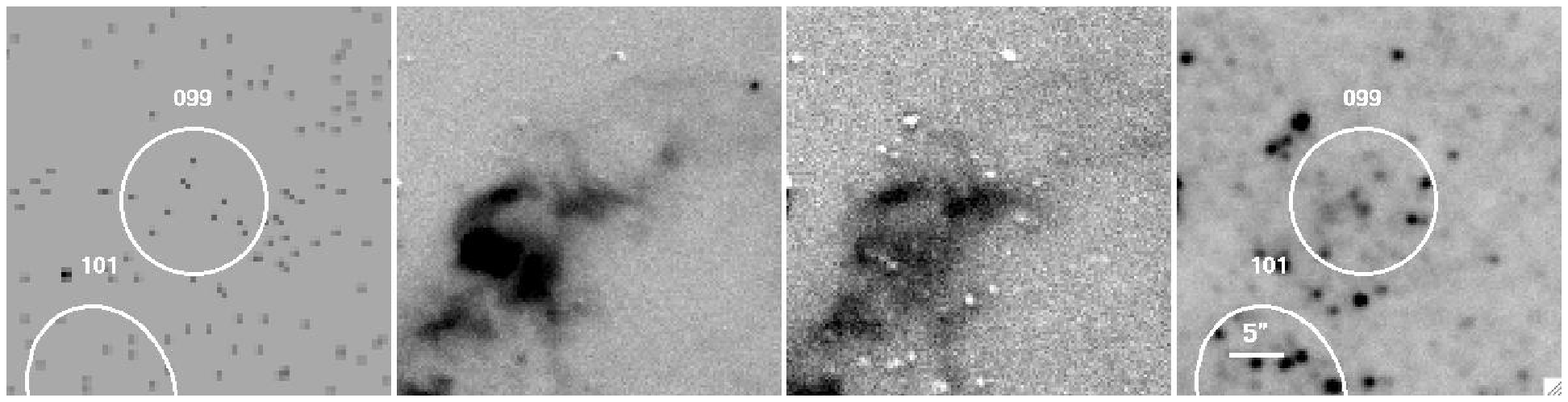}
\vspace{0.05in}
%\plotone{fig_atlas_GKL76.eps}
\plotone{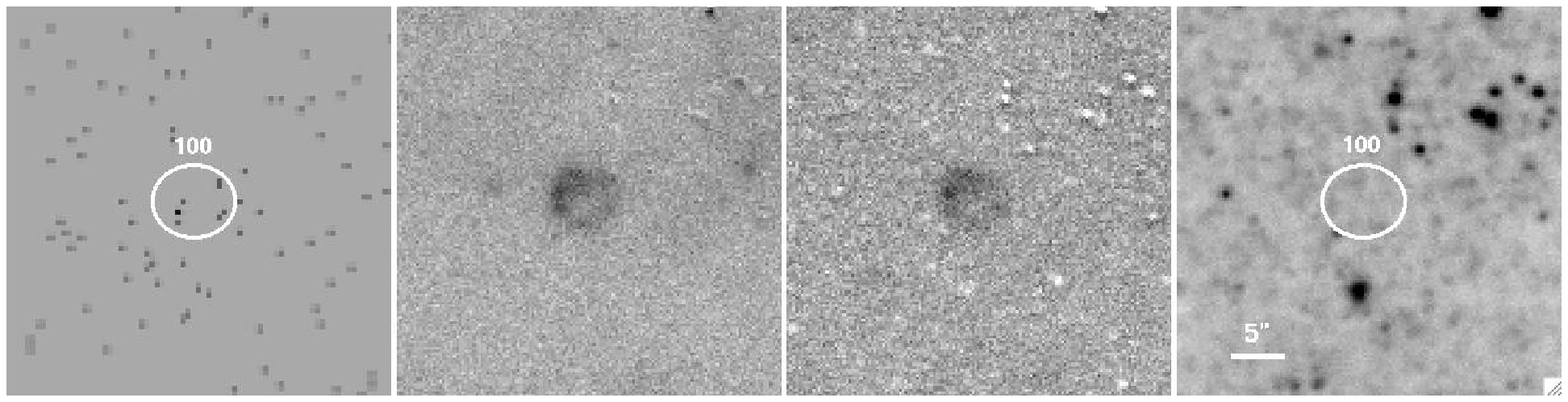}
\figcaption{Images from top to bottom of G98-72,  G98-74,  G98-75,  G98-76.  The format is identical to Fig.\ \ref{fig_atlas01}.  \label{fig_atlas24}  }
\end{figure}

\begin{figure}
%\plotone{fig_atlas_GKL77.eps}
\plotone{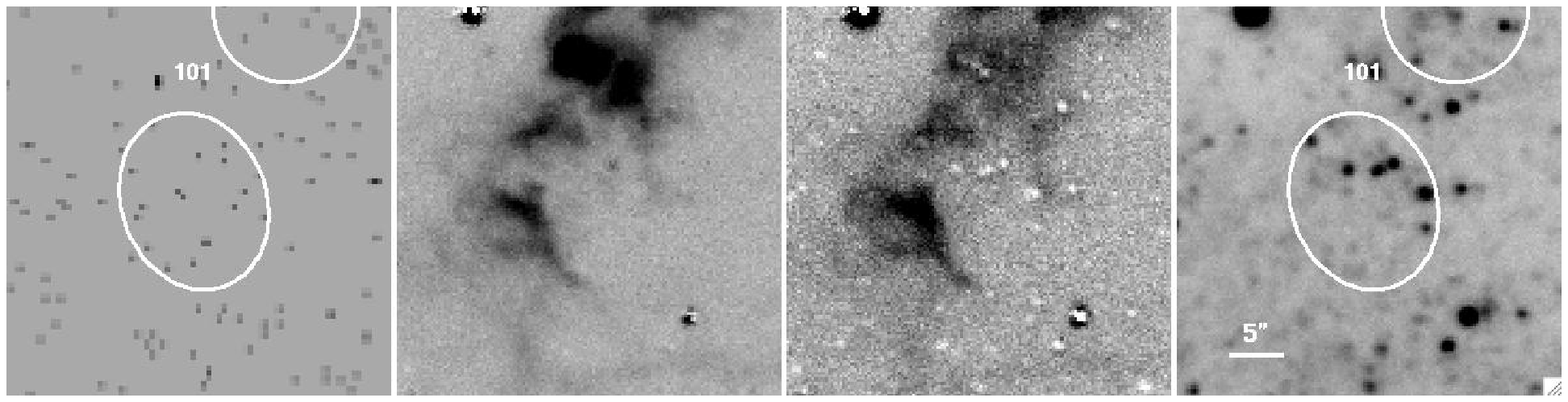}
\vspace{0.05in}
%\plotone{fig_atlas_GKL80.eps}
\plotone{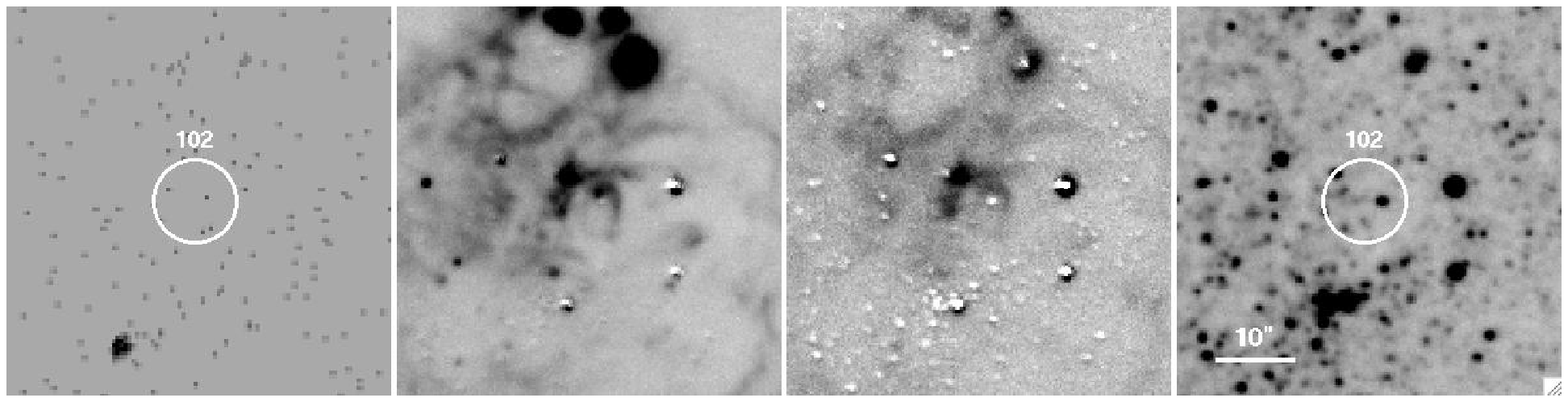}
\vspace{0.05in}
%\plotone{fig_atlas_GKL79.eps}
\plotone{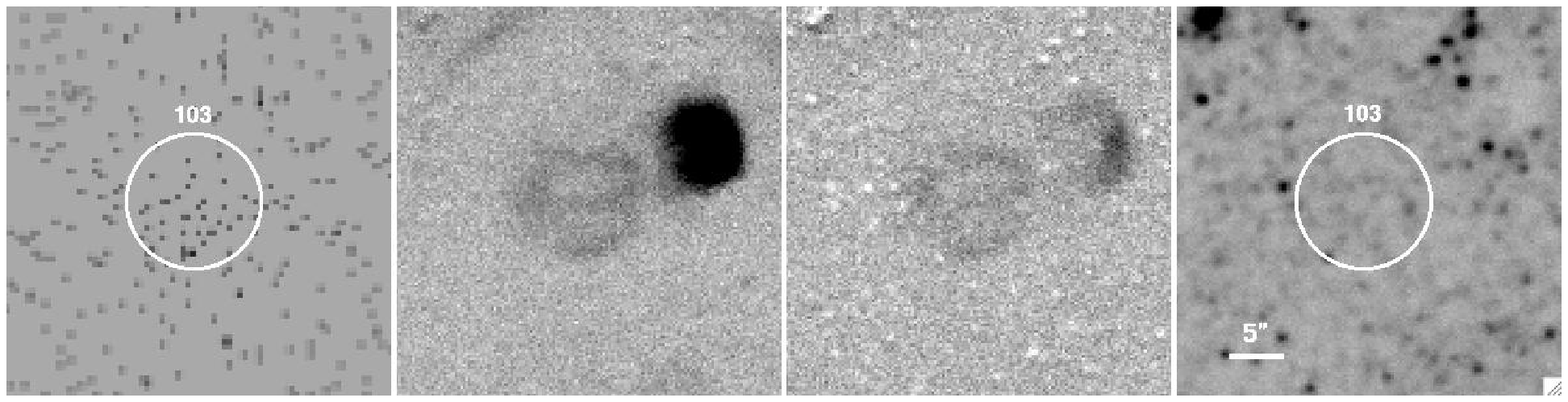}
\vspace{0.05in}
%\plotone{fig_atlas_GKL81.eps}
\plotone{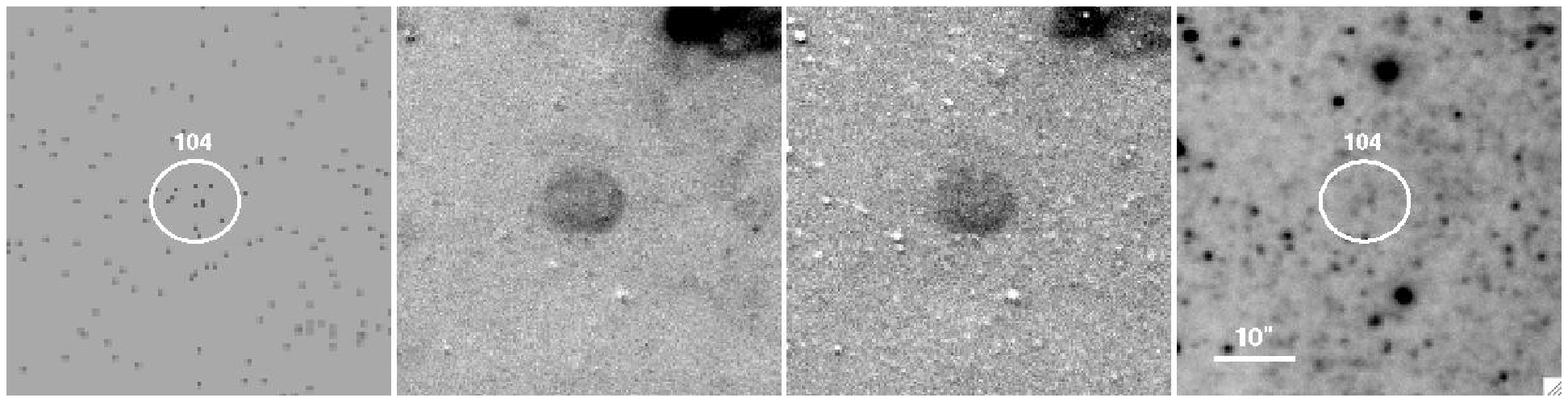}
\figcaption{Images from top to bottom of G98-77,  G98-80,  G98-79,  G98-81.  The format is identical to Fig.\ \ref{fig_atlas01}.  \label{fig_atlas25}  }
\end{figure}

\begin{figure}
%\plotone{fig_atlas_GKL78.eps}
\plotone{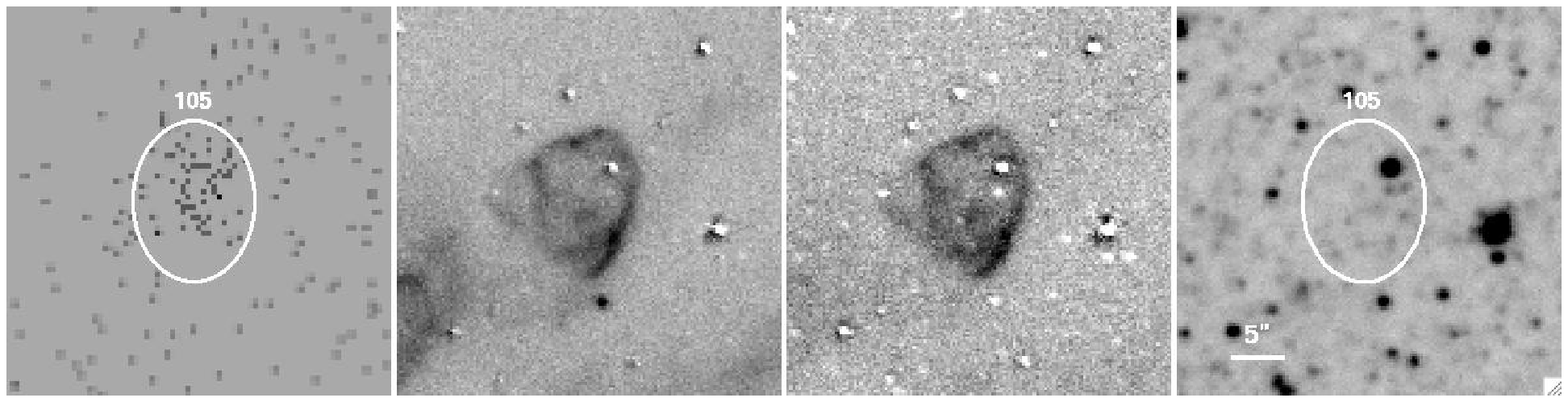}
\vspace{0.05in}
%\plotone{fig_atlas_EM50.eps}
\plotone{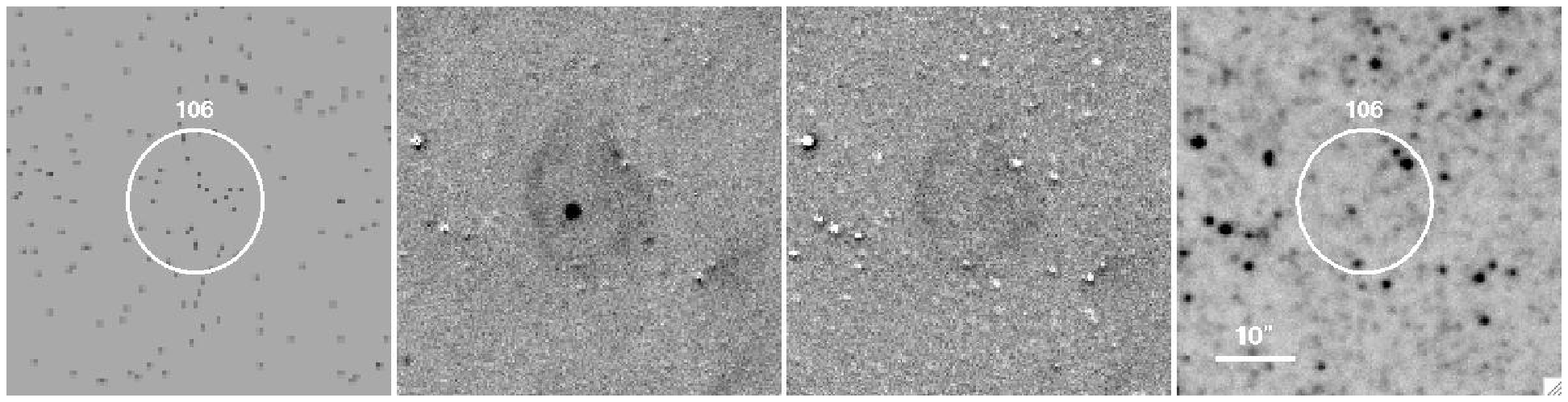}
\vspace{0.05in}
%\plotone{fig_atlas_GKL82.eps}
\plotone{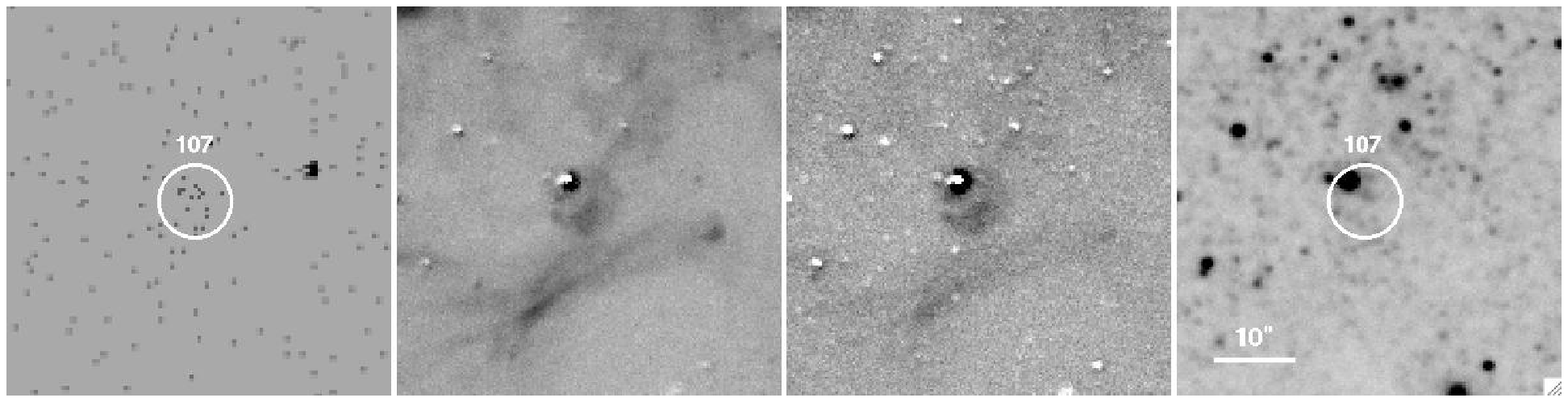}
\vspace{0.05in}
%\plotone{fig_atlas_GKL84.eps}
\plotone{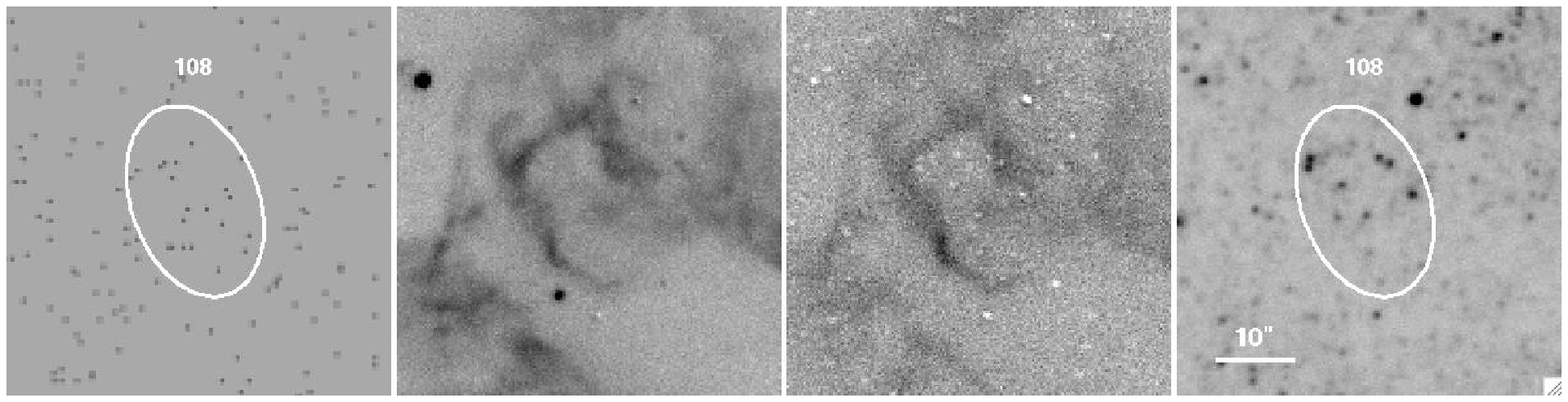}
\figcaption{Images from top to bottom of G98-78,  L10-106,  G98-82,  G98-84.  The format is identical to Fig.\ \ref{fig_atlas01}. \label{fig_atlas26}   }
\end{figure}

\begin{figure}
%\plotone{fig_atlas_GKL83.eps}
\plotone{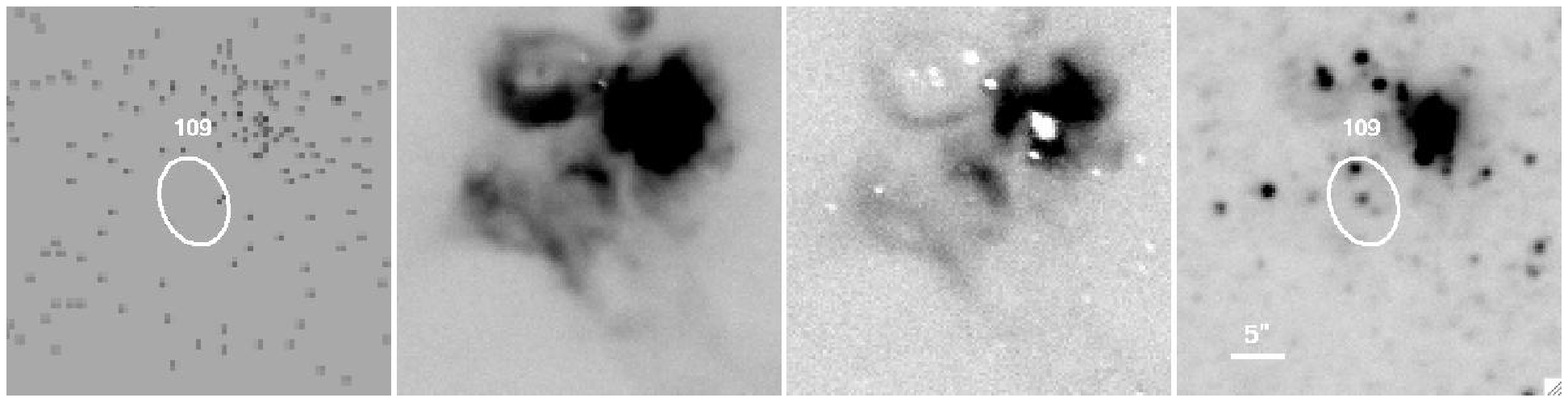}
\vspace{0.05in}
%\plotone{fig_atlas_EM17.eps}
\plotone{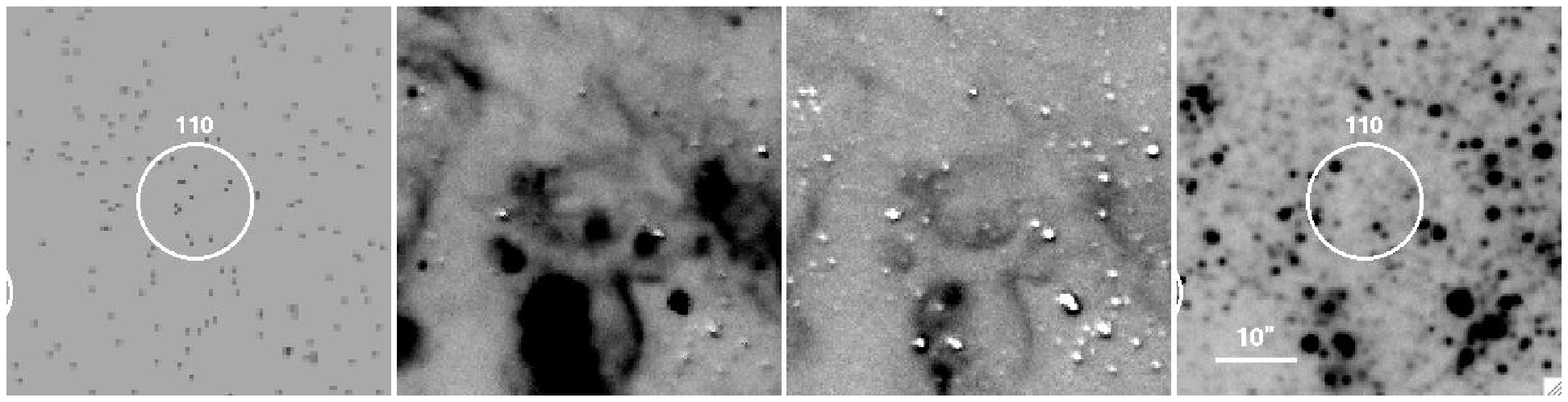}
\vspace{0.05in}
%\plotone{fig_atlas_GKL85.eps}
\plotone{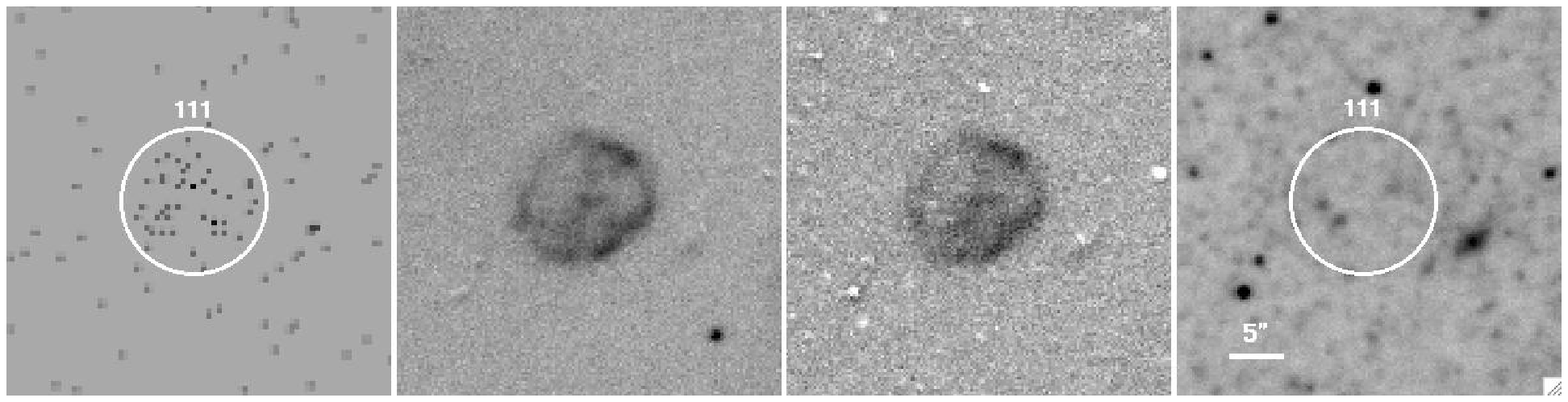}
\vspace{0.05in}
%\plotone{fig_atlas_EM62.eps}
\plotone{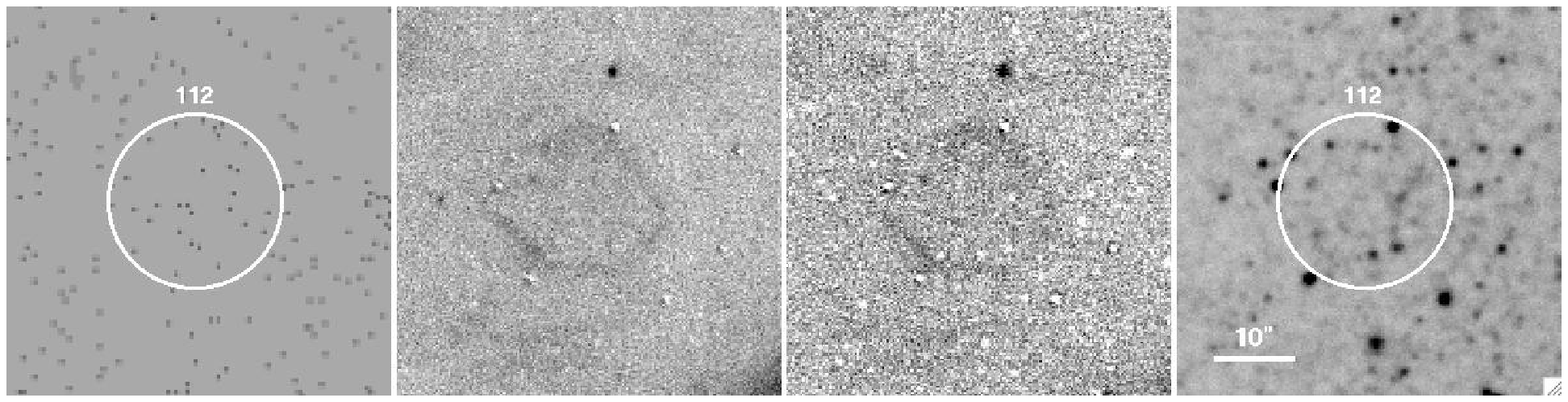}
\figcaption{Images from top to bottom of G98-83,  L10-110,  G98-85,  L10-112.  The format is identical to Fig.\ \ref{fig_atlas01}.  \label{fig_atlas27}  }
\end{figure}

\begin{figure}
%\plotone{fig_atlas_GKL86.eps}
\plotone{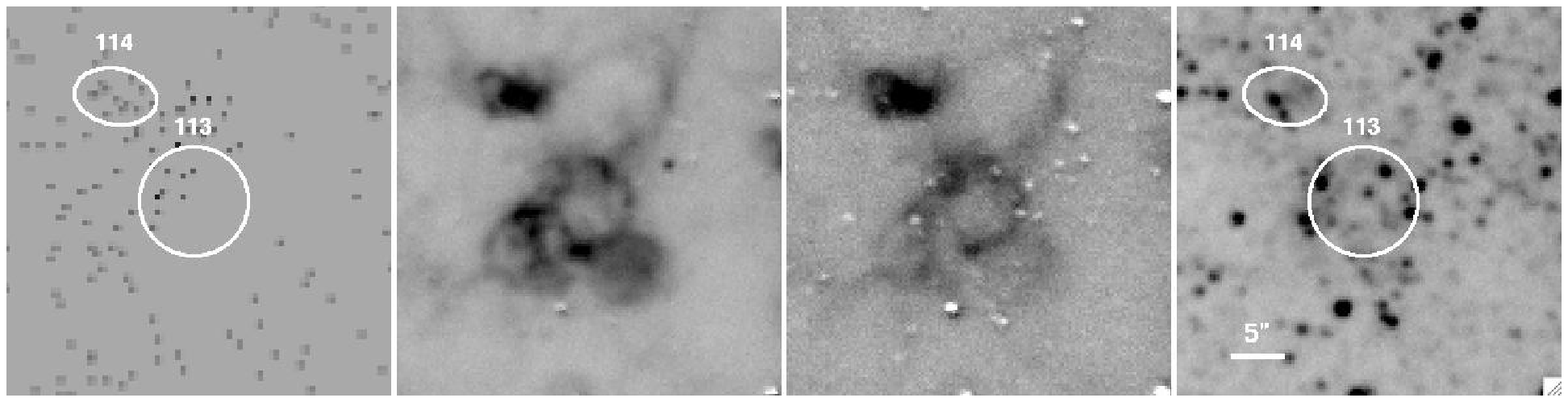}
\vspace{0.05in}
%\plotone{fig_atlas_GKL87.eps}
\plotone{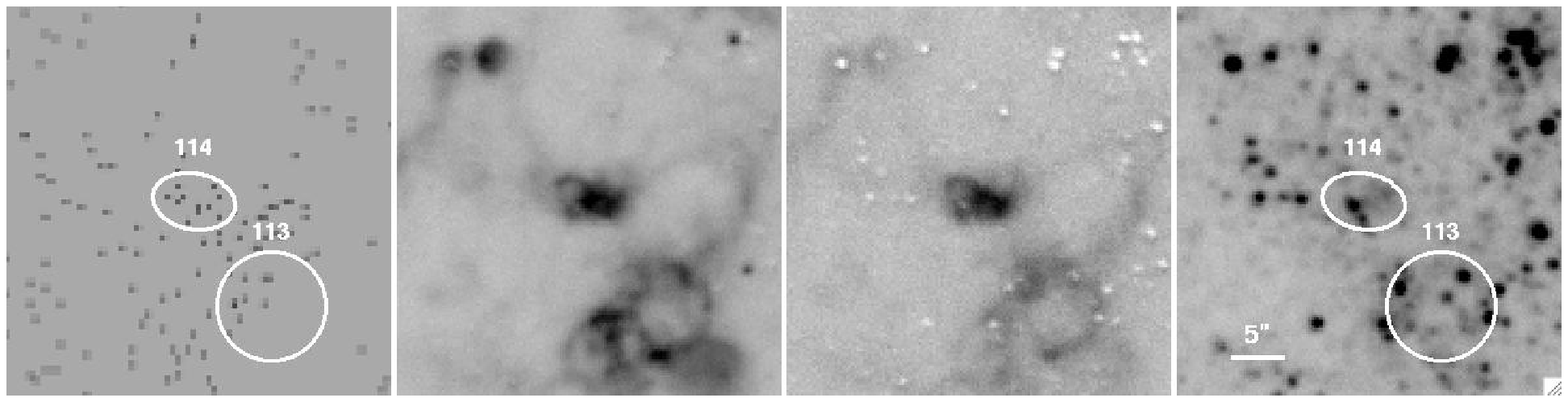}
\vspace{0.05in}
%\plotone{fig_atlas_GKL88.eps}
\plotone{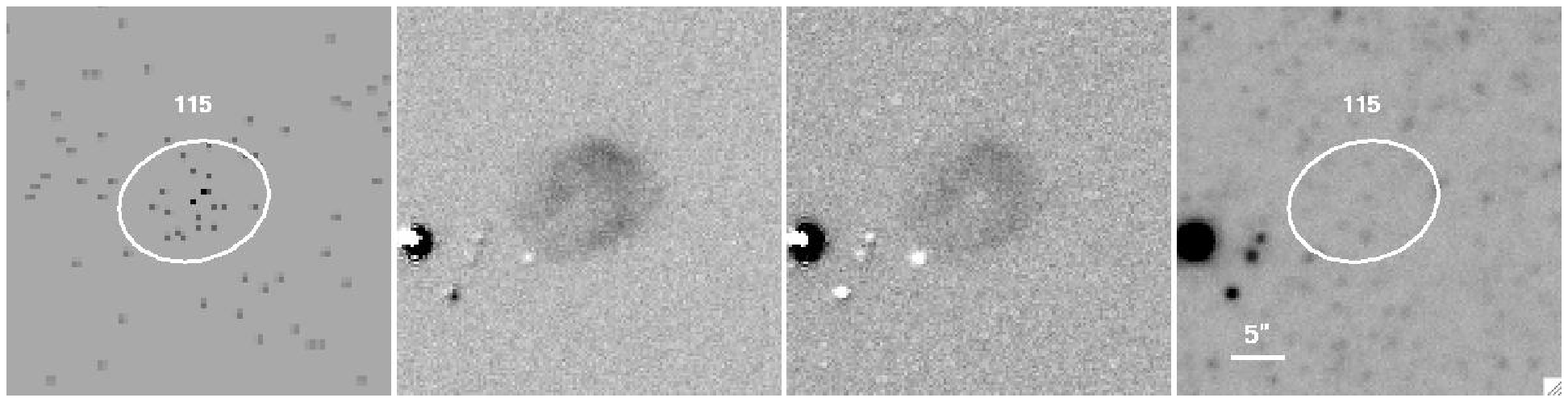}
\vspace{0.05in}
%\plotone{fig_atlas_XMM270.eps}
\plotone{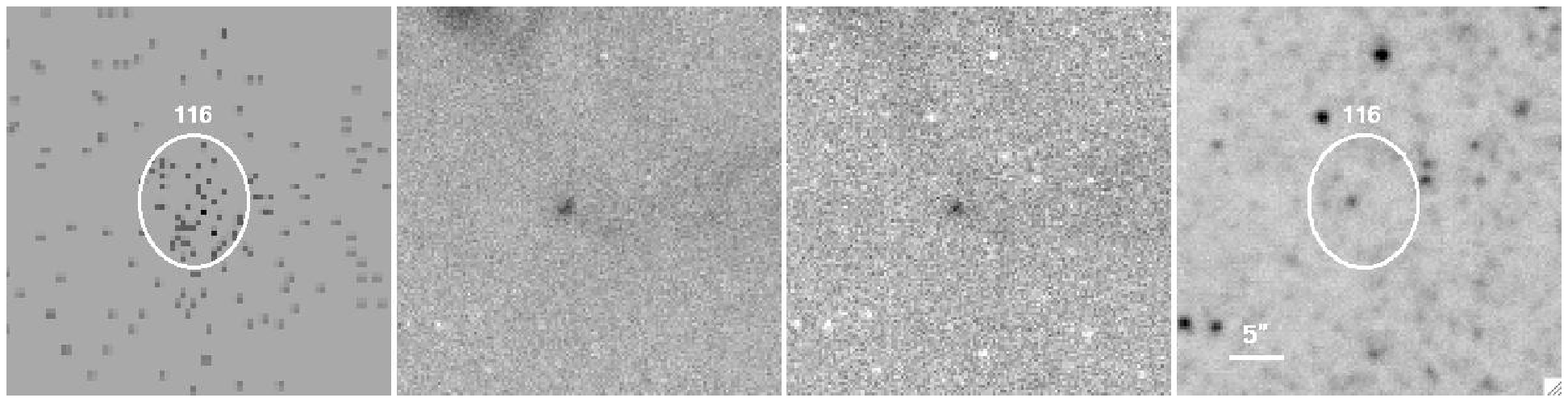}
\figcaption{Images from top to bottom of G98-86,  G98-87,  G98-88,  XMM270.  The format is identical to Fig.\ \ref{fig_atlas01}. \label{fig_atlas28}    }
\end{figure}

\begin{figure}
%\plotone{fig_atlas_GKL89.eps}
\plotone{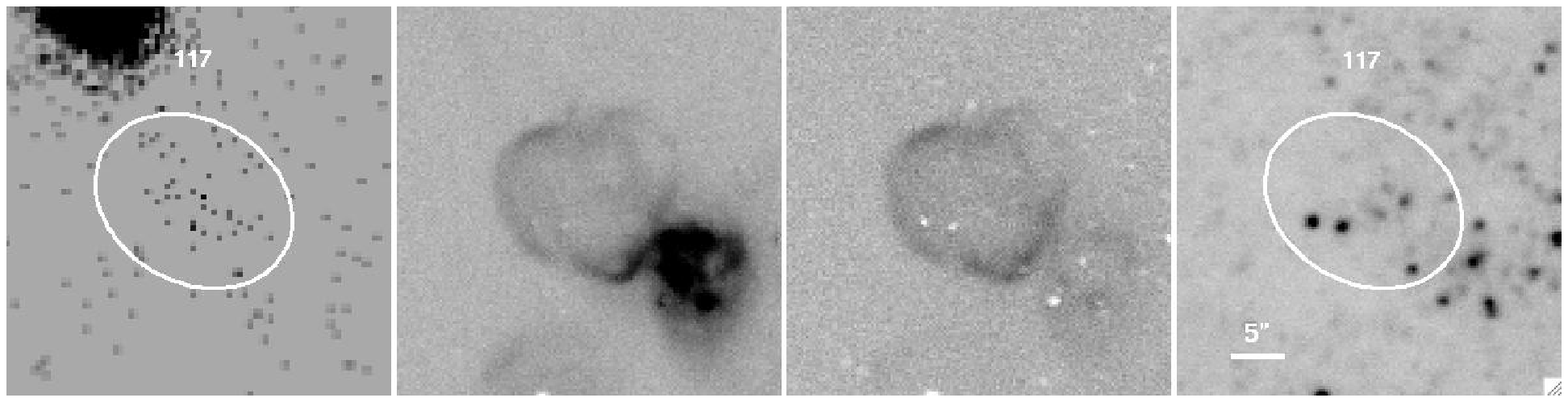}
\vspace{0.05in}
%\plotone{fig_atlas_GKL90.eps}
\plotone{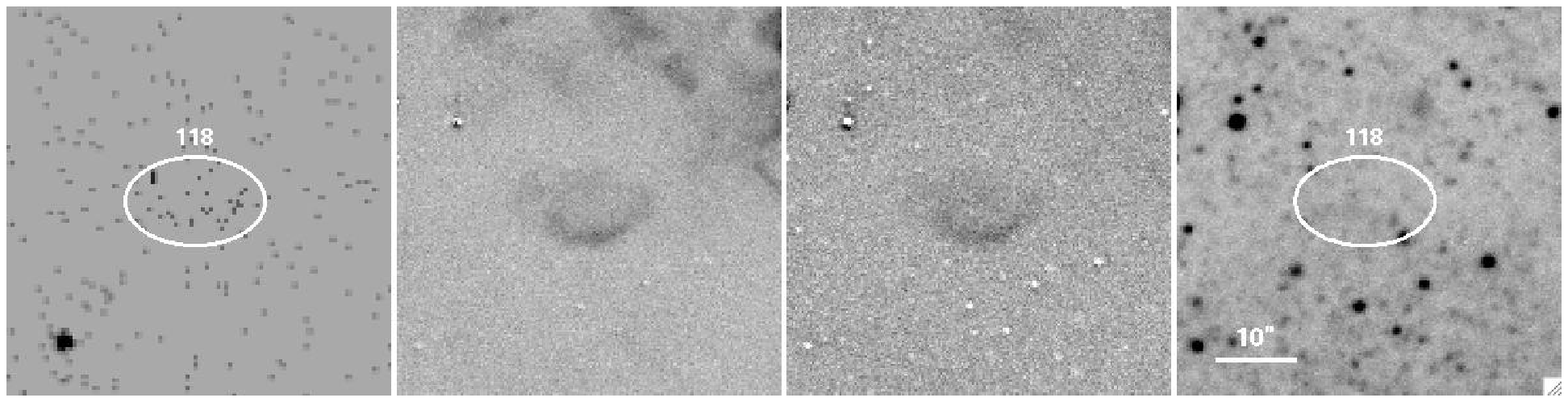}
\vspace{0.05in}
%\plotone{fig_atlas_FL312.eps}
\plotone{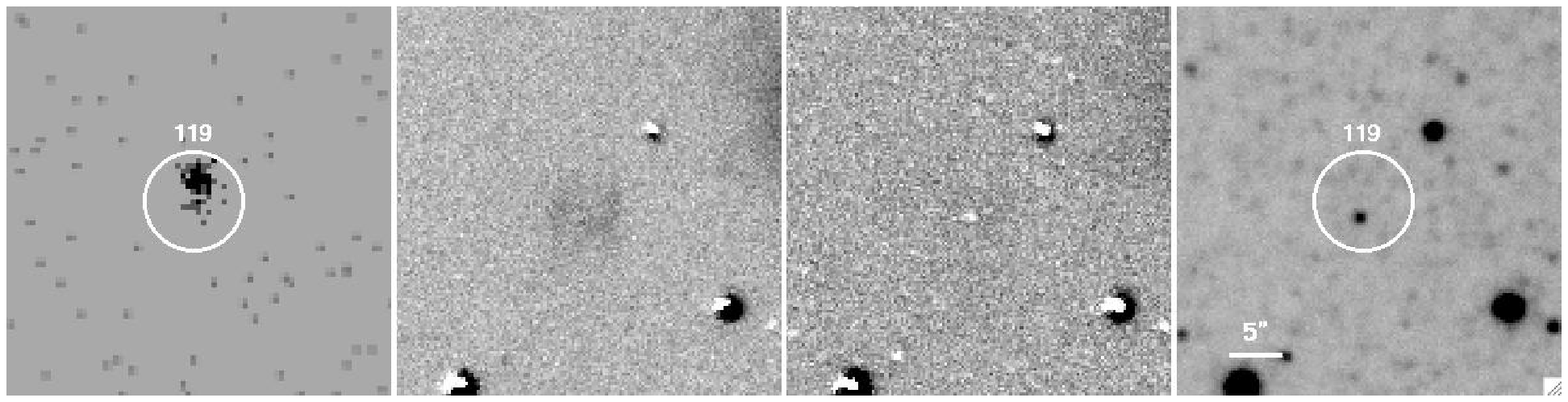}
\vspace{0.05in}
%\plotone{fig_atlas_GKL91.eps}
\plotone{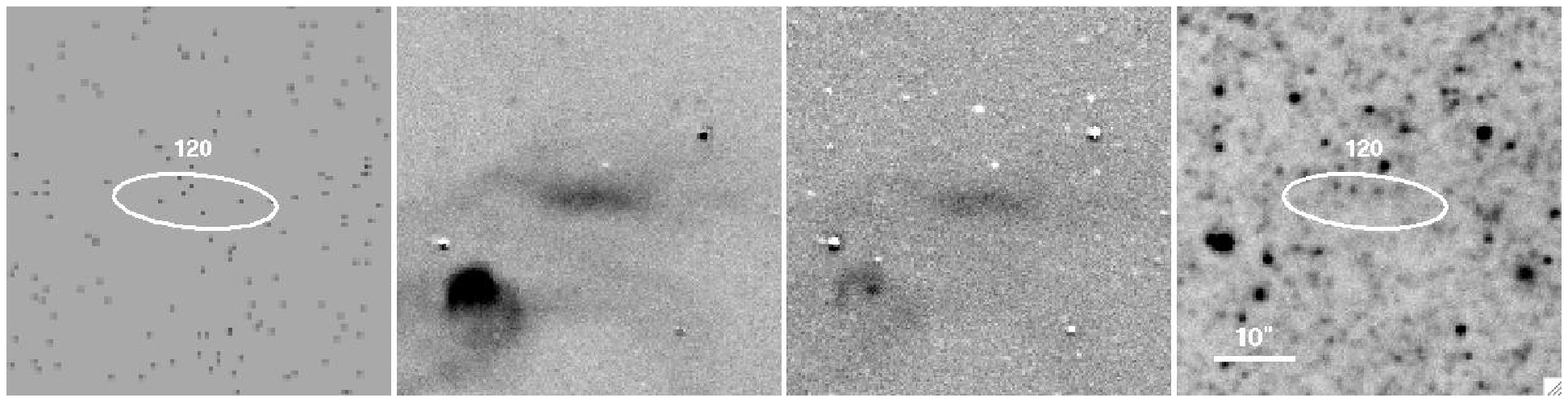}
\figcaption{Images from top to bottom of G98-89,  G98-90,  FL312,  G98-91.  The format is identical to Fig.\ \ref{fig_atlas01}. \label{fig_atlas29}   }
\end{figure}

\begin{figure}
%\plotone{fig_atlas_GKL92.eps}
\plotone{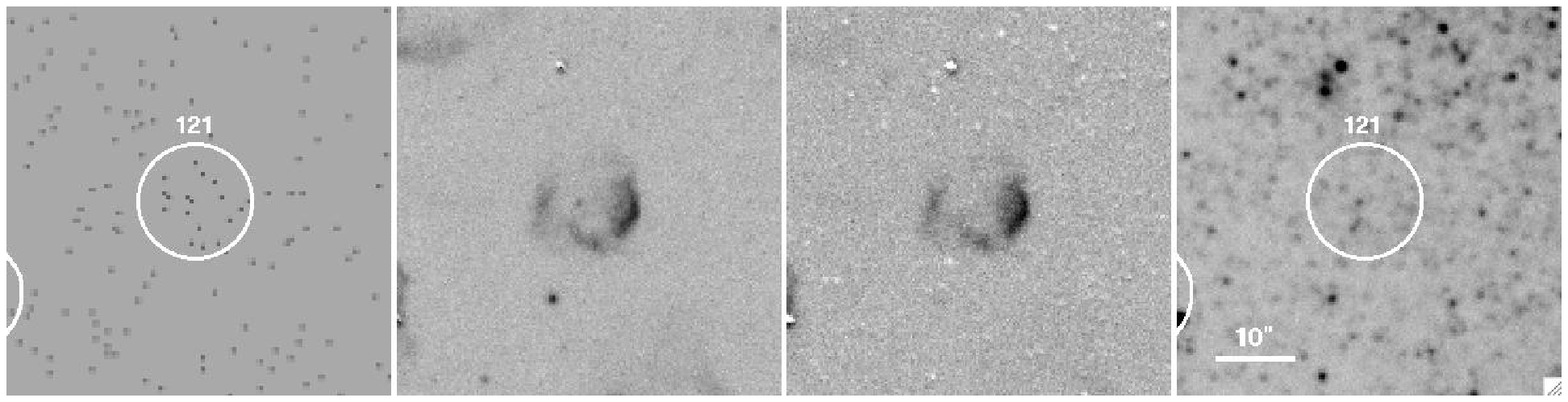}
\vspace{0.05in}
%\plotone{fig_atlas_EM34.eps}
\plotone{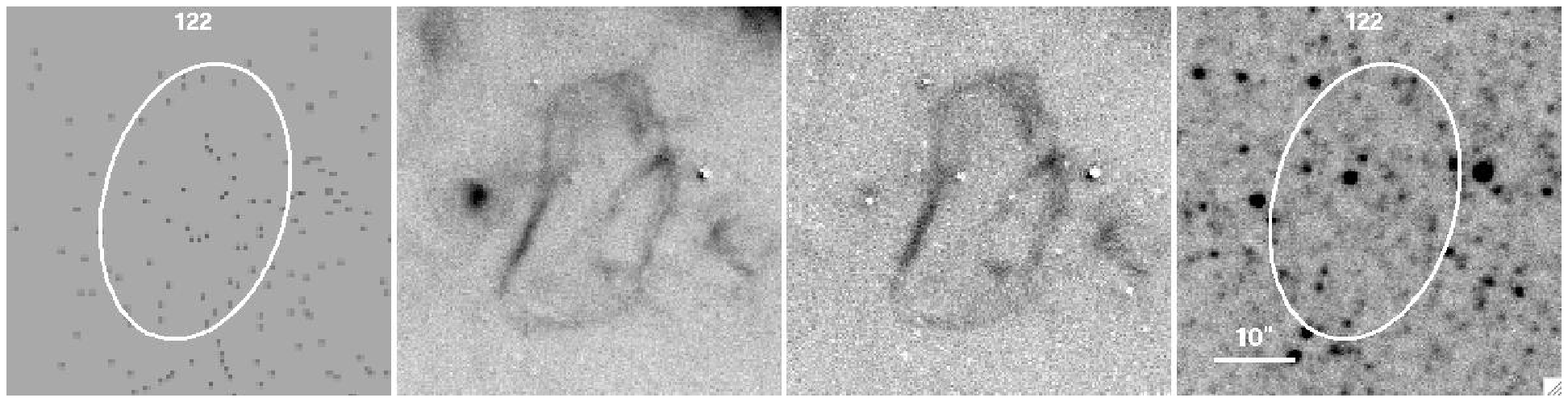}
\vspace{0.05in}
%\plotone{fig_atlas_GKL93.eps}
\plotone{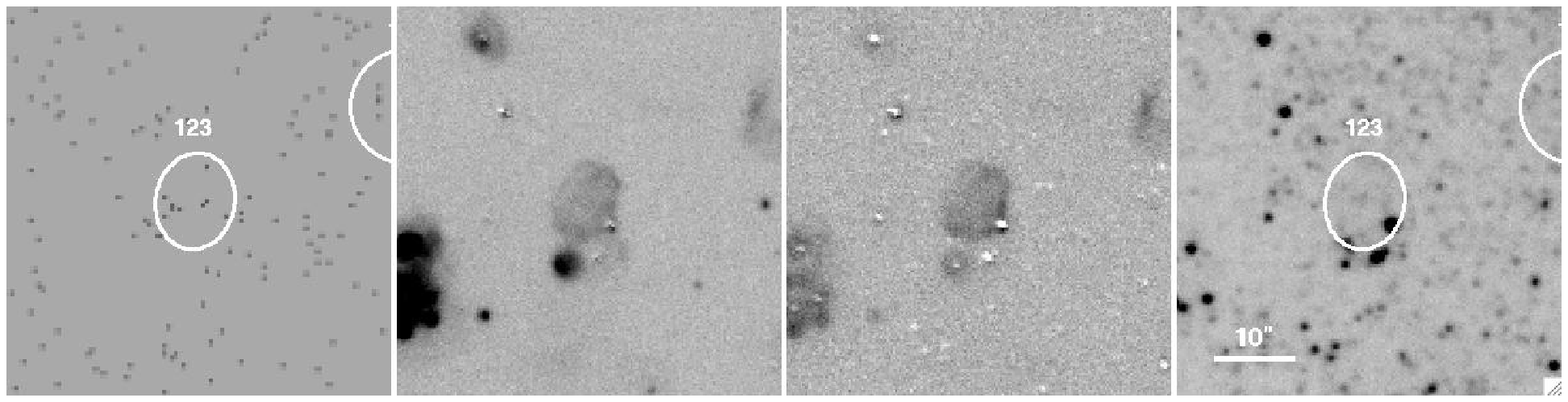}
\vspace{0.05in}
%\plotone{fig_atlas_GKL94.eps}
\plotone{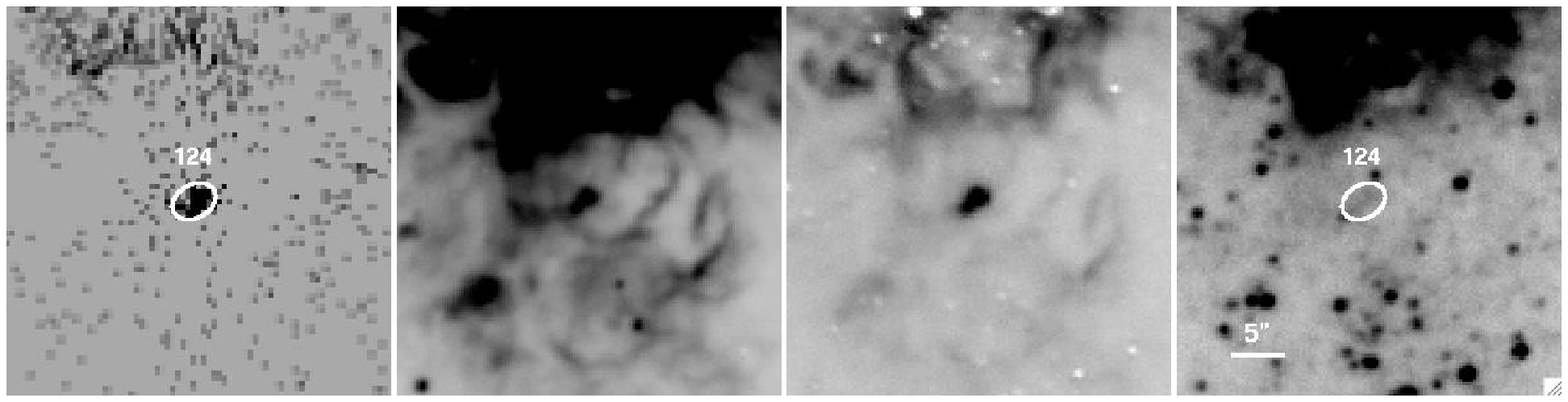}
\figcaption{Images from top to bottom of G98-92,  L10-122,  G98-93,  G98-94.  The format is identical to Fig.\ \ref{fig_atlas01}.  \label{fig_atlas30}  }
\end{figure}

\begin{figure}
%\plotone{fig_atlas_kip-E.eps}
\plotone{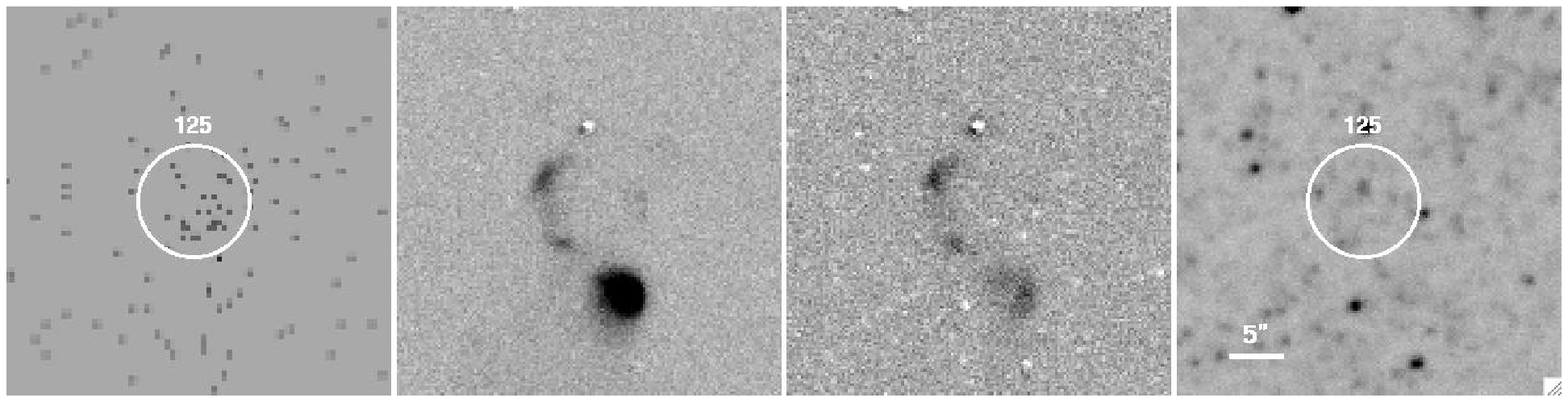}
\vspace{0.05in}
%\plotone{fig_atlas_GKL95.eps}
\plotone{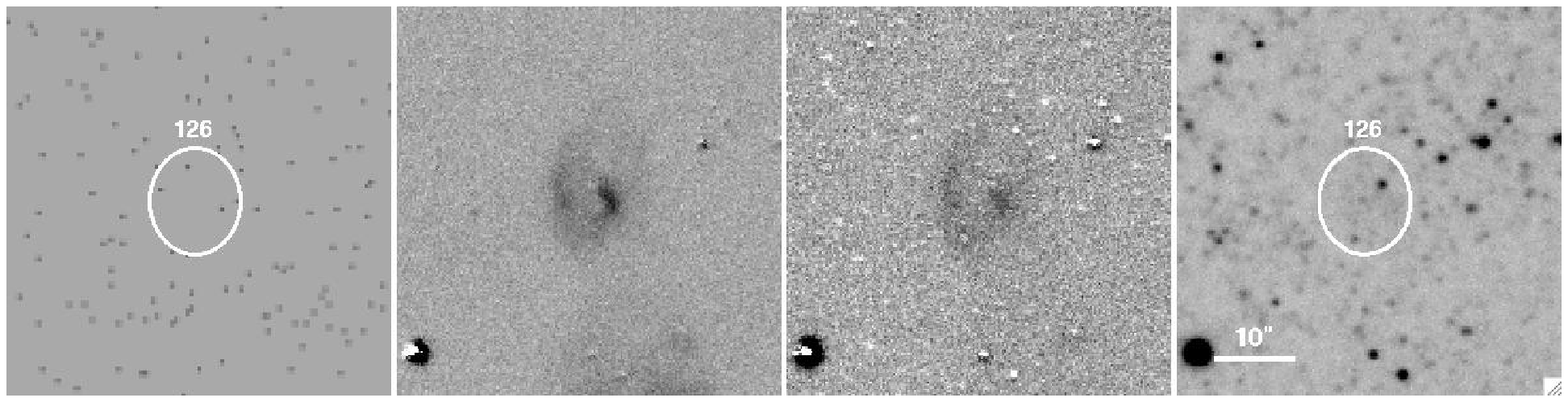}
\vspace{0.05in}
%\plotone{fig_atlas_GKL96.eps}
\plotone{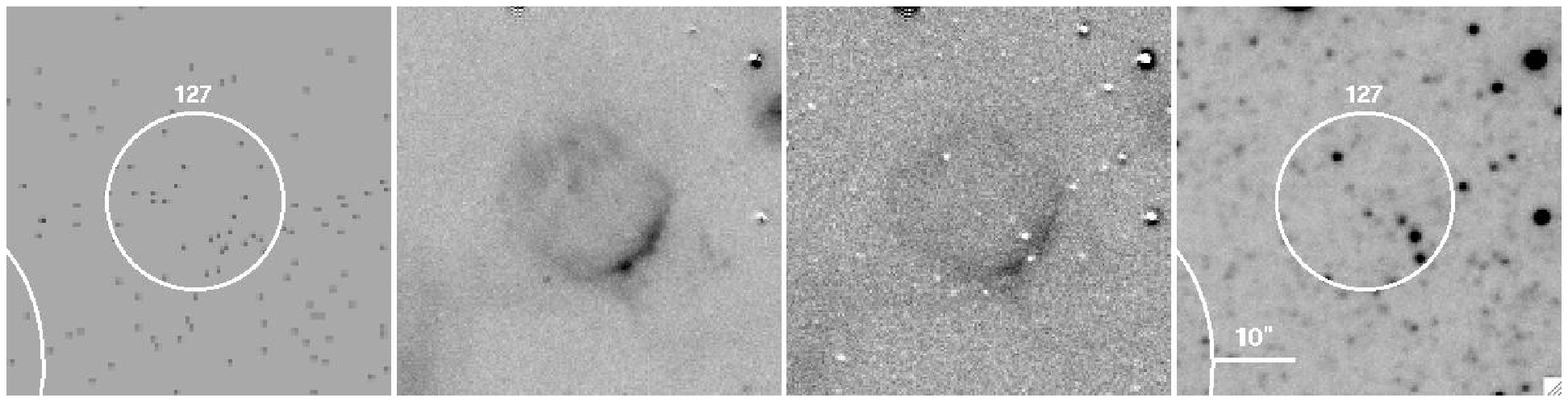}
\vspace{0.05in}
%\plotone{fig_atlas_GKL97AB.eps}
\plotone{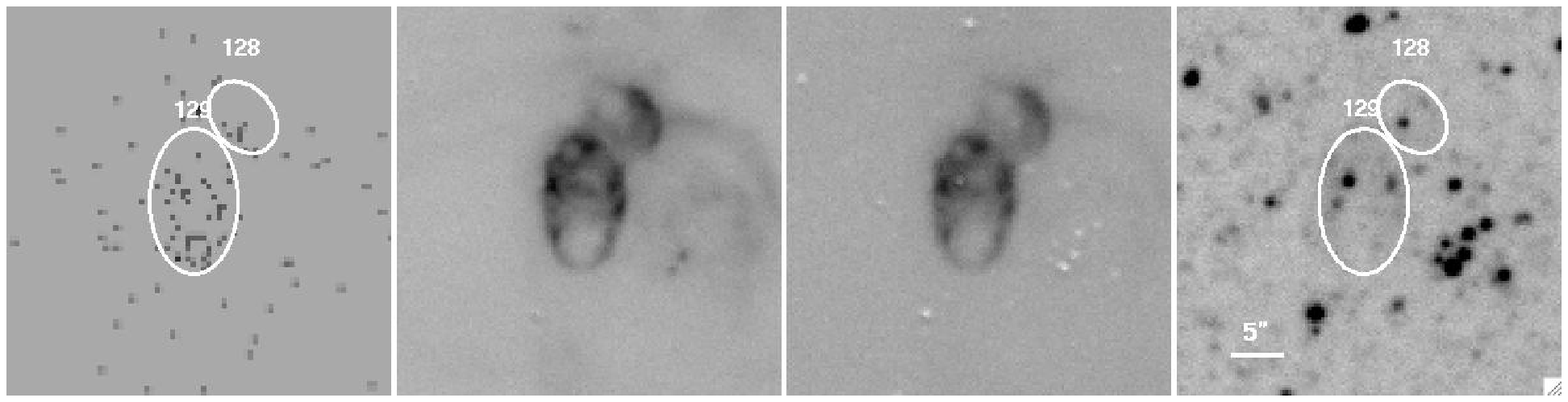}
\figcaption{Images from top to bottom of L10-125,  G98-95,  G98-96,  G98-97 A and B.  Note that G08-97 A=L10-129 is the larger of the two objects in the bottom panel. The format is identical to Fig.\ \ref{fig_atlas01}. \label{fig_atlas31}   }
\end{figure}

\begin{figure}
%\plotone{fig_atlas_WPB1.eps}
\plotone{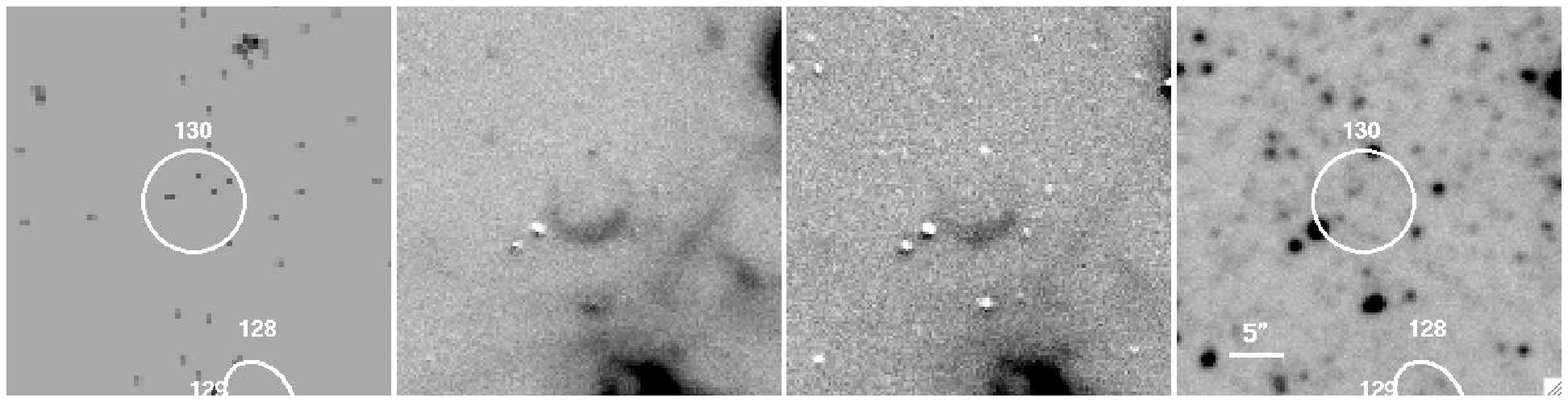}
\vspace{0.05in}
%\plotone{fig_atlas_EM36.eps}
\plotone{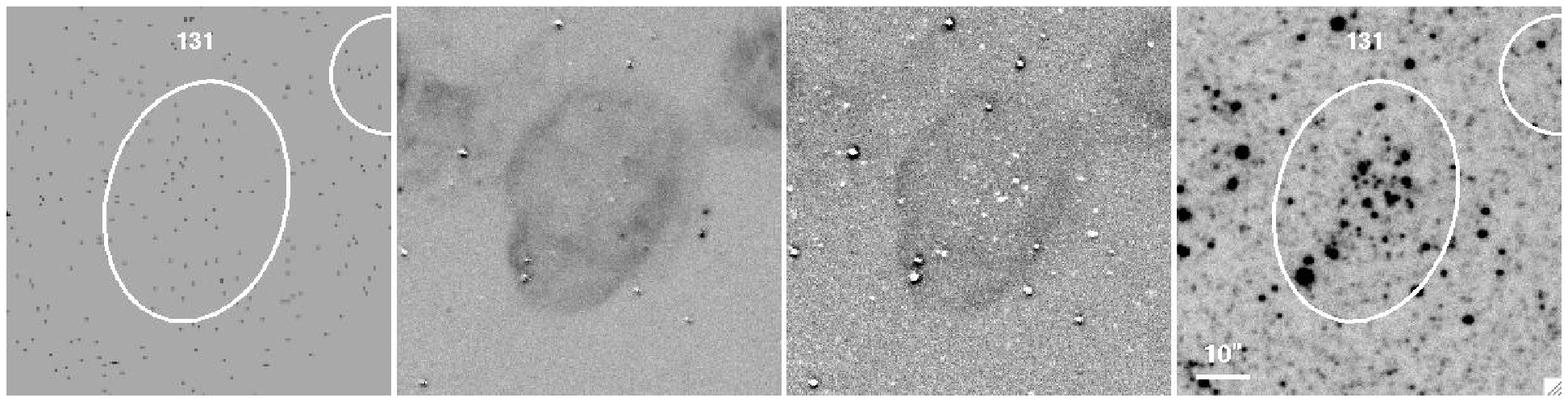}
\vspace{0.05in}
%\plotone{fig_atlas_GKL98.eps}
\plotone{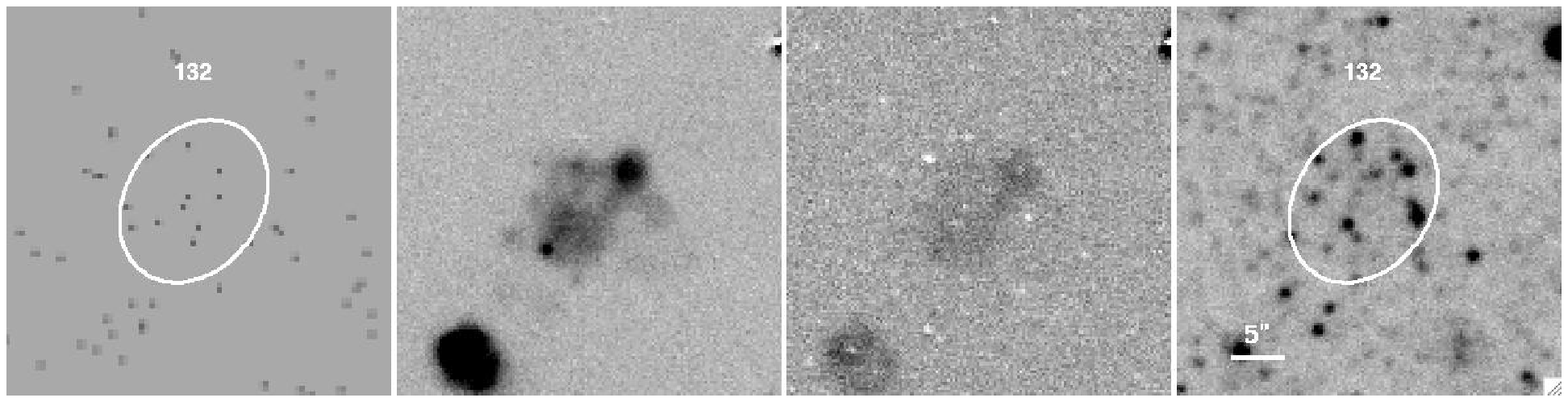}
\vspace{0.05in}
%\plotone{fig_atlas_EM52.eps}
\plotone{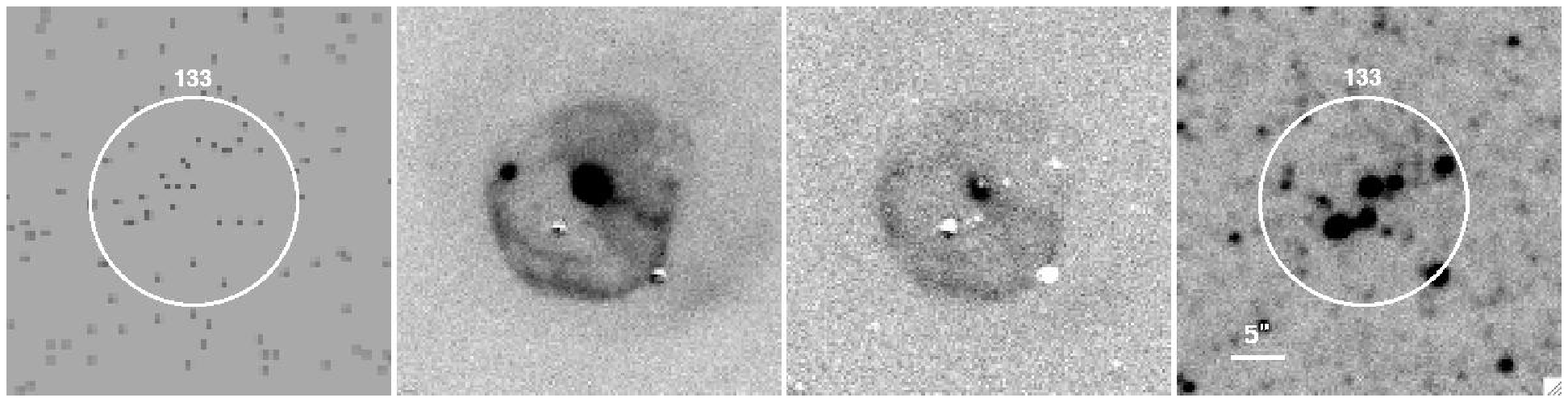}
\figcaption{Images from top to bottom of L10-130,  L10-131,  G98-98,  L10-133.  The format is identical to Fig.\ \ref{fig_atlas01}. \label{fig_atlas32}    }
\end{figure}

\begin{figure}
%\plotone{fig_atlas_EM48.eps}
\plotone{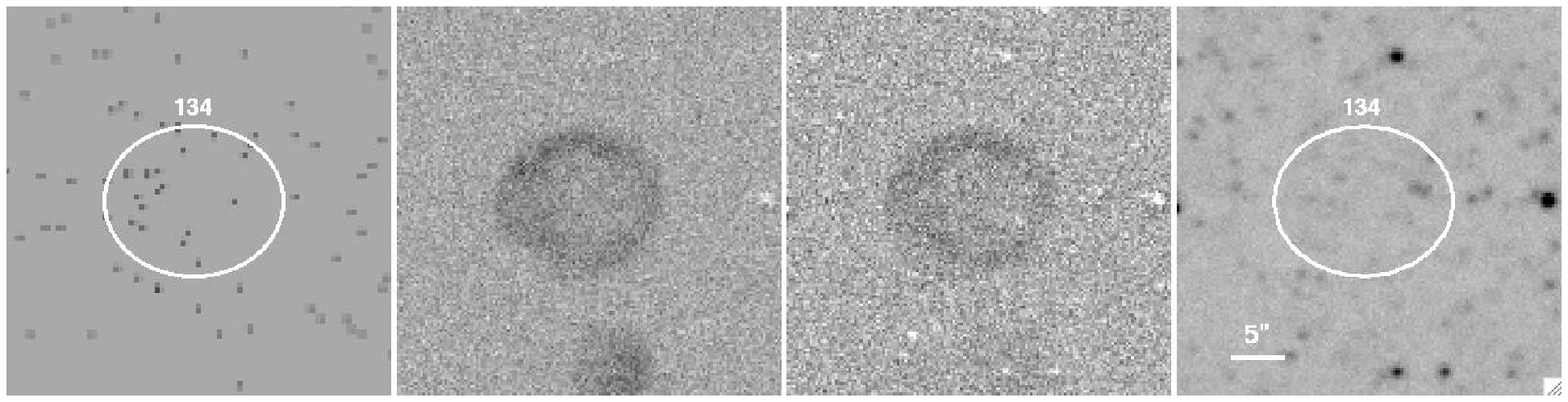}
\vspace{0.05in}
%\plotone{fig_atlas_EM24.eps}
\plotone{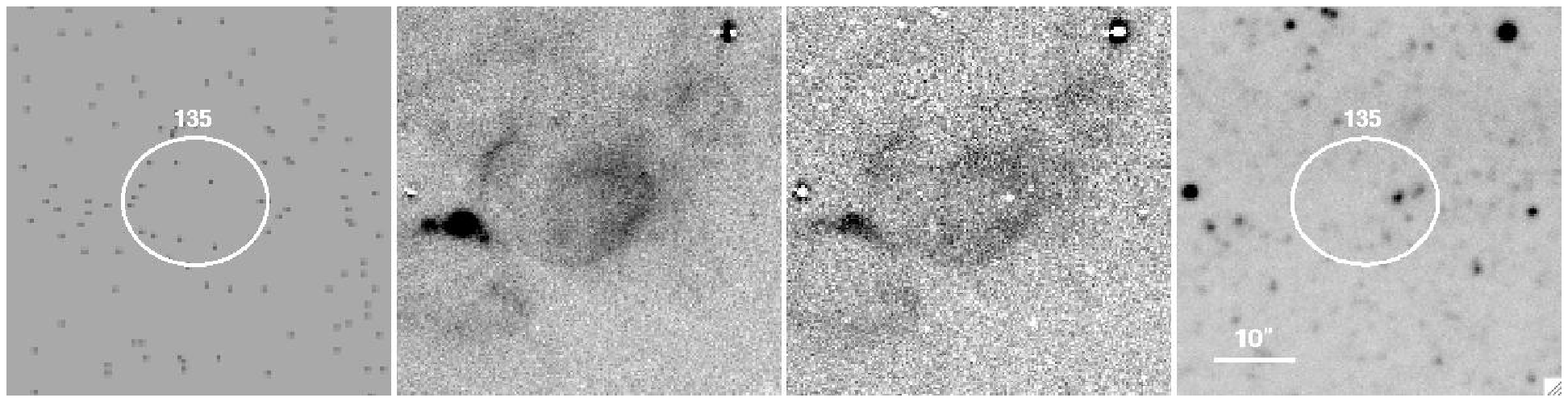}
\vspace{0.05in}
%\plotone{fig_atlas_EM26.eps}
\plotone{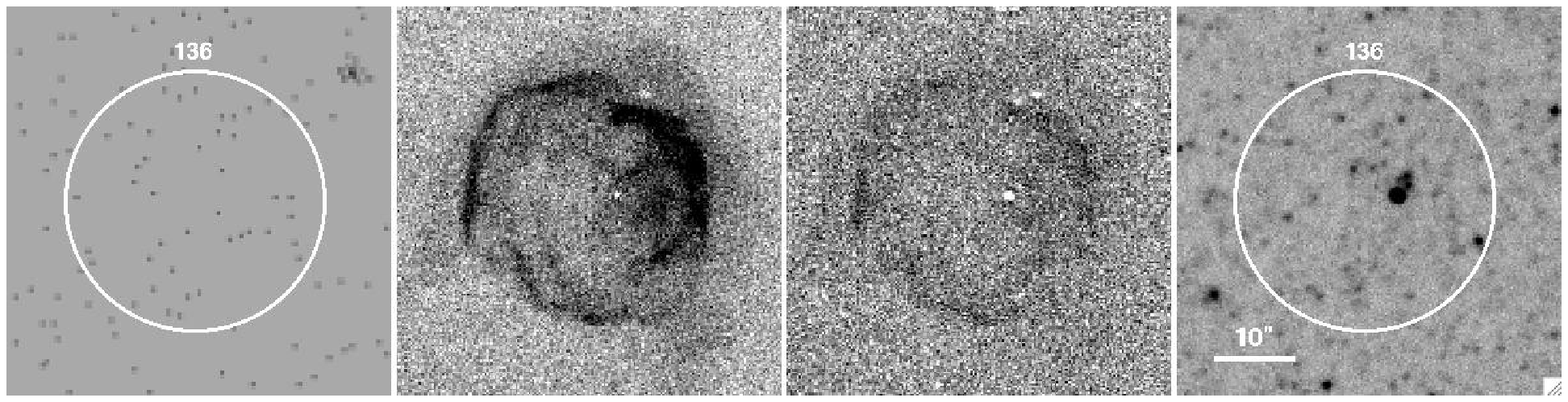}
\vspace{0.05in}
%\plotone{fig_atlas_EM25.eps}
\plotone{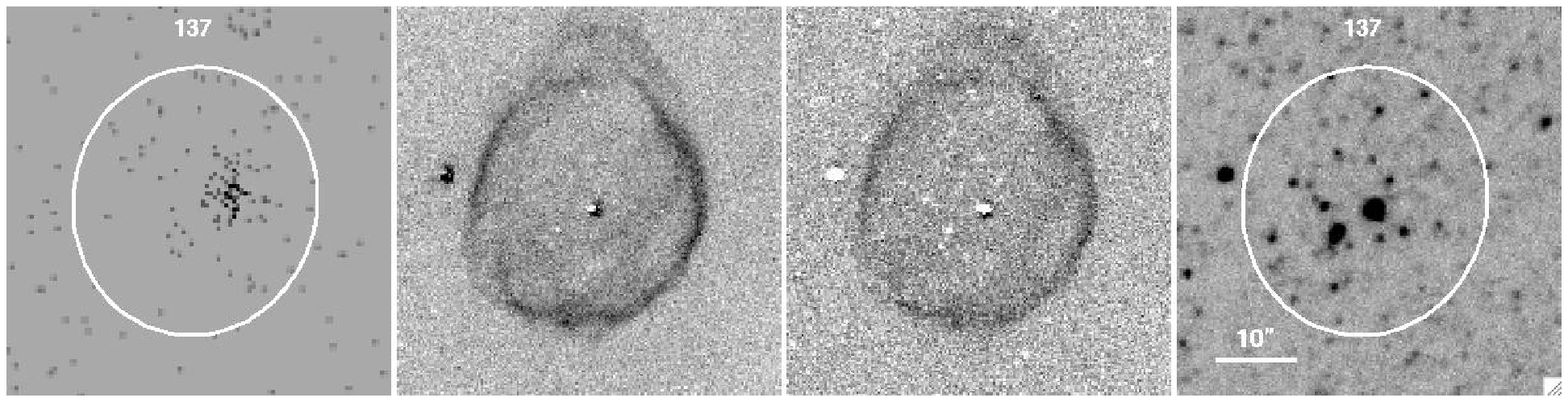}
\figcaption{Images from top to bottom of L10-134,  L10-135,  L10-136,  L10-137.  The format is identical to Fig.\ \ref{fig_atlas01}.\label{fig_atlas33}    }
\end{figure}

%\end{appendix}


\begin{thebibliography}{}









% Thermal and Nonthermal X-Rays from the Large Magellanic Cloud Superbubble 30 Doradus C  
%http://adsabs.harvard.edu/abs/2004ApJ...602..257B 
\bibitem[Bamba et al.(2004)]{bamba04} Bamba, A., Ueno, M., Nakajima, H., \& Koyama, K.\ 2004, \apj, 602, 257 

% A detailed observation of a LMC supernova remnant DEM L241 with XMM-Newton  
%http://adsabs.harvard.edu/abs/2006A%26A...450..585B 
\bibitem[Bamba et al.(2006)]{bamba06} Bamba, A., Ueno, M., Nakajima, H., Mori, K., \& Koyama, K.\ 2006, \aap, 450, 585 

% Radio Images of 3C 58: Expansion and Motion of Its Wisp  
%http://adsabs.harvard.edu/abs/2006ApJ...645.1180B 
\bibitem[Bietenholz(2006)]{bietenholz06} Bietenholz, M.~F.\ 2006, \apj, 645, 1180 

% Long Slit Echelle Spectroscopy of Supernova Remnants in M33  
%http://adsabs.harvard.edu/abs/1988srim.conf..193B 
\bibitem[Blair et al.(1988)]{blair88} Blair, W.~P., Chu, Y.-H., 
\& Kennicutt, R.~C.\ 1988, IAU Colloq.~101: Supernova Remnants and the Interstellar Medium, ed. by R. S. Roger and T. L. Landecker (CUP: Cambridge),193 


% Resolved structure in M33 supernova remnants  
%http://adsabs.harvard.edu/abs/1993PASP..105..494B 
\bibitem[Blair \& Davidsen(1993)]{blair93} Blair, W.~P., \& Davidsen, A.~F.\ 1993, \pasp, 105, 494 

% Identification of Supernova Remnants in the Sculptor Group Galaxies NGC 300 and NGC 7793  
%http://adsabs.harvard.edu/abs/1997ApJS..108..261B 
\bibitem[Blair \& Long(1997)]{blair97} Blair, W.~P., \& Long, K.~S.\ 1997, \apjs, 108, 261 

% A detailed optical study of Kepler's supernova remnant  
%http://adsabs.harvard.edu/abs/1991ApJ...366..484B 
\bibitem[Blair et al.(1991)]{blair91} Blair, W.~P., Long, K.~S., \& Vancura, O.\ 1991, \apj, 366, 484 

\bibitem[Borkowski et al.(2001)]{borkowski2001} Borkowski, K.~J., 
Lyerly, W.~J., \& Reynolds, S.~P.\ 2001, \apj, 548, 820 

% An atlas of supernova remnant candidates in Messier 31  
%http://adsabs.harvard.edu/abs/1993A%26AS...98..327B 
\bibitem[Braun \& Walterbos(1993)]{braun93} Braun, R., \& Walterbos, R.~A.~M.\ 1993, \aaps, 98, 327 


\bibitem[Broos et al.(2002)]{broos02} Broos, P. S., Townsley, L. K., Getman,
K., \& Bauer, F. E.\ 2002, ACIS Extract, An ACIS Point Source Extraction
Package (University Park: The Pennsylvania State Univ.)
\url{http://www.astro.psu.edu/xray/docs/TARA/ae\_users\_guide.html}

% X-rays from superbubbles in the Large Magellanic Cloud  
%http://adsabs.harvard.edu/abs/1990ApJ...365..510C 
\bibitem[Chu \& Mac Low(1990)]{chu90} Chu, Y.-H., \& Mac Low, M.-M.\ 1990, \apj, 365, 510 

\bibitem[Cox(1972)]{cox1972} Cox, D.~P.\ 1972, \apj, 178, 159 


% Possible Detection of O VI from the Large Magellanic Cloud Superbubble N70  
%http://adsabs.harvard.edu/abs/2006ApJ...646..205D 
\bibitem[Danforth \& Blair(2006)]{danforth06} Danforth, C.~W., \& Blair, W.~P.\ 2006, \apj, 646, 205 

% Excitation rate coefficients and line ratios for the optical and ultraviolet transitions in S II  
%http://adsabs.harvard.edu/abs/1993ApJS...88..329C 
\bibitem[Cai \& Pradhan(1993)]{cai93} Cai, W., \& Pradhan, A.~K.\ 1993, \apjs, 88, 329 

\bibitem[Chu \& Kennicutt (1988)]{chuken88} Chu, Y.-H., \& Kennicutt, R.~c.\ 1988, \aj, 95, 1111 

% H I in the Galaxy  
%http://adsabs.harvard.edu/abs/1990ARA%26A..28..215D 
\bibitem[Dickey \& Lockman(1990)]{dickey90} Dickey, J.~M., \& Lockman, F.~J.\ 1990, \araa, 28, 215 


% Supernova remnants in M33  
%http://adsabs.harvard.edu/abs/1978A%26A....63...63D 
% First detection
\bibitem[D'Odorico et al.(1978)]{dodorico78} D'Odorico, S., Benvenuti, P., \& Sabbadin, F.\ 1978, \aap, 63, 63 


% This list contains 17 candidates in 1980 and is the list that was mostly used until our CCD survey
% A catalogue of supernova remnant candidates in nearby galaxies  
%http://adsabs.harvard.edu/abs/1980A%26AS...40...67D 
\bibitem[D'Odorico et al.(1980)]{dodorico80} D'Odorico, S., Dopita, M.~A., \& Benvenuti, P.\ 1980, \aaps, 40, 67 

% The evolution of the line intensity ratios for supernova remnants in the Galaxy and in the Large Magellanic Cloud  
%http://adsabs.harvard.edu/abs/1976A%26A....53..443D 
\bibitem[D'Odorico \& Sabbadin(1976)]{dodorico76} D'Odorico, S., \& Sabbadin, F.\ 1976, \aap, 53, 443 

% The VLA-WSRT survey of M33 - Statistical properties of a sample of optically selected supernova remnants  
%http://adsabs.harvard.edu/abs/1993A%26AS...99..217D 
\bibitem[Duric et 
al.(1993)]{duric93} Duric, N., Viallefond, F., Goss, W.~M., \& van der Hulst, J.~M.\ 1993, \aaps, 99, 217 

% Cas A reddening
\bibitem[Eriksen et al.(2009)]{eriksen09} Eriksen, K.~A., Arnett, D., McCarthy, D.~W., \&Young, P.\ 2009, \apj, 697, 29

% Hectospec, the MMT's 300 Optical Fiber-Fed Spectrograph  
%http://adsabs.harvard.edu/abs/2005PASP..117.1411F 
\bibitem[Fabricant et al.(2005)]{fabricant05} Fabricant, D., et al.\ 2005, \pasp, 117, 1411 


% The Crab Nebula. I - Spectrophotometry of the filaments  
%http://adsabs.harvard.edu/abs/1982ApJ...258....1F 
\bibitem[Fesen \& Kirshner(1982)]{fesen82} Fesen, R.~A., \& Kirshner, R.~P.\ 1982, \apj, 258, 1 

% Hubble Space Telescope WFPC2 Imaging of Cassiopeia A  
%http://adsabs.harvard.edu/abs/2001AJ....122.2644F 
\bibitem[Fesen et al.(2001)]{fesen01} Fesen, R.~A., Morse, J.~A., Chevalier, R.~A., Borkowski, K.~J., Gerardy, C.~L., 
Lawrence, S.~S., \& van den Bergh, S.\ 2001, \aj, 122, 2644 


% Optical Imaging and Spectroscopy of the Galactic Supernova Remnant 3C 58 (G130.7+3.1)  
%http://adsabs.harvard.edu/abs/2008ApJS..174..379F 
\bibitem[Fesen et al.(2008)]{fesen08} Fesen, R., Rudie, G., Hurford, A., \& Soto, A.\ 2008, \apjs, 174, 379 






% New XMM-Newton observations of supernova remnants in the Small Magellanic Cloud  
%http://adsabs.harvard.edu/abs/2008A%26A...485...63F 
\bibitem[Filipovi{\'c} et al.(2008)]{filipovic08} Filipovi{\'c}, M.~D., et al.\ 2008, \aap, 485, 63 


% Final Results from the Hubble Space Telescope Key Project to Measure the Hubble Constant  
%http://adsabs.harvard.edu/abs/2001ApJ...553...47F 
\bibitem[Freedman et al.(2001)]{freedman01} Freedman, W.~L., et al.\ 2001, \apj, 553, 47 


% Chandra ACIS Survey of M33 (ChASeM33): X-Ray Imaging Spectroscopy of M33SNR 21, the Brightest X-Ray Supernova Remnant in M33  
%http://adsabs.harvard.edu/abs/2007ApJ...663..234G 
\bibitem[Gaetz et al.(2007)]{gaetz07} Gaetz, T.~J., et al.\ 2007, \apj, 663, 234 



% Revealing the Supernova Remnant Population of M33 with Chandra  
%http://adsabs.harvard.edu/abs/2005AJ....130..539G 
\bibitem[Ghavamian et al.(2005)]{ghavamian05} Ghavamian, P., Blair, W.~P., Long, K.~S., Sasaki, M., Gaetz, T.~J., 
\& Plucinsky, P.~P.\ 2005, \aj, 130, 539 


% A New Sample of Radio-selected and Optically Confirmed Supernova Remnants in M33  
%http://adsabs.harvard.edu/abs/1999ApJS..120..247G 
\bibitem[Gordon et al.(1999)]{gordon99} Gordon, S.~M., Duric, N., Kirshner, R.~P., Goss, W.~M., 
\& Viallefond, F.\ 1999, \apjs, 120, 247 

% The Discovery of a Supernova Remnant Embedded in a Giant H II Region of M33  
%http://adsabs.harvard.edu/abs/1993ApJ...418..743G 
\bibitem[Gordon et al.(1993)]{gordon93} Gordon, S.~M., Kirshner, R.~P., Duric, N., \& Long, K.~S.\ 1993, \apj, 418, 743 


% A New Optical Sample of Supernova Remnants in M33  
%http://adsabs.harvard.edu/abs/1998ApJS..117...89G 
\bibitem[Gordon et al.(1998)]{gordon98} Gordon, S.~M., Kirshner, R.~P., Long, K.~S., Blair, W.~P., Duric, N., 
\& Smith, R.~C.\ 1998, \apjs, 117, 89 



% Radio observations of three supernova remnants in M33  
%http://adsabs.harvard.edu/abs/1980MNRAS.193..901G 
\bibitem[Goss et al.(1980)]{goss80} Goss, W.~M., Ekers, R.~D., Danziger, I.~J., 
\& Israel, F.~P.\ 1980, \mnras, 193, 901 

% A ROSAT PSPC catalogue of X-ray sources in the LMC region  
%http://adsabs.harvard.edu/abs/1999A%26AS..139..277H 
\bibitem[Haberl \& Pietsch(1999)]{haberl99} Haberl, F., \& Pietsch, W.\ 1999, \aaps, 139, 277 



% X-ray line emission from supernova remnants. I - Models for adiabatic remnants  
%http://adsabs.harvard.edu/abs/1983ApJS...51..115H 
\bibitem[Hamilton et al.(1983)]{hamilton83} Hamilton, A.~J.~S., Chevalier, R.~A., \& Sarazin, C.~L.\ 1983, \apjs, 51, 115 

% Ejecta Detection in Middle-Aged Large Magellanic Cloud Supernova Remnants 0548-70.4 and 0534-69.9  
%http://adsabs.harvard.edu/abs/2003ApJ...593..370H 
\bibitem[Hendrick et al.(2003)]{hendrick} Hendrick, S.~P., Borkowski, K.~J., \& Reynolds, S.~P.\ 2003, \apj, 593, 370 

% The form of abundance gradients in three nearby spiral galaxies: M33, M81 and M101  
%http://adsabs.harvard.edu/abs/1995ApJ...438..170H 
\bibitem[Henry \& Howard(1995)]{henry95} Henry, R.~B.~C., \& Howard, J.~W.\ 1995, \apj, 438, 170 

% Iron-rich Ejecta in the Supernova Remnant DEM L71  
%http://adsabs.harvard.edu/abs/2003ApJ...582L..95H 
\bibitem[Hughes et al.(2003)]{hughes03} Hughes, J.~P., Ghavamian, P., Rakowski, C.~E., 
\& Slane, P.~O.\ 2003, \apjl, 582, L95 
% DS9
\bibitem[Joye 
\& Mandel(2003)]{joye03} Joye, W.~A., \& Mandel, E.\ 2003, Astronomical Data Analysis Software and Systems XII, 295, 489 

% Large scale structure of the ionized gas in the magellanic clouds  
%http://adsabs.harvard.edu/abs/1995AJ....109..594K 
\bibitem[Kennicutt et al.(1995)]{kennicutt95} Kennicutt, R.~C., Jr., Bresolin, F., Bomans, D.~J., Bothun, G.~D., 
\& Thompson, I.~B.\ 1995, \aj, 109, 594 

\bibitem[Kirshner et al.(1987)]{kirshner87} Kirshner, R., Winkler, P.~F., \& Chevalier, R.~A.\ 1987, \apjl, 315, L135 


% A High-Resolution Study of Nonthermal Radio and X-Ray Emission from Supernova Remnant G347.3-0.5  
%http://adsabs.harvard.edu/abs/2004ApJ...602..271L 
\bibitem[Lazendic et al.(2004)]{lazendic04} Lazendic, J.~S., Slane, P.~O., Gaensler, B.~M., Reynolds, S.~P., Plucinsky, 
P.~P., \& Hughes, J.~P.\ 2004, \apj, 602, 271 

% Spectrophotometry in the galactic supernova remnants RCW 86, 103 and Kepler  
%http://adsabs.harvard.edu/abs/1983MNRAS.204..273L 
\bibitem[Leibowitz \& Danziger(1983)]{leibowitz83} Leibowitz, E.~M., \& Danziger, I.~J.\ 1983, \mnras, 204, 273 

\bibitem[Levenson et al.(1995)]{levenson95} Levenson, N.~A., Kirshner, R.~P., Blair, W.~P. \&
	Winkler, P.~F.\ 1995, \aj, 110, 739

% An atlas of confirmed and candidate supernova remnants in M33  
%http://adsabs.harvard.edu/abs/1990ApJS...72...61L 
\bibitem[Long et al.(1990)]{long90} Long, K.~S., Blair, W.~P., Kirshner, R.~P., \& Winkler, P.~F.\ 1990, \apjs, 72, 61 

% A Deep X-Ray Image of M33  
%http://adsabs.harvard.edu/abs/1996ApJ...466..750L 
\bibitem[Long et al.(1996)]{long96} Long, K.~S., Charles, P.~A., Blair, W.~P., \& Gordon, S.~M.\ 1996, \apj, 466, 750 



% Observations of the X-ray sources in the nearby SC galaxy M33  
%http://adsabs.harvard.edu/abs/1981ApJ...246L..61L 
\bibitem[Long et al.(1981)]{long81} Long, K.~S., D'Odorico, S., Charles, P.~A., \& Dopita, M.~A.\ 1981, \apjl, 246, L61 


% Supernova remnants in the Large Magellanic Cloud  
%http://adsabs.harvard.edu/abs/1979ApJ...234L..77L 
\bibitem[Long \& Helfand(1979)]{long79} Long, K.~S., \& Helfand, D.~J.\ 1979, \apjl, 234, L77 

% Superbubbles in disk galaxies  
%http://adsabs.harvard.edu/abs/1988ApJ...324..776M 
\bibitem[Mac Low \& McCray(1988)]{maclow88} Mac Low, M.-M., \& McCray, R.\ 1988, \apj, 324, 776 

% X-ray observations of M33 with the high resolution imager on the Einstein Observatory  
%http://adsabs.harvard.edu/abs/1983ApJ...275..571M 
\bibitem[Markert \& Rallis(1983)]{markert83} Markert, T.~H., \& Rallis, A.~D.\ 1983, \apj, 275, 571 

% New supernova remnant candidates in M 31.  
%http://adsabs.harvard.edu/abs/1995A%26AS..114..215M 
\bibitem[Magnier et 
al.(1995)]{magnier95} Magnier, E.~A., Prins, S., van Paradijs, J., Lewin, W.~H.~G., Supper, R., Hasinger, G., Pietsch, W., \& Truemper, J.\ 1995, \aaps, 114, 215 



% Spectrophotometric standards  
%http://adsabs.harvard.edu/abs/1988ApJ...328..315M 
\bibitem[Massey et al.(1988)]{massey88} Massey, P., Strobel, K., Barnes, J.~V., \& Anderson, E..\ 1988, \apj, 328, 315 

%A Survey of Local Group Galaxies Currently Forming Stars. III. A Search for Luminous Blue Variables and Other H? Emission-Line Stars
%http://adsabs.harvard.edu/abs/2007AJ....134.2474M
\bibitem[Massey et al.(2007)]{massey07} Massey, P., McNeill, R.~T., Olsen, K.~A.~G., Hodge, P.~W.,  Blaha, C., Jacoby, G.~H., Smith, R.~C., \& Strong, S.~B.\  2007, \aj, 134, 2474 

% A Survey of Local Group Galaxies Currently Forming Stars. I. UBVRI Photometry of Stars in M31 and M33  
%http://adsabs.harvard.edu/abs/2006AJ....131.2478M 
\bibitem[Massey et al.(2006)]{massey06} Massey, P., Olsen, K.~A.~G., Hodge, P.~W., Strong, S.~B., Jacoby, G.~H., 
Schlingman, W., \& Smith, R.~C.\ 2006, \aj, 131, 2478 

% Supernova remnants in the Large Magellanic Cloud.  
%http://adsabs.harvard.edu/abs/1973ApJ...180..725M 
\bibitem[Mathewson \& Clarke(1973)]{mathewson73} Mathewson, D.~S., \& Clarke, J.~N.\ 1973, \apj, 180, 725 



% Optically Identified Supernova Remnants in the Nearby Spiral Galaxies: NGC 5204, NGC 5585, NGC 6946, M81, and M101  
%http://adsabs.harvard.edu/abs/1997ApJS..112...49M 
\bibitem[Matonick \& Fesen(1997)]{matonick97} Matonick, D.~M., \& Fesen, R.~A.\ 1997, \apjs, 112, 49 


% An Optical Search for Supernova Remnants in NGC 2403  
%http://adsabs.harvard.edu/abs/1997ApJS..113..333M 
\bibitem[Matonick et al.(1997)]{matonick97b} Matonick, D.~M., Fesen, R.~A., Blair, W.~P., 
\& Long, K.~S.\ 1997, \apjs, 113, 333 

% The violent interstellar medium  
%http://adsabs.harvard.edu/abs/1979ARA%26A..17..213M 
\bibitem[McCray \& Snow(1979)]{mccray79} McCray, R., \& Snow, T.~P., Jr.\ 1979, \araa, 17, 213 

\bibitem[McNeil(2006)]{mcneil06} McNeil, E.K. 2006, unpublished B.A. thesis, Middlebury College


% An XMM-Newton survey of the Local Group galaxy M 33 - variability of the detected sources  
%http://adsabs.harvard.edu/abs/2006A%26A...448.1247M 
\bibitem[Misanovic et 
al.(2006)]{misanovic06} Misanovic, Z., Pietsch, W., Haberl, F., Ehle, M., Hatzidimitriou, D., \& Trinchieri, G.\ 2006, \aap, 448, 1247 

% Neutral hydrogen and spiral structure in M33  
%http://adsabs.harvard.edu/abs/1980MNRAS.190..689N 
\bibitem[Newton(1980)]{newton80} Newton, K.\ 1980, \mnras, 190, 689 

% Nucleosynthesis in type II supernovae  
%http://adsabs.harvard.edu/abs/1997NuPhA.616...79N 
\bibitem[Nomoto et al.(1997)]{nomoto97} Nomoto, K., Hashimoto, M., Tsujimoto, T., Thielemann, F.-K., Kishimoto, N., Kubo, 
Y., \& Nakasato, N.\ 1997, Nuclear Physics A, 616, 79 

% The H I Environment of Three Superbubbles in the Large Magellanic Cloud  
%http://adsabs.harvard.edu/abs/2002AJ....123..255O 
\bibitem[Oey et al.(2002)]{oey02} Oey, M.~S., Groves, B., Staveley-Smith, L., \& Smith, R.~C.\ 2002, \aj, 123, 255 



% A Search for Chandra-detected X-Ray Counterparts to Optically Identified and Candidate Radio Supernova Remnants in Five Nearby Face-on Spiral Galaxies  
%http://adsabs.harvard.edu/abs/2007AJ....133.1361P 
\bibitem[Pannuti et al.(2007)]{pannuti07} Pannuti, T.~G., Schlegel, E.~M., \& Lacey, C.~K.\ 2007, \aj, 133, 1361 

% SNR N49 in the LMC: X-ray emission from multi-phase shock and neutron star  
%http://adsabs.harvard.edu/abs/2004AdSpR..33..409P 
\bibitem[Park et al.(2004)]{park04} Park, S., Burrows, D.~N., Garmire, G.~P., Nousek, J.~A., Hughes, J.~P., 
\& Williams, R.~M.\ 2004, Advances in Space Research, 33, 409 

% 0103-72.6: A New Oxygen-rich Supernova Remnant in the Small Magellanic Cloud  
%http://adsabs.harvard.edu/abs/2003ApJ...598L..95P 
\bibitem[Park et al. (2003)]{park03}  
Park, S.  Hughes, J. P., Burrows, D. N., Slane, P. O., Nousek, J. A., \& Garmire, G. P.  2003, ApJ, 598, 95

% Detection of Magnesium-rich Ejecta in the Middle-aged Supernova Remnant N49B  
%http://adsabs.harvard.edu/abs/2003ApJ...592L..41P 
\bibitem[Park et al.(2003b)]{park03b} Park, S., Hughes, J.~P., Slane, P.~O., Burrows, D.~N., Warren, J.~S., Garmire, 
G.~P., \& Nousek, J.~A.\ 2003, \apjl, 592, L41 



%Multi-frequency study of extragalactic supernova remnants and H II regions. Sculptor group Sd galaxy NGC 300  
%http://adsabs.harvard.edu/abs/2004A%26A...425..443P 
\bibitem[Payne et 
al.(2004)]{payne04} Payne, J.~L., Filipovi{\'c}, M.~D., Pannuti, T.~G., Jones, P.~A., Duric, N., White, G.~L., \& Carpano, S.\ 2004, \aap, 425, 443 



% Long-slit optical spectroscopy of Large Magellanic Cloud radio supernova remnants  
%http://adsabs.harvard.edu/abs/2008MNRAS.383.1175P 
\bibitem[Payne et al.(2008)]{payne08} Payne, J.~L., White, G.~L., \& Filipovi{\'c}, M.~D.\ 2008, \mnras, 383, 1175 



% ROSAT observation of a new supernova remnant in the constellation Scorpius.  
%http://adsabs.harvard.edu/abs/1996rftu.proc..267P 
\bibitem[Pfeffermann \& Aschenbach(1996)]{pfeffermann96} Pfeffermann, E., \& Aschenbach, B.\ 1996, Roentgenstrahlung from the Universe, 267 

% An XMM-Newton survey of M 31  
%http://adsabs.harvard.edu/abs/2005A%26A...434..483P 
\bibitem[Pietsch et al.(2005)]{pietsch05} Pietsch, W., Freyberg, M., \& Haberl, F.\ 2005, \aap, 434, 483 


% XMM-Newton survey of the Local Group galaxy <ASTROBJ>M 33</ASTROBJ>  
%http://adsabs.harvard.edu/abs/2004A%26A...426...11P 
\bibitem[Pietsch et 
al.(2004)]{pietsch04} Pietsch, W., Misanovic, Z., Haberl, F., Hatzidimitriou, D., Ehle, M., \& Trinchieri, G.\ 2004, \aap, 426, 11 



% Chandra ACIS Survey of M33 (ChASeM33): A First Look  
%http://adsabs.harvard.edu/abs/2008ApJS..174..366P 
\bibitem[Plucinsky et al.(2008)]{plucinsky08} Plucinsky, P.~P., et al.\ 2008, \apjs, 174, 366 

\bibitem[Rauscher et al.(2002)]{rauscher2002} Rauscher, T., Heger, 
A., Hoffman, R.~D., \& Woosley, S.~E.\ 2002, \apj, 576, 323 


% A Spitzer Space Telescope Infrared Survey of Supernova Remnants in the Inner Galaxy  
%http://adsabs.harvard.edu/abs/2006AJ....131.1479R 
\bibitem[Reach et al.(2006)]{reach06} Reach, W.~T., et al.\ 2006, \aj, 131, 1479 

%Cas A distance
\bibitem[Reed et al.(1995)]{reed95} Reed, J.~E., Hester, J.~J., Fabian, A.~C., \& Winkler, P.~F.\ 1995, \apj, 440, 706


% A radio search for Crab-like nebulae in M33  
%http://adsabs.harvard.edu/abs/1987ApJ...322..673R 
\bibitem[Reynolds \& Fix(1987)]{reynolds87} Reynolds, S.~P., \& Fix, J.~D.\ 1987, \apj, 322, 673 

%\bibitem[Saul et al.(2010)]{saul10} Saul, D. et al.\ 2010, in preparation

% X-ray survey of the Small Magellanic Cloud  
%http://adsabs.harvard.edu/abs/1981ApJ...243..736S 
\bibitem[Seward \& Mitchell(1981)]{seward81} Seward, F.~D., \& Mitchell, M.\ 1981, \apj, 243, 736 

% The Bell Laboratories H I survey  
%http://adsabs.harvard.edu/abs/1992ApJS...79...77S 
\bibitem[Stark et al.(1992)]{stark92} Stark, A.~A., Gammie, C.~F., Wilson, R.~W., Bally, J., Linke, R.~A., Heiles, C., 
\& Hurwitz, M.\ 1992, \apjs, 79, 77 

% The thermal and non-thermal gaseous halo of NGC 5775  
%http://adsabs.harvard.edu/abs/2000A%26A...364L..36T 
\bibitem[T{\"u}llmann et 
al.(2000)]{tuellmann00} T{\"u}llmann, R., Dettmar, R.-J., Soida, M., Urbanik, M., \& Rossa, J.\ 2000, \aap, 364, L36 

% The Chandra ACIS Survey of M33 (ChASeM33): Investigating the Hot Ionized Medium in NGC 604  
%http://adsabs.harvard.edu/abs/2008ApJ...685..919T 
\bibitem[T{\"u}llmann et al.(2008)]{tuellmann08} T{\"u}llmann, R., et al.\ 2008, \apj, 685, 919 

% Chandra ACIS Survey of M33 (ChASeM33): The Enigmatic X-Ray Emission from IC131  
%http://adsabs.harvard.edu/abs/2009ApJ...707.1361T 
\bibitem[T{\"u}llmann et al.(2009)]{tuellmann09} T{\"u}llmann, R., et al.\ 2009, \apj, 707, 1361 

% This is the point source paper
\bibitem[T{\"u}llmann et al.(2010)]{tuellmann10} T{\"u}llmann, R., et al.\ 2010, in preparation 

% Synoptic study of the SMC SNRs using XMM-Newton  
%http://adsabs.harvard.edu/abs/2004A%26A...421.1031V 
\bibitem[van der Heyden et 
al.(2004)]{vanderHeyden04} van der Heyden, K.~J., Bleeker, J.~A.~M., \& Kaastra, J.~S.\ 2004, \aap, 421, 1031 



% Star formation in M 33: Spitzer photometry of discrete sources  
%http://adsabs.harvard.edu/abs/2007A%26A...476.1161V 
\bibitem[Verley et al.(2007)]{verley07} Verley, S., Hunt, L.~K., Corbelli, E., \& Giovanardi, C.\ 2007, \aap, 476, 1161 

\bibitem[Verner et al.(1996)]{verner1996} Verner, D.~A., Ferland,
G.~J., Korista, K.~T., \& Yakovlev, D.~G.\ 1996, \apj, 465, 487

% An Overview of the Performance and Scientific Results from the Chandra X-Ray Observatory  
%http://adsabs.harvard.edu/abs/2002PASP..114....1W 
\bibitem[Weisskopf et al.(2002)]{weisskopf02} Weisskopf, M.~C., Brinkman, B., Canizares, C., Garmire, G., Murray, S.,  \& Van Speybroeck, L.~P.\ 2002, \pasp, 114, 1 

% Spitzer Sage Survey of the Large Magellanic Cloud. III. Star Formation and ~1000 New Candidate Young Stellar Objects  
%http://adsabs.harvard.edu/abs/2008AJ....136...18W 
\bibitem[Whitney et al.(2008)]{whitney08} Whitney, B.~A., et al.\ 2008, \aj, 136, 18 

% The Small Magellanic Cloud in the far infrared.  II. Global properties  
%http://adsabs.harvard.edu/abs/2004A%26A...414...69W 
\bibitem[Wilke et 
al.(2004)]{wilke04} Wilke, K., Klaas, U., Lemke, D., Mattila, K., Stickel, M., \& Haas, M.\ 2004, \aap, 414, 69 

% A Potential Supernova Remnant-X-Ray Binary Association in M31  
%http://adsabs.harvard.edu/abs/2005ApJ...634..365W 
\bibitem[Williams et al.(2005)]{williamsb05} Williams, B.~F., Barnard, R., Garcia, M.~R., Kolb, U., Osborne, J.~P., 
\& Kong, A.~K.~H.\ 2005, \apj, 634, 365 



% Supernova Remnants in the Magellanic Clouds. VI. The DEM L316 Supernova Remnants  
%http://adsabs.harvard.edu/abs/2005ApJ...635.1077W 
\bibitem[Williams \& Chu(2005)]{williams05} Williams, R.~M., \& Chu, Y.-H.\ 2005, \apj, 635, 1077 

% Supernova Remnants in the Magellanic Clouds. IV. X-Ray Emission from the Largest Supernova Remnant in the Large Magellanic Cloud  
%http://adsabs.harvard.edu/abs/2004ApJ...613..948W 
\bibitem[Williams et al.(2004)]{williams04} Williams, R.~M., Chu, Y.-H., Dickel, J.~R., Gruendl, R.~A., Shelton, R., 
Points, S.~D., \& Smith, R.~C.\ 2004, \apj, 613, 948 



% Supernova Remnants in the Magellanic Clouds. III. an X-Ray Atlas of Large Magellanic Cloud Supernova Remnants  
%http://adsabs.harvard.edu/abs/1999ApJS..123..467W 
\bibitem[Williams et al.(1999)]{williams99} Williams, R.~M., Chu, Y.-H., Dickel, J.~R., Petre, R., Smith, R.~C., 
\& Tavarez, M.\ 1999, \apjs, 123, 467 

\bibitem[Wilms et al.(2000)]{wilms2000} Wilms, J., Allen, A., 
\& McCray, R.\ 2000, \apj, 542, 914 

% Kinematics and composition of H II regions in spiral galaxies. I - M33  
%http://adsabs.harvard.edu/abs/1989AJ.....97...97Z 
\bibitem[Zaritsky et al.(1989)]{zaritsky89} Zaritsky, D., Elston, R., \& Hill, J.~M.\ 1989, \aj, 97, 97 



\end{thebibliography}
\end{document}